\documentclass[10pt,aps,prd,amsmath,amssymb,superscriptaddress,onecolumn,nofootinbib,showpacs,preprintnumbers,amsfonts, notitlepage]{revtex4-1}

\usepackage[usenames,dvipsnames,svgnames,table]{xcolor} 
\usepackage{graphicx}% Include figure files
\usepackage{dcolumn}% Align table columns on decimal point
\usepackage{bm}% bold math
\usepackage{hyperref}% add hypertext capabilities
\hypersetup{ 
    pdfnewwindow=true,      % links in new window
    colorlinks=true,       % false: boxed links; true: colored links
    allcolors=[RGB]{31 119 180}
} 
\usepackage{cleveref}
\usepackage{dsfont}
\usepackage{placeins}
\usepackage{todonotes}
\usepackage[utf8]{inputenc}
\usepackage{gensymb}
\usepackage{multirow}
\usepackage{amsmath}
\usepackage{amsfonts}
\usepackage{amssymb}
\usepackage{enumitem,amssymb}
\usepackage{array}   % for \newcolumntype macro
\newcolumntype{L}{>{$}l<{$}} % math-mode version of "l" column type
\newcolumntype{R}{>{$}r<{$}}
\newcolumntype{C}{>{$}c<{$}}
\usepackage{xspace}
\usepackage{braket}
\usepackage{units}
\usepackage{slashed}
\usepackage[normalem]{ulem}
\usepackage[format=plain,justification=RaggedRight,singlelinecheck=false]{caption}
\usepackage[format=plain,justification=centering,singlelinecheck=false]{subcaption}
\usepackage{tabularx}
\usepackage{nameref}

\usepackage{xpatch}% http://ctan.org/pkg/xpatch
\makeatletter
\xpatchcmd{\@ssect@ltx}{\@xsect}{\protected@edef\@currentlabelname{#8}\@xsect}{}{}% Patch \<section>*
\xpatchcmd{\@sect@ltx}{\@xsect}{\protected@edef\@currentlabelname{#8}\@xsect}{}{}% Patch \<section>
\makeatother
\usepackage{hyperref}% http://ctan.org/pkg/hyperref

\graphicspath{{figures/}}

\newcommand{\jpsi}{\ensuremath{J/\psi}\xspace}
\newcommand{\sig}{\ensuremath{\sigma/f_0(500)}\xspace}
\newcommand{\fzero}{\ensuremath{f_0(980)}\xspace}

\newcommand{\KSKS}{\ensuremath{K_S^0 K_S^0}\xspace}
\newcommand{\pizpiz}{\ensuremath{\pi^0 \pi^0}\xspace}
\newcommand{\h}{\ensuremath{h}\xspace}
\newcommand{\eg}{{\it e.g.}\xspace}
\newcommand{\cf}{{\it cf.}\xspace}

\newcommand{\ie}{{\it i.e.}\xspace}

\newcommand{\mevnospace}{\ensuremath{{\mathrm{\,Me\kern -0.1em V}}}}
\newcommand{\gevnospace}{\ensuremath{{\mathrm{\,Ge\kern -0.1em V}}}}
\newcommand{\tevnospace}{\ensuremath{{\mathrm{\,Te\kern -0.1em V}}}}
\newcommand{\mev}{\mevnospace\xspace}
\newcommand{\gev}{\gevnospace\xspace}
\newcommand{\gevsq}{\ensuremath{\gevnospace^2}}

\newcommand{\mevp}{\ensuremath{(\!\mevnospace)}}

\newcommand{\KCDD}{$K$-matrix$\big/$CDD poles\xspace}

\newlist{todolist}{itemize}{2}
\setlist[todolist]{label=$\square$}
\usepackage{pifont}
\usepackage{mfirstuc} % to uppercase the name
\newcommand{\addReviewer}[2]{
  \expandafter\newcommand\csname #1\endcsname[1]{{\bf \color{#2} \capitalisewords{#1}:\,##1}}
  \expandafter\newcommand\csname #1cor\endcsname[2]{{\color{#2} \capitalisewords{#1}:\,\st{##1}{\bf ##2}}}
  \expandafter\newcommand\csname #1cite\endcsname{{\color{#2} \capitalisewords{#1}:[Add citation]}}
  \expandafter\newcommand\csname #1color\endcsname{#2}
}

\usepackage{tikz,marginnote} % margin notes with circles
\newcommand{\checkedby}[1]{
\ifdefined\CROSSCHECKS
  \marginnote{
    \begin{tikzpicture}
      \foreach \x [count=\xi] in {#1} {
         \node[shape=circle,inner sep=0mm,
         minimum size=2mm,
         fill=\csname \x color\endcsname] at (\xi*3mm,0) {};
       }
    \end{tikzpicture}
  }
\else
\fi
}

\usepackage{soul}

\definecolor{chromeyellow}{rgb}{1.0, 0.65, 0.0}
\definecolor{DodgeBlue}{rgb}{0.118, 0.565,1.000}
\definecolor{asparagus}{rgb}{0.53, 0.66, 0.42}
\definecolor{cadmiumgreen}{rgb}{0.0, 0.42, 0.24}
\definecolor{cfrorange}{rgb}{1.0, 0.6, 0.4}

\addReviewer{adam}{red}
\addReviewer{ale}{chromeyellow}
\addReviewer{vincent}{brown}
\addReviewer{misha}{cadmiumgreen}
\addReviewer{cesar}{cfrorange}
\addReviewer{astrid}{cyan}
\addReviewer{arkaitz}{asparagus}
\addReviewer{wyatt}{Plum}
\addReviewer{miguel}{ForestGreen}
\allowdisplaybreaks

\begin{document}

\title{\boldmath Scalar and tensor resonances in $\jpsi$ radiative decays}

%----------------------------------------------------
% Authors
%----------------------------------------------------

\author{A.~Rodas}
\email{arodas@wm.edu}\affiliation{\wm}\affiliation{\jlab}
\author{A.~Pilloni}
\email{pillaus@jlab.org}
\affiliation{\rome}
\affiliation{\mift}
\affiliation{\catania}
\author{M.~Albaladejo}
\affiliation{\ific}
\author{C.~Fern\'andez-Ram\'irez}
\affiliation{\icn}
\author{V. Mathieu}
\affiliation{\ub}
\affiliation{\ucm}
\author{A.~P.~Szczepaniak}
\affiliation{\jlab}\affiliation{\indiana}\affiliation{\ceem}

\collaboration{\jpac}
\noaffiliation

%----------------------------------------------------
% Affiliations
%----------------------------------------------------

\newcommand{\bonn}{Universit\"at Bonn, 
Helmholtz-Institut f\"ur Strahlen- und Kernphysik, 
53115 Bonn, Germany}
\newcommand{\cern}{CERN, 1211 Geneva 23, Switzerland}
\newcommand{\ceem}{Center for  Exploration  of  Energy  and  Matter,  
Indiana  University,  
Bloomington,  IN  47403,  USA}
\newcommand{\ectstar}{European Centre for Theoretical Studies in Nuclear Physics and Related
Areas (ECT$^*$) and Fondazione Bruno Kessler,
I-38123 Villazzano (TN), Italy}
\newcommand{\ghent}{Department of Physics and Astronomy, Ghent University, Ghent 9000, Belgium}
\newcommand{\icn}{Instituto de Ciencias Nucleares, 
Universidad Nacional Aut\'onoma de M\'exico, 
Ciudad de M\'exico 04510, Mexico}
\newcommand{\indiana}{Physics  Department,  
Indiana  University,  
Bloomington,  IN  47405,  USA}
\newcommand{\jlab}{Theory Center,
Thomas  Jefferson  National  Accelerator  Facility, 
Newport  News,  VA  23606,  USA}
\newcommand{\mainz}{Institut f\"ur Kernphysik \& PRISMA Cluster of Excellence, 
Johannes Gutenberg Universit\"at, 
D-55099 Mainz, Germany}
\newcommand{\murcia}{Departamento de F\'isica, 
Universidad de Murcia, 
E-30071 Murcia, Spain}
\newcommand{\ucm}{Departamento de F\'isica Te\'orica, Universidad Complutense de Madrid and IPARCOS, E-28040 Madrid, Spain}
\newcommand{\ub}{Departament de F\'isica Qu\`antica i Astrof\'isica and Institut de Ci\`encies del Cosmos, Universitat de Barcelona, E08028, Spain}
\newcommand{\icsup}{Institute of Computer Science, Pedagogical University of Cracow, 30-084 Krak\'ow, Poland}
\newcommand{\wm}{Department of Physics, College of William and Mary, Williamsburg, VA 23187, USA}
\newcommand{\jpac}{Joint Physics Analysis Center}
\newcommand{\genova}{INFN Sezione di Genova, Genova, I-16146, Italy}
\newcommand{\rome}{INFN Sezione di Roma, I-00185 Roma, Italy}
\newcommand{\mift}{Dipartimento di Scienze Matematiche e Informatiche, Scienze Fisiche e Scienze della Terra,
Universit\`a degli Studi di Messina, I-98166 Messina, Italy}
\newcommand{\catania}{INFN Sezione di Catania, I-95123 Catania, Italy}
\newcommand{\uned}{Departamento de F\'isica Interdisciplinar, 
Universidad Nacional de Educaci\'on a Distancia, Madrid E-28040, Spain}
\newcommand{\ific}{Instituto de F\'isica Corpuscular (IFIC), Centro Mixto CSIC-Universidad de Valencia, Institutos de Investigaci\'on de Paterna, Aptd. 22085, E-46071 Valencia, Spain}

\preprint{JLAB-THY-21-3506}
\begin{abstract}
We perform a systematic analysis of the 
$\jpsi \to \gamma \pi^0\pi^0$ and $\to \gamma K_S^0 K_S^0$ partial waves measured by BESIII. We use a large set of amplitude parametrizations to reduce the model bias. We determine the physical properties of seven scalar and tensor resonances in the $1$--$2.5\gev$ mass range.  These include the well known $f_0(1500)$ and $f_0(1710)$, that are considered to be the primary glueball candidates. The hierarchy of resonance couplings determined from this analysis favors the 
 latter as the one with the largest  glueball component.
\end{abstract}
\maketitle

%====================================================================
\section{Introduction}
\label{sec:intro}

The  vast majority of observed mesons can be understood as simple $q\bar q$ bound states, although in principle strong interactions permit a more complex spectrum. 
In a pure Yang-Mills theory, massive gluon bound states
(named ``glueballs") populate the spectrum, as shown for example in lattice calculations.
The lightest glueball is expected to have
$J^{PC} = 0^{++}$, and a mass between 1.5 and 2\gev~\cite{Bali:1993fb,Patel:1986vv,Albanese:1987ds,Michael:1988jr,Sexton:1995kd,Morningstar:1999rf,Szczepaniak:2003mr,Chen:2005mg,Athenodorou:2020ani}. 
An enhanced glueball production is expected in OZI--suppressed processes,  \ie when the the quarks of the initial state annihilate into gluons.
For example, this is the case for central exclusive  production in $pp$ collisions (where mesons are produced by Pomeron---\ie gluon ladder---fusion),
or for \jpsi radiative decays, the $c \bar c$ must annihilate to gluons before hadronizing into the final state.  
In QCD, the mixing between glueballs and $q\bar q$ isoscalar  mesons makes the identification of a glueball candidate challenging, both theoretically and experimentally. The simplest argument for the existence of a glueball component is the presence of a supernumerary state with respect to how many are predicted by the quark model~\cite{Mathieu:2008me,Ochs:2013gi,Llanes-Estrada:2021evz}. 
It is thus of key importance to 
have a precise determination of the number and properties of the resonances seen in data. 

The most recent edition of Particle Data Group (PDG) identifies nine  isoscalar-scalar resonances. The two lightest ones, the \sig and \fzero, have been extensively studied in recent years, and are by now very well established~\cite{Ananthanarayan:2000ht,Colangelo:2001df,GarciaMartin:2011cn,Moussallam:2011zg,Caprini:2005zr,GarciaMartin:2011nna,Pelaez:2015qba}.
Quark model predicts other two scalars below 2\gev, but the three $f_0(1370)$, $f_0(1500)$ and $f_0(1710)$ are observed. This stimulated an intense work to identify one of them as the long-sought glueball~\cite{Chanowitz:1980gu,Amsler:1995td,Amsler:1995tu,Lee:1999kv,Giacosa:2005zt,Giacosa:2005qr,Albaladejo:2008qa,Janowski:2014ppa}. 
The existence of the $f_0(1370)$ is still debated. 
It seems to couple strongly to $4\pi$~\cite{Abele:2001js,Abele:2001pv}, while the analyses of two-body final states led to contradictory results. While some analysis claim to find this resonance in either $\pi \pi \to K \bar K$ or $\eta \eta$ scattering~\cite{Amsler:1992rx,Anisovich:1994bi,Amsler:1995bz,Gaspero:1992gu,Lanaro:1993km,Amsler:1994rv,Cohen:1980cq,Etkin:1982se} other analyses coming from meson-meson reactions do not find it~\cite{Hyams:1973zf,Grayer:1974cr,Hyams:1975mc,Estabrooks:1978de,Adolph:2015tqa}.
The $f_0(1500)$  and  $f_0(1710)$  are instead well established.
They have been determined from $\pi\pi$ production from fixed target experiments~\cite{Hyams:1973zf,Grayer:1974cr,Hyams:1975mc,clas:2017vxx}, and from heavy meson decays~\cite{Ablikim:2013hq,Dobbs:2015dwa,Ablikim:2018izx,dArgent:2017gzv,Lees:2012kxa,Ropertz:2018stk}, with the $f_0(1710)$  coupling mainly to kaon pairs~\cite{Barberis:1999am,Uehara:2013mbo,Ablikim:2018izx}.
Discerning which of the three is (or has the largest component of) the glueball, is an even harder task.
Since photons do not couple directly to gluons, the scarce production of $f_0(1500)$ in $\gamma\gamma$ collisions suggests it may be dominantly a   glueball. On the other hand, arguments based on the chiral suppression of the  perturbative matrix element of a scalar glueball to a $q\bar q$ pair, point to the $f_0(1710)$ as a better candidate~\cite{Chanowitz:2005du,Albaladejo:2008qa}. Although the argument does not necessarily hold nonperturbatively~\cite{Chao:2005si,Chanowitz:2007ma}, it seems to be supported by a quenched Lattice QCD calculation~\cite{Sexton:1995kd}.
The spectrum of scalars above 2\gev is even more confusing. The PDG currently lists $f_0(2020)$, $f_0(2100)$, $f_0(2200)$,  and $f_0(2330)$, but none of them is marked as established. The first one has been 
recently confirmed by a reanalysis of the $B_s^0 \to \jpsi \,\pi\pi$ and $\to \jpsi K \bar K$ decays~\cite{Ropertz:2018stk}.
The $f_0(2100)$ and $f_0(2200)$ appear to decay to only pions or kaons, respectively.
Since their resonance parameters are not dramatically different, they might originate from a single physical resonance (\cf Ref.~\cite{JPAC:2018zyd}).
Finally, the $f_0(2330)$ was seen in $p\bar p$ annihilations fifteen years ago~\cite{Anisovich:2000ut,Bugg:2004rj}, and was recently confirmed by a global reanalysis of reactions where isoscalar-scalar mesons appear~\cite{Sarantsev:2021ein}. 

The isoscalar-tensor sector appears to be better understood.
The $f_2(1270)$ and $f_2'(1525)$  are 
identified as ordinary $u\bar u + d\bar d$ and $s\bar s$ mesons, respectively.
 Indeed, the former couples largely to $\pi \pi$, and  the latter to $K\bar  K$~\cite{GarciaMartin:2011cn,Pelaez:2018qny,Pelaez:2020gnd}. Both resonances are relatively narrow and 
  have also been extracted from lattice QCD with a high degree of accuracy~\cite{Briceno:2017qmb}.\footnote{Alternative interpretations for the $f_2(1270)$ were discussed in~\cite{Molina:2008jw,Gulmez:2016scm,Geng:2016pmf,Du:2018gyn,Molina:2019rai}.}
 The status of  the other four resonances in the mass range up to 2\gev, the $f_2(1810)$, $f_2(1910)$, $f_2(1950)$, and $f_2(2010)$, is not as clear. The $f_2(2010)$ was seen in final states with strangeness only, $K\bar K$ and $\phi\phi$, suggesting the $s\bar s$ assignment. 
The other decay predominantly to multibody channels, making their identification more complicated. Above 2\gev, the PDG reports the $f_2(2150)$ and two more tensors, the  $f_2(2300)$ and the $f_2(2340)$. It is worth noting that a $2^{++}$ glueball is also expected at about $2.5\gev$~\cite{Morningstar:1999rf}. We summarize the status 
 of the isoscalar-scalar and -tensor resonances in Table~\ref{tab:pdgpoles}.

With more of high precision data coming from present and future experiments, including multibody final states,  in order to make further progress in identification of the resonance,   it is necessary to develop adequate amplitude analysis methods. For example,  dispersive techniques that rely on fundamental $S$-matrix principles  have played a key role in determining properties of the lightest scalar resonances~\cite{Caprini:2005zr,DescotesGenon:2006uk,GarciaMartin:2011nna,Hoferichter:2011wk,Moussallam:2011zg,Ditsche:2012fv,Pelaez:2020uiw,Pelaez:2020gnd}.  Their  application, however, has so far been limited to roughly the region below $1\gev$. At higher energies, other  approaches, such as Pad\'e approximants~\cite{Masjuan:2013jha,Masjuan:2014psa,Caprini:2016uxy,Pelaez:2016klv}, Laurent-Pietarinen expansion~\cite{Svarc:2014sqa}, or the Schlessinger point method~\cite{Tripolt:2016cya,Tripolt:2018xeo,Binosi:2019ecz} have been used. However, these methods often require as input an analytic parametrization of the data. But --- unlike men--- not all parameterizations are created equal, and the ones that fulfill as many $S$-matrix principles as possible should be considered more trustworthy.

\begin{table}
\begin{ruledtabular}
\begin{tabular}{c|ccccc}
 & Mass \mevp& Width \mevp & $\mathcal{B} (f\to\pi\pi)$ ($\%$)  & $\mathcal{B} (f\to K\bar{K})$  ($\%$)& $\mathcal{B} (f \to 4\pi )$  ($\%$) \\ 
\hline
$f_0(1370)$ & $1200$--$1500$& $300$--$500$&$<10$ \cite{Ochs:2013gi}& $35\pm13$ \cite{Bugg:1996ki}& $>72$ \cite{Gaspero:1992gu}\\ 
$f_0(1500)$ &$1506\pm6$ &$112\pm9$ & $34.5\pm2.2$& $8.5\pm 1.0$& $48.9\pm 3.3$\\ 
$f_0(1710)$ & $1704\pm12$&$123\pm18$ &$3.9^{+3.0}_{-2.4}$ \cite{Longacre:1986fh}& $36\pm12$ \cite{Albaladejo:2008qa}& $-$\\ 
$\left[f_0(2020)\right]$ & $1992\pm16$&$442\pm60$ &$-$ & $-$&$-$\\
$\left[f_0(2200)\right]$ & $2187\pm 14$& $207\pm40$&$-$ &$-$ & $-$\\
$\left[f_0(2330)\right]$ & $2324\pm 35$& $195\pm71$&$-$ &$-$ & $-$\\
\hline
$f_2(1270)$ & $1275.5\pm0.8$&$186.7^{+2.2}_{-2.5}$ &$84.2^{+2.9}_{-0.9}$ & $4.6^{+0.5}_{-0.4}$ & $10.4^{+1.6}_{-3.7}$ $^{(a)}$\\
$\left[f_2(1430)\right]$ & $\approx 1430$& $-$ &$-$ & $-$& $-$\\
$f_2'(1525)$ & $1517.4\pm2.5$& $86\pm5$ & $0.83\pm0.16$ &$87.6\pm2.2$ & $-$ \\
$\left[f_2(1565)\right]$ & $1542\pm19$& $122\pm 13$&$-$ & $-$& $-$\\
$\left[f_2(1640)\right]$ & $1639\pm 6$& $99^{+60}_{-40}$& $-$& $-$& $-$\\
$\left[f_2(1810)\right]$ & $1815\pm12$& $197\pm 22$& $21^{+2}_{-3}$ \cite{Longacre:1986fh}&$2\times 0.3^{+1.9}_{-0.2}$ \cite{Longacre:1986fh}&$-$\\
$\left[f_2(1910)\right]$ & $1900\pm9$ $^{(b)}$& $167\pm 21$ $^{(b)}$& $-$& $-$& $-$\\
$f_2(1950)$ & $1936\pm12$& $464\pm24$&$-$ &$-$ &$-$\\
$f_2(2010)$ & $2011^{+62}_{-76}$& $202^{+67}_{-62}$& $-$ & $-$&$-$\\
$\left[f_2(2150)\right]$ & $2157\pm12$& $152\pm30$& $-$& $-$& $-$\\
$\left[f_2(2300)\right]$ & $2297\pm28$& $149\pm41$& $-$& $-$& $-$\\
\end{tabular}
\end{ruledtabular}
\caption{\label{tab:pdgpoles}Summary of scalar and tensor resonances in the $1$--$2.5\gev$ region listed in the PDG~\cite{pdg}. Resonances in square brackets are not well established.  $^{(a)}$ Combination of entries on  $\pi^+\pi^- 2\pi^0$, $2\pi^+2\pi^-$, and $4\pi^0$, errors added linearly due to being asymmetric.
$^{(b)}$ Mass and width from the $\omega\omega$ decay mode.} 
\end{table}
In this paper we extract the scalar and tensor resonances from the partial waves of $\jpsi\to\gamma \pi^0\pi^0$ and $\to \gamma \KSKS$ 
determined by BESIII~\cite{Ablikim:2015umt,Ablikim:2018izx}.
We use a number of different parametrizations that satisfy unitarity and analyticity, in order to put under control large model dependencies. At the energies 
 of interest, the number of available open channels makes the complete rigorous analysis unfeasible. We start by considering the $\pi\pi$ and $K\bar K$ final states only. Implementing unitarity on a subset of available channels does not affect seriously the resonant parameters, provided that resonances are sufficiently separated from each other~\cite{JPAC:2017dbi,JPAC:2018zyd}. This is definitely not the case here. We find that 2-channel fits fail to reproduce some of the details of the resonant peaks and the interference patterns in the regions between nearest resonances. This can bias the pole determination. 
 In addition to $\pi\pi$ and $K\bar K$, the PDG lists at least three  other decay channels for these resonances, \ie $\eta \eta$, $\eta \eta'$, $4 \pi$, with larger coupling to $4\pi$. The channels $\jpsi \to \gamma \,(2\pi^+2\pi^-,\pi^+\pi^-2\pi^0)$ were seen in the experiments done  in the 80s~\cite{Baltrusaitis:1985nd,Bisello:1988as} and later by BES~\cite{Bai:1999mm}. BESIII also measured $J/\psi \to \gamma \eta \eta$~\cite{BESIII:2013qqz}. However these analyses do not provide mass-independent partial wave extractions, and are not comparable in statistics and quality with the most recent ones that we use. For this reason, we decided to add an effective  third channel, which without loss of generality we may interpret as  $\rho\rho$, that is  however, not  constrained by any other  data. 
 Finally, the statistical uncertainties are determined via bootstrap~\cite{recipes,EfroTibs93,Landay:2016cjw}.

The rest of the paper is organized as follows. A brief description of the data and 
of our selection of the fit region is discussed in Section~\ref{sec:data}. We describe our set of parametrizations in Section~\ref{sec:model}. The 2-channel fits are described in Section~\ref{sec:2charesults}, and in Section~\ref{sec:3charesults} we study the role of the third channel and perform the statistical analysis. 
The summary of results we obtain for the resonant poles are detailed in  Section~\ref{sec:poles} and our conclusions are given in Section~\ref{sec:summary}.

%====================================================================
\section{The Dataset}
\label{sec:data}

\begin{figure}[t]
{\centering
\includegraphics[scale=.8]{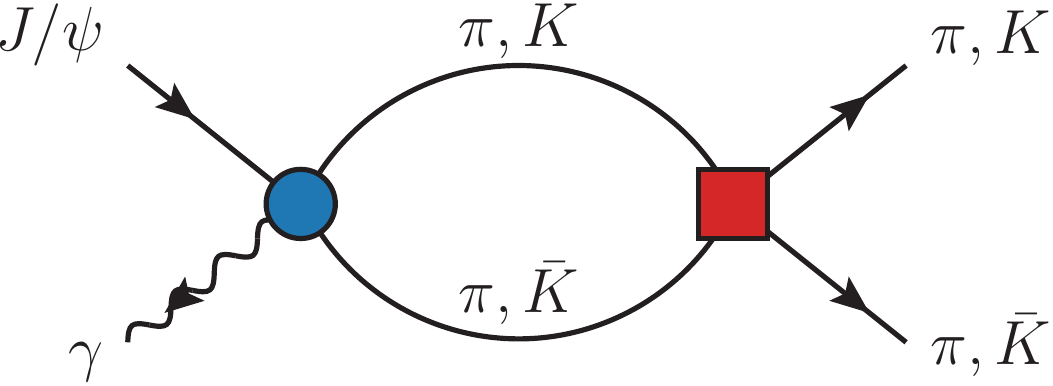} \hspace{0.5cm} \includegraphics[scale=.8]{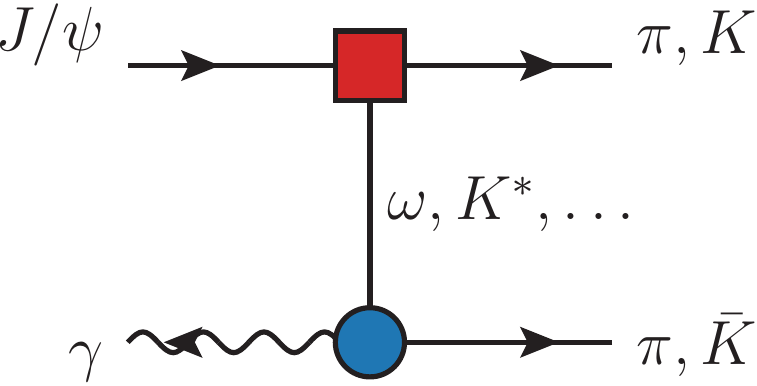}}
\caption{Processes contributing to $\jpsi \to \gamma \,\h\bar \h$, with $\h = \pi,K$. Left panel: \jpsi decays through short-range $c\bar c \to \gamma gg$ (blue disk), then resonances are seen emerging from final state interaction (red square). Right panel: \jpsi decays to $V\,\h$ through the short-range $c\bar c \to ggg$ (red square), then the resonance $V$ decays radiatively to $\gamma \bar \h$ (blue disk).}
\label{fig:proces}
\end{figure}

We consider the data from the 
  mass independent analysis of $J/\psi$ 
   radiative decays, 
    $\jpsi \to \gamma\pi^0\pi^0$~\cite{Ablikim:2015umt} and $\to \gamma \KSKS$~\cite{Ablikim:2018izx} 
     by BESIII. 
Bose symmetry requires the two pseudoscalars to have $J^{PC}=\text{(even)}^{++}$; moreover the isospin zero amplitude is dominant.\footnote{In fact, for the $\pi^0\pi^0$ system, isospin one is forbidden, and the decay into isospin two  would be higher order in the isospin breaking. Since $I=2$ has no resonances, there would be no dynamical mechanism that could enhance it. For the \KSKS system, isospin one is allowed, and exhibits a rich resonant structure, that includes, for example, the $a_0(980)$ and the $a_2(1320)$. However, the production of isovector is OZI-suppressed, since the topology  $c\bar c\to \gamma gg$ couples to isoscalars only. }
The mass independent $J=0$ and $J=2$ partial waves are given in the multipole basis~\cite{sebastian:1992xq}. 
  The latter is visible in three different multipoles, $E1$, $M2$, and $E3$.\footnote{We use the standard notation,  $Xj_\gamma$, in which 
   $j_\gamma \ge 1$ is the angular momentum carried by the electromagnetic field, and $X=E$ ($X=M$) if the parities of initial and final state satisfy (or not) $P_\text{in} P_\text{fin} = (-)^{j_\gamma}$. The values allowed for $j_\gamma$ are $|J_\psi - J| \le j_\gamma \le J_\psi + J$.} The three intensities look very similar up to the overall normalization with,  $E1 > M2 > E3$. 
  Considering that the quark model predicts each multipole to scale as $(E_\gamma)^{j_\gamma}$, $E_\gamma$ being the photon energy, the observed 
      hierarchy is consistent with theoretical expectation, at  least close to threshold. 
  Intensities and
phase differences determined with respect to the $2^{++}\,E1$ are given in 15\mev invariant mass bins, from
threshold up to 3\gev. In order to make use of the information on the relative phase, we should analyze simultaneously the $S$- and all the $D$-waves, however, 
 because of the dominance of the lowest multipoles,
   we focus on $0^{++}$ and $2^{++}\,E1$. 

\begin{figure}[t]
{\centering
\includegraphics[width=0.5\textwidth]{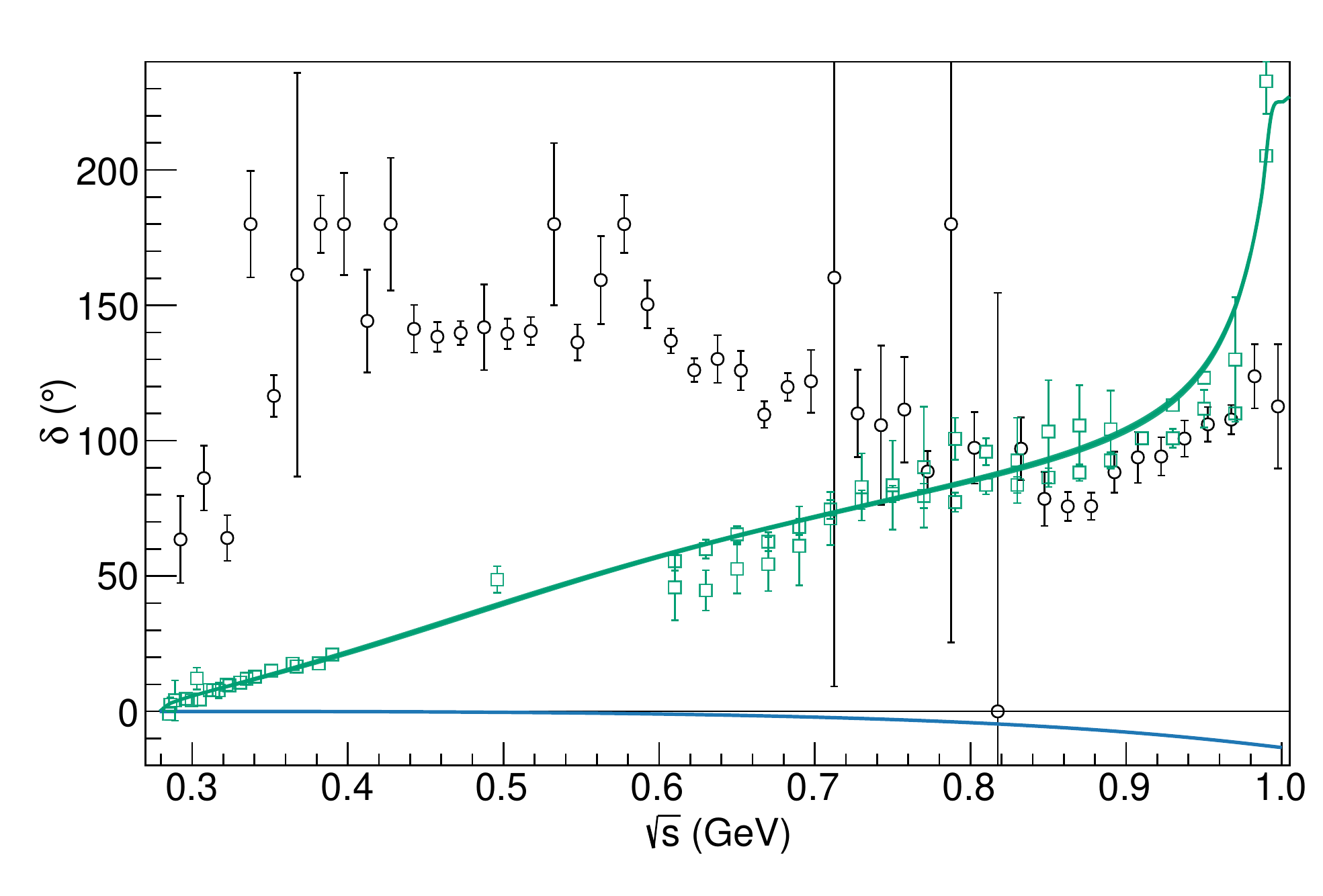} 
}
\caption{$S$-wave $\pi\pi$ phase shifts in the elastic region. Black empty circles show the $0^{++} - 2^{++}\,E1$  
phase difference of the radiative $\jpsi\to\gamma\pi\pi$ decays by BESIII~\cite{Ablikim:2015umt}. The well known $\pi \pi$ scattering data is shown in green empty squares~\cite{Hyams:1973zf,Grayer:1974cr,Hyams:1975mc,Cohen:1980cq,Kaminski:1996da,Batley:2010zza}. The dispersive fits of~\cite{GarciaMartin:2011cn,Pelaez:2019eqa} are shown in green for the $S$-wave, and blue for (minus) the $D$-wave phase shift, which is basically zero at these energies. Radiative data are incompatible with the dispersive result by more than $9\sigma$, that reduce to $\sim 7\sigma$ if one allows for a constant shift.}
\label{fig:upto1GeV}
\end{figure}
 
\begin{table}[b]
\caption{Branching ratios of resonances appearing in the $\gamma \bar\h$ channel, compared to the total branching ratio. The largest contribution is given by the $\omega$. However, it is removed in~\cite{Ablikim:2015umt} by vetoing the events within 50\mev from the nominal $\omega$ mass.
}
\begin{tabular}{l | c c}
\hline\hline
$\mathcal{B}\!\left(\jpsi \to \gamma \pi^0\pi^0\right)$ & $(11.5 \pm 0.5) \times 10^{-4}$ &\cite{Ablikim:2015umt}\\\hline
$\mathcal{B}\!\left(\jpsi \to \omega\pi^0 \to \gamma \pi^0\pi^0\right)$ & $(3.8 \pm 0.4) \times 10^{-5}$ &\multirow{3}{*}{\cite{pdg}}\\
$\mathcal{B}\!\left(\jpsi \to \rho\pi^0 \to \gamma \pi^0\pi^0\right)$ & $(2.6 \pm 0.5) \times 10^{-6}$ &\\
$\mathcal{B}\!\left(\jpsi \to b_1(1232)\pi^0 \to \gamma \pi^0\pi^0\right)$ & $(3.6 \pm 1.3) \times 10^{-6}$ & \\\hline\hline
$\mathcal{B}\!\left(\jpsi \to \gamma \KSKS\right)$ & $(8.1 \pm 0.4) \times 10^{-4}$ &\cite{Ablikim:2018izx}\\\hline
$\mathcal{B}\!\left(\jpsi \to K^{*}(892)^0 K_S^0 \to \gamma \KSKS\right)$ & $(6.3 \pm 0.6) \times 10^{-6}$ &\multirow{2}{*}{\cite{Ablikim:2018izx}} \\
$\mathcal{B}\!\left(\jpsi \to K_1(1270)^0 K_S^0 \to \gamma \KSKS\right)$ & $(8.5 \pm 2.5) \times 10^{-7}$& \\\hline\hline
\end{tabular}
\label{tab:tchannel}
\end{table}

The dynamics underlying these radiative decays can be represented by the diagrams in Fig.~\ref{fig:proces}. In the left diagram, the \jpsi decay is mediated by the short-range process, for example,  $c\bar c \to \gamma g g$ and resonances originate from rescattering of the two mesons. On the right diagram, the \jpsi decays through another short-range process, {\it e.g.} $c\bar c \to g g g$ to a state containing  an intermediate resonance $V$ and a bachelor meson,  $\h=\pi,K$. The resonance $V$ then decays radiatively to $\gamma \bar \h$.\footnote{Charge conjugation is understood.} 
The latter class of reactions introduce  a nontrivial background to the processes  we are interested in. 
These intermediate resonances appear as peaks in the $\gamma \bar \h$ invariant mass, but their contribution is mostly flat when projected onto the $\h\bar \h$ direction.
Morevover, the region within 50\mev from the dominant  exchange of the of the $\omega$, that appears as a narrow  peak in the $\gamma \bar{h}$ Dalitz plot
 has  been removed from the \pizpiz dataset~\cite{Ablikim:2015umt}. The effect of $K^*(892)$ and $K_1(1270)$ on the \KSKS spectrum was estimated to be negligible~\cite{Ablikim:2018izx}. Indeed,  looking at the branching ratios given in Table~\ref{tab:tchannel}, one can appreciate how small the contribution of these resonances is, even more so when spread over the two-meson invariant mass. While in principle these resonances can still affect the partial waves through 3-body rescattering~\cite{Niecknig:2012sj,*Gan:2020aco,*JPAC:2020umo}, it is expected that these corrections are small for large phase spaces like the ones considered here. 
 We thus restrict the dynamics of the right diagram of Fig.~\ref{fig:proces} to possible heavier resonances that lie outside of the Dalitz plot region.    
The first consequence is that it is expected that  Watson's theorem holds from $\pi\pi$ to $K\bar K$ threshold~\cite{Watson:1952ji}. 
 Specifically, the phase of the $0^{++}\,E1$ multipole of $\jpsi \to \gamma \pi\pi$ should match 
 that of the $S$-wave elastic $\pi\pi$ scattering. The latter is well established~\cite{Hyams:1973zf,Grayer:1974cr,Hyams:1975mc,Cohen:1980cq,Kaminski:1996da,Batley:2010zza,Ananthanarayan:2000ht,Colangelo:2001df,GarciaMartin:2011cn,Moussallam:2011zg,Pelaez:2019eqa} and 
Fig.~\ref{fig:upto1GeV} compares the two. 
Even if one reconsidered the effect of 3-body rescattering, it would be impossible for crossed channel resonances to produce such a fast phase motion, in particular close to the $\pi\pi$ threshold.
Since the focus of this work is on the higher resonances, we shall not consider further  the data below the $K\bar K$ threshold. Moreover, no significant structure appears in data above 2.5\gev. Since the high energy region would require a  different approach~\cite{Bibrzycki:2021rwh}, we also drop it from this analysis. 

The extraction of partial waves from data suffers from Barrelet ambiguities~\cite{Barrelet:1971pw}. For $\jpsi \to \gamma \h\bar \h$ truncated to $J=2$, there are two possible solutions in each  channel, as shown in Appendix~\ref{app:ambi}.
While the nominal ones have a roughly vanishing $2^{++}\,E1-2^{++}\,M2$ phase difference, the alternative solutions display rapid motion, in particular at $\sim 1.5\gev$.  It is well known that the $f_2(1270)$ and $f_2'(1525)$ are mostly elastic and dominate the $\pi\pi$ and $K\bar K$ channels, respectively. Inelasticities contribute to $\lesssim 15\%$ to the width of each resonance. 
In this case, 
Watson's theorem requires that the phase difference between two $2^{++}$ multipoles vanishes in this region.
Hence, the phase motions observed in the alternative solutions are not
justified. These solutions also exhibit phase motion at both low and high masses, where no resonances are expected to contribute. Incidentally, the mass dependent fit of $K\bar K$ in~\cite{Ablikim:2018izx} clearly favors the nominal solution. For these reasons, in our analysis we consider the nominal solutions only.

To summarize, we will perform a coupled-channel analysis of the $0^{++}\,E1$ and $2^{++}\,E1$ intensities and relative phase in the invariant mass region between $1$ and $2.5\gev$ using as data input the nominal solutions. In the following, we will refer to these two multipoles as $S$- and $D$-waves. In total, we fit 606 data points.

%====================================================================
\section{Amplitude models}
\label{sec:model}
We describe here the sets of models used to fit the data. We consider several possible variations, in order to perform a thorough study of the systematic uncertainties of our results. In the analysis of $\eta^{(\prime)}\pi$ COMPASS data~\cite{JPAC:2018zyd}, we chose one model as the nominal one, and the differences with other models were quoted as systematic uncertainties. However, here the spread of the results is too large to permit this strategy, and we will simply list the results of each model without selecting a preferred one.

We parametrize the partial wave amplitudes following the coupled-channel $N/D$
formalism~\cite{Chew:1960iv,Bjorken:1960zz,Aitchison:1972ay,Oller:1998hw},
\begin{equation}
\label{eq:amplitude}
 a^J_i(s) = E_\gamma\, p_i^{J} \, \sum_k n^J_k(s) \left[ {D^J(s)}^{-1} \right]_{ki}\,,
\end{equation}
with the index $i=\h\bar \h=\pi\pi$, $K\bar K$, and later $\rho\rho$; as customary, $s$ is the $\h\bar \h$ invariant mass squared, 
$p_i=\sqrt{s- 4m_{i}^2}/2$
is the breakup momentum in the $h\bar h$ rest frame. One power of photon energy
$E_\gamma=(m_{\jpsi}^2-s)/(2\sqrt{s})$ for E1 transitions is required by gauge invariance. The intensities are calculated as $I^J_i(s) = \mathcal{N} p_i \left|a^J_i(s)\right|^2$, with $\mathcal{N}$ a normalization factor. The $n^J_k(s)$ incorporate exchange forces ({\it cf.} the right diagram in 
  Fig.~\ref{fig:proces}) 
     in the production process
and are smooth functions of $s$ in the physical region. The matrix 
$D^J(s)$  represents the $\h \bar \h \to \h \bar \h$ final state
interactions, and contains cuts only on the real axis above
thresholds (right hand cuts), which are constrained by
unitarity. For the numerator $n^J_k(s)$, we use an effective polynomial expansion,
\begin{equation}
    n^J_k(s)=\sum_{n=0}^{n_\text{max}}{a^{J,k}_n  T_n\left[\omega(s)\right]},
    \label{eq:production}
\end{equation}

where $T_n$ are the Chebyshev polynomials of order $n$. For systematic studies, we consider three different choices of   $\omega(s)$,
\begin{subequations}
\begin{align}
\omega(s)_\text{pole}&=\frac{s}{s+s_0}\,, \label{eq:omegapole} \\
\omega(s)_\text{scaled}&=2\frac{s-s_\text{min}}{s_\text{max}-s_\text{min}}-1\,, \label{eq:omegascaled}\\
\omega(s)_\text{pole+scaled} &=2\frac{\omega(s)_\text{pole}-\omega(s_\text{min})_\text{pole}}{\omega(s_\text{min})_\text{pole}-\omega(s_\text{max})_\text{pole}}-1\,, \label{eq:omegapolescaled}
\end{align}
\end{subequations}
where $s_0 = 1\gev^2$ is an effective scale parameter that controls the position of the left-hand singularities in Eqs.~\eqref{eq:omegapole} and~\eqref{eq:omegapolescaled}, and reflects the short range nature of production. Instead, Eq.~\eqref{eq:omegascaled} has no singularity, which corresponds to neglecting completely the right diagram in Fig.~\ref{fig:proces}. 
Eqs.~\eqref{eq:omegascaled} and~\eqref{eq:omegapolescaled} exploit the orthogonality of Chebyshev polynomials in the $[-1,1]$ interval in order to reduce correlations, being $\left[s_\text{min},s_\text{max}\right] = \left[\left(1\gev\right)^2,\left(2.5\gev\right)^2\right]$ the fitting region.  

A customary parametrization of the denominator is given by~\cite{Aitchison:1972ay}
\begin{equation}
\label{eq:Dsol}
D^J_{ki}(s) =  \left[ {K^J(s)}^{-1}\right]_{ki} - \frac{s}{\pi}\int_{4m_{k}^2}^{\infty}ds'\frac{\rho N^J_{ki}(s') }{s'(s'-s - i\epsilon)}, 
\end{equation}
where 
\begin{subequations}
\begin{align}
\rho N^J_{ki}(s')_\text{nominal}  &= \delta_{ki} \,\frac{(2p_i)^{2J+1}}{\left(s'+s_L\right)^{2J+\alpha}}, \label{eq:rhoN}
\end{align}
that is an effective description of the left-hand singularities in scattering 
controlled by the $s_L$ parameter, which we vary between $0$ and $1\gev^2$. The parameter $\alpha$ controls the asymptotic behavior of the integrand. 
As an alternative model, we consider the projection of a cross-channel exchange of mass squared $s_L$,
\begin{equation}
\label{eq:QCM}
\rho N^J_{ki}(s')_\text{$Q$-model} = \delta_{ki}\, \frac{Q_J(z_{s'})}{2p_i^2},
\end{equation}
\end{subequations}
where $Q_J(z_{s'})$ is the second kind Legendre function, and $z_{s'}=1+s_L/2p_i^2$. 
This function behaves asymptotically as $\log(s')/ s'$,  and has a left hand cut starting at $s'= 4m_i^2 - s_L$. For the $K$-matrix, we consider 
\begin{subequations}
\begin{align}
K^J_{ki}(s)_\text{nominal} &= \sum_R \frac{g^{J,R}_k g^{J,R}_i}{m_R^2 - s} + c^J_{ki} + d^J_{ki} \,s,\label{eq:Kmatrix}
\end{align}
with $c^J_{ki}= c^J_{ik}$ and $d^J_{ki}= d^J_{ik}$.
Alternatively, we parametrize the inverse of the $S$-wave $K$-matrix as a sum of CDD poles~\cite{Castillejo:1955ed,JPAC:2017dbi},
\begin{equation}
\left[K^J(s)^{-1}\right]_{ki}^\text{CDD} = c^J_{ki} - d^J_{ki} \,s - \sum_R \frac{g^{J,R}_k g^{J,R}_i}{m_R^2 - s} \, , \label{eq:CDD}
\end{equation}
\end{subequations}
where $c^J_{ki} = c^J_{ik}$ and $d^J_{ki} = d^J_{ik}$ are constrained to be positive.
For single channel, this choice ensures that no poles can appear on the first Riemann sheet. Even in the case of coupled channels, their occurrence is scarce, and when they do occur they are deep in the complex plane, far from the physical region. 
No CDD-like denominator will be used for the $D$-wave, as its structure looks much simpler. Ideally, the natural extension of the single channel CDD parametrization would be the inclusion of positive defined matrices for each  term in Eq.~\eqref{eq:CDD}, however this is expensive to compute from a numerical point of view, and not so simple to implement in our fits~\cite{Bedlinskiy:2014tvi}.

%====================================================================
\section{2-channel Results}
\label{sec:2charesults}

\begin{figure}
\centering
\includegraphics[width=0.32\textwidth]{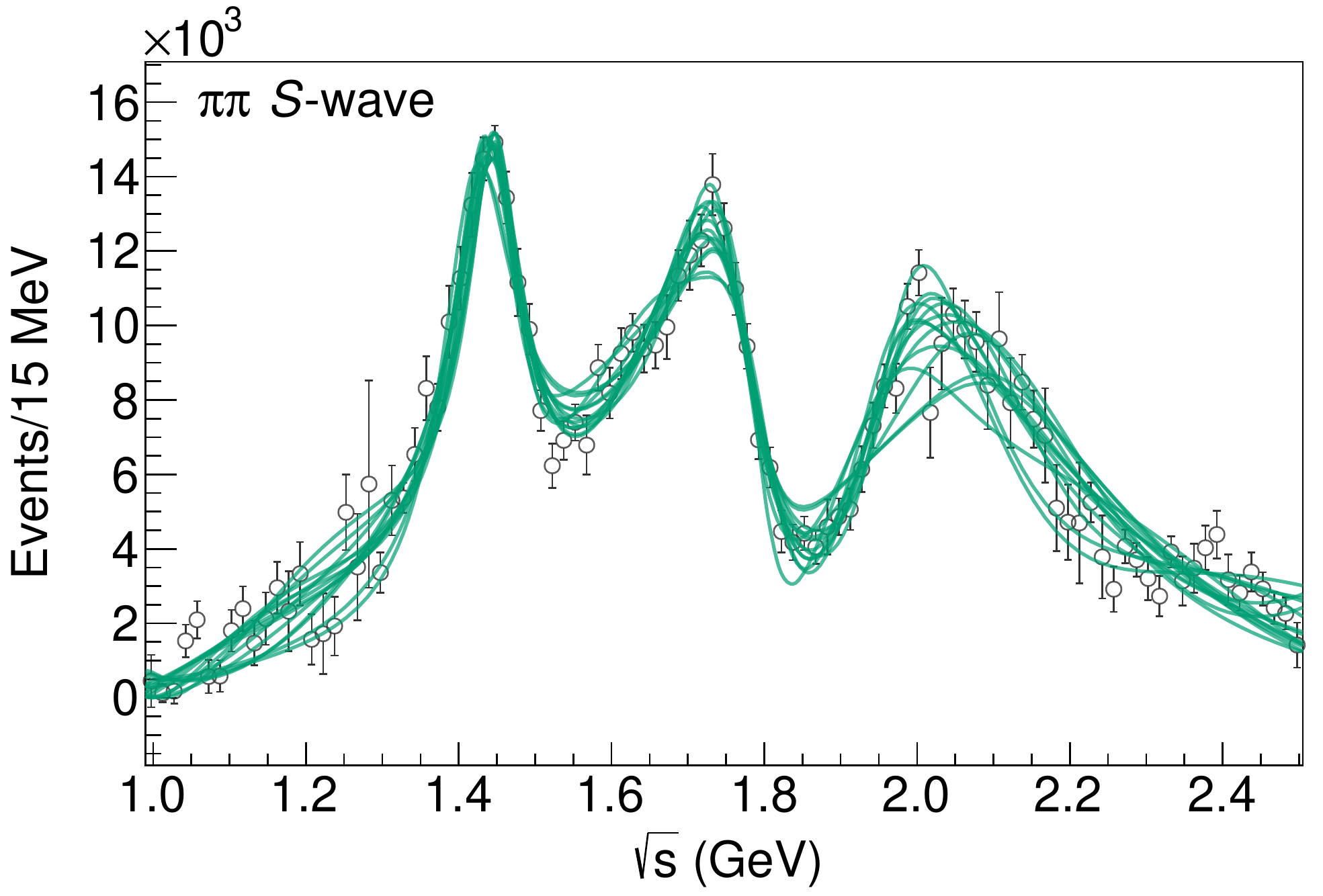} \includegraphics[width=0.32\textwidth]{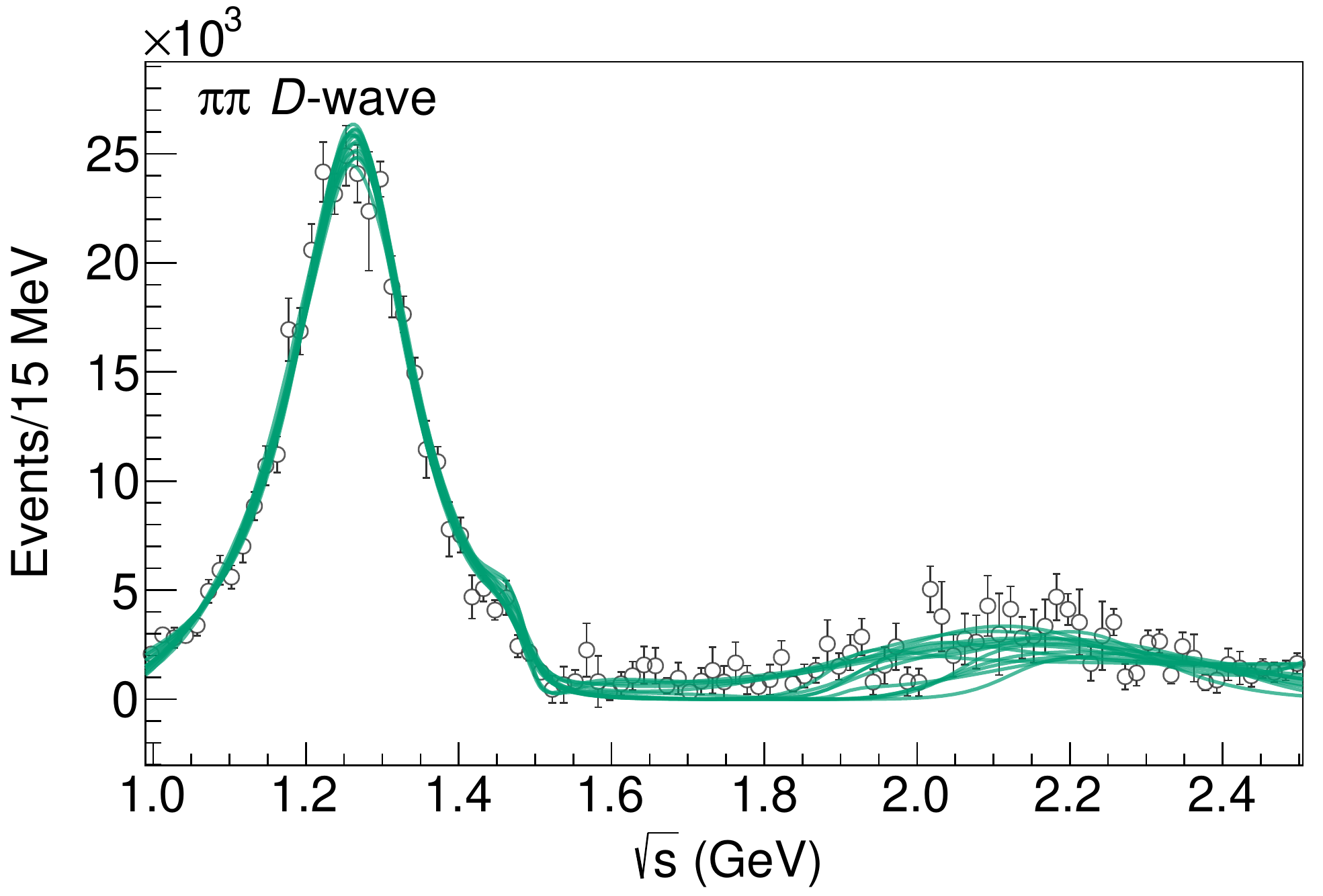} \includegraphics[width=0.32\textwidth]{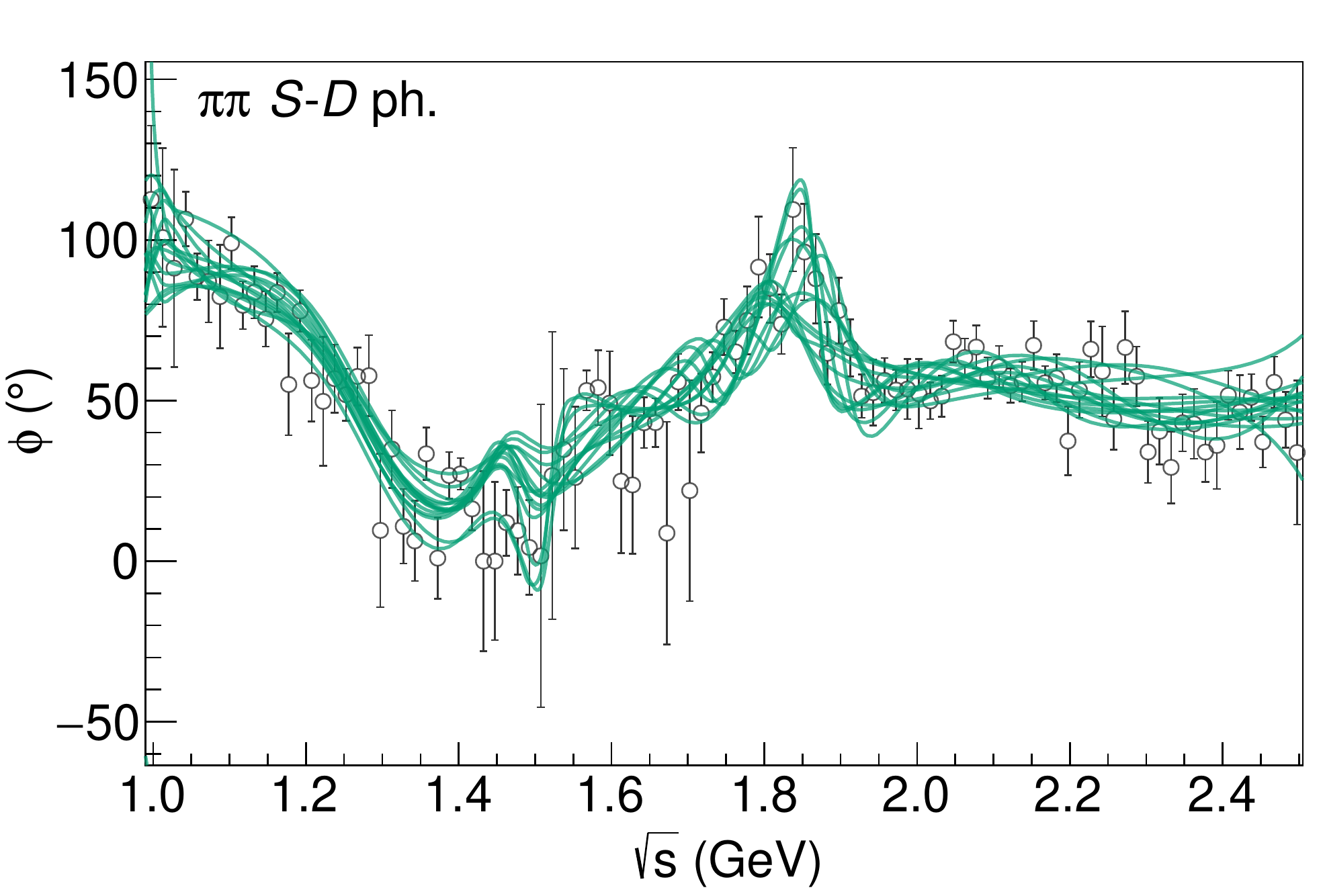} \\
\includegraphics[width=0.32\textwidth]{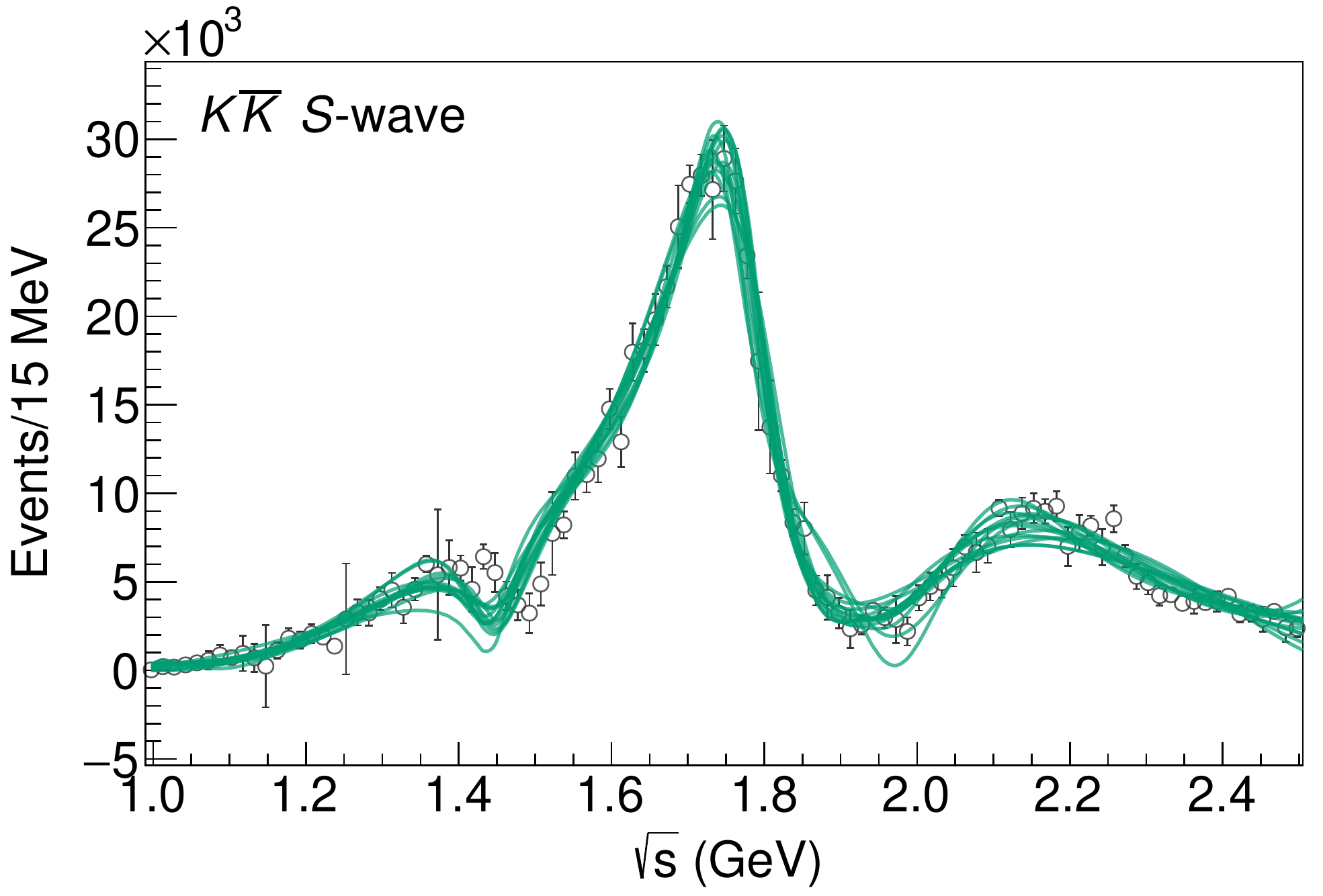} \includegraphics[width=0.32\textwidth]{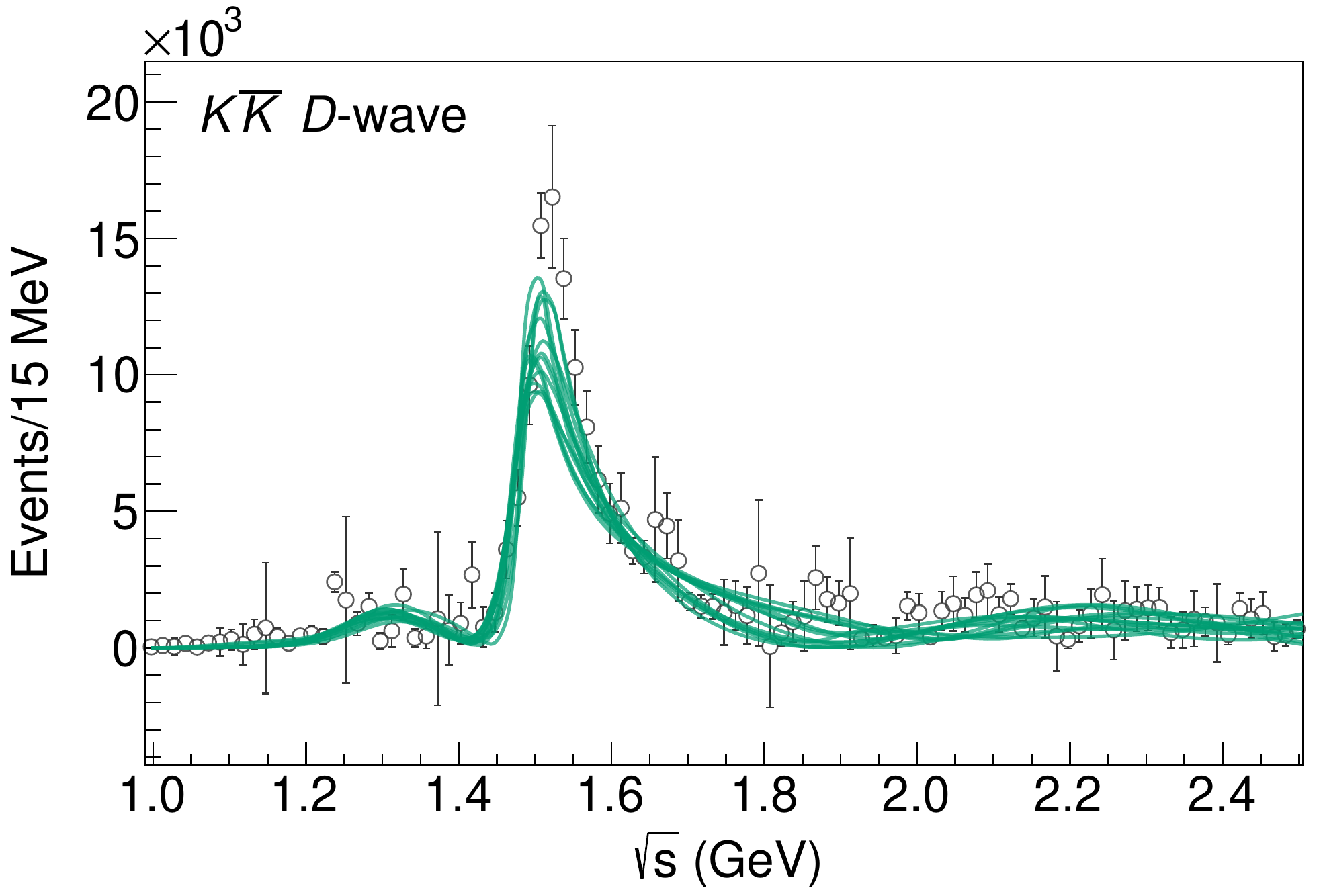} \includegraphics[width=0.32\textwidth]{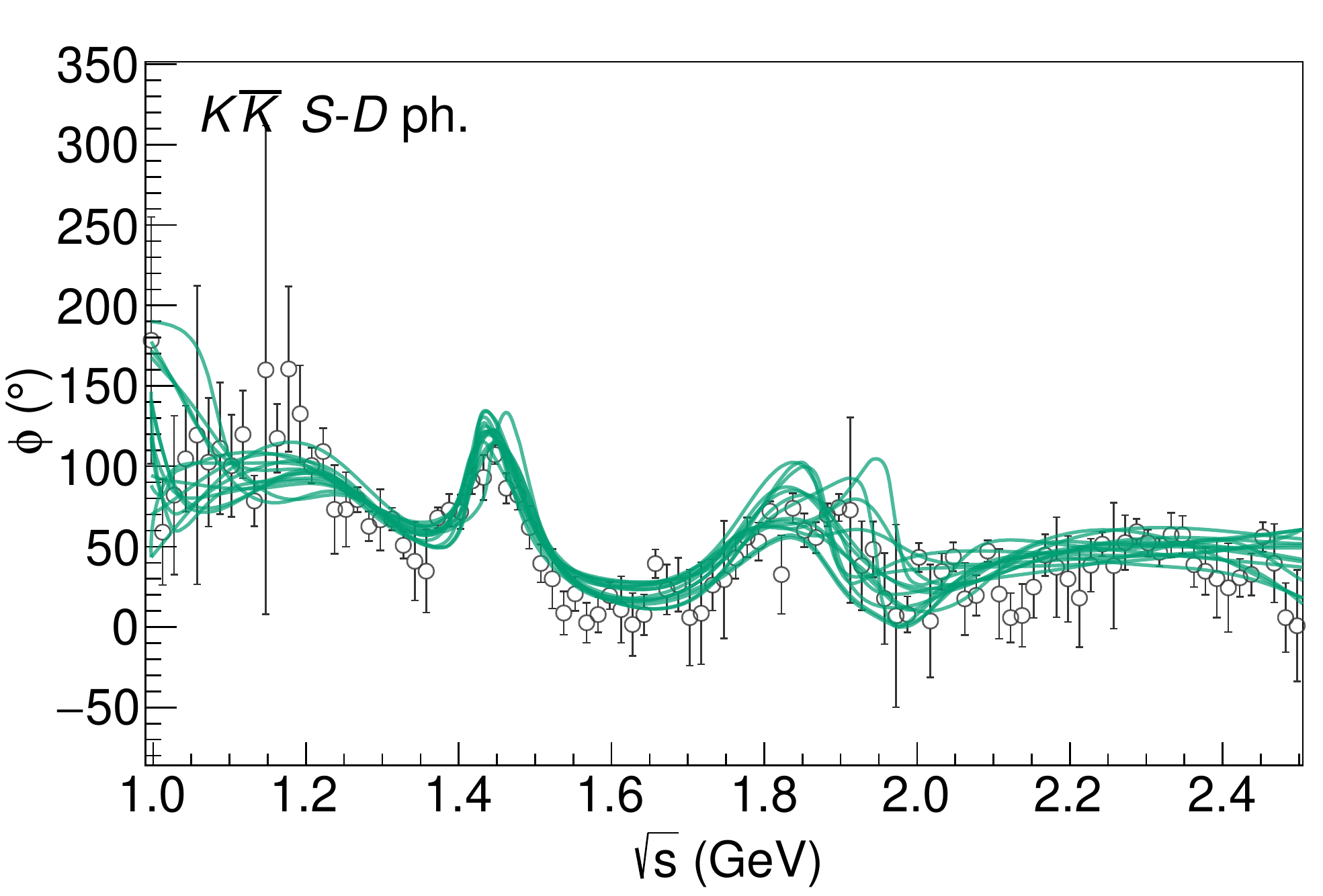}
\caption{\label{fig:2channelfits} Best 2-channel fits to $\pi \pi$ (top) and $K \bar K$ (second row) final states. The intensities for the $S$- (left), $D$-wave (center), and their relative phase (right) are shown. The green and red lines denote the fit results. We remark that the model variations in the 2-channel fits are much larger than the statistical uncertainties, in particular for the phases. All these fits produce $\chi^2/\text{dof}\sim 1.7$--$2$. 
}
\end{figure}

We first explore the 2-channel fits. In total, considering the various possibilities discussed in Section~\ref{sec:model}, we could fit 27 different amplitude choices, without considering further variations of the fixed parameters (\eg the position of the left hand cut, the number of \KCDD, the order of the  polynomial in the numerator $n^J(s)$, \dots), which would amount to thousands of different possibilities. 
For phase space functions, we consider both Eq.~\eqref{eq:QCM} and Eq.~\eqref{eq:rhoN} for $\alpha\leq 1$. 
The choice of $\alpha$ is motivated by the asymptotic behavior of the phase space: if $\alpha> 1$, the integral is oversubtracted, making other subtractions redundant. Even if those fits produce similar results and fit quality, they tend to produce narrow unphysical $1^\text{st}$ sheet poles. Thus we restrict $\alpha=0$ for the final best fits. 
For the denominator, we vary the order of the background terms. We also tried to increase the \KCDD from the nominal 3 to 5 to see if extra resonances are produced. These fits do not produce noticeable differences and the additional poles produced are unstable and far from the fitted region, effectively merging with the background. 
Finally, we consider the different numerator variables listed in Eq.~\eqref{eq:production}. We also vary the order of the production polynomial between $2^\text{nd}$ and $3^\text{rd}$ order. Lower orders are not able to reproduce the data, in particular the relative phases would be heavily affected.

We select 14 models that do not produce noticeable unphysical behaviors, as  $1^\text{st}$-sheet poles narrower than $1\gev$. 
Summarizing, we have 3--4 parameters per wave per channel for the numerator polynomial, 2 couplings and a mass for the six bare resonances, 3--6 per wave for the background polynomial in the denominator.
Depending on the specific choices, they amount to 
40--44 parameters fitted to data, by performing a $\chi^2$ minimization with {\tt MINUIT}~\cite{minuit}.\footnote{This requires systematic uncertainties and correlations between partial waves to be negligible, as found in~\cite{Schluter:2012mep}. Correlations can actually be relevant, in particular in the high energy region, as shown in~\cite{Bibrzycki:2021rwh}.}
For each model, the fits are initialized by randomly choosing $O(10^5)$ different sets of values for the parameters. The best fits that do not produce any unphysical behavior have $\chi^2/\text{dof} \sim 1.7$--$2$. 
We show in Fig.~\ref{fig:2channelfits} the various 2-channel fits selected as best choices. Notice that none of them can reproduce the bump at $\sim 2.4\gev$ in the $S$-wave. Moreover, some of the dips between the peaks in the intensities are poorly described, with some local  $\chi^2/\text{bins}\gtrsim 4$.
As we anticipated in the \nameref{sec:intro}, some of the resonances in the fitted region have sizeable coupling to a $4\pi$ channel, which is not included in the two channel fits. Absence of a channel may be responsible 
 of producing tension between model and the data.  In particular, most of the 2-channel fits fail to describe the $f_2'(1525)$ lineshape properly, and our assumption that this state is saturated by $\pi\pi$ and $K\bar K$ only seems far too rigid, in particular considering that its coupling to $\pi\pi$ is negligible. This is the main reason why we expect the opening a third channel to improve the description of data. .

\begin{figure}
{\centering
\includegraphics[width=0.49\textwidth]{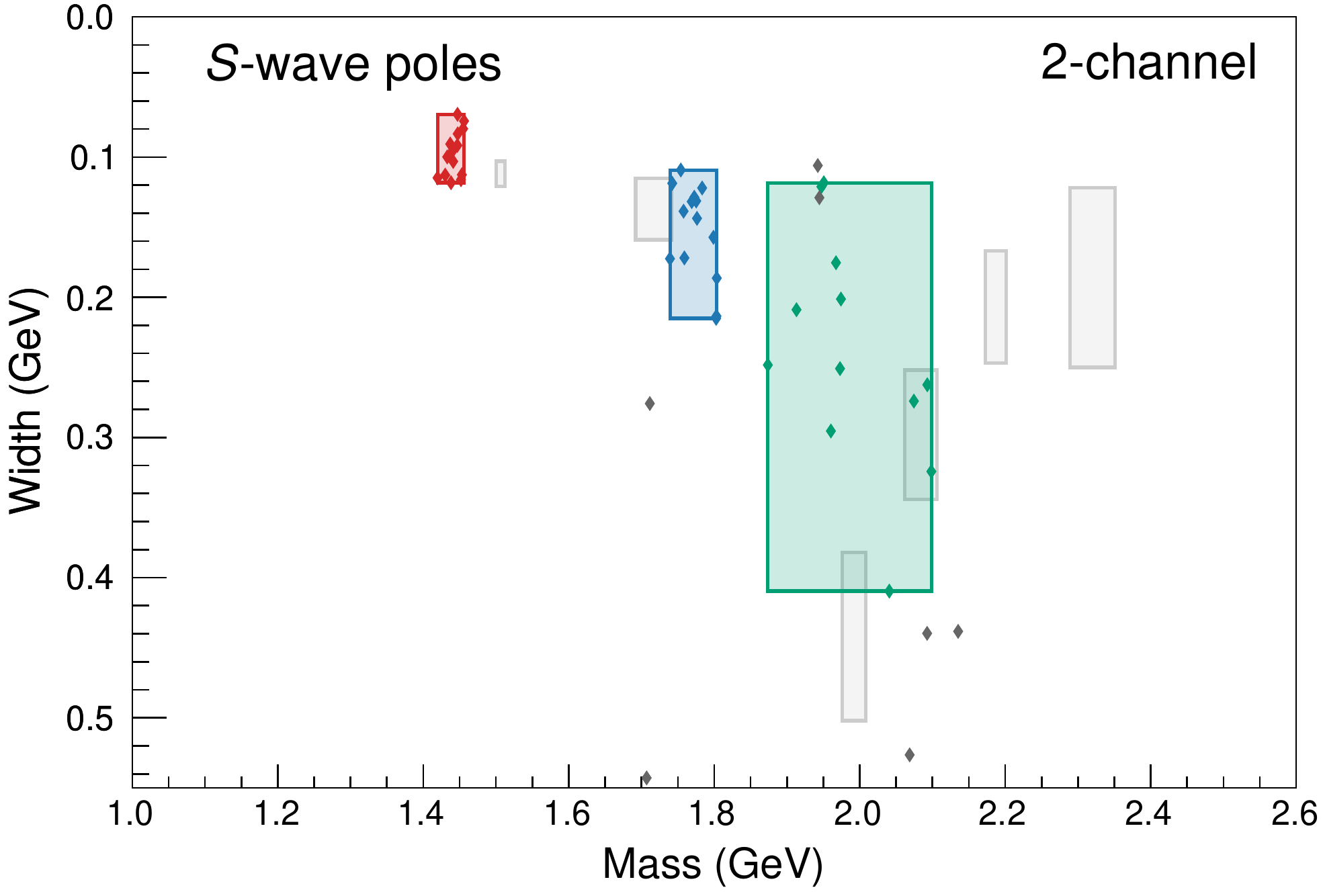}  \includegraphics[width=0.49\textwidth]{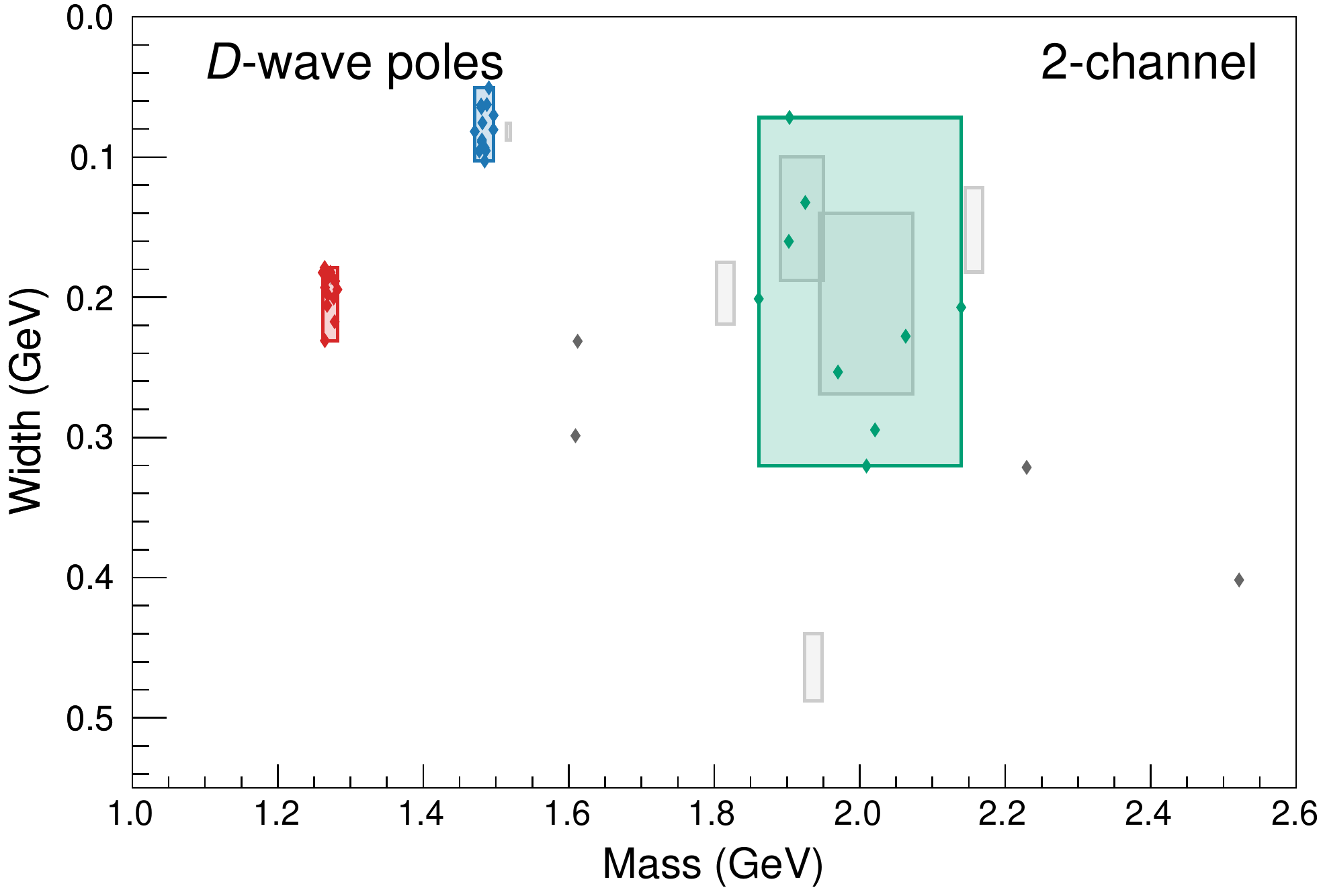}} \\
{\centering
\includegraphics[width=0.49\textwidth]{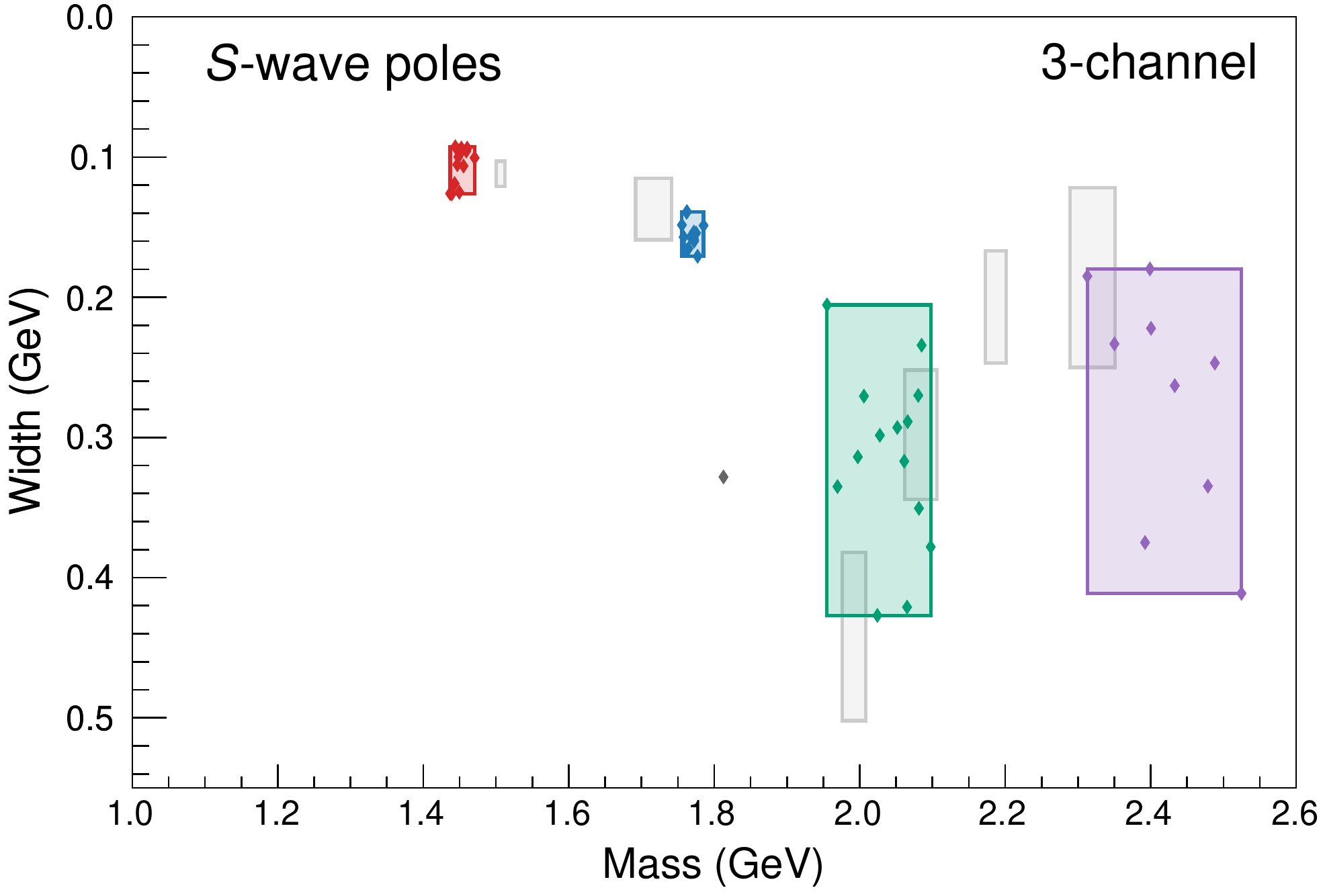}  \includegraphics[width=0.49\textwidth]{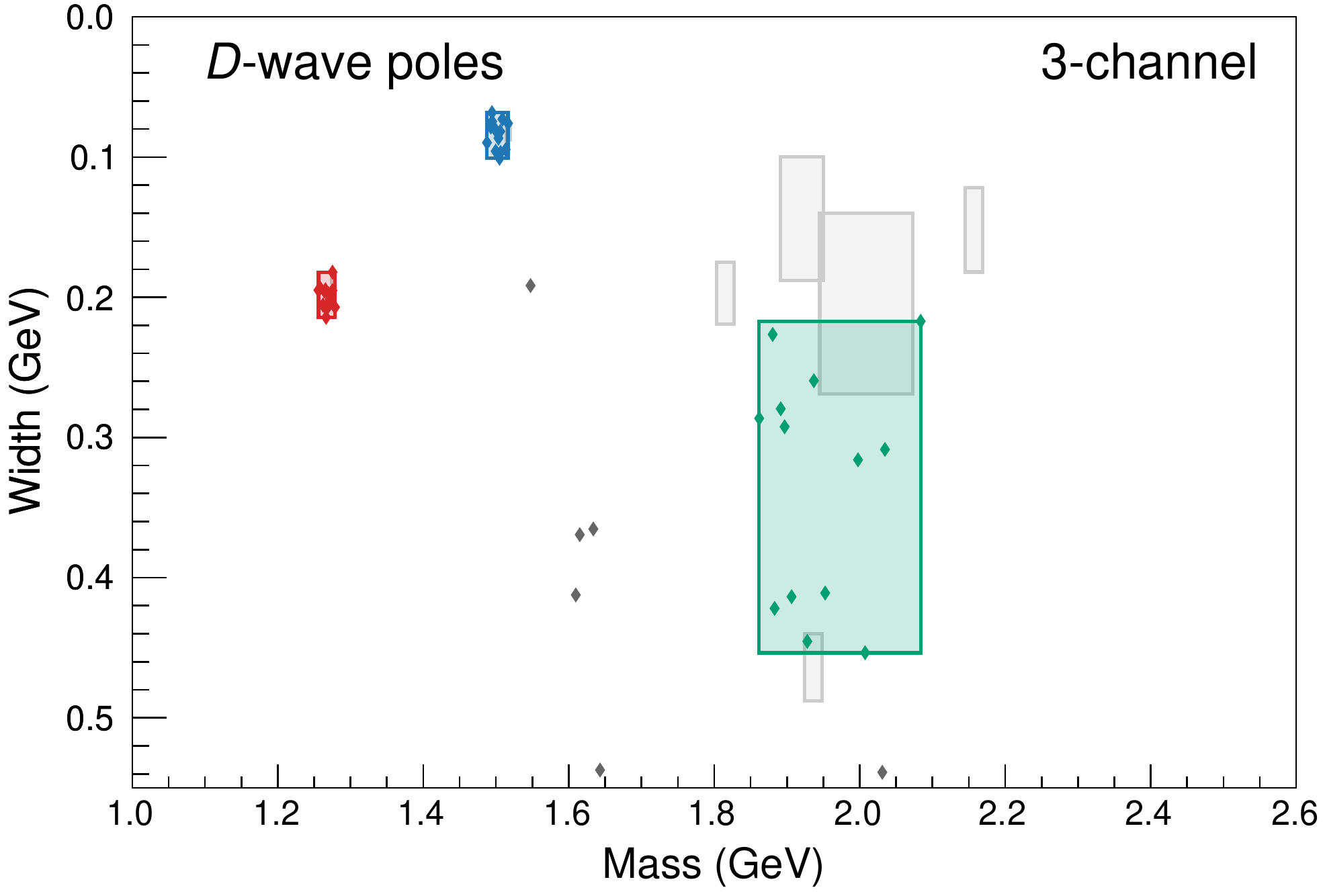}}
\caption{\label{fig:poles2channel} Pole position of the various candidates for the 2- and 3-channel $\pi \pi, K \bar K$ fits, for the 14 systematics considered. We show in the left panels the $S$-wave with the $f_0(1500), f_0(1710)$, $f_0(2020)$ and a possible $f_0(2330)$ resonances. In the right panels the $D$-wave is shown, with the $f_2(1270), f_2'(1525)$ and a possible $f_2(1950)$. Identified poles are represented by colored markers, unidentified ones by gray ones. The colored rectangles represent the maximum spread in mass and width among the 14 models. For comparison, we show as gray rectangles the mass and widths (with uncertainties) of the 12 resonances listed in the PDG. We remark that the PDG lists mostly Breit-Wigner parameters, rather than pole positions in the complex plane. The 3-channel fits show a general improvement of the pole spreads. A new $f_0(2330)$ is found, while the $f_2(1950)$ is pushed deeper into the complex plane.
}
\end{figure}

In these exploratory 2-channel studies no detailed statistical analysis is  performed. Nevertheless, we discuss the results on the pole positions. 
We show in Fig.~\ref{fig:poles2channel} the poles that appear on the Riemann sheets closest to the physical axis. Firstly, it is worth noting that not all fits produce the same number of resonant poles. Secondly, some of the fits produce additional ``spurious'' poles nearby, unstable upon variations of the model.  As mentioned in the \nameref{sec:intro}, the PDG lists five   $S$- and seven  $D$-wave resonances in this energy region, respectively ({\it cf.} Table~\ref{tab:pdgpoles}).
Grouping in clusters the poles obtained from the fits of the various models  that can be identified with physical resonances is not a simple task, especially for the heavier broad resonances that have large  uncertainties. 
Out of the 12 PDG resonances, we can identify only 6. As said, increasing the number of \KCDD does not help. 
The four lower mass clusters do not spread much and can be easily recognized. We note that the $f_0(1500)$ is clearly lighter than what is listed in the PDG average, whereas the $f_0(1710)$ is systematically heavier. Both $f_2(1270)$ and $f'_2(1525)$ seem to have masses close to those of the PDG, although the latter's width is not very well determined in  the fits.
Two heavier mass clusters seem to exist, each spreading over at least two states listed in the PDG. We identify them as the $f_0(2020)$ and the $f_2(1950)$. Some models produce a fourth narrow $S$-wave pole at around $\sim 2\gev$. One might wonder whether this cluster should be identified as the $f_0(2020)$, or as an additional state with almost the same mass. Most of the models produce the broader pole only. For those parametrizations that produce both, the narrower state has a much smaller total coupling, and decays preferably to the $K \bar K$ state. As we will see later, when including a third  channel this narrow pole disappears. 
Finally, we note that there is no pole that could be identified with the $f_0(1370)$, even when an {\em ad hoc} $K$-matrix$\big/$CDD pole is added. However, this is not unexpected, as the $f_0(1370)$ couples mostly to $4 \pi$. Phenomenologically, little mixing is expected between this resonance and the scalar glueball~\cite{Giacosa:2005zt,Giacosa:2005qr,Albaladejo:2008qa,Janowski:2014ppa}, which would additionally suppress its production in $\jpsi$ radiative decays. Its broad width would make its identification even more complicated. We conclude that, although we do not find evidence for this resonance in our analysis, its existence is not challenged.

The intervals of mass and width where the six resonances appear in the best models are shown in Fig.~\ref{fig:poles2channel} and summarized in Tab.~\ref{tab:poles2channel}. It is worth noting, as shown in Fig.~\ref{fig:poles2channel}, that the spreads for the heavier poles are compatible with several different resonances listed in the PDG. We remark again that we have intentionally conducted no statistical analysis here.
Summarizing, even though 2-channel fits describe data reasonably overall, they miss local features that affect the determination of some resonances.

\begin{table}[b]
\caption{Poles positions of the 2- and 3-channel fits. The intervals summarize the spread of results among the 15 best models. Statistical uncertainties are not taken into account. }
\begin{ruledtabular}
\begin{tabular}{l l  c  c  c c  c c c}
&  &$f_0(1500)$ & $f_0(1710)$ & $f_0(2020)$ & $f_0(2330)$ & $f_2(1270)$ & $f_2'(1525)$ & $f_2(1950)$ \\ \hline
\multirow{2}{*}{2-ch.} & Mass \mevp & $[1420,1456]$ & $[1739,1803]$ & $[1874,2098]$ & $-$ & $[1262,1282]$ & $[1471,1497]$ & $[1861,2139]$ \\ 
& Width \mevp & $[70,118]$ & $[109,215]$ & $[118,410]$ & $-$ & $[179,231]$ & $[51,103]$ & $[72,320]$ \\ \hline
\multirow{2}{*}{3-ch.} & Mass \mevp & $[1437,1471]$ & $[1756,1785]$ & $[1955,2098]$ & $[2313,2525]$ & $[1256,1279]$ & $[1488,1517]$ & $[1862,2084]$ \\ 
& Width \mevp & $[93,126]$ & $[139,171]$ & $[206,427]$ & $[180,411]$ & $[182,214]$ & $[68,101]$ & $[217,539]$ \\ \end{tabular}
\end{ruledtabular}
\label{tab:poles2channel}
\end{table}

%====================================================================
\section{3-channel Results}
\label{sec:3charesults}
\begin{figure}
\centering
\includegraphics[width=0.32\textwidth]{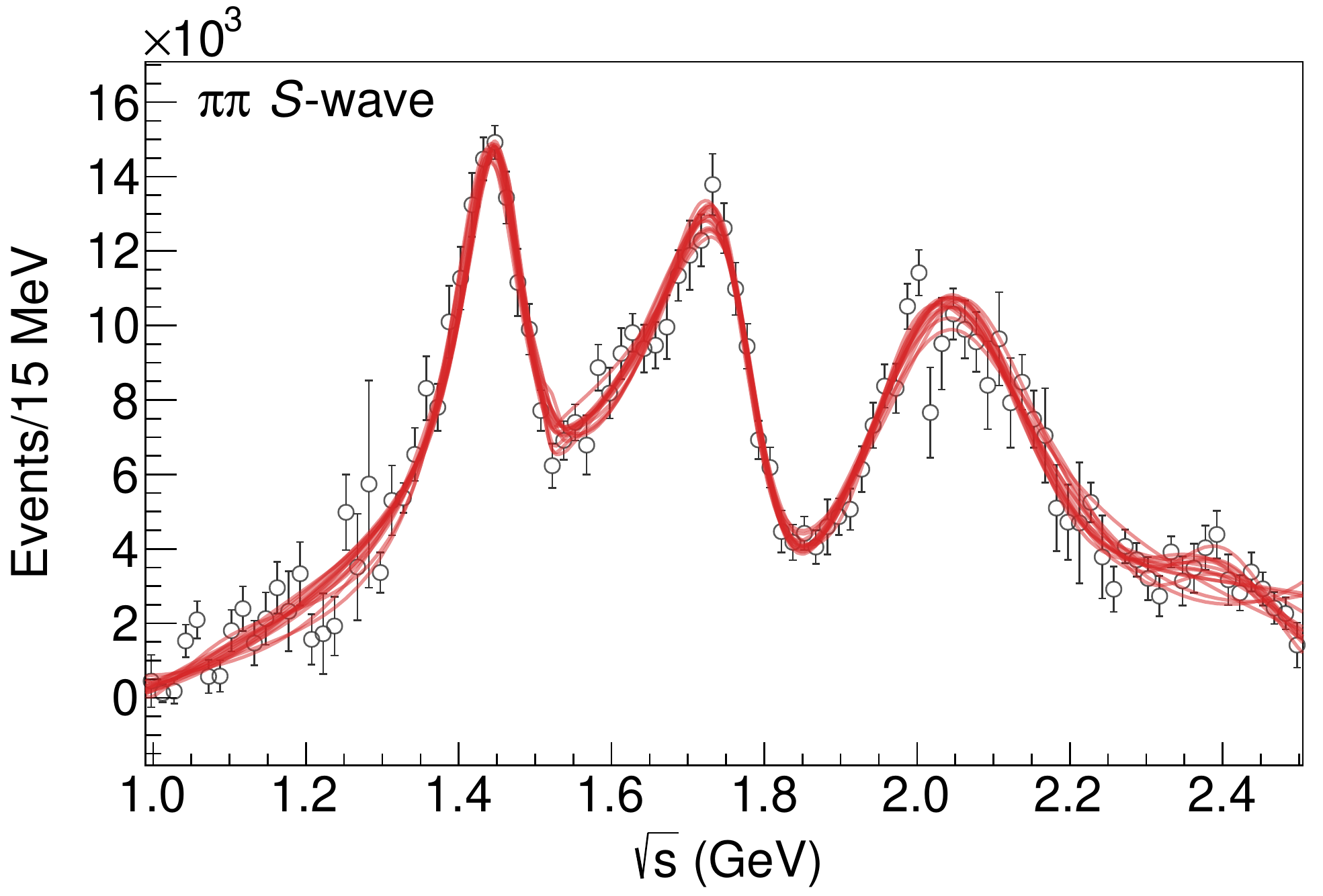} \includegraphics[width=0.32\textwidth]{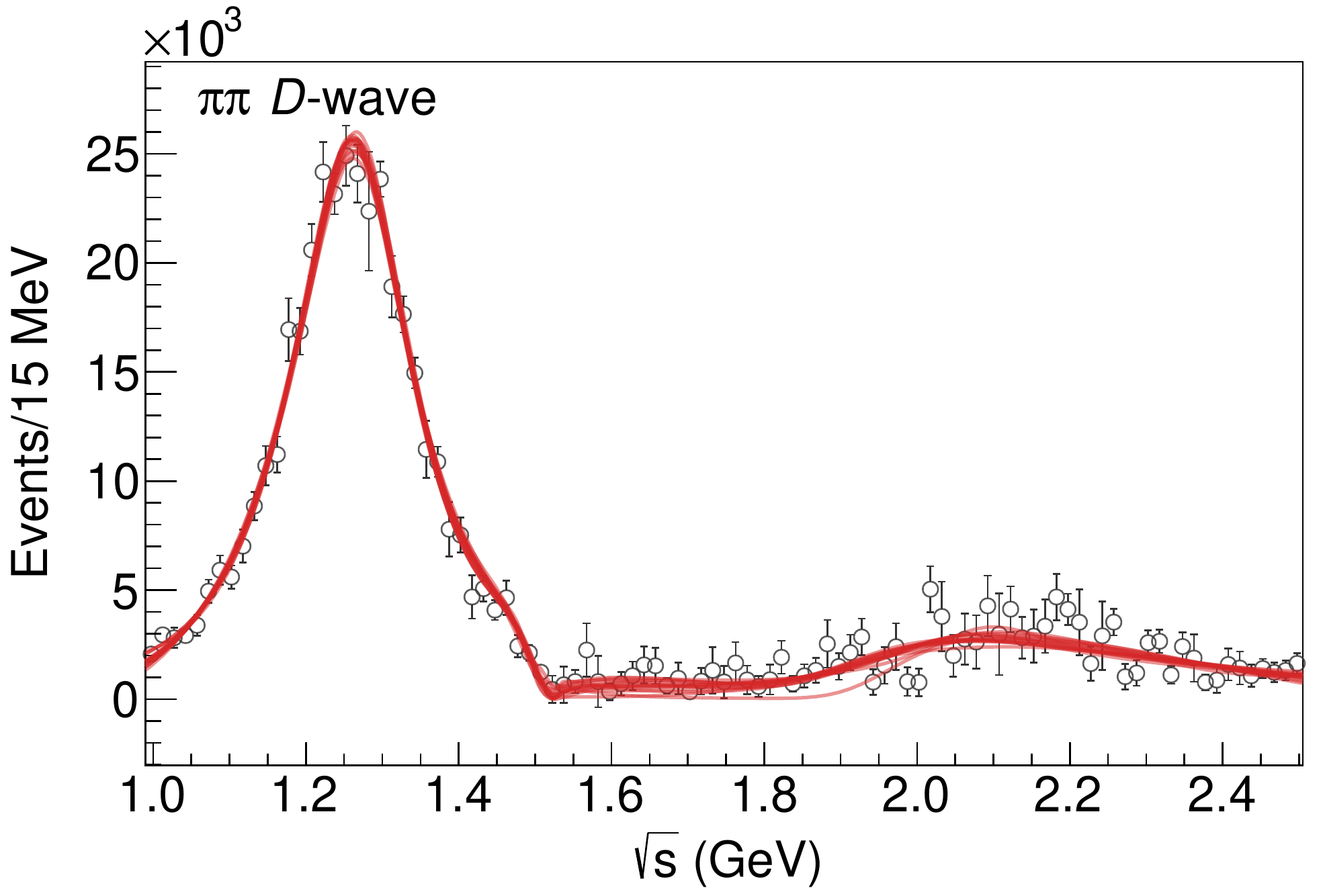} \includegraphics[width=0.32\textwidth]{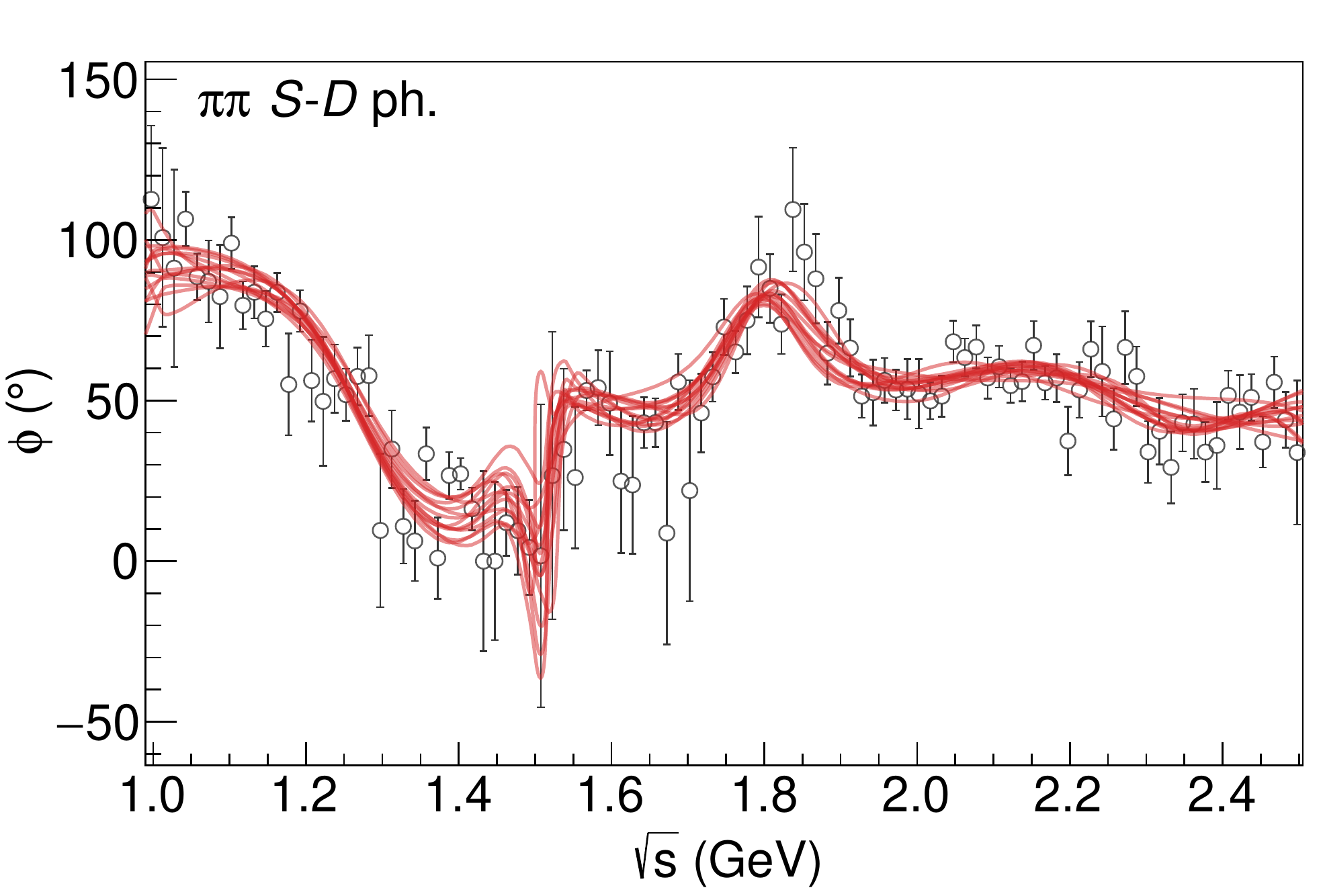} \\
\includegraphics[width=0.32\textwidth]{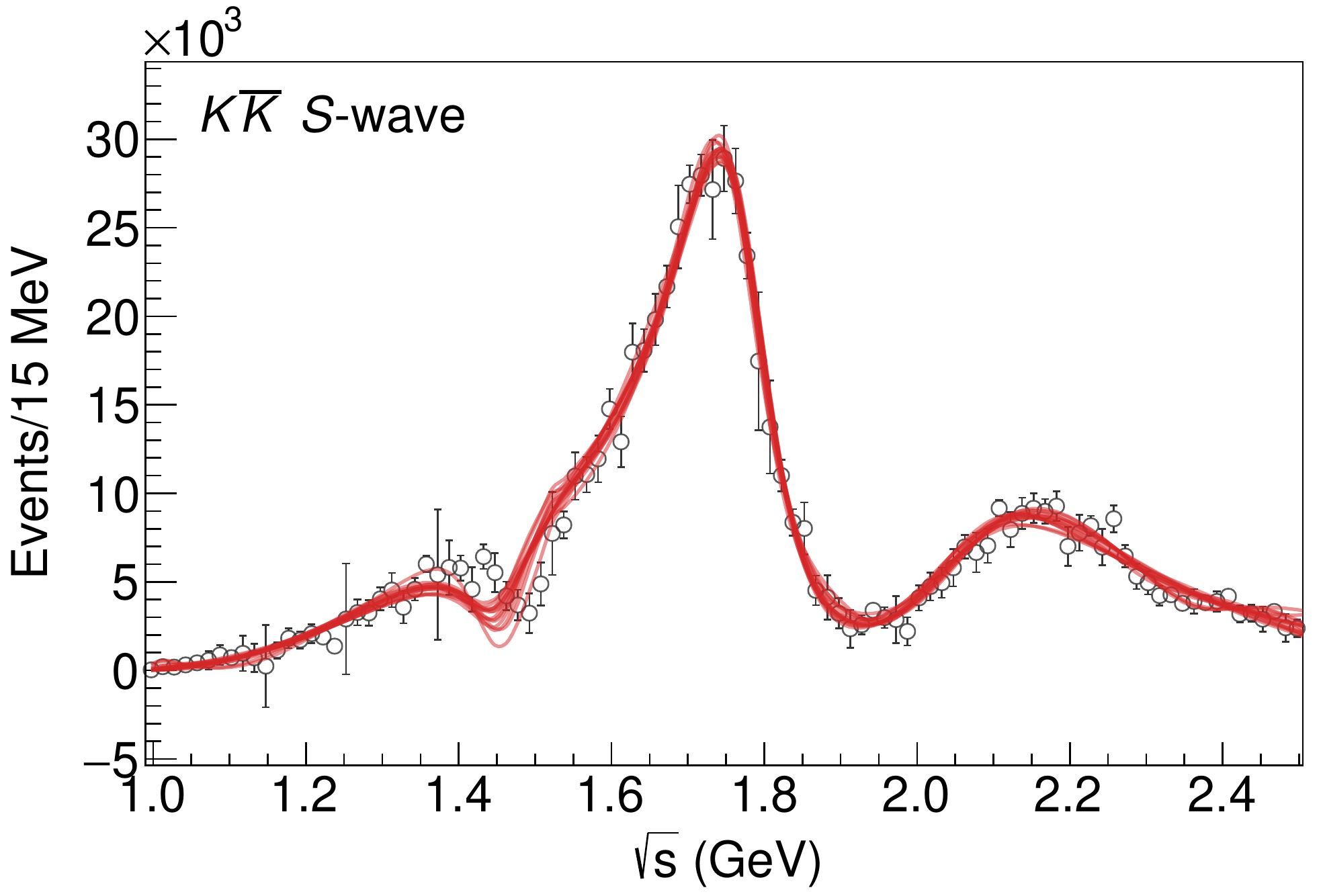} \includegraphics[width=0.32\textwidth]{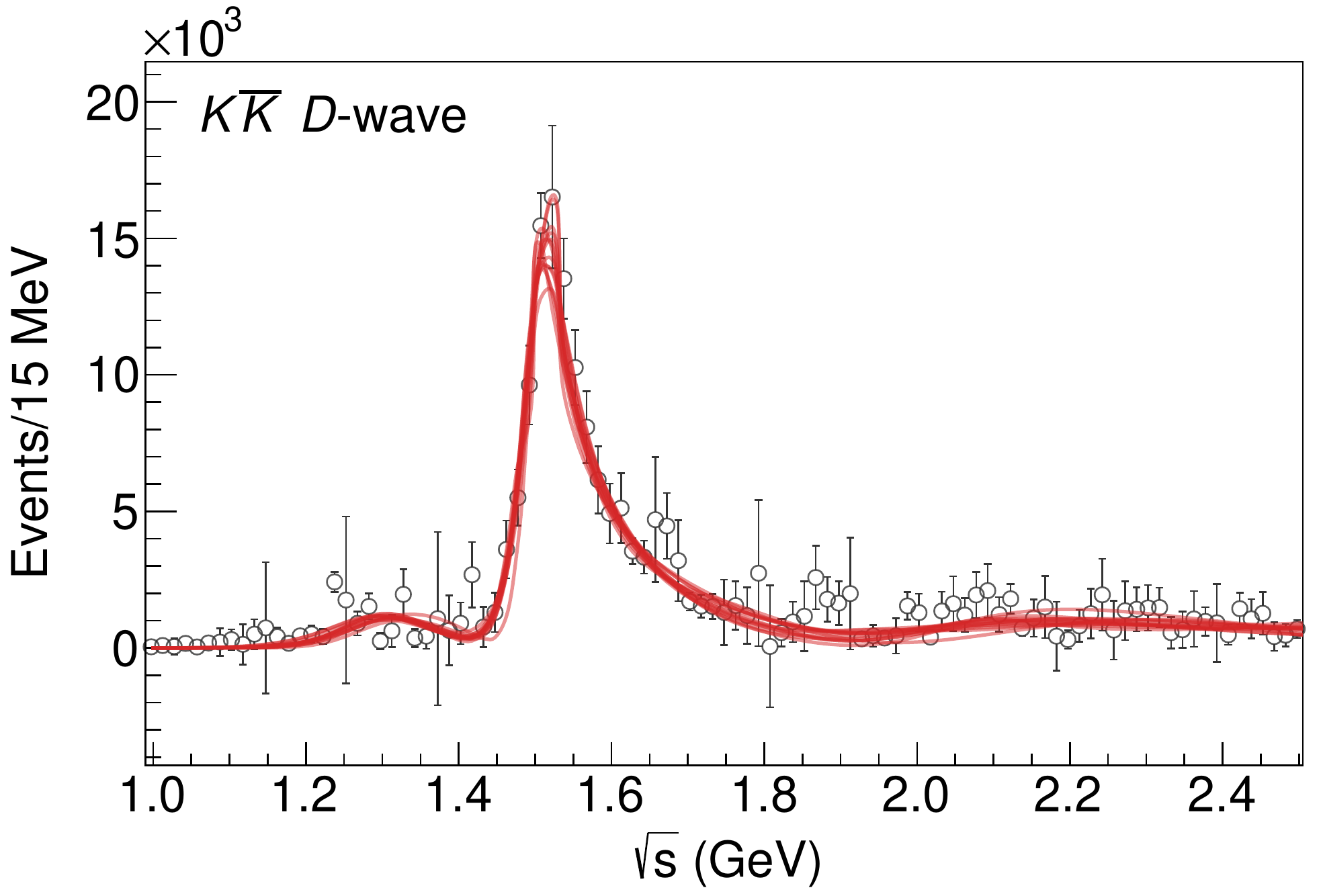} \includegraphics[width=0.32\textwidth]{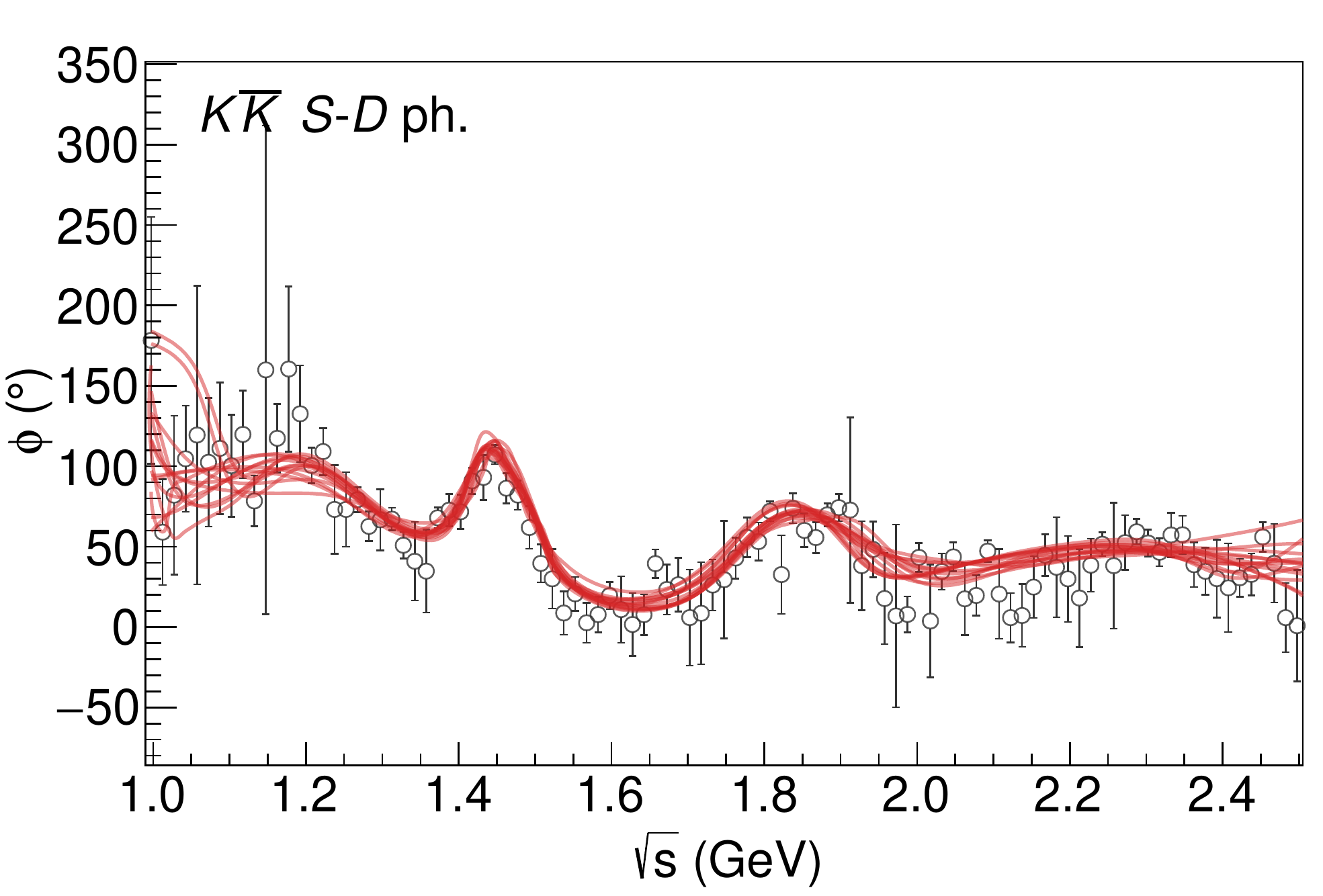} 
\caption{\label{fig:3channelfits} Best 3-channel fits to $\pi \pi$ (top) and $K \bar K$ (second row) final states. The intensities for the $S$- (left), $D$-wave (center), and their relative phase (right) are shown. The red lines denote the fit results. All these produce $\chi^2/\text{dof}\sim 1.1$--$1.2$.
 }
\end{figure}
We extend our model to include a third channel corresponding to an effective $4\pi$ final state. Since we are not sensitive to the details of the dynamics populating it, we approximate it as a stable $\rho\rho$ channel, with $m_\rho = 762\mev$~\cite{GarciaMartin:2011nna}.  Indeed, including the $\rho$ width does not improve the fit sizably, but makes the analytic continuation extremely complicated~\cite{JPAC:2018zwp}.\footnote{Alternatively, one could use approximate methods for analytic continuation, for example Pad\'e approximants, as in~\cite{Ropertz:2018stk}.} 
Restarting the fits from scratch with an additional unconstrained channel is unfeasible.
Instead we use the best 2-channel fits of the models of Section~\ref{sec:2charesults}, and use their parameters as starting point for the new 3-channel fits, to obtain more stable results. Since the 2-channel fits have reasonable quality already, we expect the contribution of the third channel to be small. To reduce the number of parameters, the numerator coefficients $a^{J,\rho\rho}_n$ are set to zero. Moreover, in the coefficients of the polynomial in $K^J(s)^{(-1)}$, 
we set the cross terms between the first two and the third channel to zero. The total number of parameters increases to 53--56, depending on the specific model.

\begin{figure}
\centering
\includegraphics[width=0.32\textwidth]{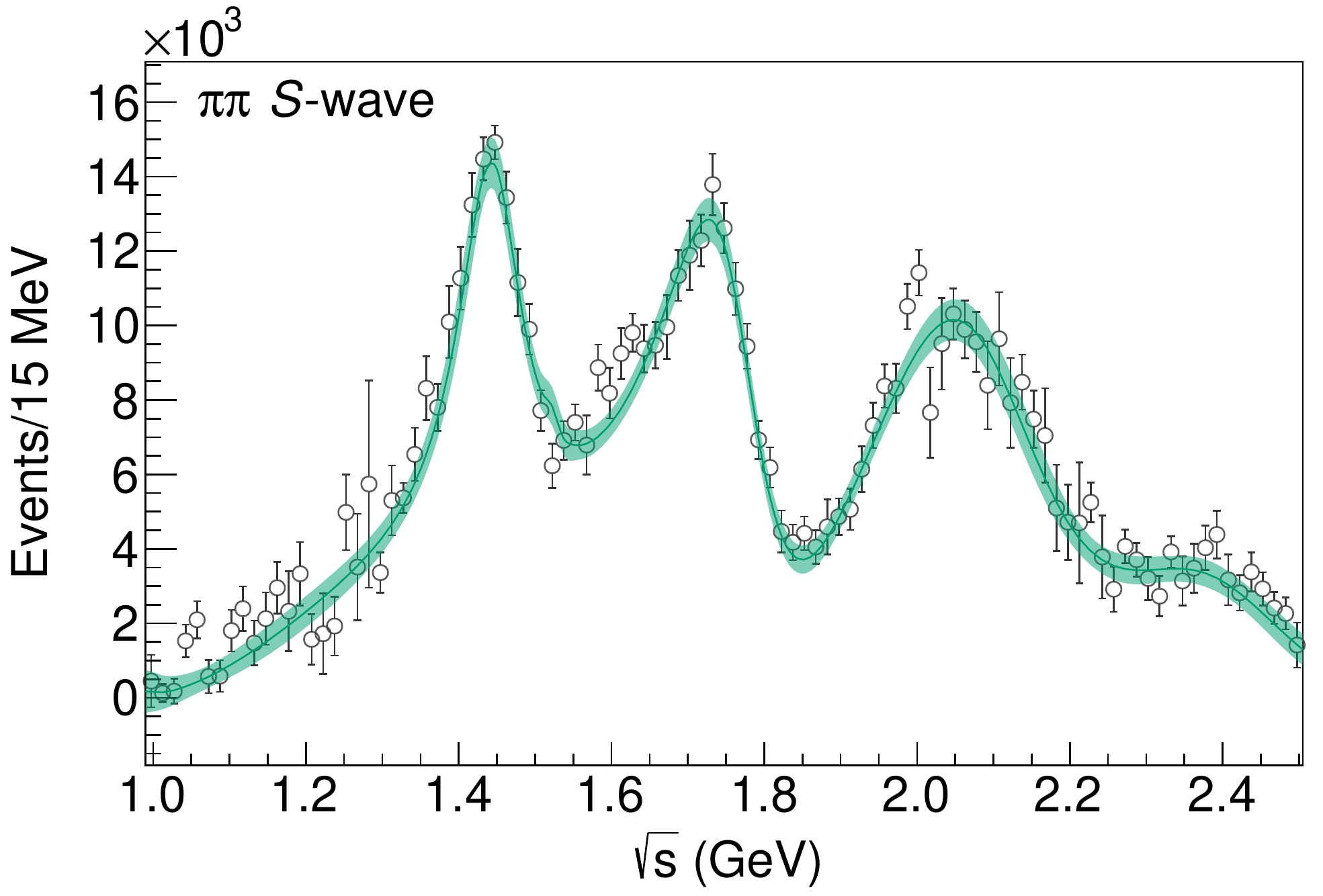} \includegraphics[width=0.32\textwidth]{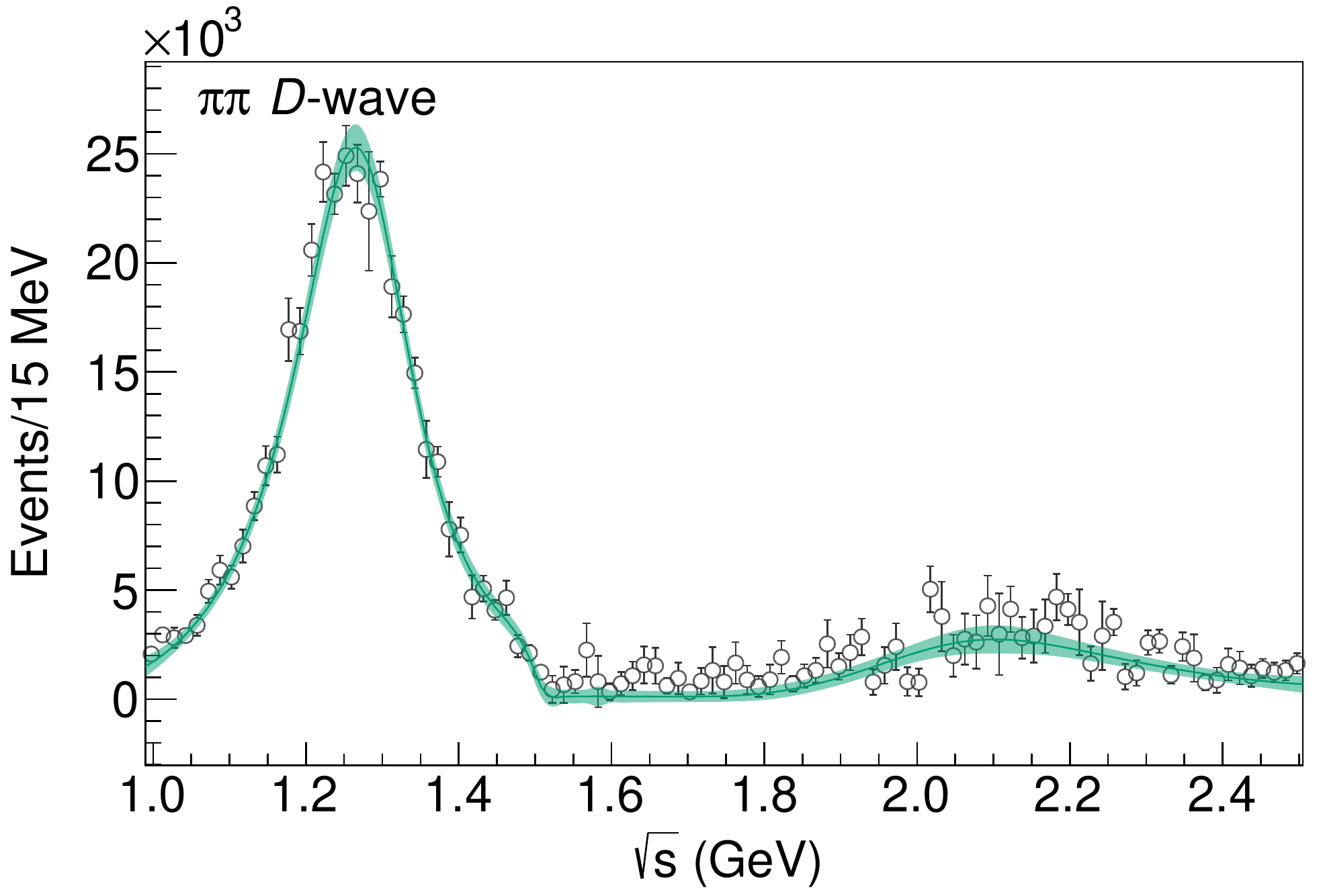} \includegraphics[width=0.32\textwidth]{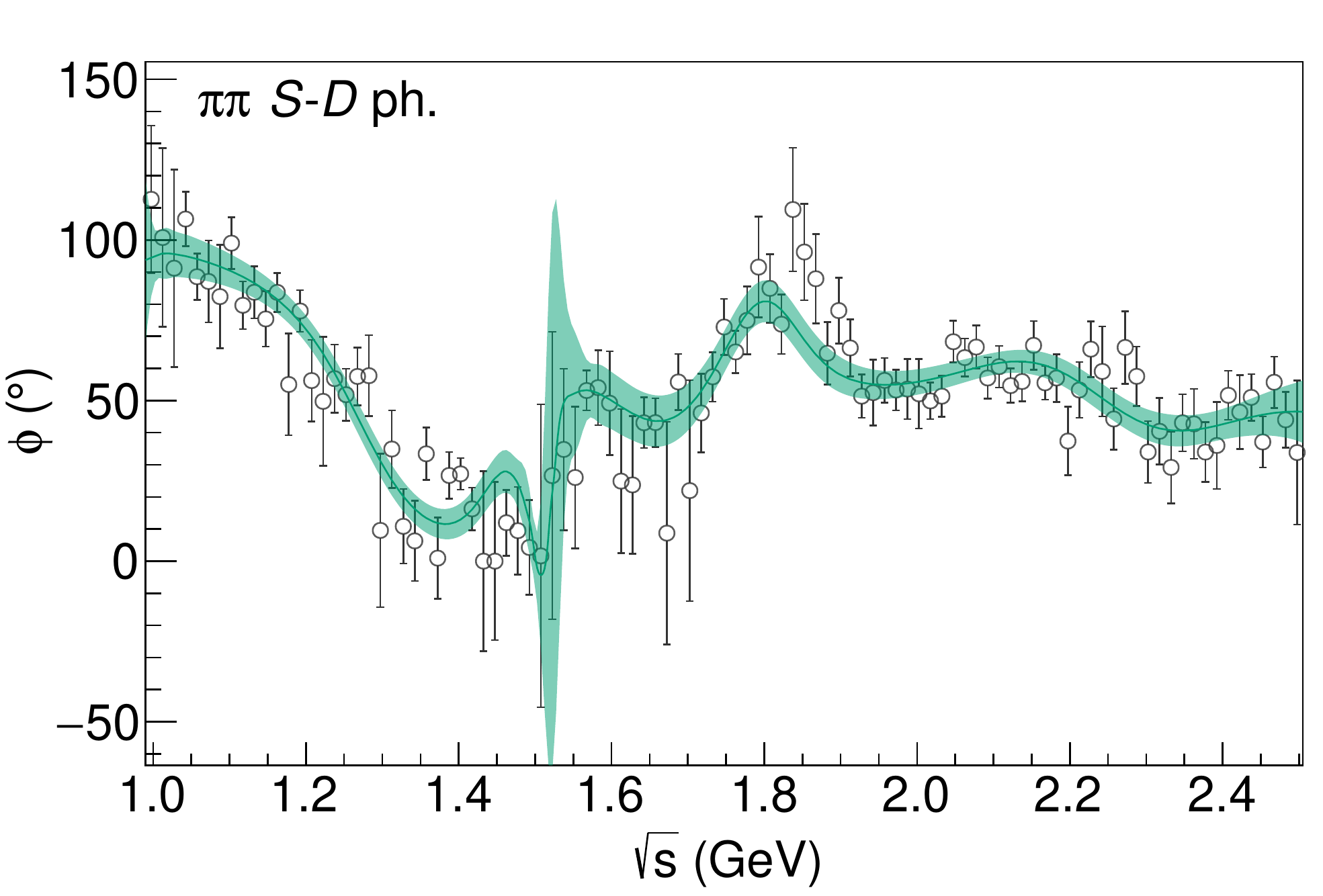} \\
\includegraphics[width=0.32\textwidth]{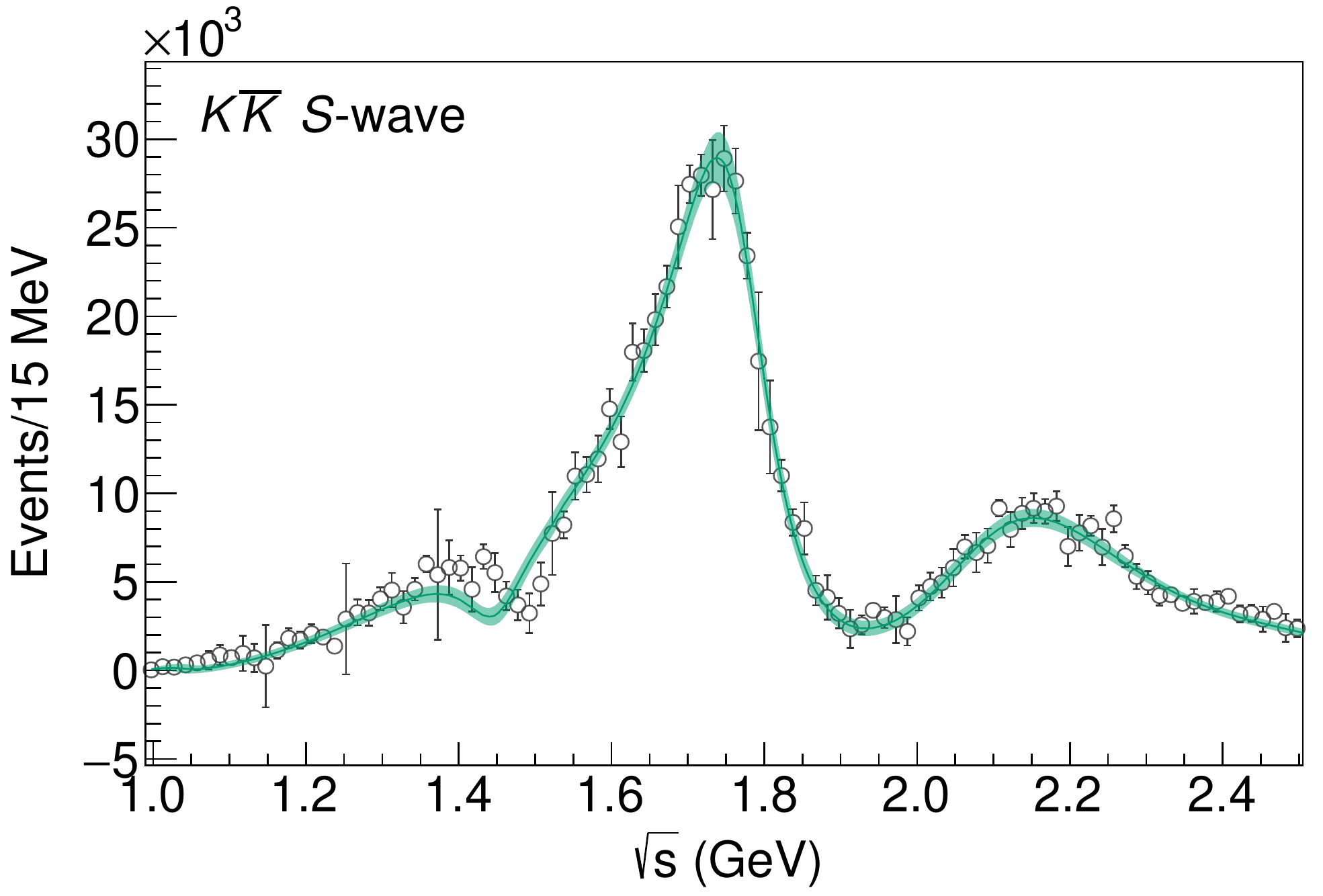} \includegraphics[width=0.32\textwidth]{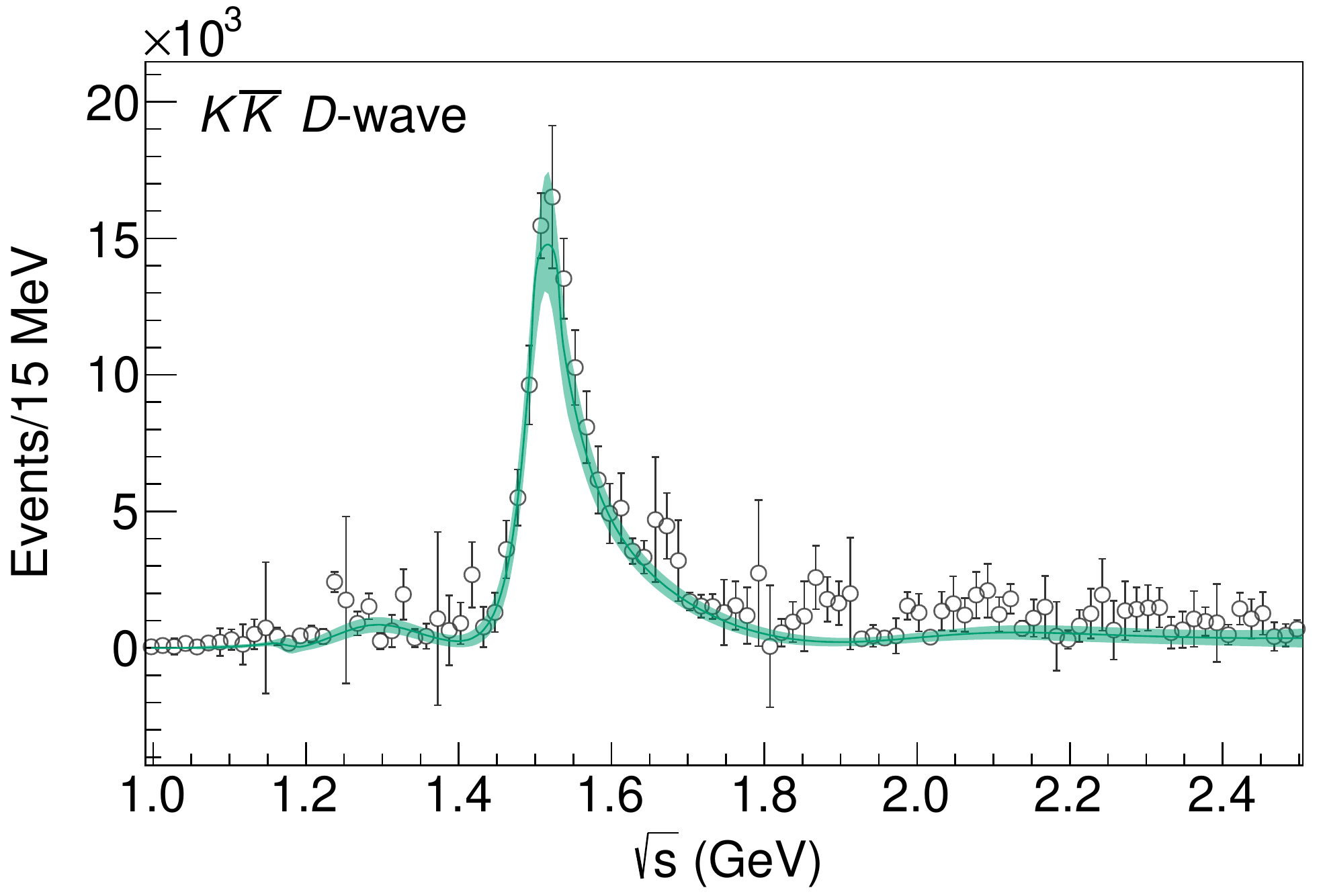} \includegraphics[width=0.32\textwidth]{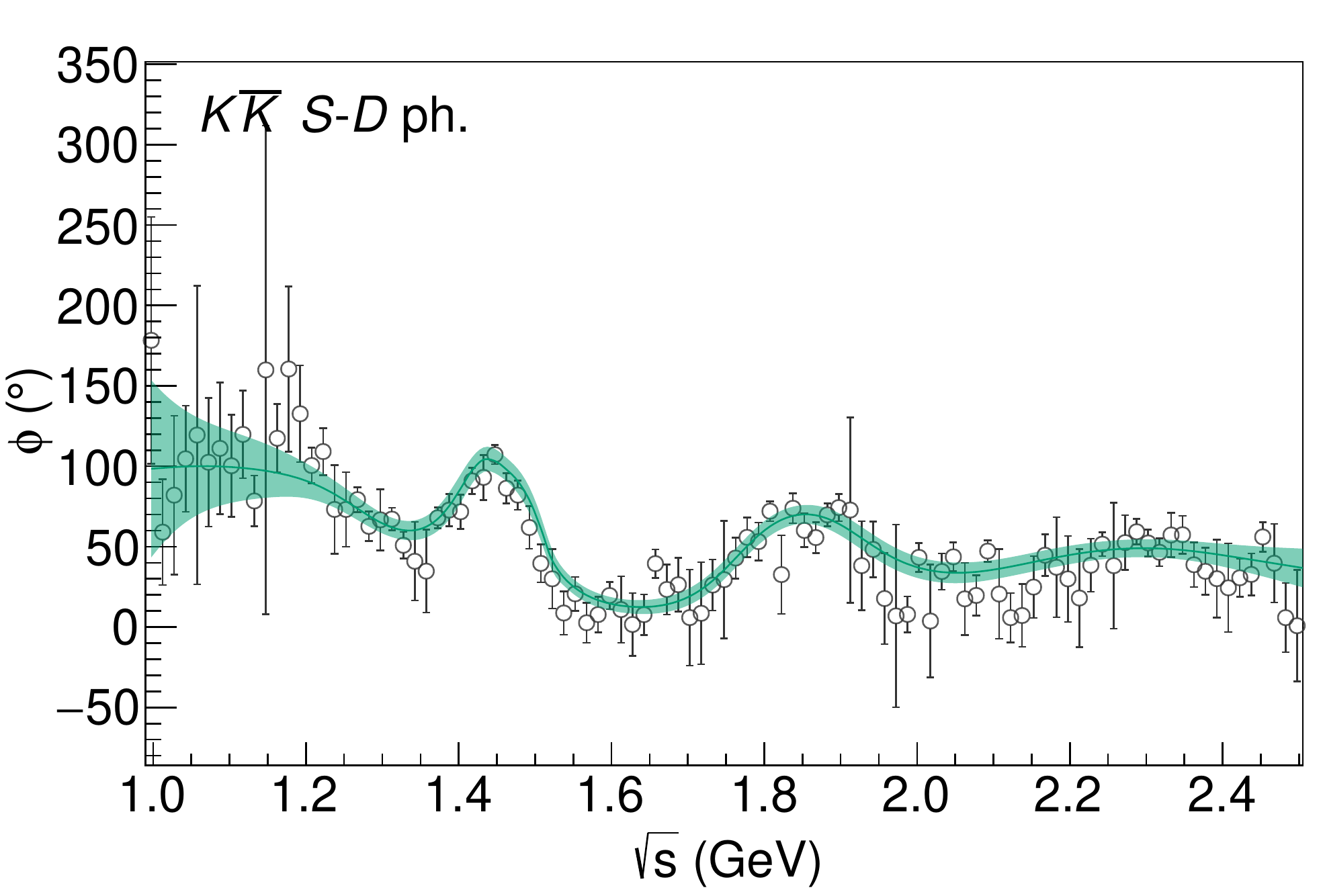}
\caption{One of the final 3-channel fits, with statistical uncertainties included. The solid line and green band show the central value and the $1\sigma$ confidence level provided by the bootstrap analysis, calculated for $O(10^4)$ samples. }
\label{fig:bootstrapbestfit}
\end{figure}

The full list of plots and fit parameters for the 14 models is available in the \nameref{sup:supp-material}. 
In Fig.~\ref{fig:3channelfits} we show the results for the 14 best  3-channel models. These can be compared to the 2-channel fits in Fig.~\ref{fig:2channelfits}. It is evident that the fits improve: the average $\chi^2/\text{dof}$ drops from $\sim 1.7$--$2$ to $\sim 1.1$--$1.2$. 
More importantly, the local description of the relative phases and of the regions around the peaks are much more accurate. The effect can be seen in Fig.~\ref{fig:poles2channel},  where it is evident that poles are determined more precisely when the new channel is added.

Most of the models lead to similar results, except for some deviation of the $K \bar K$ phase close to threshold. By construction our models respect Watson's theorem, which means that at the $K \bar K$ threshold  the $\pi \pi$ and the $K \bar K$ phases are identical. However, in some of our fits the $K \bar K$ phase moves rapidly just above threshold, because of peculiar cancellations between large numerators. Another interesting feature is the ``quasi-zero'' behavior on the $\pi \pi$ $D$-wave around 1.5\gev, which is evident in the intensity and seems to produce a sharp motion in the relative phase. A simple interpretation is that, if the $D$-waves are almost elastic, one expects a zero to appear between two resonances. If one assumes the coupling of the $f_2'(1525)$ to $\pi \pi$ to be almost zero, then this behavior could be explained by the interference between the $f_2(1270)$ and a heavier resonance coupling strongly to $\pi \pi$. This matches the behavior shown in the $D$-wave intensity, where the $f_2(1950)$ candidate produces a small peak. Moreover, the $K \bar K$ $D$-wave does not show any rapid motion, suggesting that, were a heavy resonance to exist, it would couple mostly to the other channels.

The structure of the $S$-waves is much richer. There are four clearly visible peaks in $\pi \pi$, and three  in $K \bar K$. It is worth noticing how different the values at the peak intensities look when comparing the same resonance in both final states. In particular, in $K \bar K$ the peak associated to the $f_0(1710)$ is roughly six times stronger than the $f_0(1500)$ one. We will show later that this is reflected in a much larger coupling of the $f_0(1710)$ to this channel. As can be seen in Fig.~\ref{fig:3channelfits}, 
our best fit reproduces all intensity peaks with high accuracy. There is a slightly larger local $\chi^2$ value around the $1.5\gev$ region in the $K \bar K$ $S$-wave. This region is below the $\rho\rho$ open channel, which seems to prevent our fit from fully reproducing the peak and interference. We are aware that the $f_0(1500)$ resonance couples to $4\pi$, but the local description is nevertheless reasonable. We thus conclude that a third channel is not strictly needed to describe such behavior. Ideally, the $\rho \rho$ channel should include  both $\ell =0$ and $\ell =2$  contributions. However, the latter is suppressed at threshold, and having no data to fit makes it impossible to distinguish the two. Nonetheless, we performed some alternative fits including just an $\ell =2$ channel to asses our systematics. We get $\chi^2/\text{dof}\sim 1.4$, not as good as in the $\ell =0$ case, being the channel suppressed as  mentioned. The pole positions calculated this way are compatible with the $\ell =0$ models (see below), and we do not see much variation in the $S$-wave, as expected. We do not consider these fits any further.
Even for these 3-channel fits, there is no evidence for more resonances than the 
seven ones  discussed above. Fits with additional \KCDD do not improve the data description, and the additional poles are far and unstable. 

The statistical uncertainties are determined via bootstrap~\cite{recipes,EfroTibs93,Landay:2016cjw}. We generate $O(10^4)$
pseudodatasets: each data point is resampled from a gaussian distribution having by mean and standard deviation its value and uncertainty; to avoid unphysical negative intensities, data points compatible with zero within $2\sigma$ are instead resampled from a Gamma distribution (see Appendix~\ref{app:gamma}). Each 
pseudodataset is refitted to the original model, and the (co)variance of the population of the fit parameters provides an estimate of their statistical uncertainties and correlations. In Fig.~\ref{fig:bootstrapbestfit} we show as an example the uncertainties for one of the models.

%====================================================================
\section{Resonant poles}
\label{sec:poles}

\begin{table}[b] 
\caption{List of final pole position and uncertainties resulting from the combination of the 14 different final fits to the data. The errors have been obtained as the variance of the full samples, by assuming that the spread of results for each pole, shown in Fig.~\ref{fig:finalpoles}, resembles a Gaussian distribution.}
\begin{ruledtabular}
\begin{tabular}{c c c c}
$S$-wave  & $\sqrt{s_p}$ \mevp & $D$-wave & $\sqrt{s_p}$ \mevp \\ \hline
\rule[-0.2cm]{-0.1cm}{.55cm} $f_0 (1500)$ &  $(1450 \pm 10) - i (106 \pm 16)/2$  &  $f_2 (1270)$ &  $(1268 \pm 8) - i (201 \pm 11)/2$ \\
\rule[-0.2cm]{-0.1cm}{.55cm} $f_0 (1710)$ &  $(1769 \pm 8) - i (156 \pm 12)/2$  &  $f_2 (1525)$ &  $(1503 \pm 11) - i (84 \pm 15)/2$ \\
\rule[-0.2cm]{-0.1cm}{.55cm} $f_0 (2020)$ &  $(2038 \pm 48) - i (312 \pm 82)/2$  &  $f_2 (1950)$ &  $(1955 \pm 75) - i (350 \pm 113)/2$ \\
\rule[-0.2cm]{-0.1cm}{.55cm} $f_0 (2330)$ &  $(2419 \pm 64) - i (274 \pm 94)/2$  &   &  \\
\end{tabular}
\end{ruledtabular}

\label{tab:polesfinal}
\end{table}

\begin{figure}[t]
\centering
\includegraphics[width=\textwidth]{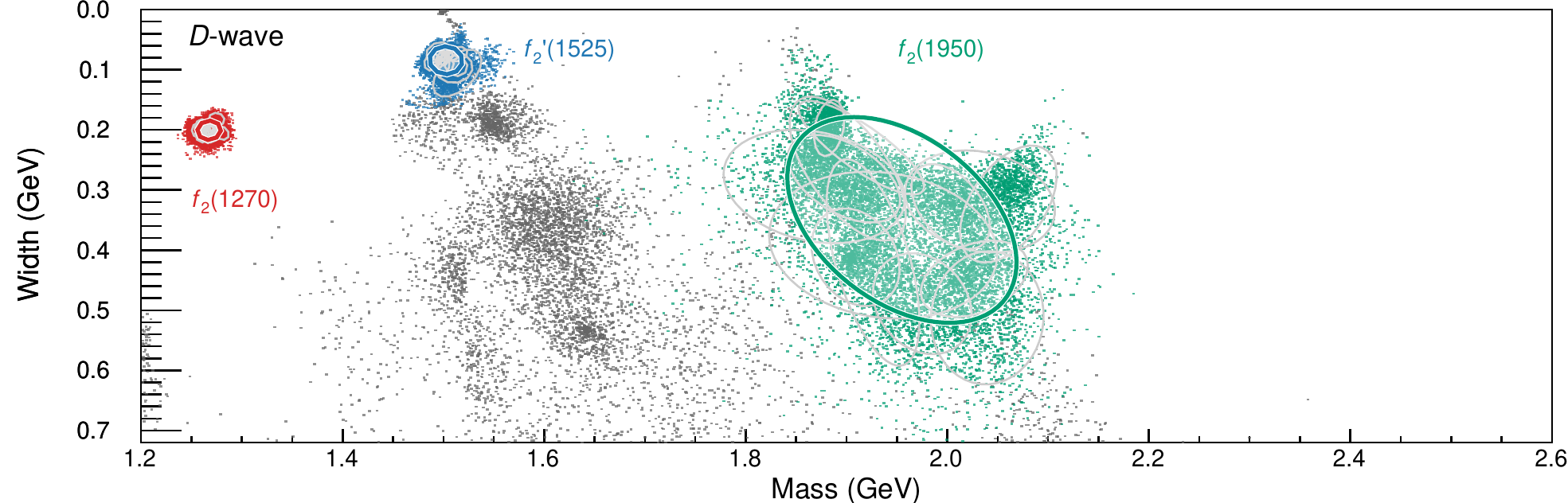} \\ \includegraphics[width=\textwidth]{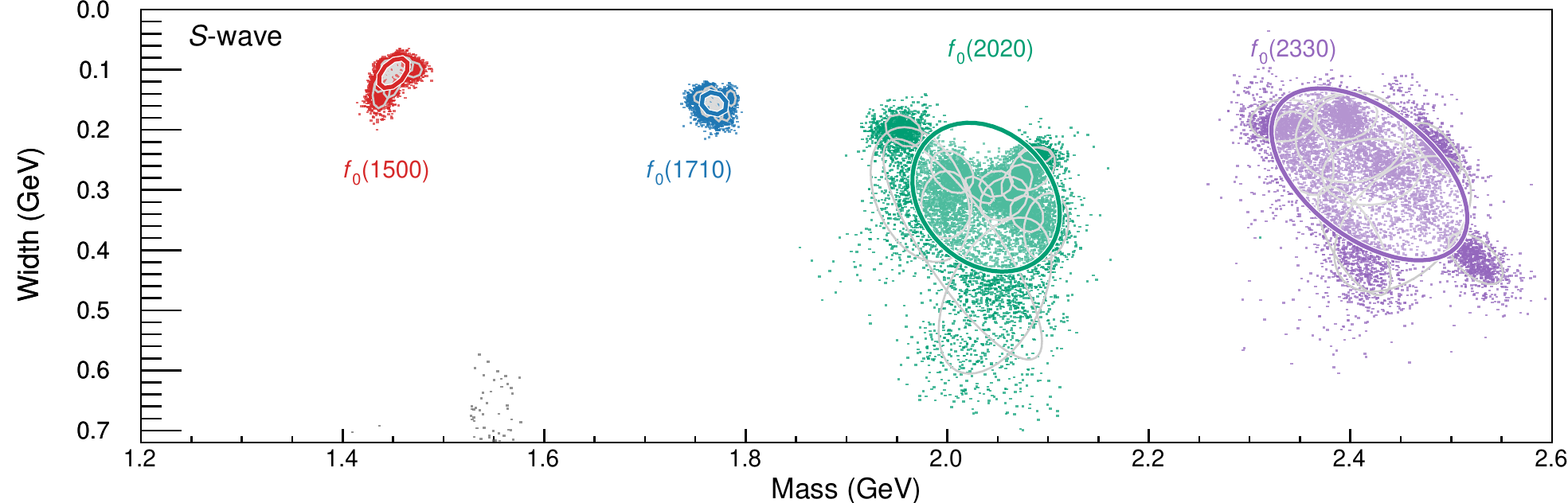} 
\caption{Results for the pole positions of the 3-channel fits, superimposed for the 14 models. A point is drawn for each pole found in each of the $O(10^4)$ pseudodatasets  generated by the bootstrap analysis. Colored points represent poles identified as a physical resonance, gray points are spurious. For the physical resonance, gray ellipses show the $68\%$ confidence region of each systematic. Colored ellipses show the average of all 14 systematics, as explained in the text.
}
\label{fig:finalpoles}
\end{figure}
As already discussed, it is not possible to fix a priori the number of poles that appear on the proximal Riemann sheets. In general, there is no one-to-one correspondence between the poles of the amplitude and the number of \KCDD in coupled channel problems. This relation becomes even more complicated because of the additional background polynomial. 
Moreover, the simple left-hand cut parametrizations in $\rho N^J_{ki}(s')$ also tend to generate additional broad poles close to threshold~\cite{JPAC:2017dbi}.
Some of the poles capture the real features of the amplitude, and are associated with the physical resonances. Other poles are mere artefacts of the model implemented, and are unstable upon bootstrap and model variations. Therefore, a sound statistical analysis and a large set of systematic variations are required to filter out the spurious singularities and identify the remaining ones with the physical resonances. 
The pole positions for the systematic variations of amplitudes studied here are plotted in Fig.~\ref{fig:finalpoles}, while the separate plots for each systematic are left in the \nameref{sup:supp-material}. 
For each model, the statistical uncertainties are determined via bootstrap, as explained in Section~\ref{sec:3charesults}. While in~\cite{JPAC:2018zyd} we were able to identify a nominal model, and explored how model variation affected the central values, here the clusters of poles, in particular the heaver ones, move too much to make this strategy feasible. In order to quote 
an average of masses and widths obtained by the 14 models, we calculate the mean and (co)variance of the pole positions among the $14\times O(10^4)$ 
pseudodatasets from the bootstrap analysis for all the models at once. 

In addition to the pole positions, one can extract the residues of the amplitude. The residues of $a_i^J(s)$ can be associated with the couplings of the resonance $f$ to the initial $\jpsi\,\gamma$ and final $\h \bar \h$ states. We remark that we do not include all the possible open channels involved at these energies. However, since the unconstrained third $\rho\rho$ channel  can effectively reabsorb the presence of other channels, we believe that the relative size of the $\pi\pi$ and $K\bar K$ coupling provides reliable information. One can also study the residues of the $D^J(s)^{-1}$ matrix, that are connected to the scattering couplings $\h \bar \h \to f \to \h' \bar \h'$, albeit not rigorously.\footnote{To get the full scattering amplitude, the  $D^J(s)^{-1}$ matrix should be multiplied by the appropriate $N^J(s)$ that satisfies an integral equation that depends on the left-hand cuts of the scattering process. However, since $N^J(s)$ is smooth, we believe it should not affect much the relative size of the couplings, that we discuss here.} Since we are not fitting scattering data, the residues of $D^J(s)^{-1}$ are mostly unconstrained, and have large uncertainties~\cite{Briceno:2021xlc}.

The lightest two $D$-wave poles are very well determined. They correspond to the $f_2(1270)$ and $f_2'(1525)$ resonances, and decay almost elastically to $\pi \pi$ and $K \bar K$ respectively. The $f_2$ peak in $\pi \pi$ and the $f_2'$ peak in $K \bar K$ is very well described by all models. The $f_2'$ lies close to the $\rho\rho$ threshold, so we have to ensure that the poles that form the cluster appear always on the proximal Riemann sheet. For all the models, the $f_2'$ is always centered below this threshold. 
Even though we do not fit scattering data directly, these resonances are so well behaved that the scattering couplings have reasonable ratios:
\begin{align}
f_2(1270): \quad r_{\pi\pi}\Big/\sqrt{r^2_{\pi\pi} + r^2_{K \bar K}} &=82^{+6}_{-8}\% \,, &
f_2'(1525): \quad r_{K\bar K}\Big/\sqrt{r^2_{\pi\pi} + r^2_{K \bar K}}&=95 ^{+3}_{-5}\% \,,
\end{align}
where $r_{\h\bar \h}$ are the absolute values of the residues of the $D^J(s)^{-1}$ matrix in the elastic $\h\bar \h \to \h\bar \h$ channel. These are reasonably close to the PDG estimates~\cite{pdg}. 
Some of the fits produce a second broader cluster in  $D$-wave behind the $f'_2(1525)$. As can be seen in the \nameref{sup:supp-material}, this second pole appears in most of the $K$-matrix parametrizations, often with very large spread, but not in the CDD ones. 
Furthermore, when the pole appears the local $\chi^2$ in that region does not improve. For these reasons, the existence of an additional resonance is not compelling in data.

Moving to the $S$-wave, our result for the $f_0(1500)$ is perfectly compatible with~\cite{Ropertz:2018stk}, even though  we have a $f_0(1710)$  close by, which could easily affect its pole position. The $f_0(1500)$ turns out to be rather narrow, and produces a simple phase motion for the $S$-wave phases. The $f_0(1710)$ is noticeably broader, but nevertheless very well determined. The mass we find for the 
 $f_0(1710)$ is considerably larger than the PDG average, however, it is still compatible with many of the determinations listed in the PDG. All the four scalar resonances we found are roughly compatible with those identified  in~\cite{Sarantsev:2021ein}, although what we call $f_0(1710)$ and $f_0(2020)$ seem to correspond to their $f_0(1770)$ and $f_0(2100)$. 

\begin{figure}[t]
\centering
\includegraphics[width=0.48\textwidth]{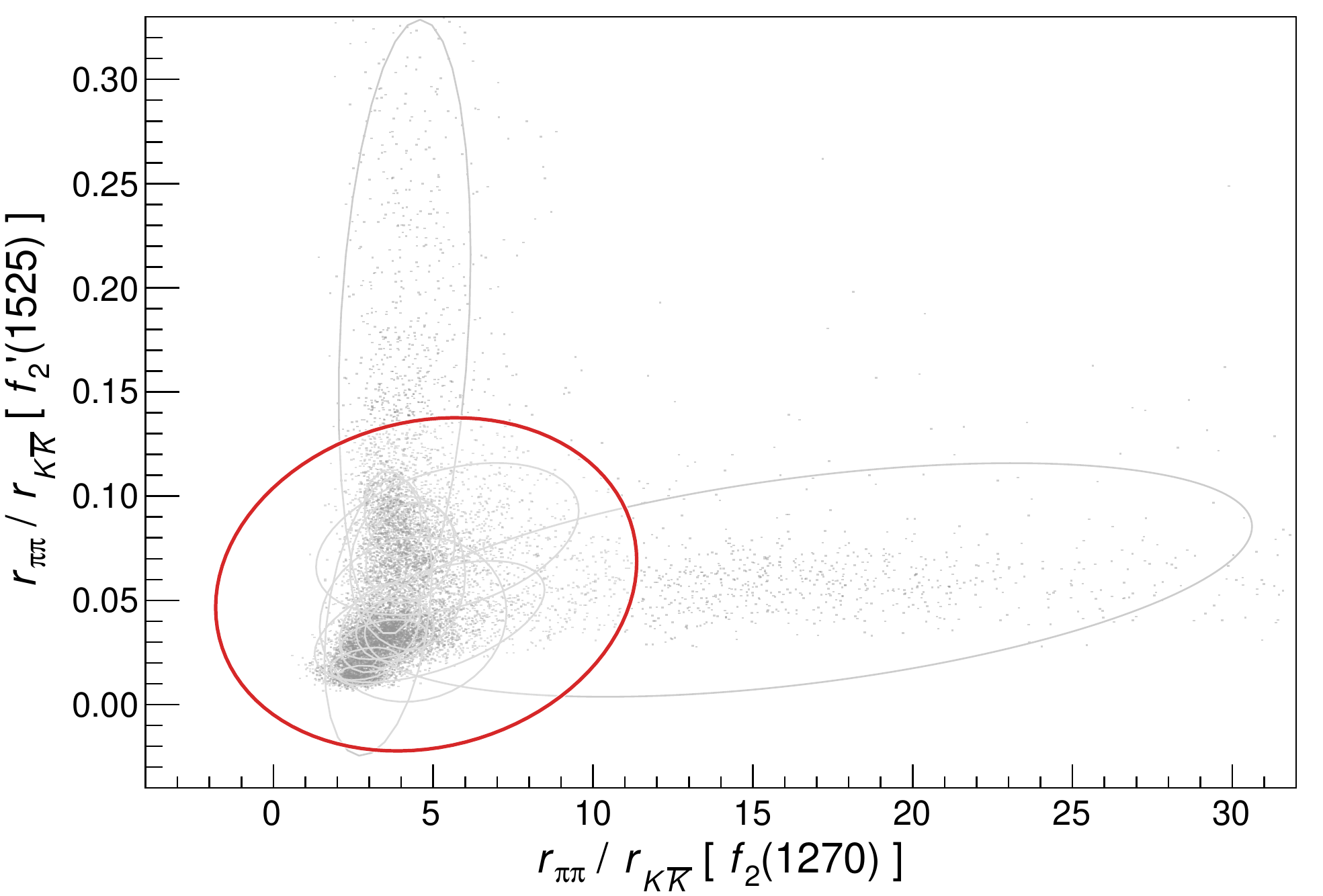}  \includegraphics[width=0.48\textwidth]{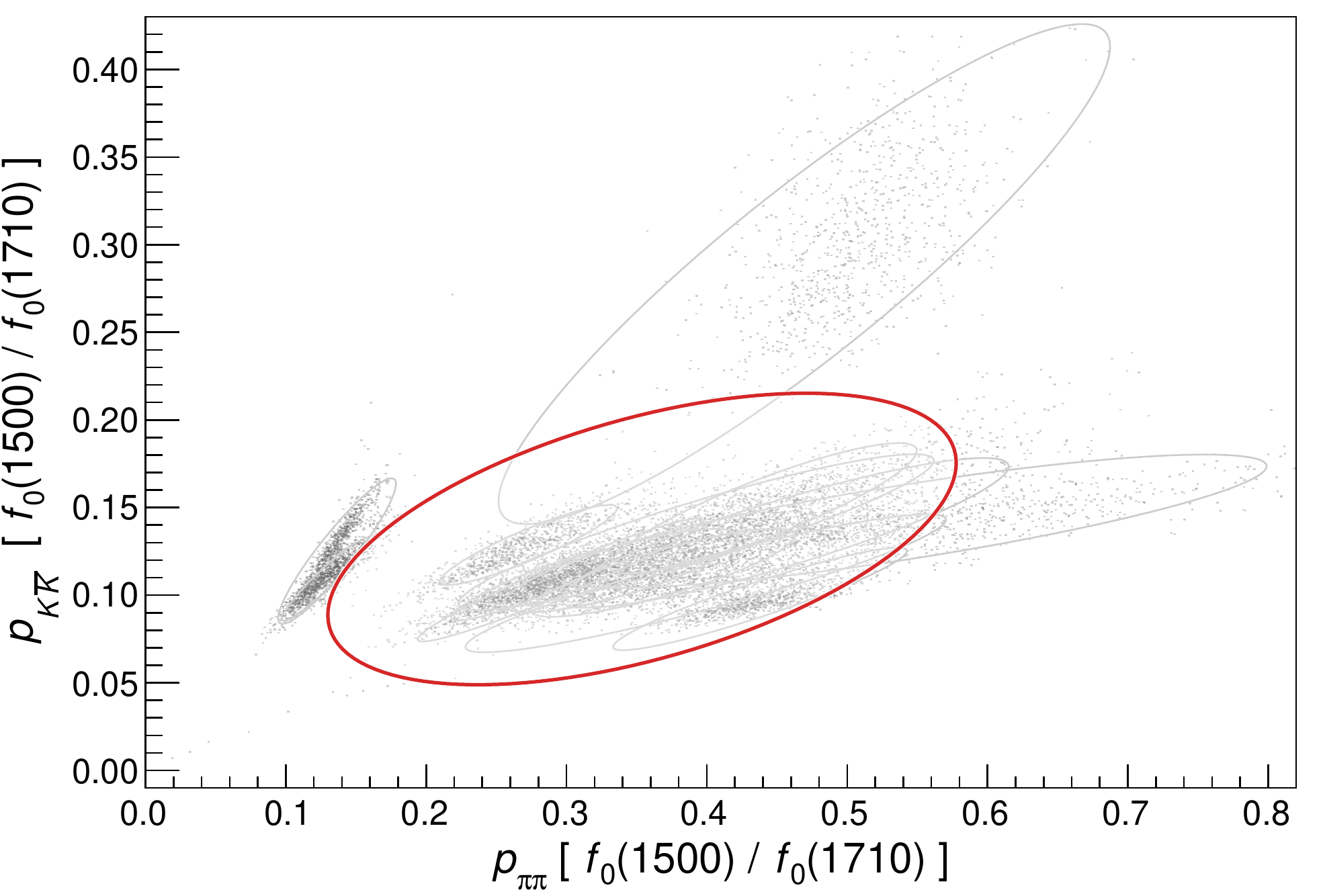} 
\caption{Left panel: ratio of absolute values of the scattering residues  of $\pi\pi$ and $K\bar K$ final states, for the $f_2(1270)$ against the $f_2'(1525)$. Right panel: ratio of absolute values of decay residues of $f_0(1500)$ and $f_0(1710)$, for $\pi\pi$ against $K\bar K$ final state.A point is drawn for each pole found in each of the $O(10^4)$ pseudodatasets  generated by the bootstrap analysis. Gray ellipses show the $68\%$ confidence region of each systematic. The colored ellipses represents the $68\%$ confidence region of all the systematics at once. Ratios are nongaussian positive-defined quantities, and the results of each systematics scarcely overlap, so this ellipse cannot be taken literally, but nevertheless provides a crude idea of average values, errors and correlations. 
}
\label{fig:residues}
\end{figure}

When comparing the $f_0(1500)$ and $f_0(1710)$ couplings of the full $\jpsi \to \gamma f_0 \to \gamma \h \bar \h$ process, we find that the heavier one couples more strongly to both final states. In particular the coupling of the $f_0(1710)$ 
 to $K \bar K$ is roughly eight times larger than that of the $f_0(1500)$  and roughly three times larger in $\pi\pi$, as can be seen in Fig.~\ref{fig:residues}. It is worth noting that the values of the residues change substantially  under amplitude variations,
 which makes us cautious about strong claims regarding a precise determination of these ratios. However, all determinations agree qualitatively: the heavier resonance is stronger in $\jpsi$ radiative decays, and in particular in the $K \bar K$ channel. As we mentioned in the \nameref{sec:intro}, these arguments favor the interpretation for the $f_0(1710)$ to have a sizeable glueball component.

The final set of poles that can be identified as physical ones is shown in Fig.~\ref{fig:finalpoles}, and the mean values and uncertainties are listed in Table~\ref{tab:polesfinal}. It is worth noting that our poles are compatible with the ones on the BESIII $J/\psi \to \gamma \eta\eta$ decay~\cite{BESIII:2013qqz}, even if we do not include this channel. This supports our choice of including the most relevant high-statistics $\pi \pi$ and $K \bar K$ channels only.

%====================================================================
\section{Summary}
\label{sec:summary}
We presented a detailed analysis of the isoscalar-scalar and -tensor resonances in the $1$-$2.5\gev$ mass region. We study the BESIII mass-independent partial waves from $\jpsi \to \gamma \pi^0\pi^0$ and $\to \gamma K_S^0 K_S^0$ radiative decays~\cite{Ablikim:2015umt,Ablikim:2018izx}. 
Data were published in two equivalent solutions in the full kinematic range. However, the region below the $K\bar K$ threshold is not compatible with Watson's theorem expectation, which made us select one of the two solutions, and to restrict to the $1$-$2.5\gev$ mass region.
To assess the model dependence realistically, we explored a large number of amplitude parametrizations that respect the $S$-matrix principles as much as possible, and discuss the results for 14 of them. We first enforce unitarity strictly on the two channels considered, which turns out to be too rigid to describe data, in particular between the resonant peaks. We then extend our models to include a third unconstrained $\rho\rho$ channel, which is known to contribute substantially to the resonances in this region. Fit quality is excellent for all the parametrizations studied. Despite the large systematic uncertainties, we can identify four scalar and three tensor states. 

The four lightest resonances are determined with great accuracy, which allows us to study their couplings. We find that the $f_2(1270)$ and $f_2'(1525)$ couple largely to $\pi\pi$ and $K\bar K$, respectively, as expected by their quark model assignments. The couplings ratios are compatible with the branching fractions reported in the PDG. In the scalar sector, it seems that the $f_0(1710)$ appears in $\jpsi \to \gamma f_0$ more strongly than the $f_0(1500)$. This affinity of the $f_0(1710)$ to the gluon-rich initial state, together with a coupling to $K\bar K$ larger by one order of magnitude, are hints for a sizeable glueball component.

\begin{acknowledgments}
We thank Daniele Binosi and Ralf-Arno Tripolt for crucial comments to the preliminary results of this work. 
This work was supported by the U.S. Department of Energy under Grants No. DE-AC05-06OR23177, under which Jefferson Science Associates, LLC, manages and operates Jefferson Lab, No.~DE-AC05-06OR23177 and No.~DE-FG02-87ER40365. 
This project has received funding from the European Union’s Horizon 2020 research and innovation programme under grant agreement No.~824093. %ECT*
AP has received funding from the European Union's Horizon 2020 research and innovation programme under the Marie Sk{\l}odowska-Curie grant agreement No.~754496. AR acknowledges the financial support of the U.S. Department of Energy contract DE-SC0018416 at the College of William \& Mary. 
CFR acknowledges the financial support of 
PAPIIT-DGAPA (UNAM, Mexico) Grant No.~IN106921 and
CONACYT (Mexico) Grant No.~A1-S-21389.
VM is a Serra Húnter fellow and acknowledges support from the Spanish national Grant No. PID2019–106080 GB-C21 and PID2020-118758GB-I00. %Vincent
MA is supported by Generalitat Valenciana Grant No. CIDEGENT/2020/002,
and by the Spanish Ministerio de Econom\'ia y Competitividad, Ministerio de Ciencia e Innovaci\'on under Grants No. PID2019-105439G-C22, No. PID2020-112777GB-I00 (Ref. 10.13039/501100011033). %Miguel

\end{acknowledgments}

%%%%%%%%%%%%%%%%%%%%%%%%%%%%%%%%%%%%%%%%%%%%%%%%%%%%%%%%%%%%%%%%%%%%%
\appendix

%====================================================================
%============================================
% Ambiguities
%============================================
\section{\boldmath $J/\psi\to \gamma\pi \pi$ Ambiguities}
\label{app:ambi}

\begin{figure}[t]
{\centering
\includegraphics[width=0.4\textwidth]{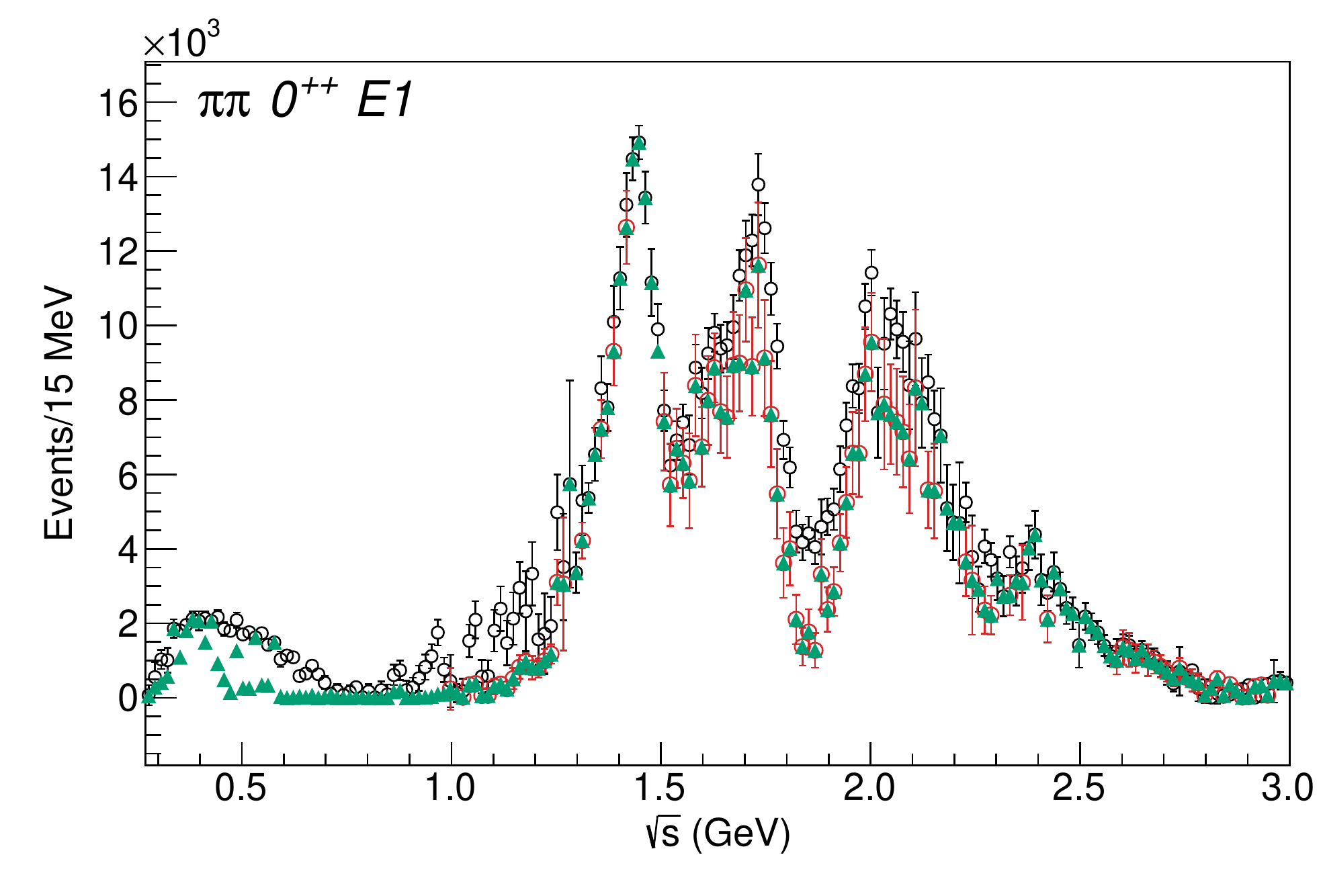} \includegraphics[width=0.4\textwidth]{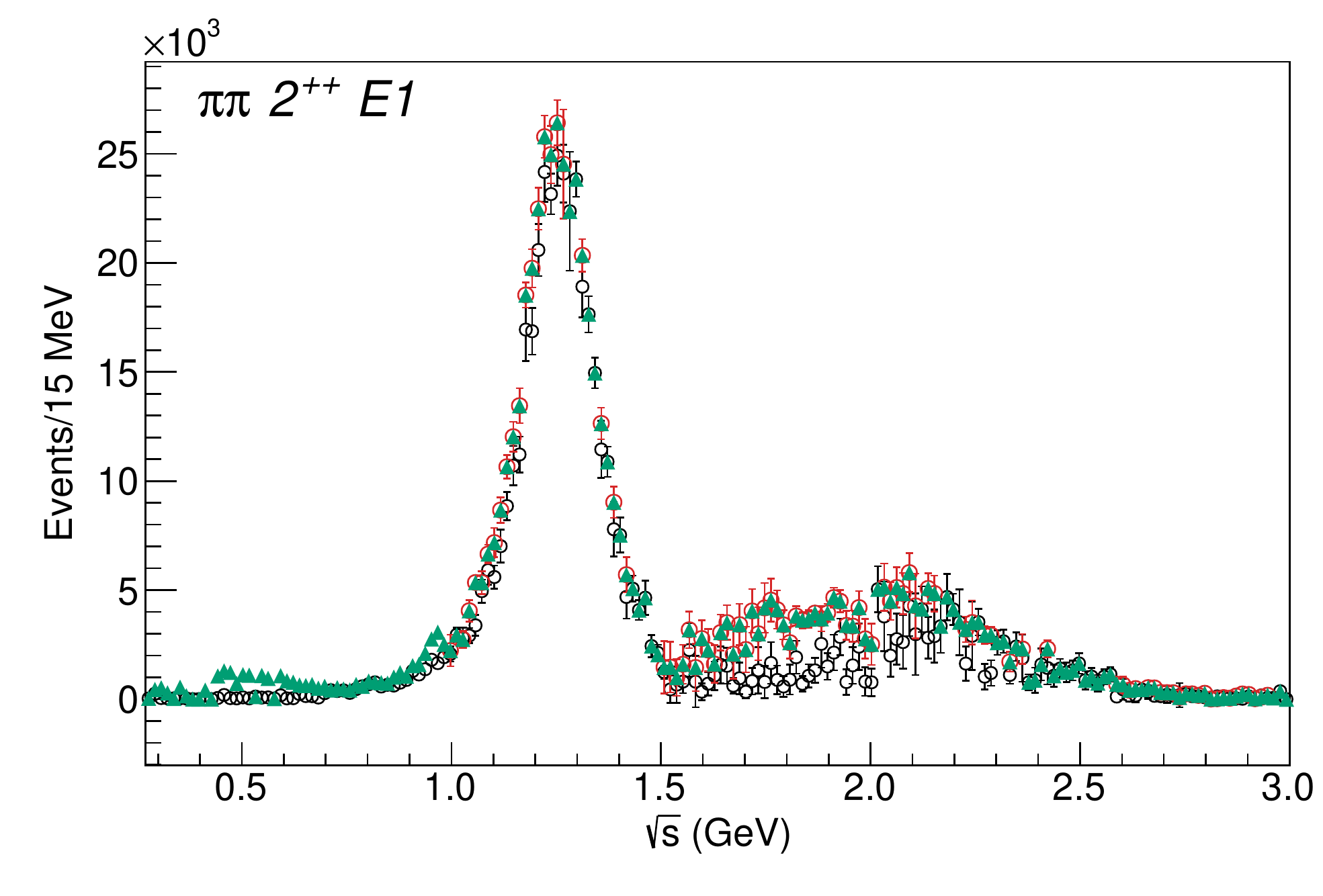} \\
\includegraphics[width=0.4\textwidth]{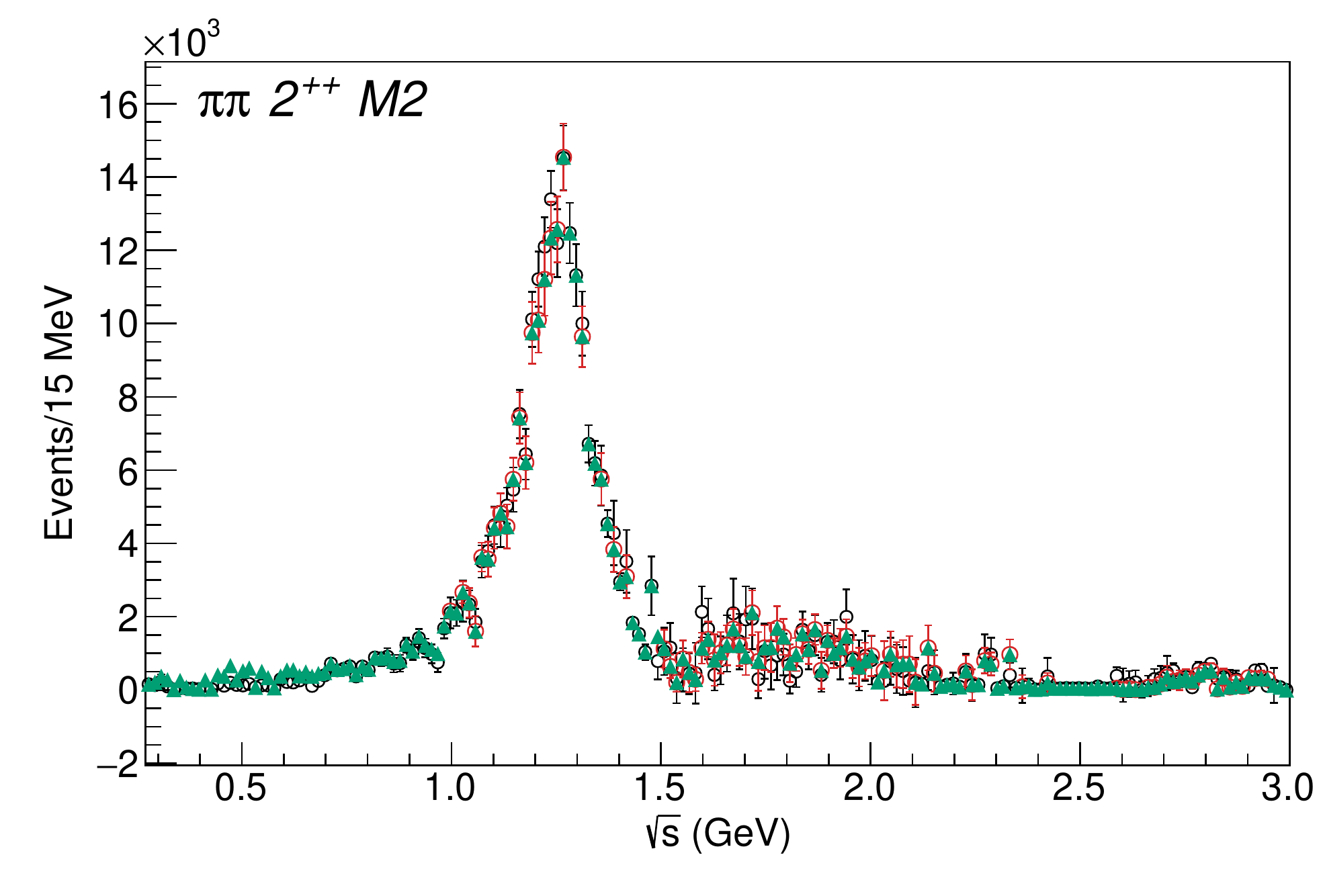} 
\includegraphics[width=0.4\textwidth]{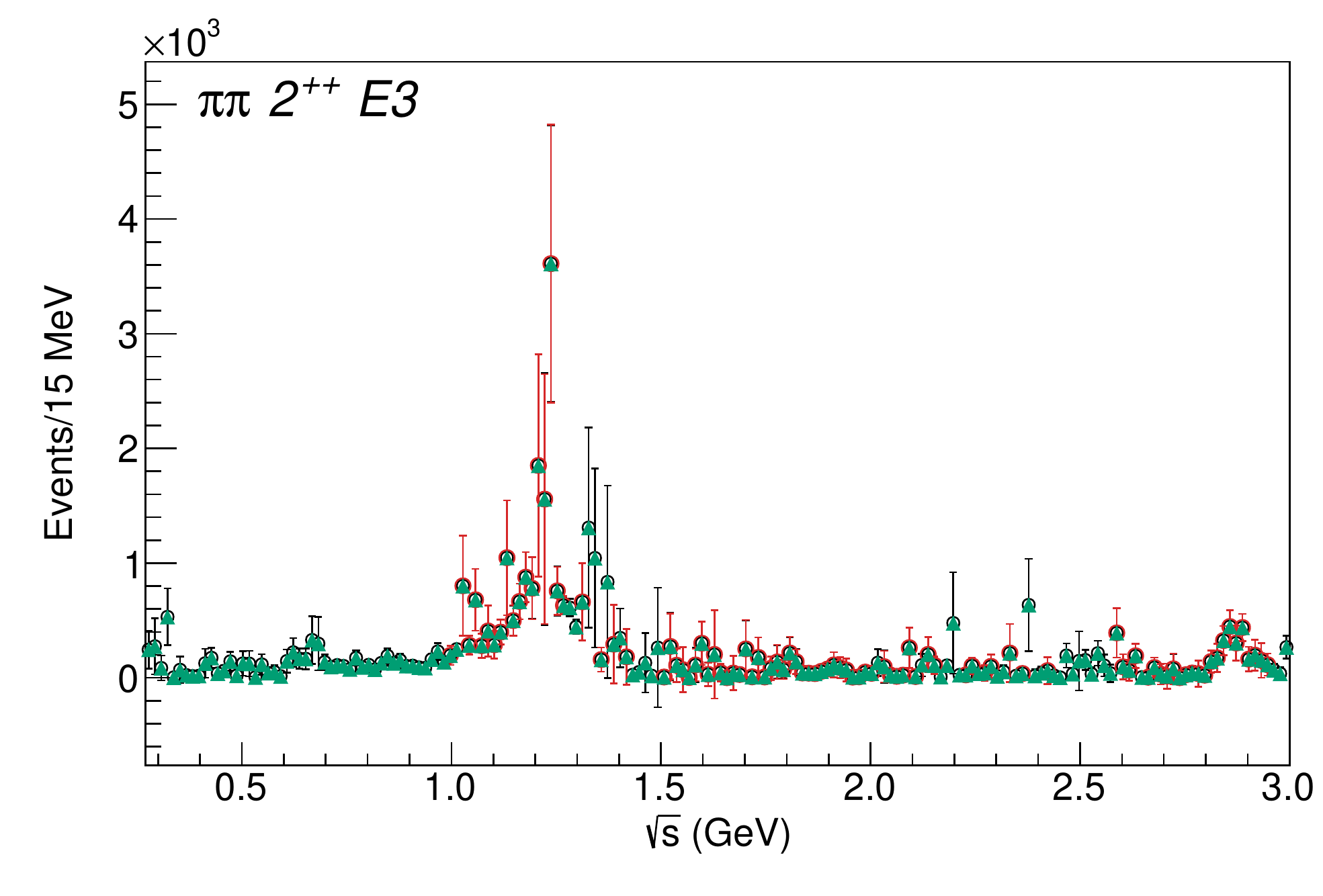} 
}
\caption{Comparison between the nominal (black) and ambiguous (red) solutions for the intensities extracted in~\cite{Ablikim:2015umt}. The prediction for the latter is shown in green using the relations derived in the experimental paper, and extended below the $K \bar K$ threshold.
}
\label{fig:ambi-intensities}
\end{figure}

\begin{figure}[t]
{\centering
\includegraphics[width=0.4\textwidth]{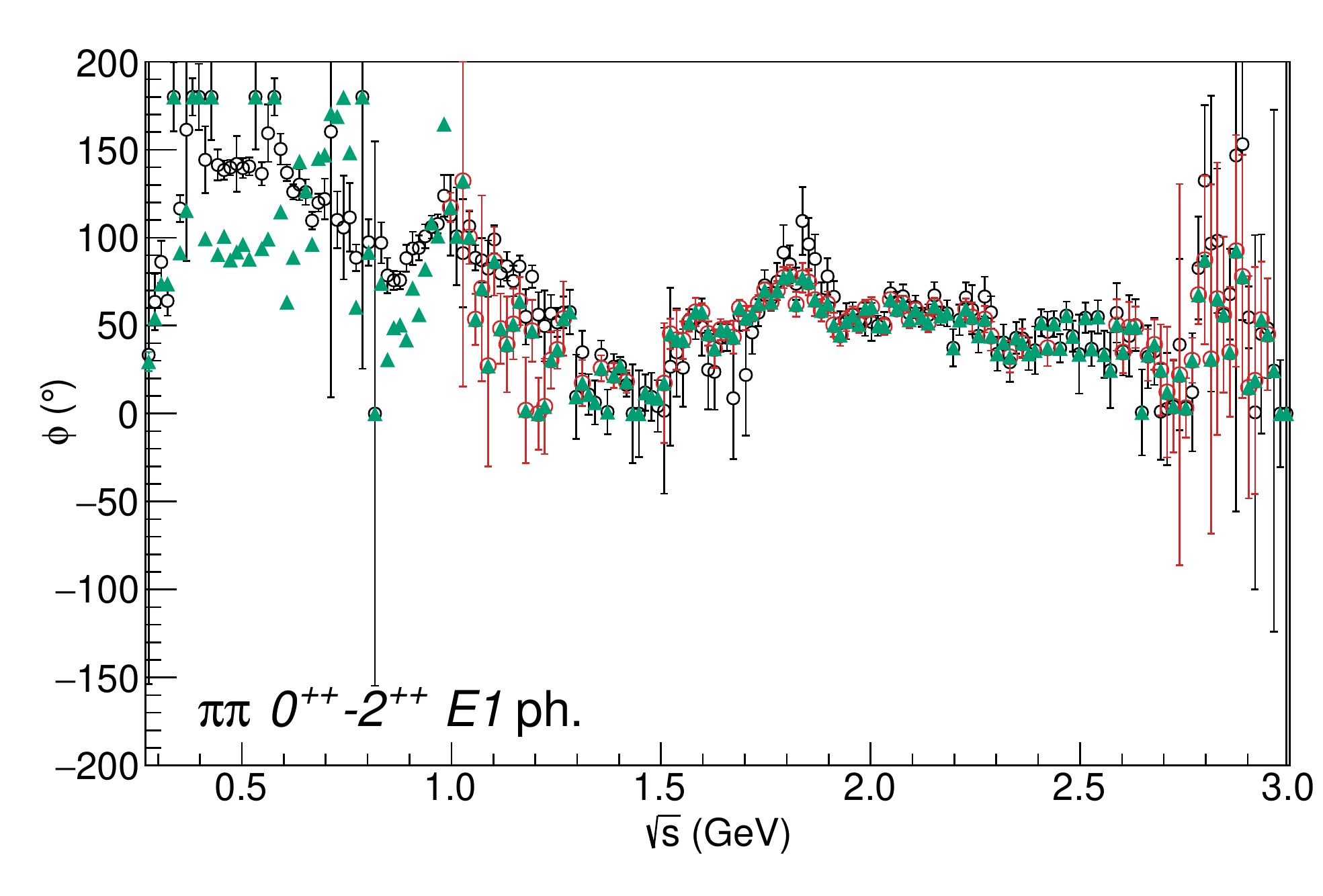}  \\
\includegraphics[width=0.4\textwidth]{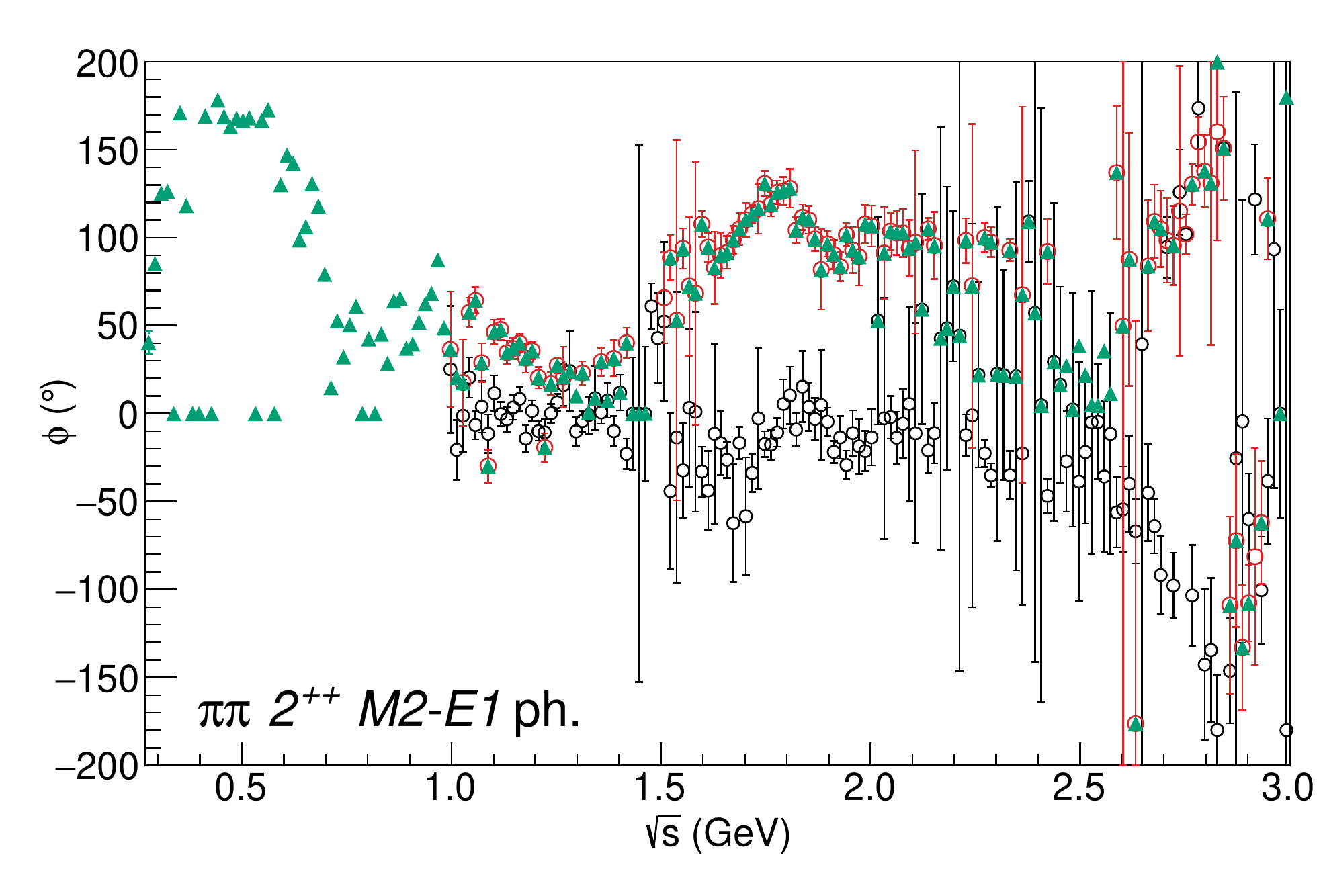} 
\includegraphics[width=0.4\textwidth]{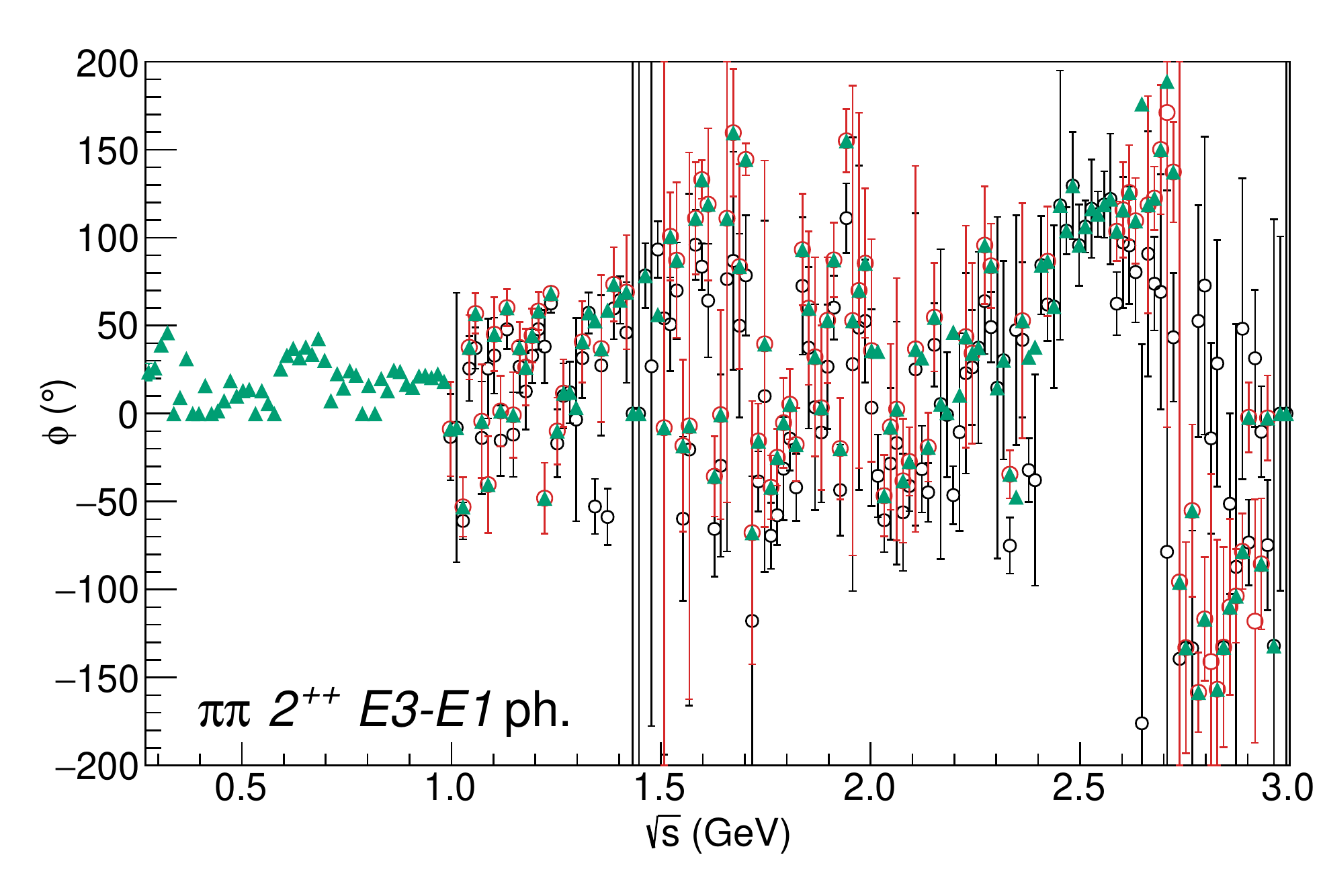} 
}
\caption{Comparison between the nominal (black) and ambiguous (red) solutions for the relative phases extracted in~\cite{Ablikim:2015umt}. The prediction for the latter is shown in green using the relations derived in the experimental paper, and extended below the $K \bar K$ threshold.}
\label{fig:ambi-phases}
\end{figure}
As mentioned in Section~\ref{sec:data}, partial wave extractions suffer from ambiguities. 
Specifically, the $J/\psi\to \gamma \h \bar \h$ radiative decays truncated to the $2^{++}$ multipoles, can have two different solutions, related mathematically~\cite{Ablikim:2015umt,Ablikim:2018izx}: in a given energy bin, one can calculate the intensities and relative phases of the four multipoles of one solution from the intensities and relative phases of the four multipoles of the other solution. Below the $K \bar K$ threshold, the experimental papers do not show the ambiguous solution: Watson's theorem is invoked in order to discard one of them. However, as we showed in Section~\ref{sec:data}, Watson's theorem also implies the $0^{++}\,E1$ phase to match the $S$-wave elastic $\pi\pi$ scattering shift, which is not the case. Based on this, and on the fact that the ambiguous solutions in $\pi\pi$ and $K\bar K$ shows some unexpected behaviour in the phases, we decided to focus on the nominal solution, and discard the region below 1\gev.

Nevertheless, we tried to see whether there is a way to make use of these data in the region where the much studied $\sig$ and the $f_0(980)$ appear.
Since the existence of ambiguities is a mathematical fact that does not  depend on unitarity arguments like Watson's theorem, we can calculate the ambiguous solution of $\jpsi \to \gamma \pi^0\pi^0$ below $K\bar K$ threshold and check whether it agrees better with $\pi\pi$ scattering. The exercise is shown in Figs.~\ref{fig:ambi-intensities} and~\ref{fig:ambi-phases}. Since the relative phase of the three $2^{++}$ multipoles is set to zero below the $K \bar K$ threshold, this turns into an underestimation of the errors of the ambiguous solution, that looks very scattered (in particular for the phases) and unusable.

\begin{figure}[ht]
{\centering
\includegraphics[width=0.5\textwidth]{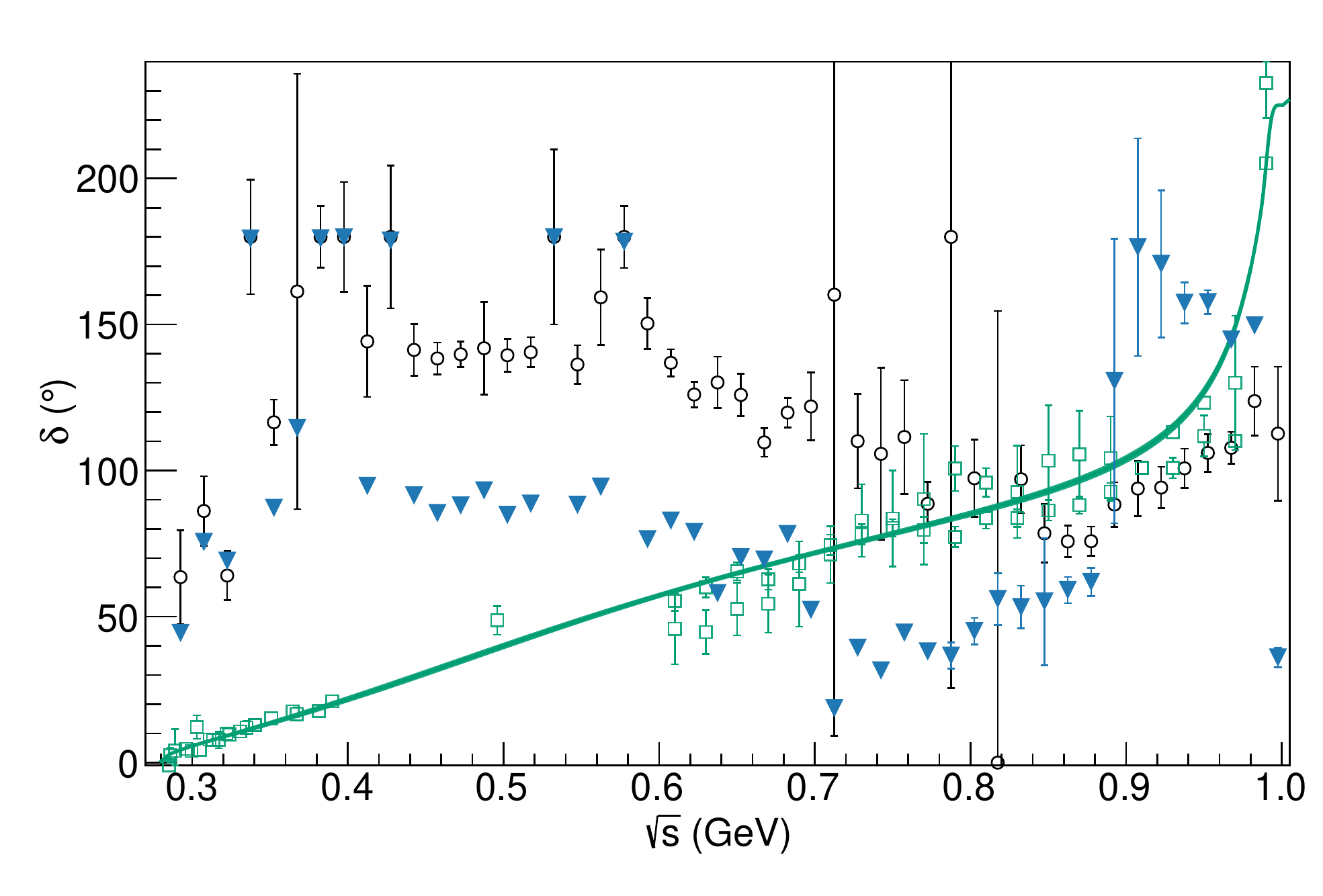}
}
\caption{Comparison between the nominal BESIII data, the elastic $\pi \pi$ phase shift from~\cite{Pelaez:2019eqa} (solid green band) and the predicted ambiguous partner of the latter.}
\label{fig:ambi-phases_new}
\end{figure}

We even tried to proceed in the opposite direction: replacing the measured $0^{++}\,E1$ phase with the known $S$-wave elastic $\pi\pi$ scattering one, we can calculate what would its  ambiguous counterpart be. The result is shown in Fig.~\ref{fig:ambi-phases_new}. This looks closer to the BESIII phase, although with some differences, most notably the sharp rise at $\sim 900\mev$.

%============================================
% Gamma dist appendix
%============================================
\section{\boldmath Bootstrap and the $\Gamma$ distribution}
\label{app:gamma}

Bootstrapping has become in the recent past a promising method to assess uncertainties in spectroscopy analyses~\cite{Landay:2016cjw,Pilloni:2016obd,JPAC:2017dbi,JPAC:2018zyd,Molina:2020qpw,Niehus:2020gmf,JPAC:2020umo,Bibrzycki:2021rwh}. In particular it allows one to map the likelihood for a given minimum, which is not accessible through simple error propagation in non-linear problems, or when the number of parameters is very large. Furthermore, this technique can also help us distinguishing between stable ``physical'' poles and spurious ones~\cite{JPAC:2018zyd,Fernandez-Ramirez:2019koa}, whereas simple error propagation would fail to describe in a robust way those uncertainties.

One usually assumes data points to be normally distributed. However, the intensities extracted in~\cite{Ablikim:2015umt,Ablikim:2018izx} are positive defined, and since they are not simple event counts, they are not even Poisson distributed. There are several data points compatible with zero, or even negative values, within $1\sigma$, which is not physical, and no sensible parametrization can reproduce. In this sense using a simple normal distribution to resample the data would produce artifacts in our uncertainties, which would then propagate into the pole errors.

For all intensity data points which are compatible with zero within $2 \sigma$, we assume they follow a $\Gamma$ distribution, having by mean and variance the central value and the error squared. This was used in previous spectroscopy analyses~\cite{Blin:2016dlf}. The distribution is given by
\begin{equation}
H\left(x \mid \mu, \sigma\right)=\theta(x)\,\left(\frac{x \mu}{\sigma^{2}}\right)^{\frac{\mu^{2}}{\sigma^{2}}} \frac{\exp \left(-x \mu / \sigma^{2}\right)}{x\, \Gamma\!\left(\mu^{2} / \sigma^{2}\right)}\,.
\end{equation}
This distribution is positive defined and has light tails as the gaussian, which makes it a good candidate for our purposes. Its mean and variance are $\mu$ and $\sigma^2$.

\bibliographystyle{apsrev4-2.bst}

\bibliography{quattro}

\newpage

%====================================================================
\clearpage
\section{Supplemental material}
\label{sup:supp-material}

\subsection{$\left[K^J(s)^{-1}\right]^\text{CDD} \quad\Big/\quad \omega(s)_\text{pole} \quad\Big/\quad \rho N^J_{ki}(s')_\text{Q-model} \quad\Big/\quad s_L = 0$}
\label{subsec:inputcddcm11newc3_bootstrap-out}

\begin{table}[h]
\begin{ruledtabular}
\begin{tabular}{c c c c c c}
\multicolumn{3}{c}{$\pi \pi$ channel}  & \multicolumn{3}{c}{$ K  \bar K$ channel} \\ \hline
\rule[-0.2cm]{-0.1cm}{.55cm} $a^{S,\pi\pi}_0$ &$-40.312$ & $-40.982 \pm 5.291$ & $a^{S,K\bar K}_0$ & $-7.767$ & $-8.598 \pm 9.618$ \\
\rule[-0.2cm]{-0.1cm}{.55cm} $a^{S,\pi\pi}_1$ &$156.891$ & $151.158 \pm 23.746$ & $a^{S,K\bar K}_1$ & $428.448$ & $412.786 \pm 38.076$ \\
\rule[-0.2cm]{-0.1cm}{.55cm} $a^{S,\pi\pi}_2$ &$220.025$ & $219.970 \pm 26.765$ & $a^{S,K\bar K}_2$ & $489.943$ & $484.947 \pm 39.625$ \\
\rule[-0.2cm]{-0.1cm}{.55cm} $a^{S,\pi\pi}_3$ &$99.516$ & $97.033 \pm 12.939$ & $a^{S,K\bar K}_3$ & $205.654$ & $198.653 \pm 20.792$ \\
\hline
\rule[-0.2cm]{-0.1cm}{.55cm} $a^{D,\pi\pi}_0$ &$2.724$ & $2.657 \pm 0.328$ & $a^{D,K\bar K}_0$ & $1.973$ & $2.006 \pm 0.368$ \\
\rule[-0.2cm]{-0.1cm}{.55cm} $a^{D,\pi\pi}_1$ &$-0.052$ & $-0.023 \pm 0.530$ & $a^{D,K\bar K}_1$ & $5.417$ & $5.247 \pm 0.611$ \\
\rule[-0.2cm]{-0.1cm}{.55cm} $a^{D,\pi\pi}_2$ &$1.205$ & $1.109 \pm 0.357$ & $a^{D,K\bar K}_2$ & $0.625$ & $0.684 \pm 0.385$ \\
\rule[-0.2cm]{-0.1cm}{.55cm} $a^{D,\pi\pi}_3$ &$0.616$ & $0.623 \pm 0.176$ & $a^{D,K\bar K}_3$ & $-0.649$ & $-0.718 \pm 0.150$ \\
\end{tabular}
\end{ruledtabular}
\end{table}

\begin{table}[h]
\begin{ruledtabular}
\begin{tabular}{c c c c c c c c c}
\multicolumn{3}{c}{$S$-wave couplings $\pi \pi$ channel}  & \multicolumn{3}{c}{$S$-wave couplings $K  \bar K$ channel} & \multicolumn{3}{c}{$S$-wave couplings $\rho\rho$ channel} \\ \hline
\rule[-0.2cm]{-0.1cm}{.55cm} $g^{S,1}_{\pi\pi}$ & $-0.061$ & $-0.059 \pm 0.072$ & $g^{S,1}_{K\bar K}$ & $0.776$ & $0.793 \pm 0.224$ & $g^{S,1}_{\rho\rho}$ & $-0.667$ & $-0.653 \pm 0.092$ \\
\rule[-0.2cm]{-0.1cm}{.55cm} $g^{S,2}_{\pi\pi}$ & $0.607$ & $0.621 \pm 0.059$ & $g^{S,2}_{K\bar K}$ & $0.952$ & $0.952 \pm 0.082$ & $g^{S,2}_{\rho\rho}$ & $0.417$ & $0.417 \pm 0.015$ \\
\rule[-0.2cm]{-0.1cm}{.55cm} $g^{S,3}_{\pi\pi}$ & $4.410$ & $4.435 \pm 0.228$ & $g^{S,3}_{K\bar K}$ & $8.443$ & $8.444 \pm 0.505$ & $g^{S,3}_{\rho\rho}$ & $-0.401$ & $-0.401 \pm 0.037$ \\
\hline 
\multicolumn{3}{c}{$D$-wave couplings $\pi \pi$ channel}  & \multicolumn{3}{c}{$D$-wave couplings $K  \bar K$ channel} & \multicolumn{3}{c}{$D$-wave couplings $\rho\rho$ channel} \\ \hline
\rule[-0.2cm]{-0.1cm}{.55cm} $g^{D,1}_{\pi\pi}$ & $3.275$ & $3.327 \pm 0.579$ & $g^{D,1}_{K\bar K}$ & $1.413$ & $1.378 \pm 0.223$ & $g^{D,1}_{\rho\rho}$ & $0.283$ & $0.215 \pm 0.439$ \\
\rule[-0.2cm]{-0.1cm}{.55cm} $g^{D,2}_{\pi\pi}$ & $-0.614$ & $-0.817 \pm 0.732$ & $g^{D,2}_{K\bar K}$ & $4.596$ & $4.665 \pm 0.446$ & $g^{D,2}_{\rho\rho}$ & $1.113$ & $1.250 \pm 0.515$ \\
\rule[-0.2cm]{-0.1cm}{.55cm} $g^{D,3}_{\pi\pi}$ & $-57.319$ & $-57.530 \pm 3.561$ & $g^{D,3}_{K\bar K}$ & $52.174$ & $52.687 \pm 4.079$ & $g^{D,3}_{\rho\rho}$ & $14.094$ & $14.268 \pm 2.262$ \\
\end{tabular}
\vspace{2em}
\begin{tabular}{c c c c c c}
\multicolumn{3}{c}{$S$-wave background}  & \multicolumn{3}{c}{$D$-wave background} \\ \hline
\rule[-0.2cm]{-0.1cm}{.55cm} $c^{S,0}_{\pi\pi,\pi\pi}$ &$2.512$ & $2.527 \pm 0.120$ & $c^{D,0}_{\pi\pi,\pi\pi}$ &$-290.233$ & $-294.927 \pm 29.460$ \\
\rule[-0.2cm]{-0.1cm}{.55cm} $c^{S,0}_{\pi\pi,K\bar K}$ &$3.380$ & $3.380 \pm 0.126$ & $c^{D,0}_{\pi\pi,K\bar K}$ &$282.875$ & $287.534 \pm 20.077$ \\
\rule[-0.2cm]{-0.1cm}{.55cm} $c^{S,0}_{\pi\pi,\rho\rho}$ &$0$ & $0$ & $c^{D,0}_{\pi\pi,\rho\rho}$ &$95.463$ & $97.541 \pm 16.022$ \\
\rule[-0.2cm]{-0.1cm}{.55cm} $c^{S,0}_{K\bar K,K\bar K}$ &$4.752$ & $4.725 \pm 0.159$ & $c^{D,0}_{K\bar K,K\bar K}$ &$-228.333$ & $-234.485 \pm 30.352$ \\
\rule[-0.2cm]{-0.1cm}{.55cm} $c^{S,0}_{K\bar K,\rho\rho}$ &$0$ & $0$ & $c^{D,0}_{K\bar K,\rho\rho}$ &$-86.069$ & $-88.340 \pm 13.787$ \\
\rule[-0.2cm]{-0.1cm}{.55cm} $c^{S,0}_{\rho\rho,\rho\rho}$ &$0.040$ & $0.007 \pm 0.102$ & $c^{D,0}_{\rho\rho,\rho\rho}$ &$-25.988$ & $-27.507 \pm 10.715$ \\
\rule[-0.2cm]{-0.1cm}{.55cm} $c^{S,1}_{\pi\pi,\pi\pi}$ &$0.495$ & $0.502 \pm 0.035$ & $c^{D,1}_{\pi\pi,\pi\pi}$ &$-18.417$ & $-18.815 \pm 3.768$ \\
\rule[-0.2cm]{-0.1cm}{.55cm} $c^{S,1}_{\pi\pi,K\bar K}$ &$0.015$ & $0.020 \pm 0.066$ & $c^{D,1}_{\pi\pi,K\bar K}$ &$16.215$ & $16.762 \pm 3.339$ \\
\rule[-0.2cm]{-0.1cm}{.55cm} $c^{S,1}_{\pi\pi,\rho\rho}$ &$0$ & $0$ & $c^{D,1}_{\pi\pi,\rho\rho}$ &$0$ & $0$ \\
\rule[-0.2cm]{-0.1cm}{.55cm} $c^{S,1}_{K\bar K,K\bar K}$ &$-0.775$ & $-0.754 \pm 0.100$ & $c^{D,1}_{K\bar K,K\bar K}$ &$-20.303$ & $-21.139 \pm 3.923$ \\
\rule[-0.2cm]{-0.1cm}{.55cm} $c^{S,1}_{K\bar K,\rho\rho}$ &$0$ & $0$ & $c^{D,1}_{K\bar K,\rho\rho}$ &$0$ & $0$ \\
\rule[-0.2cm]{-0.1cm}{.55cm} $c^{S,1}_{\rho\rho,\rho\rho}$ &$0$ & $0$ & $c^{D,1}_{\rho\rho,\rho\rho}$ &$0$ & $0$ \\
\end{tabular}
\end{ruledtabular}
\end{table}

\begin{figure}[h]
\centering\includegraphics[width=0.32\textwidth]{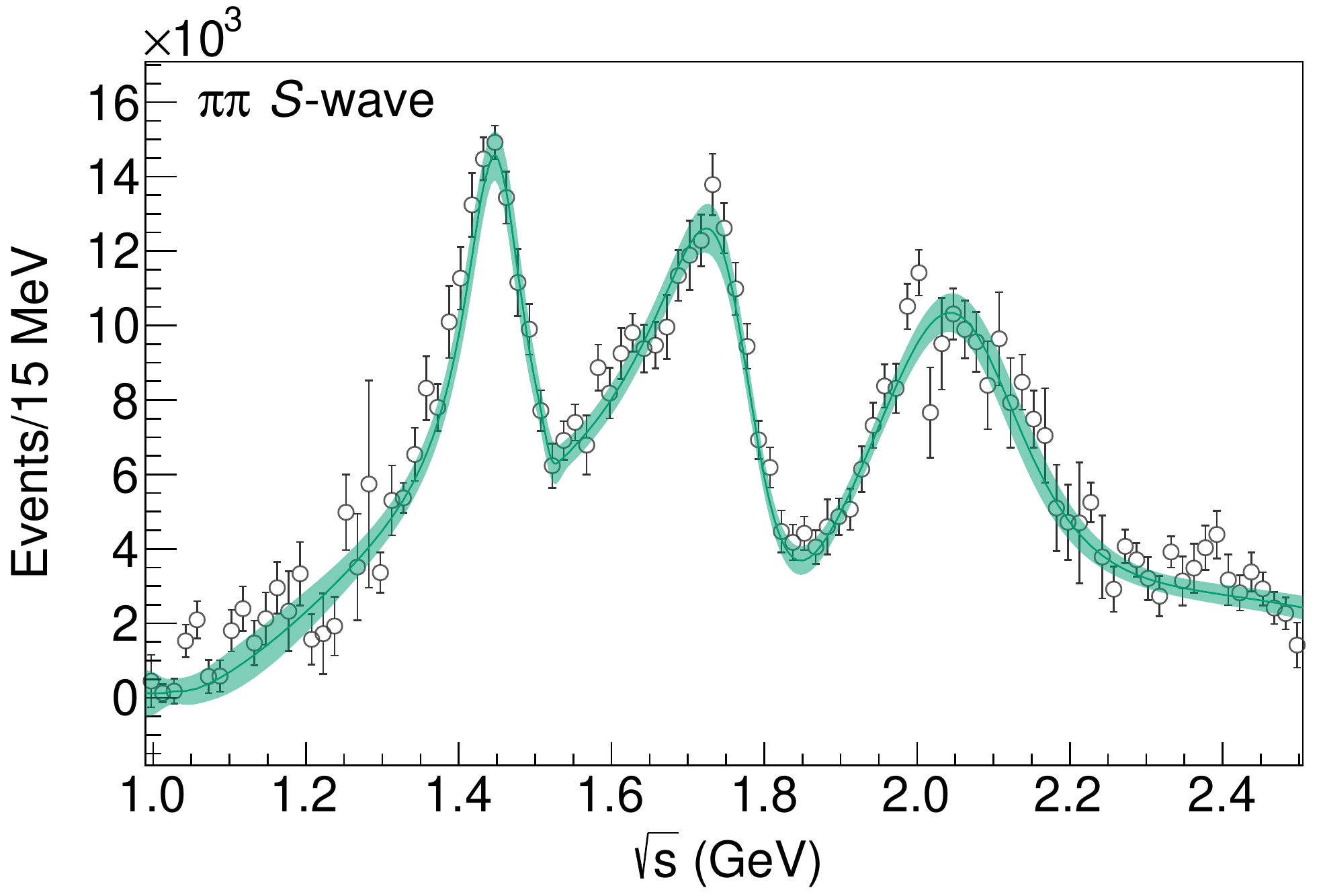} \includegraphics[width=0.32\textwidth]{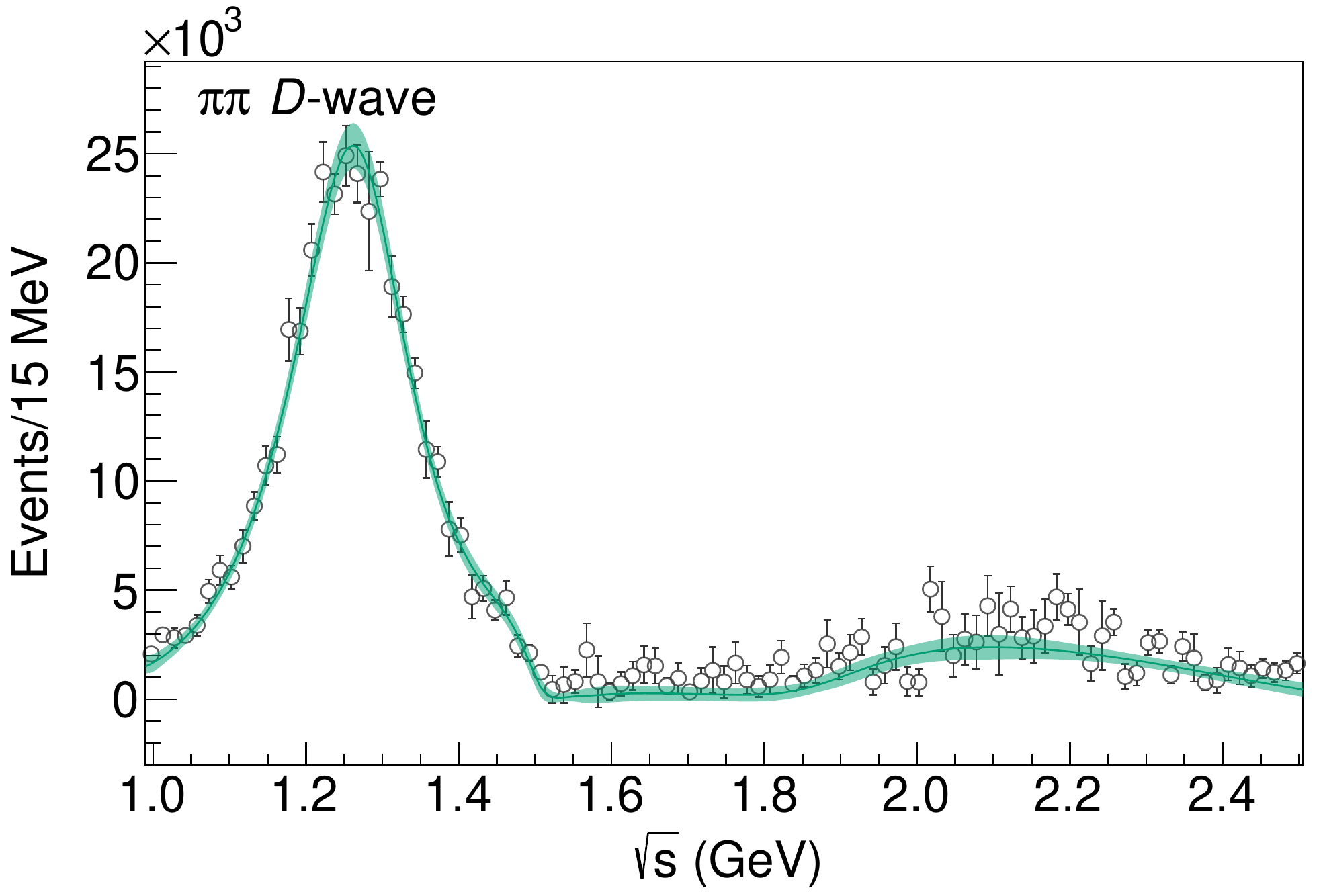} \includegraphics[width=0.32\textwidth]{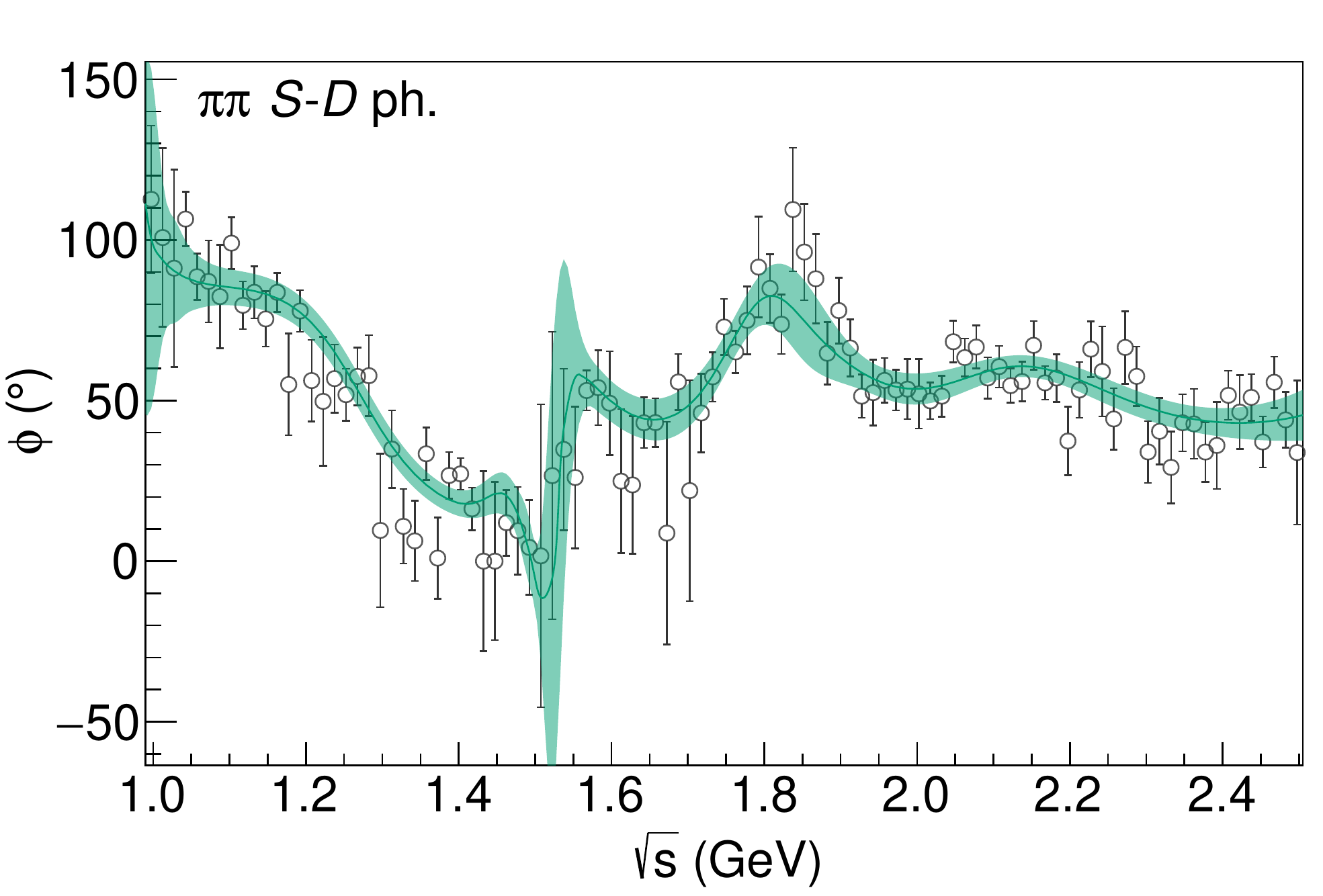}
\includegraphics[width=0.32\textwidth]{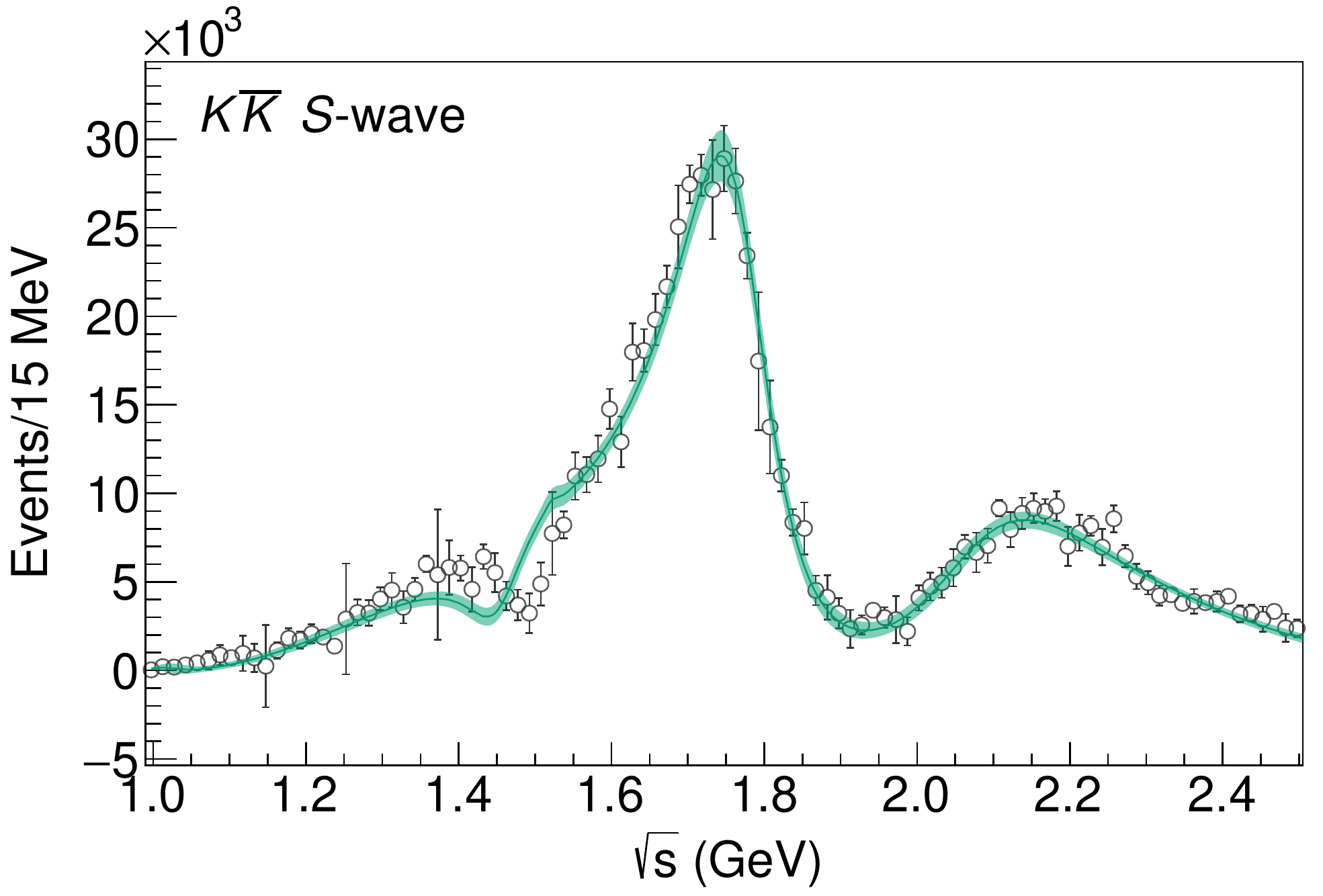} \includegraphics[width=0.32\textwidth]{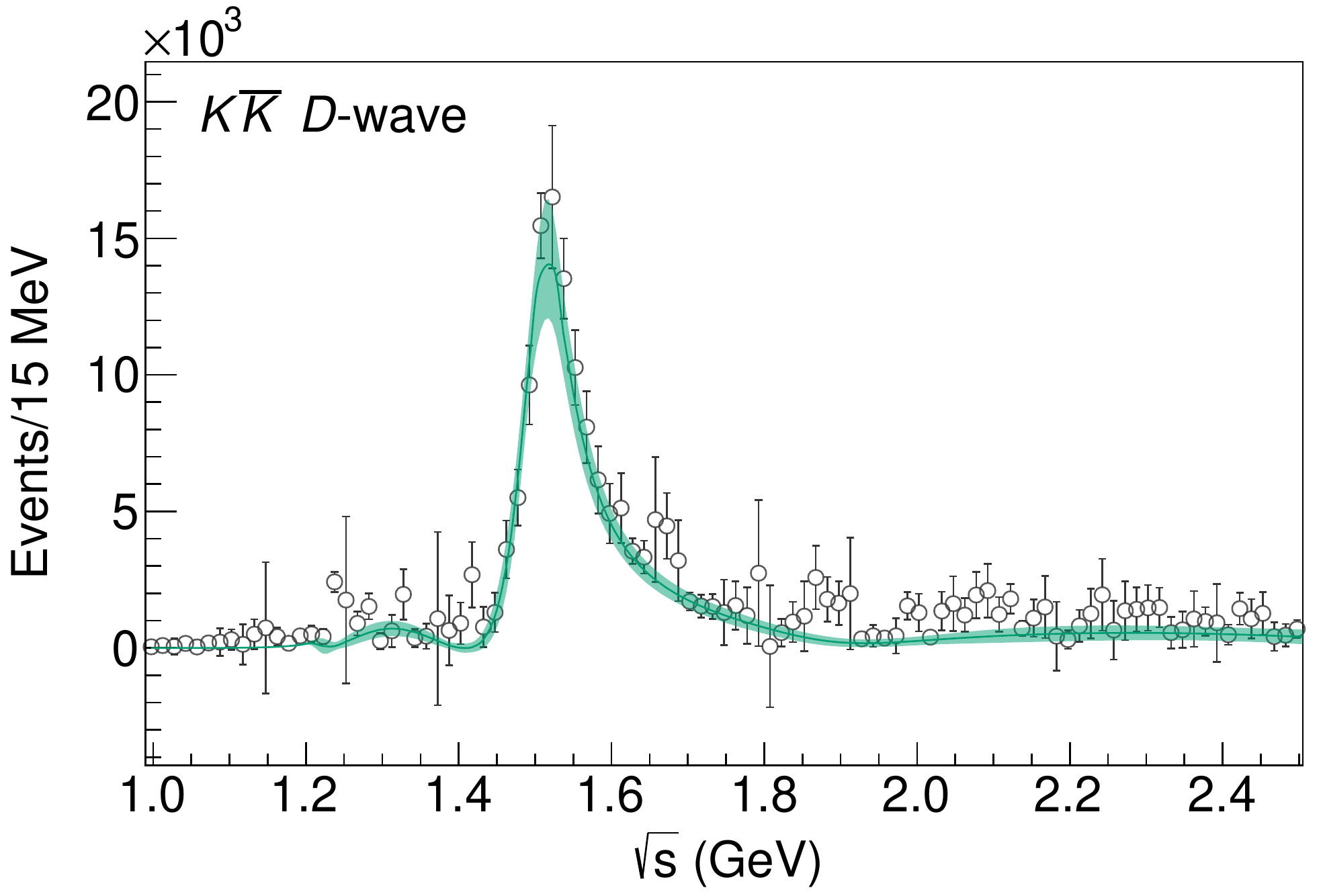} \includegraphics[width=0.32\textwidth]{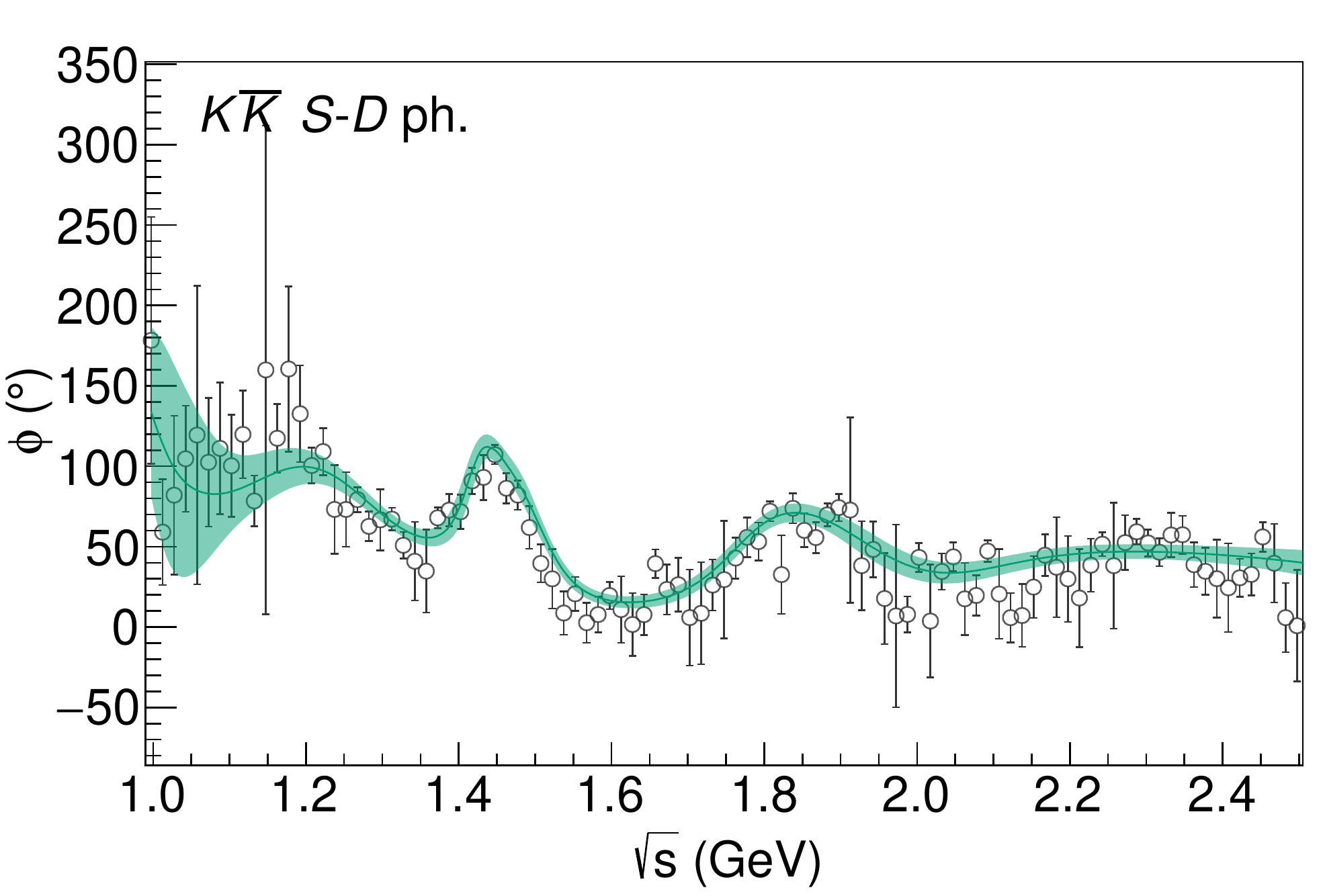}
\end{figure}

\begin{figure}[h]
\centering\includegraphics[width=0.45\textwidth]{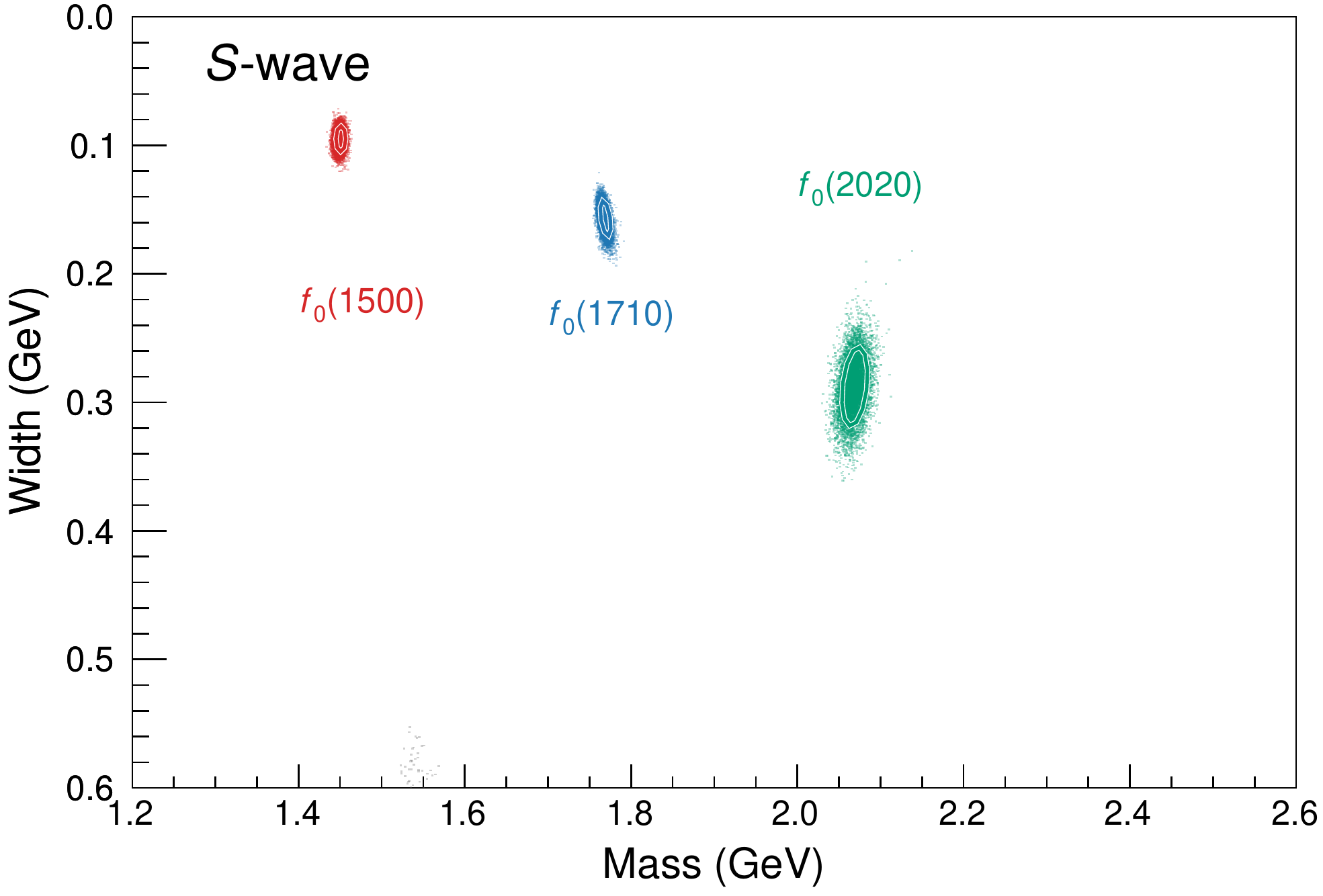} \includegraphics[width=0.45\textwidth]{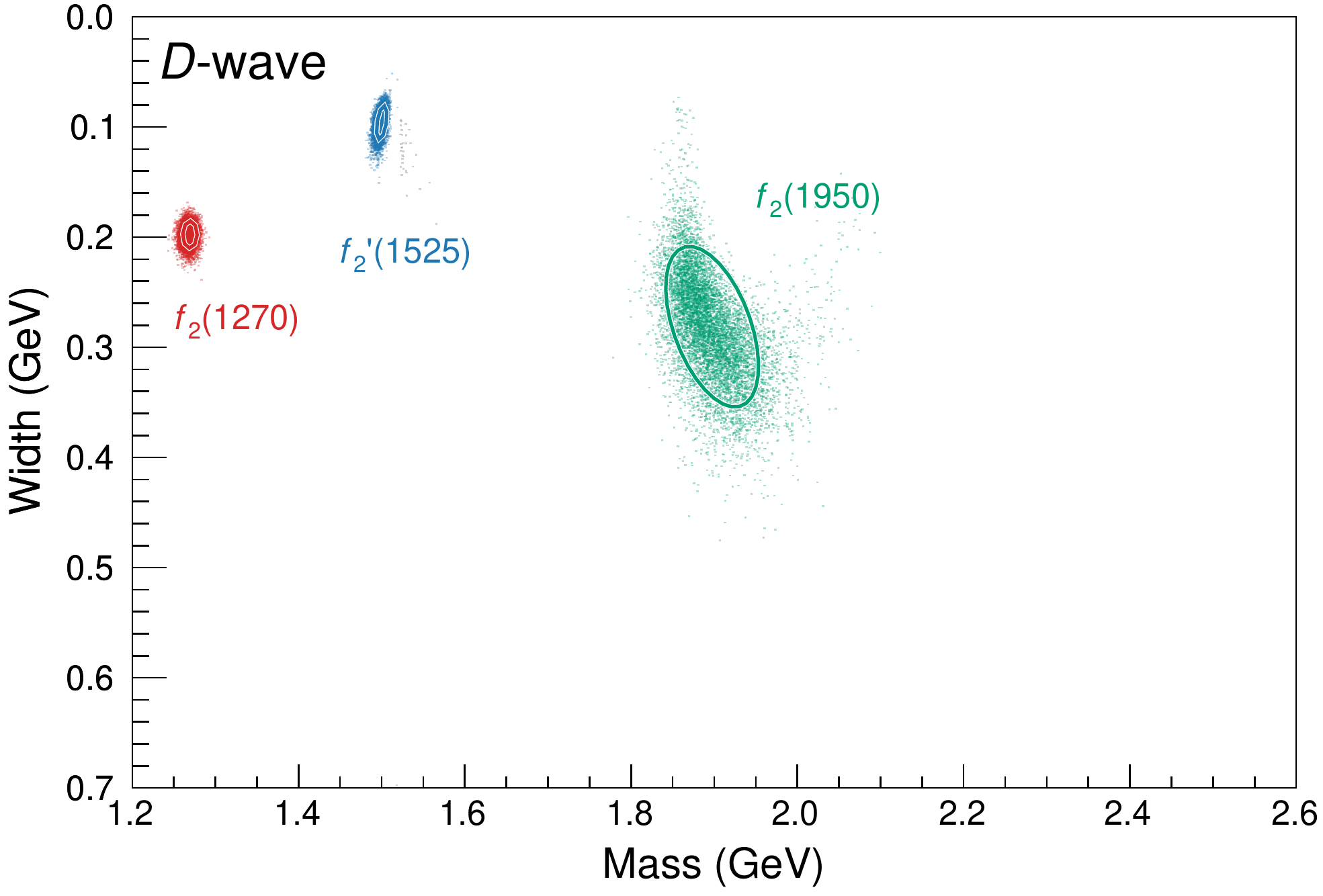}
\end{figure}

\begin{table}[h]
\begin{ruledtabular}
\begin{tabular}{c c c c}
$S$-wave  & $\sqrt{s_p}$ \mevp & $D$-wave & $\sqrt{s_p}$ \mevp \\ \hline
\rule[-0.2cm]{-0.1cm}{.55cm} $f_0(1500)$ &  $(1451 \pm 4) - i (95 \pm 6)/2$  &  $f_2(1270)$ &  $(1269 \pm 6) - i (198 \pm 8)/2$ \\
\rule[-0.2cm]{-0.1cm}{.55cm} $f_0(1710)$ &  $(1769 \pm 4) - i (157 \pm 9)/2$  &  $f_2'(1525)$ &  $(1500 \pm 4) - i (96 \pm 10)/2$ \\
\rule[-0.2cm]{-0.1cm}{.55cm} $f_0(2020)$ &  $(2069 \pm 10) - i (288 \pm 20)/2$  &  $f_2(1950)$ &  $(1897 \pm 37) - i (281 \pm 48)/2$ \\
\end{tabular}
\end{ruledtabular}
\end{table}

\clearpage

\subsection{$\left[K^J(s)^{-1}\right]^\text{CDD} \quad\Big/\quad \omega(s)_\text{scaled} \quad\Big/\quad \rho N^J_{ki}(s')_\text{Q-model} \quad\Big/\quad s_L = 0.6\gevsq$}
\label{subsec:inputcddcm3newc3_bootstrap-out}

\begin{table}[h]
\begin{ruledtabular}
\begin{tabular}{c c c c c c}
\multicolumn{3}{c}{$\pi \pi$ channel}  & \multicolumn{3}{c}{$ K  \bar K$ channel} \\ \hline
\rule[-0.2cm]{-0.1cm}{.55cm} $a^{S,\pi\pi}_0$ &$157.245$ & $154.207 \pm 34.686$ & $a^{S,K\bar K}_0$ & $150.521$ & $147.977 \pm 35.399$ \\
\rule[-0.2cm]{-0.1cm}{.55cm} $a^{S,\pi\pi}_1$ &$162.476$ & $170.138 \pm 44.823$ & $a^{S,K\bar K}_1$ & $118.193$ & $122.855 \pm 45.194$ \\
\rule[-0.2cm]{-0.1cm}{.55cm} $a^{S,\pi\pi}_2$ &$62.922$ & $66.150 \pm 23.969$ & $a^{S,K\bar K}_2$ & $54.494$ & $54.614 \pm 25.297$ \\
\rule[-0.2cm]{-0.1cm}{.55cm} $a^{S,\pi\pi}_3$ &$50.439$ & $46.958 \pm 14.090$ & $a^{S,K\bar K}_3$ & $70.733$ & $66.971 \pm 15.775$ \\
\hline
\rule[-0.2cm]{-0.1cm}{.55cm} $a^{D,\pi\pi}_0$ &$-604.559$ & $-639.498 \pm 95.786$ & $a^{D,K\bar K}_0$ & $309.966$ & $293.521 \pm 63.038$ \\
\rule[-0.2cm]{-0.1cm}{.55cm} $a^{D,\pi\pi}_1$ &$-816.649$ & $-863.980 \pm 144.188$ & $a^{D,K\bar K}_1$ & $523.716$ & $499.355 \pm 101.483$ \\
\rule[-0.2cm]{-0.1cm}{.55cm} $a^{D,\pi\pi}_2$ &$-276.267$ & $-287.779 \pm 60.369$ & $a^{D,K\bar K}_2$ & $294.291$ & $285.370 \pm 49.751$ \\
\rule[-0.2cm]{-0.1cm}{.55cm} $a^{D,\pi\pi}_3$ &$-47.735$ & $-46.296 \pm 12.500$ & $a^{D,K\bar K}_3$ & $74.811$ & $73.966 \pm 13.512$ \\
\end{tabular}
\end{ruledtabular}
\end{table}

\begin{table}[h]
\begin{ruledtabular}
\begin{tabular}{c c c c c c c c c}
\multicolumn{3}{c}{$S$-wave couplings $\pi \pi$ channel}  & \multicolumn{3}{c}{$S$-wave couplings $K  \bar K$ channel} & \multicolumn{3}{c}{$S$-wave couplings $\rho\rho$ channel} \\ \hline
\rule[-0.2cm]{-0.1cm}{.55cm} $g^{S,1}_{\pi\pi}$ & $0.651$ & $0.613 \pm 0.114$ & $g^{S,1}_{K\bar K}$ & $0.931$ & $0.961 \pm 0.087$ & $g^{S,1}_{\rho\rho}$ & $0.610$ & $0.620 \pm 0.032$ \\
\rule[-0.2cm]{-0.1cm}{.55cm} $g^{S,2}_{\pi\pi}$ & $0.223$ & $0.229 \pm 0.011$ & $g^{S,2}_{K\bar K}$ & $-0.188$ & $-0.194 \pm 0.018$ & $g^{S,2}_{\rho\rho}$ & $-0.314$ & $-0.322 \pm 0.025$ \\
\rule[-0.2cm]{-0.1cm}{.55cm} $g^{S,3}_{\pi\pi}$ & $1.219$ & $1.307 \pm 0.345$ & $g^{S,3}_{K\bar K}$ & $1.263$ & $1.329 \pm 0.355$ & $g^{S,3}_{\rho\rho}$ & $-0.292$ & $-0.271 \pm 0.049$ \\
\hline 
\multicolumn{3}{c}{$D$-wave couplings $\pi \pi$ channel}  & \multicolumn{3}{c}{$D$-wave couplings $K  \bar K$ channel} & \multicolumn{3}{c}{$D$-wave couplings $\rho\rho$ channel} \\ \hline
\rule[-0.2cm]{-0.1cm}{.55cm} $g^{D,1}_{\pi\pi}$ & $2.381$ & $2.547 \pm 0.576$ & $g^{D,1}_{K\bar K}$ & $6.949$ & $6.097 \pm 1.373$ & $g^{D,1}_{\rho\rho}$ & $61.196$ & $59.850 \pm 11.460$ \\
\rule[-0.2cm]{-0.1cm}{.55cm} $g^{D,2}_{\pi\pi}$ & $0.137$ & $0.426 \pm 0.238$ & $g^{D,2}_{K\bar K}$ & $1.463$ & $2.176 \pm 0.488$ & $g^{D,2}_{\rho\rho}$ & $-5.565$ & $-13.785 \pm 5.501$ \\
\rule[-0.2cm]{-0.1cm}{.55cm} $g^{D,3}_{\pi\pi}$ & $1.061$ & $1.047 \pm 0.029$ & $g^{D,3}_{K\bar K}$ & $0.350$ & $0.332 \pm 0.046$ & $g^{D,3}_{\rho\rho}$ & $0$ & $0$ \\
\hline 
\multicolumn{9}{c}{$S$-wave bare masses}   \\ \hline
\rule[-0.2cm]{-0.1cm}{.55cm} $m^2_{S,1}$ & $4.247$ & $4.301 \pm 0.112$ & $m^2_{S,2}$ & $2.293$ & $2.306 \pm 0.044$ & $m^2_{S,3}$ & $6.406$ & $6.455 \pm 0.130$ \\
\hline 
\multicolumn{9}{c}{$D$-wave bare masses}   \\ \hline
\rule[-0.2cm]{-0.1cm}{.55cm} $m^2_{D,1}$ & $19.512$ & $17.544 \pm 1.873$ & $m^2_{D,2}$ & $2.423$ & $3.020 \pm 0.432$ & $m^2_{D,3}$ & $1.498$ & $1.500 \pm 0.013$ \\
\end{tabular}
\vspace{2em}
\begin{tabular}{c c c c c c}
\multicolumn{3}{c}{$S$-wave background}  & \multicolumn{3}{c}{$D$-wave background} \\ \hline
\rule[-0.2cm]{-0.1cm}{.55cm} $c^{S,0}_{\pi\pi,\pi\pi}$ &$1.081$ & $0.880 \pm 0.246$ & $c^{D,0}_{\pi\pi,\pi\pi}$ &$2.643$ & $2.429 \pm 0.270$ \\
\rule[-0.2cm]{-0.1cm}{.55cm} $c^{S,0}_{\pi\pi,K\bar K}$ &$1.820$ & $1.736 \pm 0.144$ & $c^{D,0}_{\pi\pi,K\bar K}$ &$-0.723$ & $-1.001 \pm 0.466$ \\
\rule[-0.2cm]{-0.1cm}{.55cm} $c^{S,0}_{\pi\pi,\rho\rho}$ &$0$ & $0$ & $c^{D,0}_{\pi\pi,\rho\rho}$ &$0$ & $0$ \\
\rule[-0.2cm]{-0.1cm}{.55cm} $c^{S,0}_{K\bar K,K\bar K}$ &$2.063$ & $2.022 \pm 0.164$ & $c^{D,0}_{K\bar K,K\bar K}$ &$-1.287$ & $-1.177 \pm 0.972$ \\
\rule[-0.2cm]{-0.1cm}{.55cm} $c^{S,0}_{K\bar K,\rho\rho}$ &$0$ & $0$ & $c^{D,0}_{K\bar K,\rho\rho}$ &$0$ & $0$ \\
\rule[-0.2cm]{-0.1cm}{.55cm} $c^{S,0}_{\rho\rho,\rho\rho}$ &$1.185$ & $1.191 \pm 0.028$ & $c^{D,0}_{\rho\rho,\rho\rho}$ &$-820.930$ & $-920.902 \pm 173.917$ \\
\rule[-0.2cm]{-0.1cm}{.55cm} $c^{S,1}_{\pi\pi,\pi\pi}$ &$0.381$ & $0.072 \pm 0.163$ & $c^{D,1}_{\pi\pi,\pi\pi}$ &$-0.878$ & $-0.889 \pm 0.122$ \\
\rule[-0.2cm]{-0.1cm}{.55cm} $c^{S,1}_{\pi\pi,K\bar K}$ &$0.784$ & $0.726 \pm 0.055$ & $c^{D,1}_{\pi\pi,K\bar K}$ &$-0.558$ & $-0.587 \pm 0.125$ \\
\rule[-0.2cm]{-0.1cm}{.55cm} $c^{S,1}_{\pi\pi,\rho\rho}$ &$0$ & $0$ & $c^{D,1}_{\pi\pi,\rho\rho}$ &$0$ & $0$ \\
\rule[-0.2cm]{-0.1cm}{.55cm} $c^{S,1}_{K\bar K,K\bar K}$ &$-0.887$ & $-0.845 \pm 0.077$ & $c^{D,1}_{K\bar K,K\bar K}$ &$-0.629$ & $-0.710 \pm 0.242$ \\
\rule[-0.2cm]{-0.1cm}{.55cm} $c^{S,1}_{K\bar K,\rho\rho}$ &$0$ & $0$ & $c^{D,1}_{K\bar K,\rho\rho}$ &$0$ & $0$ \\
\rule[-0.2cm]{-0.1cm}{.55cm} $c^{S,1}_{\rho\rho,\rho\rho}$ &$0.534$ & $0.540 \pm 0.017$ & $c^{D,1}_{\rho\rho,\rho\rho}$ &$128.239$ & $156.655 \pm 46.467$ \\
\end{tabular}
\end{ruledtabular}
\end{table}

\begin{figure}[h]
\centering\includegraphics[width=0.32\textwidth]{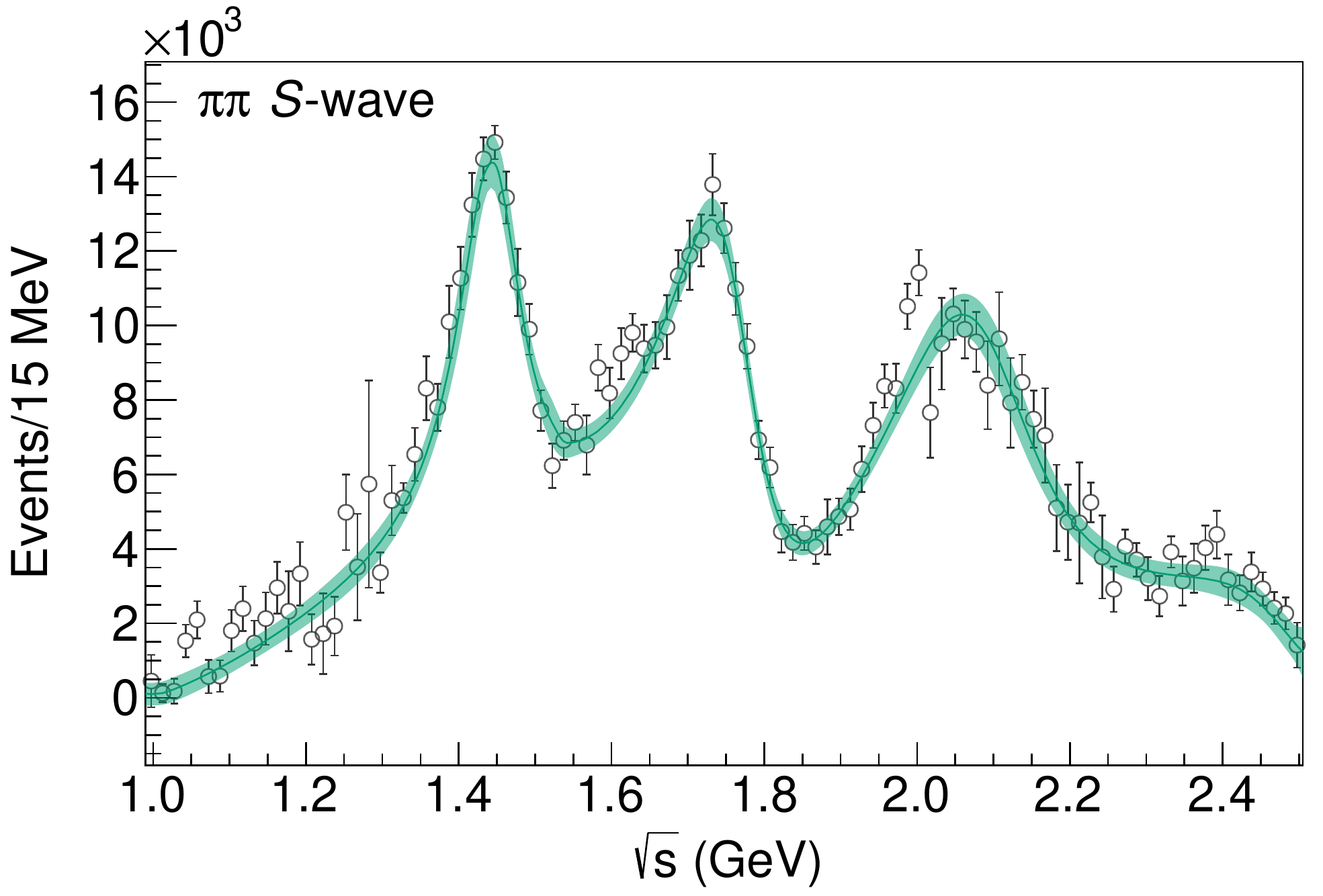} \includegraphics[width=0.32\textwidth]{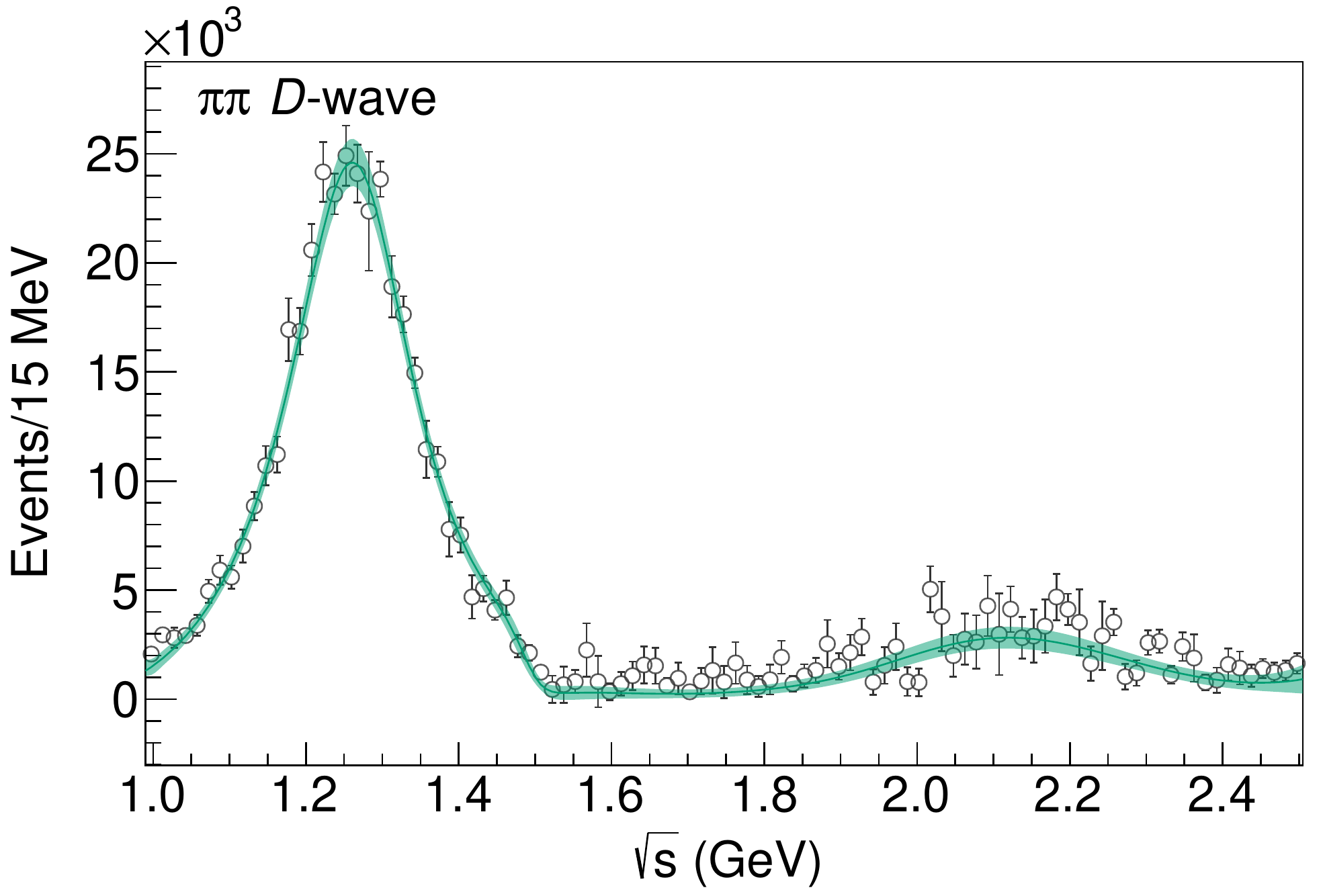} \includegraphics[width=0.32\textwidth]{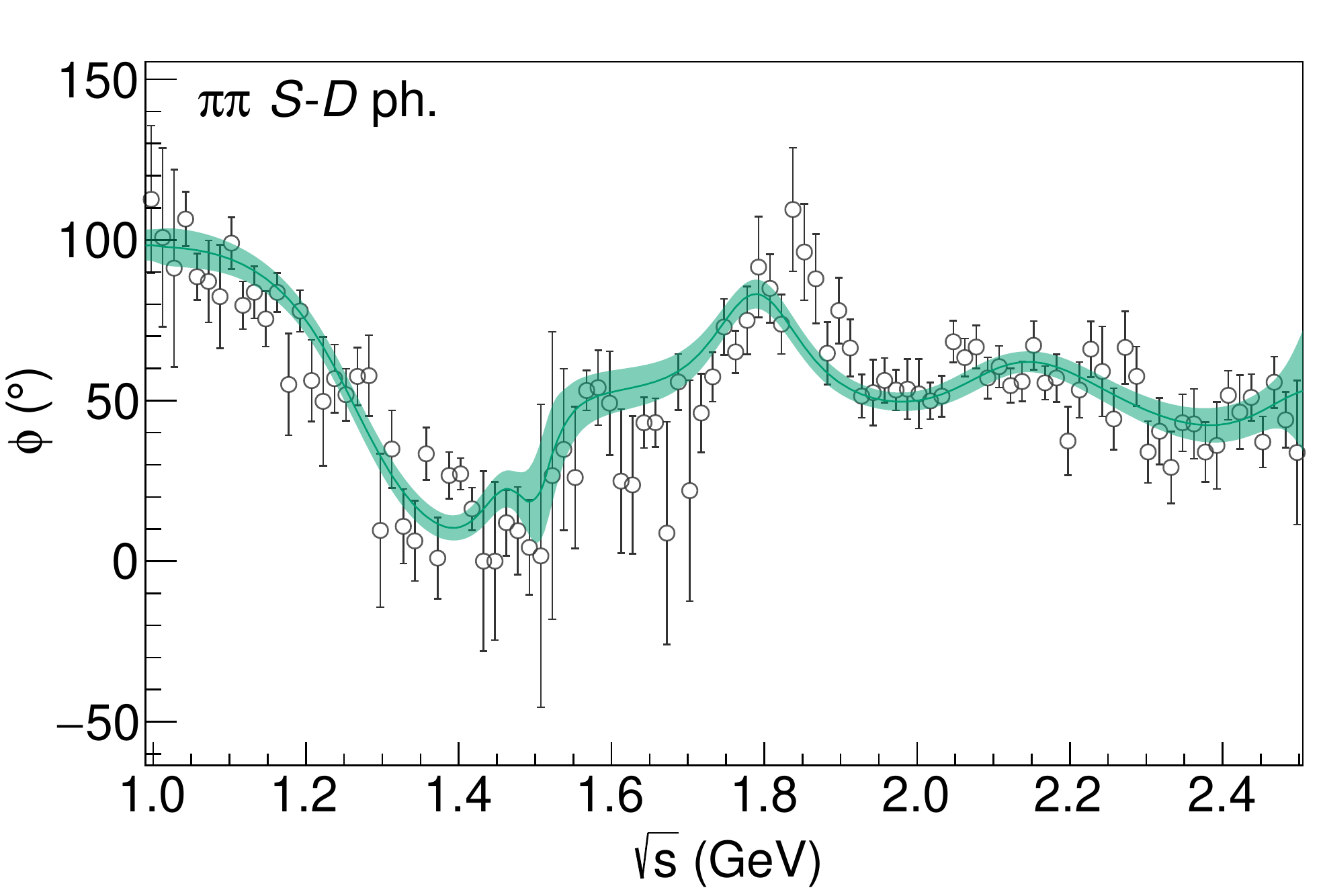}
\includegraphics[width=0.32\textwidth]{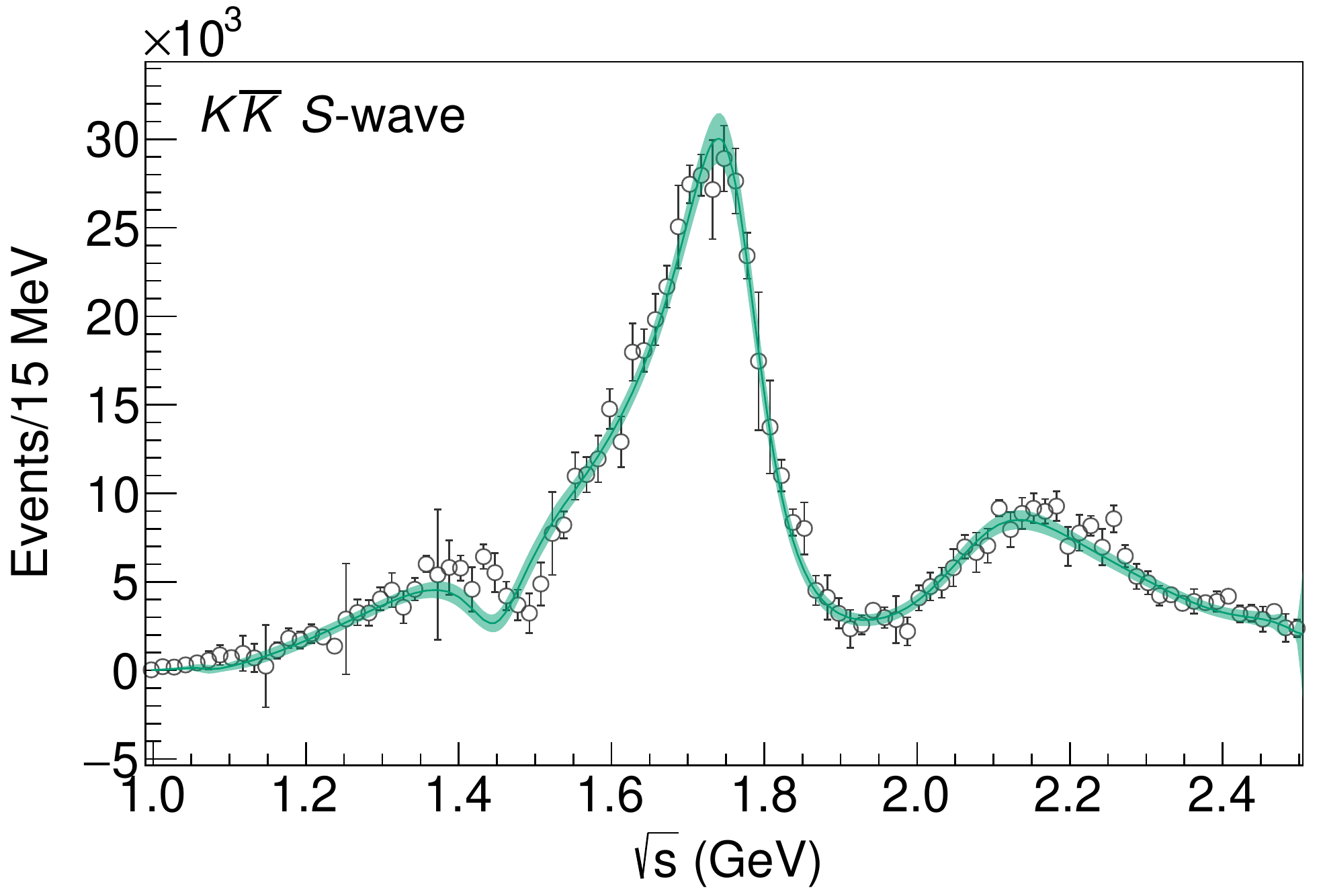} \includegraphics[width=0.32\textwidth]{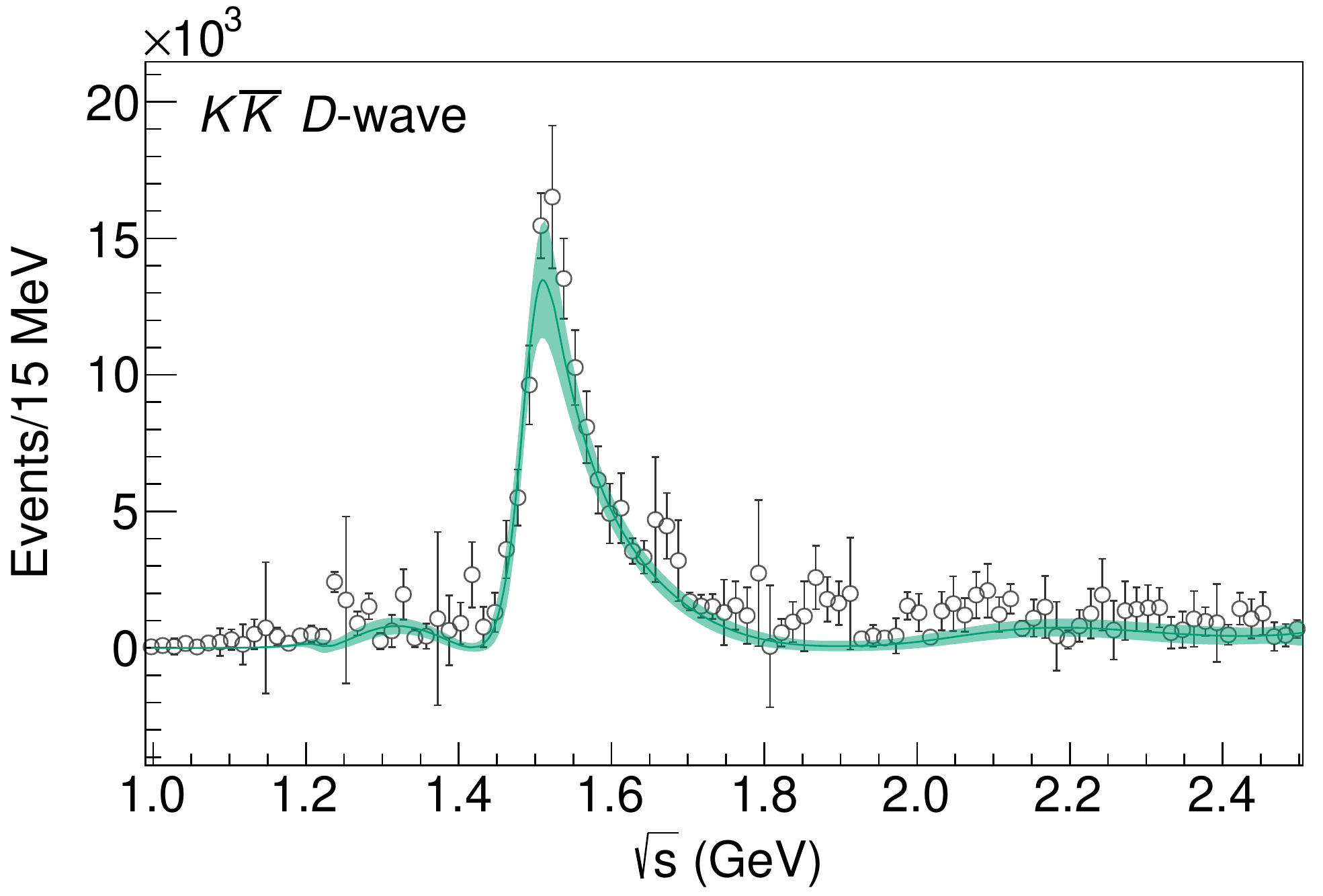} \includegraphics[width=0.32\textwidth]{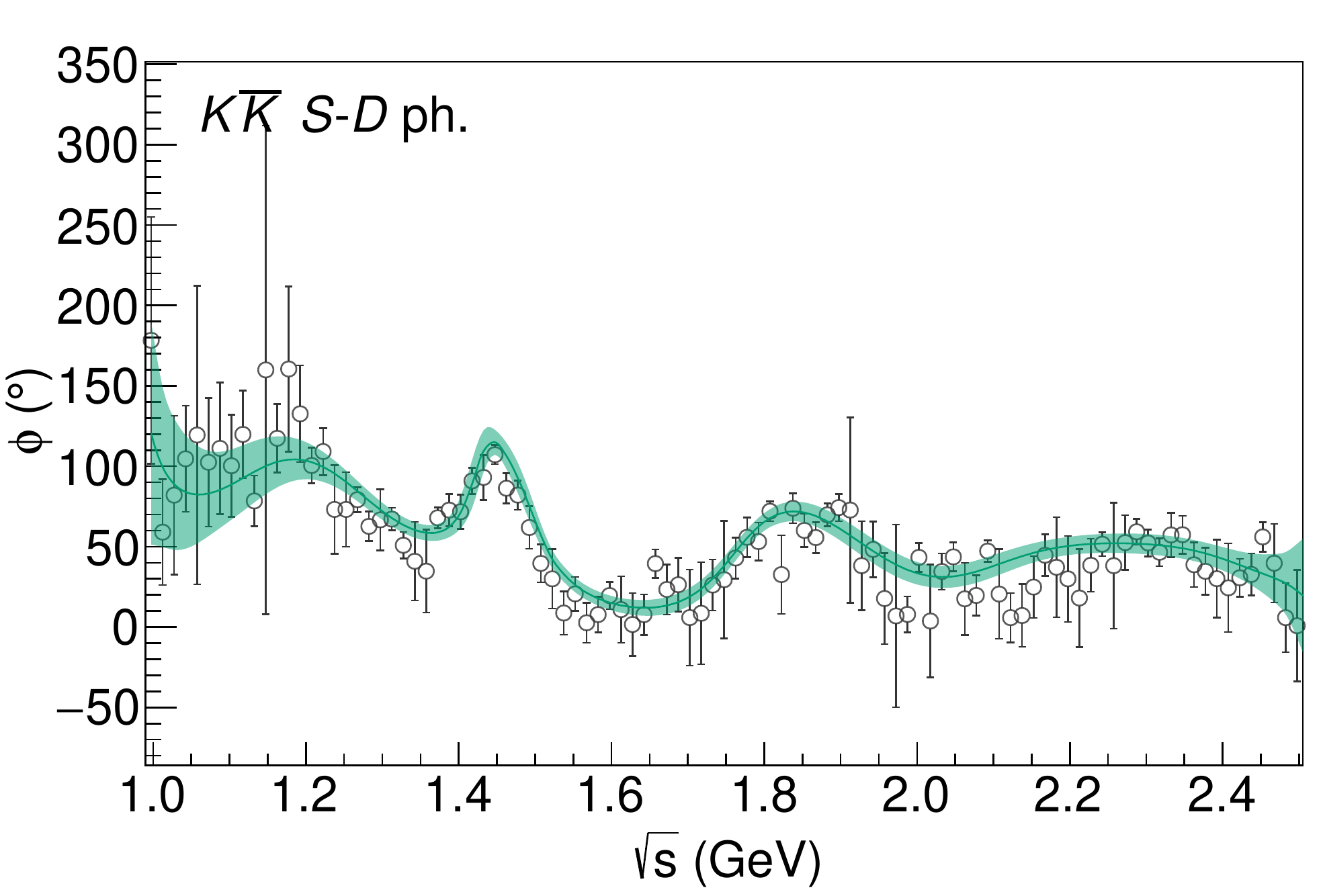}
\end{figure}

\begin{figure}[h]
\centering\includegraphics[width=0.45\textwidth]{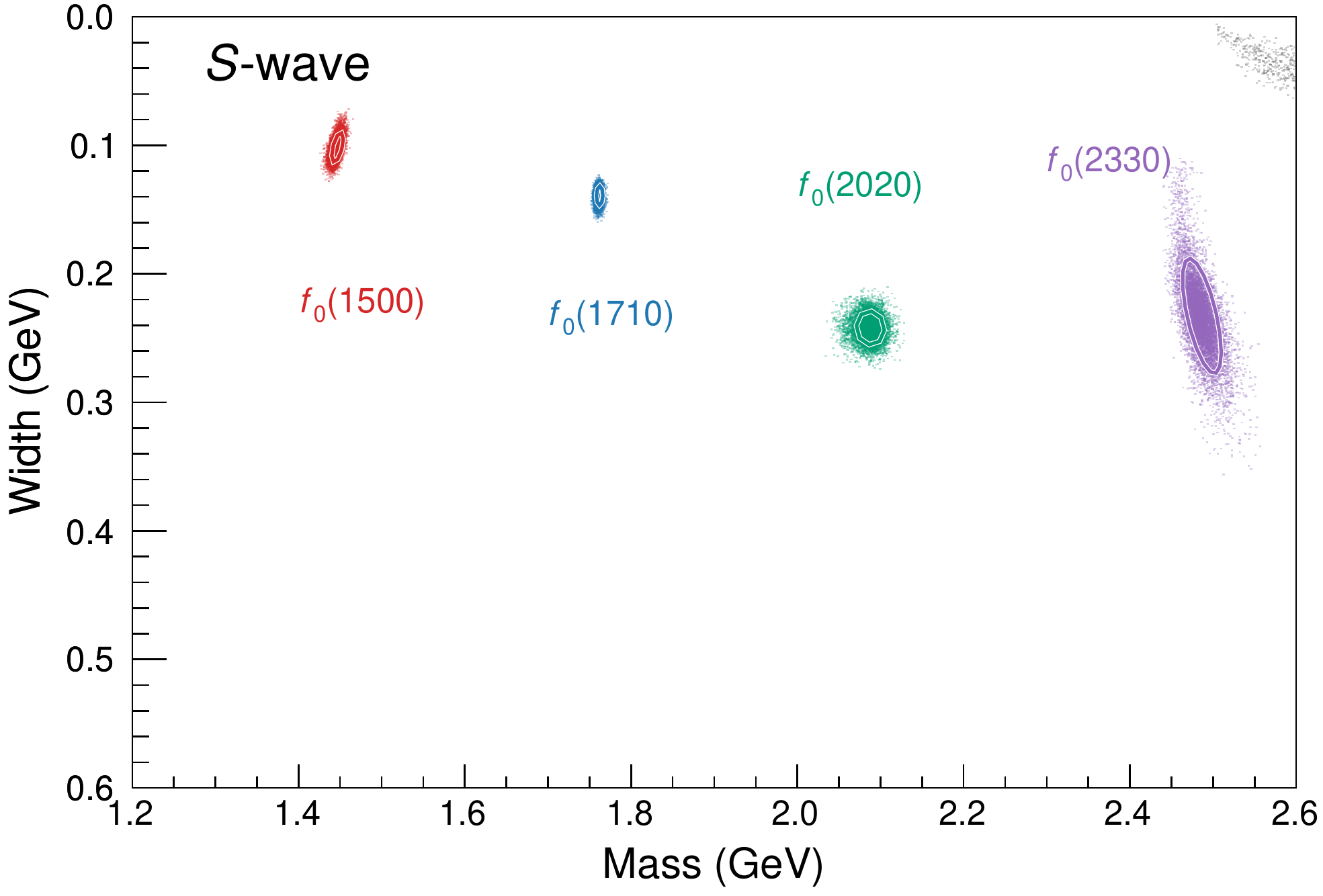} \includegraphics[width=0.45\textwidth]{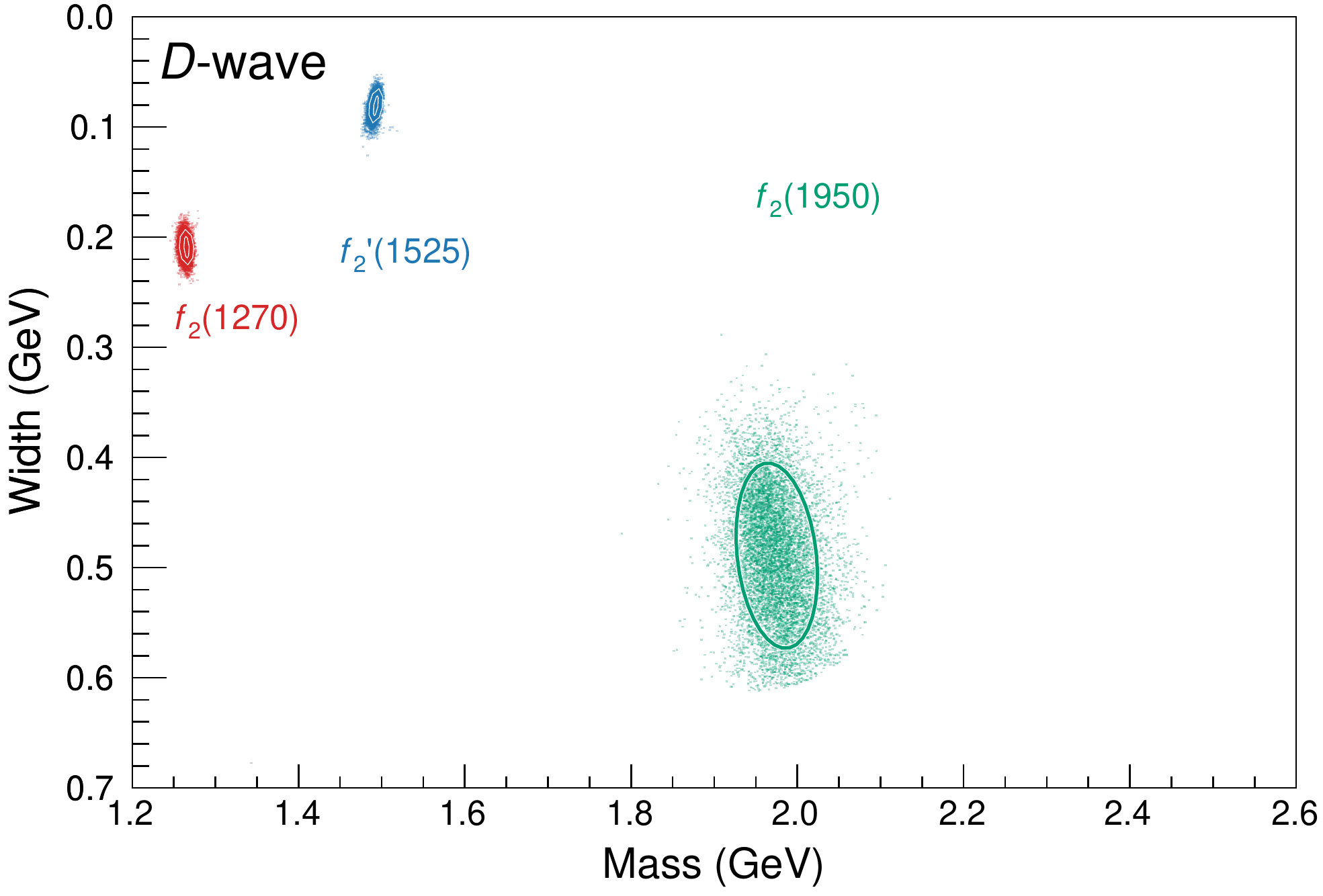}
\end{figure}

\begin{table}[h]
\begin{ruledtabular}
\begin{tabular}{c c c c}
$S$-wave  & $\sqrt{s_p}$ \mevp & $D$-wave & $\sqrt{s_p}$ \mevp \\ \hline
\rule[-0.2cm]{-0.1cm}{.55cm} $f_0(1500)$ &  $(1447 \pm 5) - i (102 \pm 7)/2$  &  $f_2(1270)$ &  $(1265 \pm 4) - i (209 \pm 9)/2$ \\
\rule[-0.2cm]{-0.1cm}{.55cm} $f_0(1710)$ &  $(1762 \pm 3) - i (139 \pm 5)/2$  &  $f_2'(1525)$ &  $(1493 \pm 4) - i (81 \pm 8)/2$ \\
\rule[-0.2cm]{-0.1cm}{.55cm} $f_0(2020)$ &  $(2088 \pm 11) - i (242 \pm 9)/2$  &  $f_2(1950)$ &  $(1975 \pm 32) - i (489 \pm 56)/2$ \\
\rule[-0.2cm]{-0.1cm}{.55cm} $f_0(2330)$ &  $(2486 \pm 15) - i (233 \pm 30)/2$  &   &   \\
\end{tabular}
\end{ruledtabular}
\end{table}

\clearpage

\subsection{$\left[K^J(s)^{-1}\right]^\text{CDD} \quad\Big/\quad \omega(s)_\text{pole+scaled} \quad\Big/\quad \rho N^J_{ki}(s')_\text{Q-model} \quad\Big/\quad s_L = 0.6\gevsq$}
\label{subsec:inputcddcm5newc3_bootstrap-out}

\begin{table}[h]
\begin{ruledtabular}
\begin{tabular}{c c c c c c}
\multicolumn{3}{c}{$\pi \pi$ channel}  & \multicolumn{3}{c}{$ K  \bar K$ channel} \\ \hline
\rule[-0.2cm]{-0.1cm}{.55cm} $a^{S,\pi\pi}_0$ &$-110.243$ & $-134.867 \pm 31.664$ & $a^{S,K\bar K}_0$ & $-151.390$ & $-187.281 \pm 48.033$ \\
\rule[-0.2cm]{-0.1cm}{.55cm} $a^{S,\pi\pi}_1$ &$-35.610$ & $-46.478 \pm 15.178$ & $a^{S,K\bar K}_1$ & $119.745$ & $115.015 \pm 23.673$ \\
\rule[-0.2cm]{-0.1cm}{.55cm} $a^{S,\pi\pi}_2$ &$59.275$ & $71.442 \pm 23.082$ & $a^{S,K\bar K}_2$ & $245.904$ & $272.587 \pm 47.872$ \\
\rule[-0.2cm]{-0.1cm}{.55cm} $a^{S,\pi\pi}_3$ &$28.277$ & $32.539 \pm 8.976$ & $a^{S,K\bar K}_3$ & $89.554$ & $96.485 \pm 18.621$ \\
\hline
\rule[-0.2cm]{-0.1cm}{.55cm} $a^{D,\pi\pi}_0$ &$-45.501$ & $-57.079 \pm 16.161$ & $a^{D,K\bar K}_0$ & $48.969$ & $49.914 \pm 5.799$ \\
\rule[-0.2cm]{-0.1cm}{.55cm} $a^{D,\pi\pi}_1$ &$-143.801$ & $-163.005 \pm 27.055$ & $a^{D,K\bar K}_1$ & $88.319$ & $88.822 \pm 10.689$ \\
\rule[-0.2cm]{-0.1cm}{.55cm} $a^{D,\pi\pi}_2$ &$-71.387$ & $-82.756 \pm 16.050$ & $a^{D,K\bar K}_2$ & $34.708$ & $36.099 \pm 5.703$ \\
\rule[-0.2cm]{-0.1cm}{.55cm} $a^{D,\pi\pi}_3$ &$-10.572$ & $-14.792 \pm 7.225$ & $a^{D,K\bar K}_3$ & $5.095$ & $5.811 \pm 3.164$ \\
\end{tabular}
\end{ruledtabular}
\end{table}

\begin{table}[h]
\begin{ruledtabular}
\begin{tabular}{c c c c c c c c c}
\multicolumn{3}{c}{$S$-wave couplings $\pi \pi$ channel}  & \multicolumn{3}{c}{$S$-wave couplings $K  \bar K$ channel} & \multicolumn{3}{c}{$S$-wave couplings $\rho\rho$ channel} \\ \hline
\rule[-0.2cm]{-0.1cm}{.55cm} $g^{S,1}_{\pi\pi}$ & $0.387$ & $0.398 \pm 0.116$ & $g^{S,1}_{K\bar K}$ & $-2.178$ & $-2.176 \pm 0.340$ & $g^{S,1}_{\rho\rho}$ & $0.584$ & $0.566 \pm 0.094$ \\
\rule[-0.2cm]{-0.1cm}{.55cm} $g^{S,2}_{\pi\pi}$ & $-0.844$ & $-0.902 \pm 0.105$ & $g^{S,2}_{K\bar K}$ & $-1.326$ & $-1.390 \pm 0.155$ & $g^{S,2}_{\rho\rho}$ & $0.383$ & $0.376 \pm 0.017$ \\
\rule[-0.2cm]{-0.1cm}{.55cm} $g^{S,3}_{\pi\pi}$ & $-3.258$ & $-3.063 \pm 0.713$ & $g^{S,3}_{K\bar K}$ & $0.648$ & $0.192 \pm 0.491$ & $g^{S,3}_{\rho\rho}$ & $-0.518$ & $-0.576 \pm 0.100$ \\
\hline 
\multicolumn{3}{c}{$D$-wave couplings $\pi \pi$ channel}  & \multicolumn{3}{c}{$D$-wave couplings $K  \bar K$ channel} & \multicolumn{3}{c}{$D$-wave couplings $\rho\rho$ channel} \\ \hline
\rule[-0.2cm]{-0.1cm}{.55cm} $g^{D,1}_{\pi\pi}$ & $1.048$ & $1.034 \pm 0.045$ & $g^{D,1}_{K\bar K}$ & $0.533$ & $0.507 \pm 0.077$ & $g^{D,1}_{\rho\rho}$ & $-1.319$ & $-1.301 \pm 0.143$ \\
\rule[-0.2cm]{-0.1cm}{.55cm} $g^{D,2}_{\pi\pi}$ & $-0.355$ & $-0.292 \pm 0.100$ & $g^{D,2}_{K\bar K}$ & $0.899$ & $0.929 \pm 0.071$ & $g^{D,2}_{\rho\rho}$ & $0.432$ & $0.392 \pm 0.145$ \\
\rule[-0.2cm]{-0.1cm}{.55cm} $g^{D,3}_{\pi\pi}$ & $0.751$ & $0.675 \pm 0.189$ & $g^{D,3}_{K\bar K}$ & $0.264$ & $0.279 \pm 0.206$ & $g^{D,3}_{\rho\rho}$ & $1.578$ & $1.565 \pm 0.139$ \\
\end{tabular}
\vspace{2em}
\begin{tabular}{c c c c c c}
\multicolumn{3}{c}{$S$-wave background}  & \multicolumn{3}{c}{$D$-wave background} \\ \hline
\rule[-0.2cm]{-0.1cm}{.55cm} $c^{S,0}_{\pi\pi,\pi\pi}$ &$-1.628$ & $-1.713 \pm 0.308$ & $c^{D,0}_{\pi\pi,\pi\pi}$ &$1.656$ & $1.498 \pm 0.255$ \\
\rule[-0.2cm]{-0.1cm}{.55cm} $c^{S,0}_{\pi\pi,K\bar K}$ &$-2.497$ & $-2.713 \pm 0.288$ & $c^{D,0}_{\pi\pi,K\bar K}$ &$3.748$ & $3.619 \pm 0.555$ \\
\rule[-0.2cm]{-0.1cm}{.55cm} $c^{S,0}_{\pi\pi,\rho\rho}$ &$0$ & $0$ & $c^{D,0}_{\pi\pi,\rho\rho}$ &$-0.446$ & $-0.510 \pm 0.207$ \\
\rule[-0.2cm]{-0.1cm}{.55cm} $c^{S,0}_{K\bar K,K\bar K}$ &$-3.630$ & $-3.847 \pm 0.373$ & $c^{D,0}_{K\bar K,K\bar K}$ &$11.214$ & $12.236 \pm 1.893$ \\
\rule[-0.2cm]{-0.1cm}{.55cm} $c^{S,0}_{K\bar K,\rho\rho}$ &$0$ & $0$ & $c^{D,0}_{K\bar K,\rho\rho}$ &$0.376$ & $0.256 \pm 0.309$ \\
\rule[-0.2cm]{-0.1cm}{.55cm} $c^{S,0}_{\rho\rho,\rho\rho}$ &$-0.360$ & $-0.362 \pm 0.115$ & $c^{D,0}_{\rho\rho,\rho\rho}$ &$1.812$ & $1.744 \pm 0.460$ \\
\rule[-0.2cm]{-0.1cm}{.55cm} $c^{S,1}_{\pi\pi,\pi\pi}$ &$0.141$ & $0.036 \pm 0.146$ & $c^{D,1}_{\pi\pi,\pi\pi}$ &$-1.288$ & $-1.140 \pm 0.253$ \\
\rule[-0.2cm]{-0.1cm}{.55cm} $c^{S,1}_{\pi\pi,K\bar K}$ &$1.078$ & $1.124 \pm 0.103$ & $c^{D,1}_{\pi\pi,K\bar K}$ &$-2.259$ & $-2.228 \pm 0.418$ \\
\rule[-0.2cm]{-0.1cm}{.55cm} $c^{S,1}_{\pi\pi,\rho\rho}$ &$0$ & $0$ & $c^{D,1}_{\pi\pi,\rho\rho}$ &$0$ & $0$ \\
\rule[-0.2cm]{-0.1cm}{.55cm} $c^{S,1}_{K\bar K,K\bar K}$ &$1.756$ & $1.837 \pm 0.149$ & $c^{D,1}_{K\bar K,K\bar K}$ &$-4.301$ & $-4.821 \pm 1.001$ \\
\rule[-0.2cm]{-0.1cm}{.55cm} $c^{S,1}_{K\bar K,\rho\rho}$ &$0$ & $0$ & $c^{D,1}_{K\bar K,\rho\rho}$ &$0$ & $0$ \\
\rule[-0.2cm]{-0.1cm}{.55cm} $c^{S,1}_{\rho\rho,\rho\rho}$ &$0$ & $0$ & $c^{D,1}_{\rho\rho,\rho\rho}$ &$0$ & $0$ \\
\end{tabular}
\end{ruledtabular}
\end{table}

\begin{figure}[h]
\centering\includegraphics[width=0.32\textwidth]{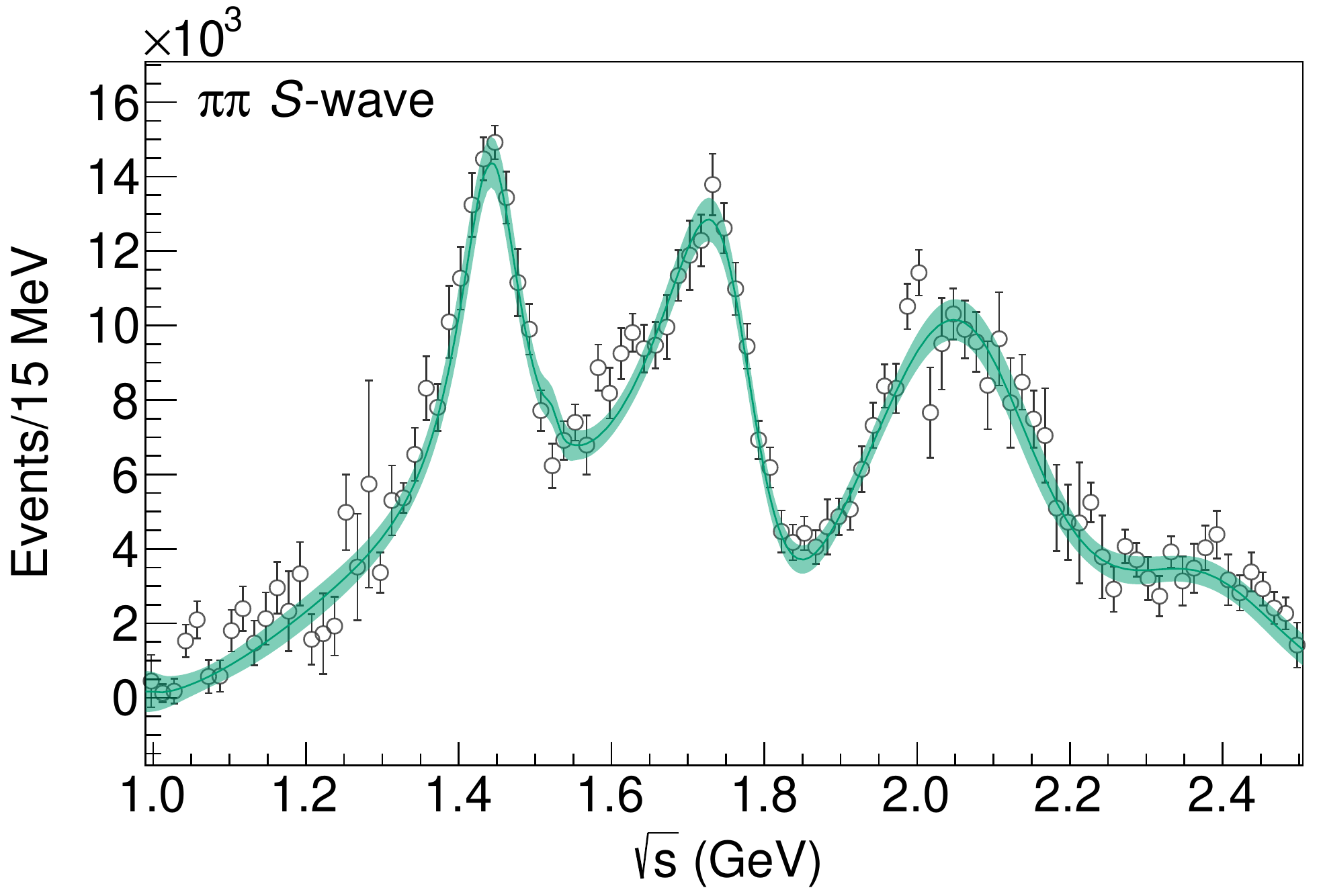} \includegraphics[width=0.32\textwidth]{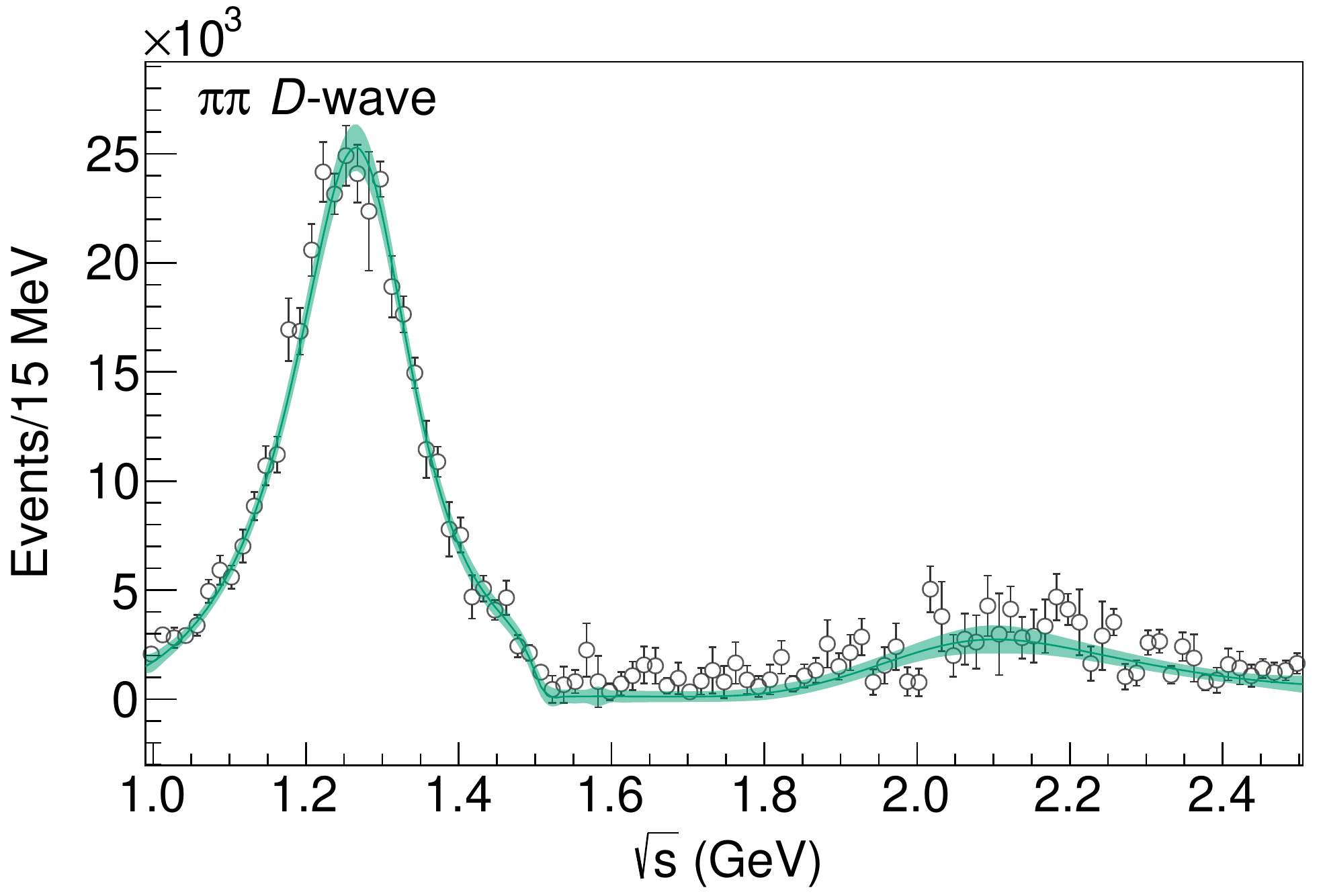} \includegraphics[width=0.32\textwidth]{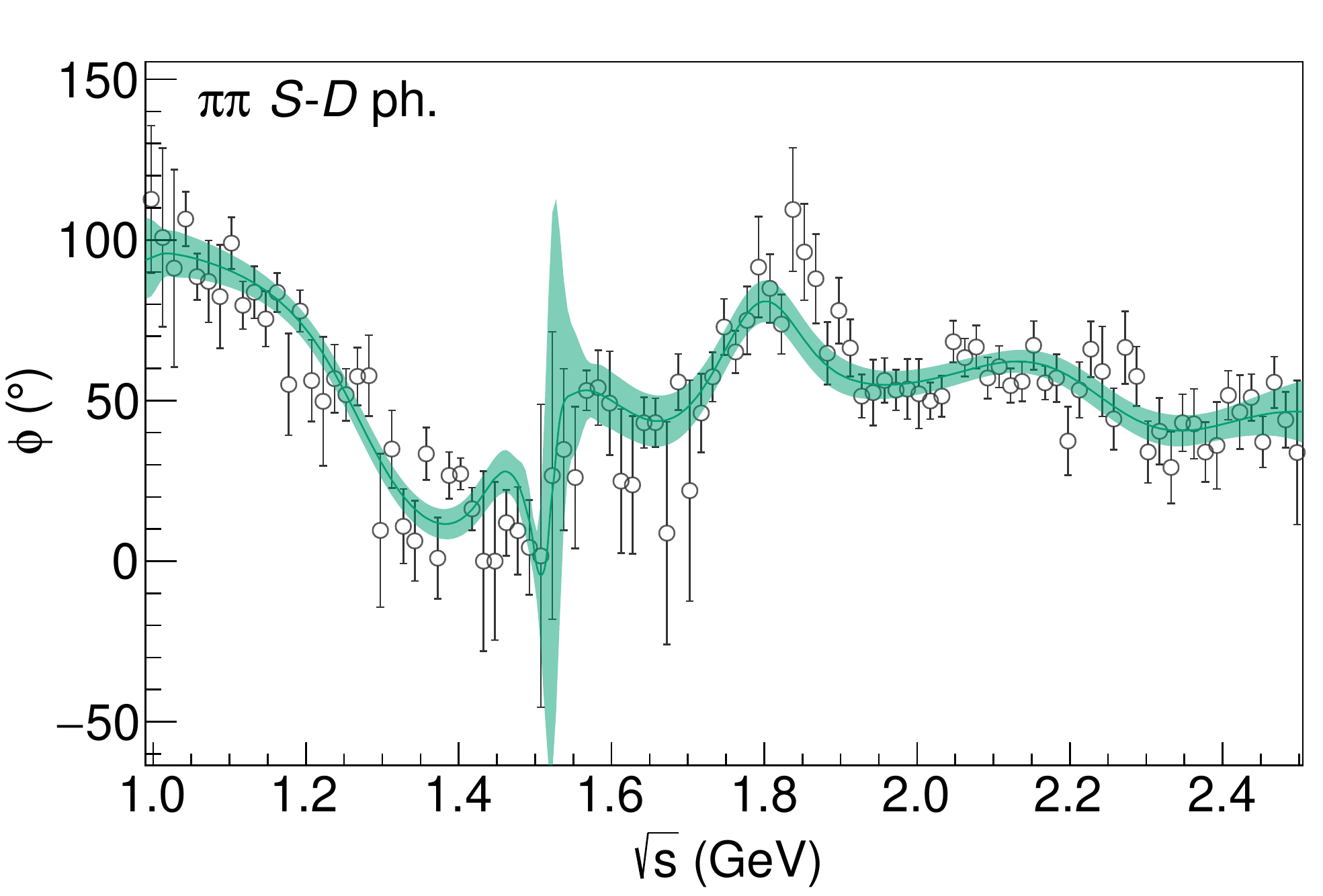}
\includegraphics[width=0.32\textwidth]{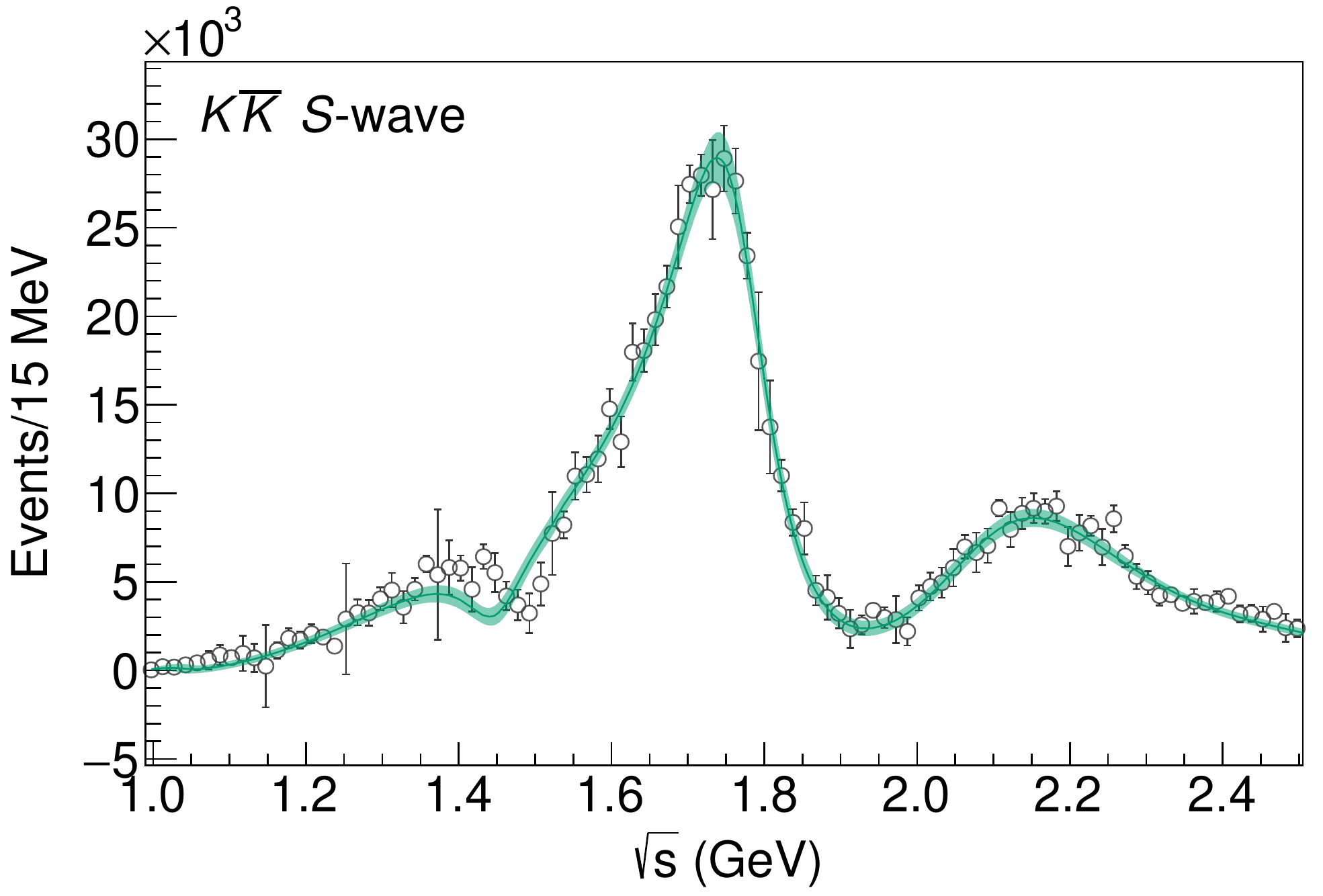} \includegraphics[width=0.32\textwidth]{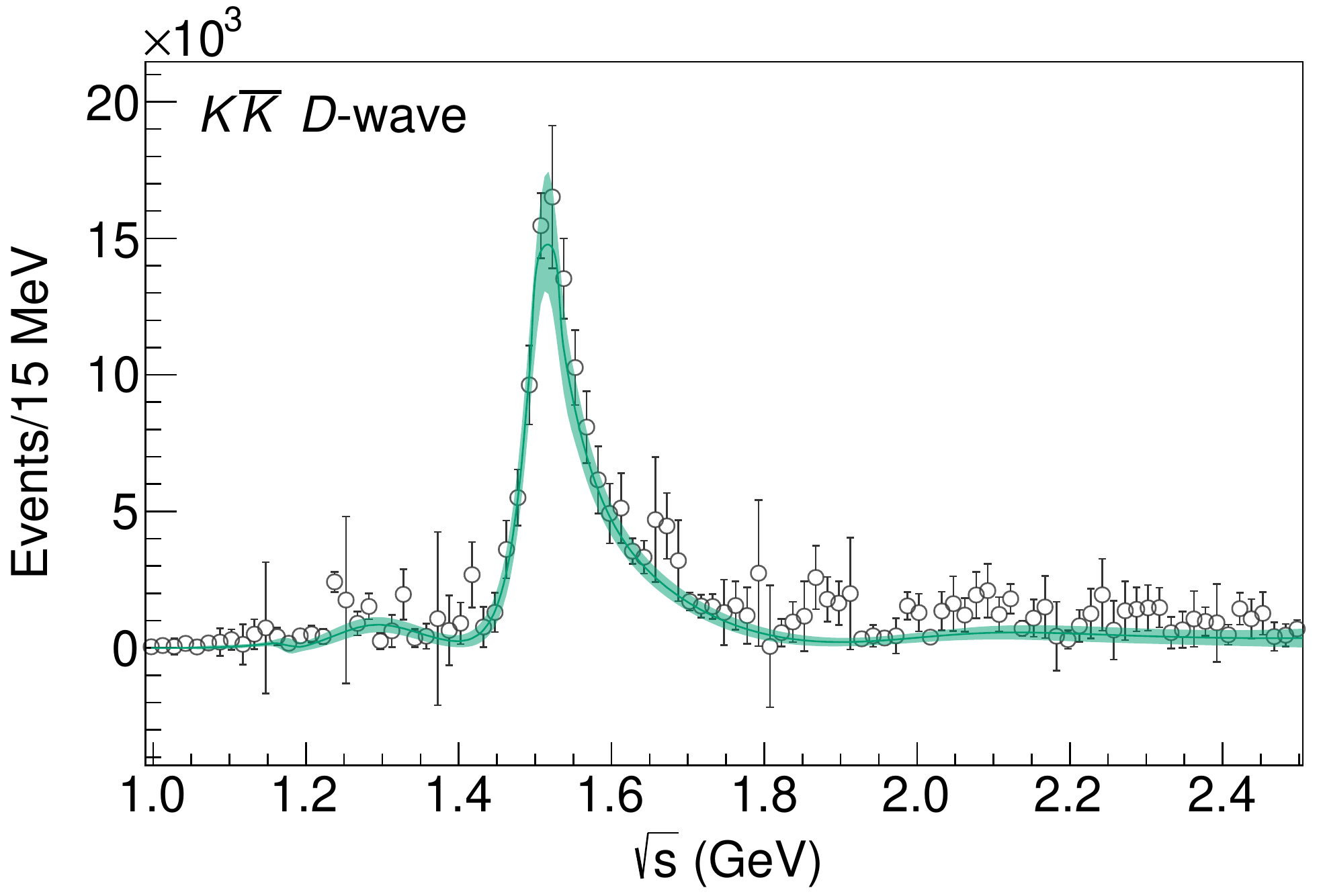} \includegraphics[width=0.32\textwidth]{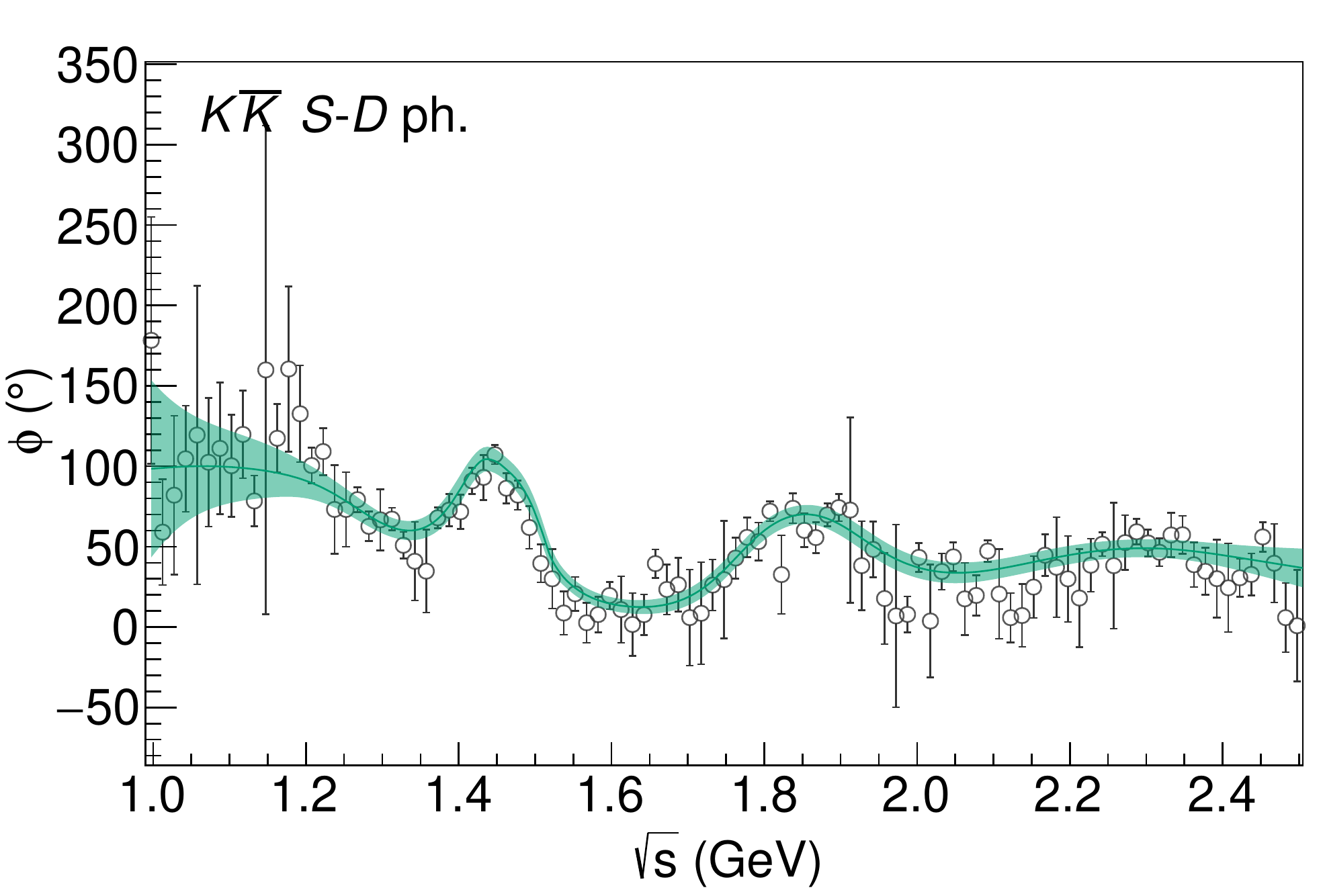}
\end{figure}

\begin{figure}[h]
\centering\includegraphics[width=0.45\textwidth]{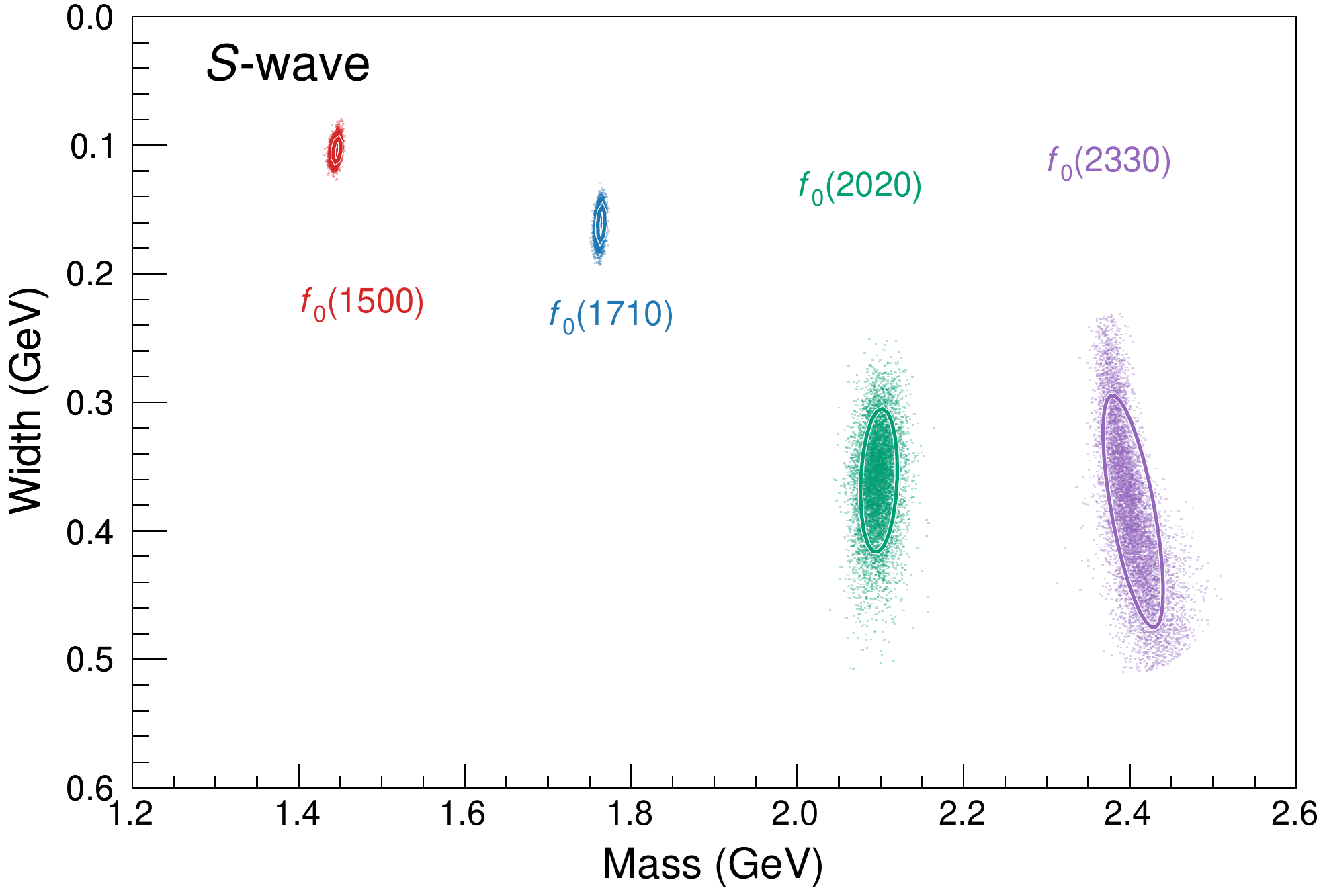} \includegraphics[width=0.45\textwidth]{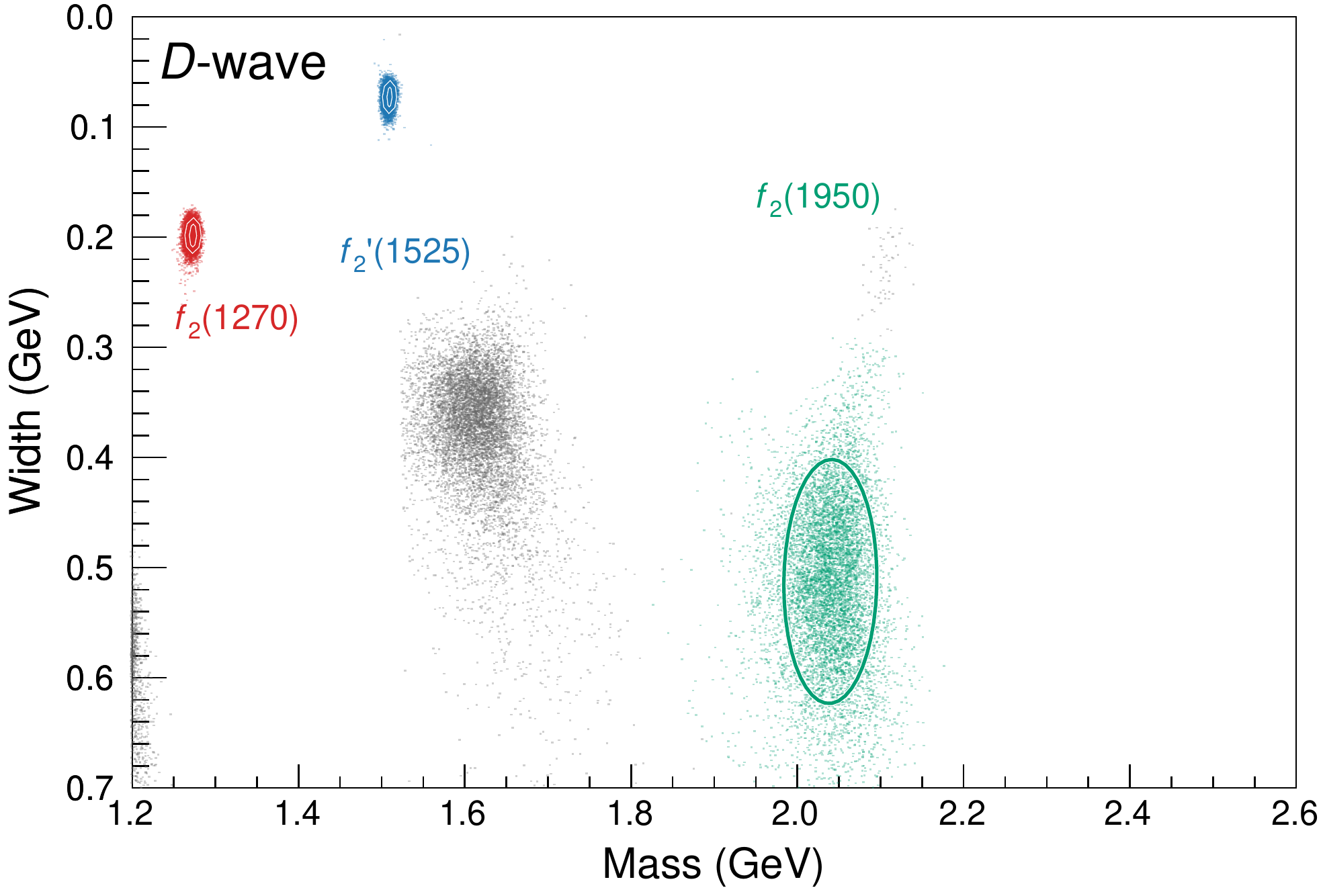}
\end{figure}

\begin{table}[h]
\begin{ruledtabular}
\begin{tabular}{c c c c}
$S$-wave  & $\sqrt{s_p}$ \mevp & $D$-wave & $\sqrt{s_p}$ \mevp \\ \hline
\rule[-0.2cm]{-0.1cm}{.55cm} $f_0(1500)$ &  $(1446 \pm 3) - i (104 \pm 6)/2$  &  $f_2(1270)$ &  $(1273 \pm 5) - i (199 \pm 9)/2$ \\
\rule[-0.2cm]{-0.1cm}{.55cm} $f_0(1710)$ &  $(1764 \pm 3) - i (161 \pm 9)/2$  &  $f_2'(1525)$ &  $(1509 \pm 4) - i (73 \pm 8)/2$ \\
\rule[-0.2cm]{-0.1cm}{.55cm} $f_0(2020)$ &  $(2098 \pm 15) - i (361 \pm 37)/2$  &  $f_2(1950)$ &  $(2040 \pm 37) - i (513 \pm 73)/2$ \\
\rule[-0.2cm]{-0.1cm}{.55cm} $f_0(2330)$ &  $(2404 \pm 24) - i (385 \pm 60)/2$  &   &   \\
\end{tabular}
\end{ruledtabular}
\end{table}

\clearpage

\subsection{$\left[K^J(s)^{-1}\right]^\text{CDD} \quad\Big/\quad \omega(s)_\text{pole+scaled} \quad\Big/\quad \rho N^J_{ki}(s')_\text{Q-model} \quad\Big/\quad s_L = 0$}
\label{subsec:inputcddcm6newc3_bootstrap-out}

\begin{table}[h]
\begin{ruledtabular}
\begin{tabular}{c c c c c c}
\multicolumn{3}{c}{$\pi \pi$ channel}  & \multicolumn{3}{c}{$ K  \bar K$ channel} \\ \hline
\rule[-0.2cm]{-0.1cm}{.55cm} $a^{S,\pi\pi}_0$ &$-44.149$ & $-44.016 \pm 7.140$ & $a^{S,K\bar K}_0$ & $-1.180$ & $-3.244 \pm 13.295$ \\
\rule[-0.2cm]{-0.1cm}{.55cm} $a^{S,\pi\pi}_1$ &$128.785$ & $118.535 \pm 23.435$ & $a^{S,K\bar K}_1$ & $433.016$ & $397.835 \pm 49.342$ \\
\rule[-0.2cm]{-0.1cm}{.55cm} $a^{S,\pi\pi}_2$ &$196.149$ & $193.429 \pm 26.093$ & $a^{S,K\bar K}_2$ & $491.470$ & $471.411 \pm 49.625$ \\
\rule[-0.2cm]{-0.1cm}{.55cm} $a^{S,\pi\pi}_3$ &$94.886$ & $93.064 \pm 12.251$ & $a^{S,K\bar K}_3$ & $211.380$ & $201.446 \pm 22.529$ \\
\hline
\rule[-0.2cm]{-0.1cm}{.55cm} $a^{D,\pi\pi}_0$ &$232.738$ & $245.804 \pm 17.326$ & $a^{D,K\bar K}_0$ & $53.519$ & $36.061 \pm 22.033$ \\
\rule[-0.2cm]{-0.1cm}{.55cm} $a^{D,\pi\pi}_1$ &$-197.597$ & $-225.738 \pm 38.642$ & $a^{D,K\bar K}_1$ & $202.272$ & $232.740 \pm 46.644$ \\
\rule[-0.2cm]{-0.1cm}{.55cm} $a^{D,\pi\pi}_2$ &$172.533$ & $185.895 \pm 19.088$ & $a^{D,K\bar K}_2$ & $25.053$ & $6.986 \pm 22.498$ \\
\rule[-0.2cm]{-0.1cm}{.55cm} $a^{D,\pi\pi}_3$ &$-34.434$ & $-43.659 \pm 11.394$ & $a^{D,K\bar K}_3$ & $5.299$ & $14.750 \pm 15.324$ \\
\end{tabular}
\end{ruledtabular}
\end{table}

\begin{table}[h]
\begin{ruledtabular}
\begin{tabular}{c c c c c c c c c}
\multicolumn{3}{c}{$S$-wave couplings $\pi \pi$ channel}  & \multicolumn{3}{c}{$S$-wave couplings $K  \bar K$ channel} & \multicolumn{3}{c}{$S$-wave couplings $\rho\rho$ channel} \\ \hline
\rule[-0.2cm]{-0.1cm}{.55cm} $g^{S,1}_{\pi\pi}$ & $-0.699$ & $-0.708 \pm 0.070$ & $g^{S,1}_{K\bar K}$ & $-1.072$ & $-1.059 \pm 0.087$ & $g^{S,1}_{\rho\rho}$ & $0.415$ & $0.414 \pm 0.017$ \\
\rule[-0.2cm]{-0.1cm}{.55cm} $g^{S,2}_{\pi\pi}$ & $-0.276$ & $-0.278 \pm 0.044$ & $g^{S,2}_{K\bar K}$ & $0.488$ & $0.471 \pm 0.117$ & $g^{S,2}_{\rho\rho}$ & $0.433$ & $0.418 \pm 0.064$ \\
\rule[-0.2cm]{-0.1cm}{.55cm} $g^{S,3}_{\pi\pi}$ & $3.613$ & $3.774 \pm 0.248$ & $g^{S,3}_{K\bar K}$ & $7.414$ & $7.672 \pm 0.566$ & $g^{S,3}_{\rho\rho}$ & $0.336$ & $0.316 \pm 0.047$ \\
\hline 
\multicolumn{3}{c}{$D$-wave couplings $\pi \pi$ channel}  & \multicolumn{3}{c}{$D$-wave couplings $K  \bar K$ channel} & \multicolumn{3}{c}{$D$-wave couplings $\rho\rho$ channel} \\ \hline
\rule[-0.2cm]{-0.1cm}{.55cm} $g^{D,1}_{\pi\pi}$ & $0.366$ & $0.279 \pm 0.140$ & $g^{D,1}_{K\bar K}$ & $0.267$ & $0.287 \pm 0.114$ & $g^{D,1}_{\rho\rho}$ & $0.526$ & $0.868 \pm 0.460$ \\
\rule[-0.2cm]{-0.1cm}{.55cm} $g^{D,2}_{\pi\pi}$ & $-0.265$ & $-0.251 \pm 0.050$ & $g^{D,2}_{K\bar K}$ & $0.992$ & $0.997 \pm 0.058$ & $g^{D,2}_{\rho\rho}$ & $0.416$ & $0.346 \pm 0.157$ \\
\rule[-0.2cm]{-0.1cm}{.55cm} $g^{D,3}_{\pi\pi}$ & $-1.341$ & $-1.426 \pm 0.376$ & $g^{D,3}_{K\bar K}$ & $0.890$ & $0.902 \pm 0.218$ & $g^{D,3}_{\rho\rho}$ & $2.648$ & $3.156 \pm 0.584$ \\
\end{tabular}
\vspace{2em}
\begin{tabular}{c c c c c c}
\multicolumn{3}{c}{$S$-wave background}  & \multicolumn{3}{c}{$D$-wave background} \\ \hline
\rule[-0.2cm]{-0.1cm}{.55cm} $c^{S,0}_{\pi\pi,\pi\pi}$ &$-2.065$ & $-2.075 \pm 0.134$ & $c^{D,0}_{\pi\pi,\pi\pi}$ &$-0.378$ & $-0.303 \pm 0.235$ \\
\rule[-0.2cm]{-0.1cm}{.55cm} $c^{S,0}_{\pi\pi,K\bar K}$ &$3.253$ & $3.253 \pm 0.123$ & $c^{D,0}_{\pi\pi,K\bar K}$ &$0.294$ & $0.413 \pm 0.225$ \\
\rule[-0.2cm]{-0.1cm}{.55cm} $c^{S,0}_{\pi\pi,\rho\rho}$ &$0$ & $0$ & $c^{D,0}_{\pi\pi,\rho\rho}$ &$2.724$ & $3.163 \pm 0.797$ \\
\rule[-0.2cm]{-0.1cm}{.55cm} $c^{S,0}_{K\bar K,K\bar K}$ &$4.694$ & $4.646 \pm 0.167$ & $c^{D,0}_{K\bar K,K\bar K}$ &$0.684$ & $0.530 \pm 0.188$ \\
\rule[-0.2cm]{-0.1cm}{.55cm} $c^{S,0}_{K\bar K,\rho\rho}$ &$0$ & $0$ & $c^{D,0}_{K\bar K,\rho\rho}$ &$-1.317$ & $-1.335 \pm 0.292$ \\
\rule[-0.2cm]{-0.1cm}{.55cm} $c^{S,0}_{\rho\rho,\rho\rho}$ &$0.004$ & $0.000 \pm 0.032$ & $c^{D,0}_{\rho\rho,\rho\rho}$ &$-0.763$ & $0.007 \pm 1.124$ \\
\rule[-0.2cm]{-0.1cm}{.55cm} $c^{S,1}_{\pi\pi,\pi\pi}$ &$0.739$ & $0.696 \pm 0.066$ & $c^{D,1}_{\pi\pi,\pi\pi}$ &$-0.430$ & $-0.540 \pm 0.161$ \\
\rule[-0.2cm]{-0.1cm}{.55cm} $c^{S,1}_{\pi\pi,K\bar K}$ &$-0.617$ & $-0.534 \pm 0.165$ & $c^{D,1}_{\pi\pi,K\bar K}$ &$0.278$ & $0.284 \pm 0.056$ \\
\rule[-0.2cm]{-0.1cm}{.55cm} $c^{S,1}_{\pi\pi,\rho\rho}$ &$0$ & $0$ & $c^{D,1}_{\pi\pi,\rho\rho}$ &$0$ & $0$ \\
\rule[-0.2cm]{-0.1cm}{.55cm} $c^{S,1}_{K\bar K,K\bar K}$ &$-1.169$ & $-1.058 \pm 0.205$ & $c^{D,1}_{K\bar K,K\bar K}$ &$-0.290$ & $-0.262 \pm 0.059$ \\
\rule[-0.2cm]{-0.1cm}{.55cm} $c^{S,1}_{K\bar K,\rho\rho}$ &$0$ & $0$ & $c^{D,1}_{K\bar K,\rho\rho}$ &$0$ & $0$ \\
\rule[-0.2cm]{-0.1cm}{.55cm} $c^{S,1}_{\rho\rho,\rho\rho}$ &$0$ & $0$ & $c^{D,1}_{\rho\rho,\rho\rho}$ &$0$ & $0$ \\
\end{tabular}
\end{ruledtabular}
\end{table}

\begin{figure}[h]
\centering\includegraphics[width=0.32\textwidth]{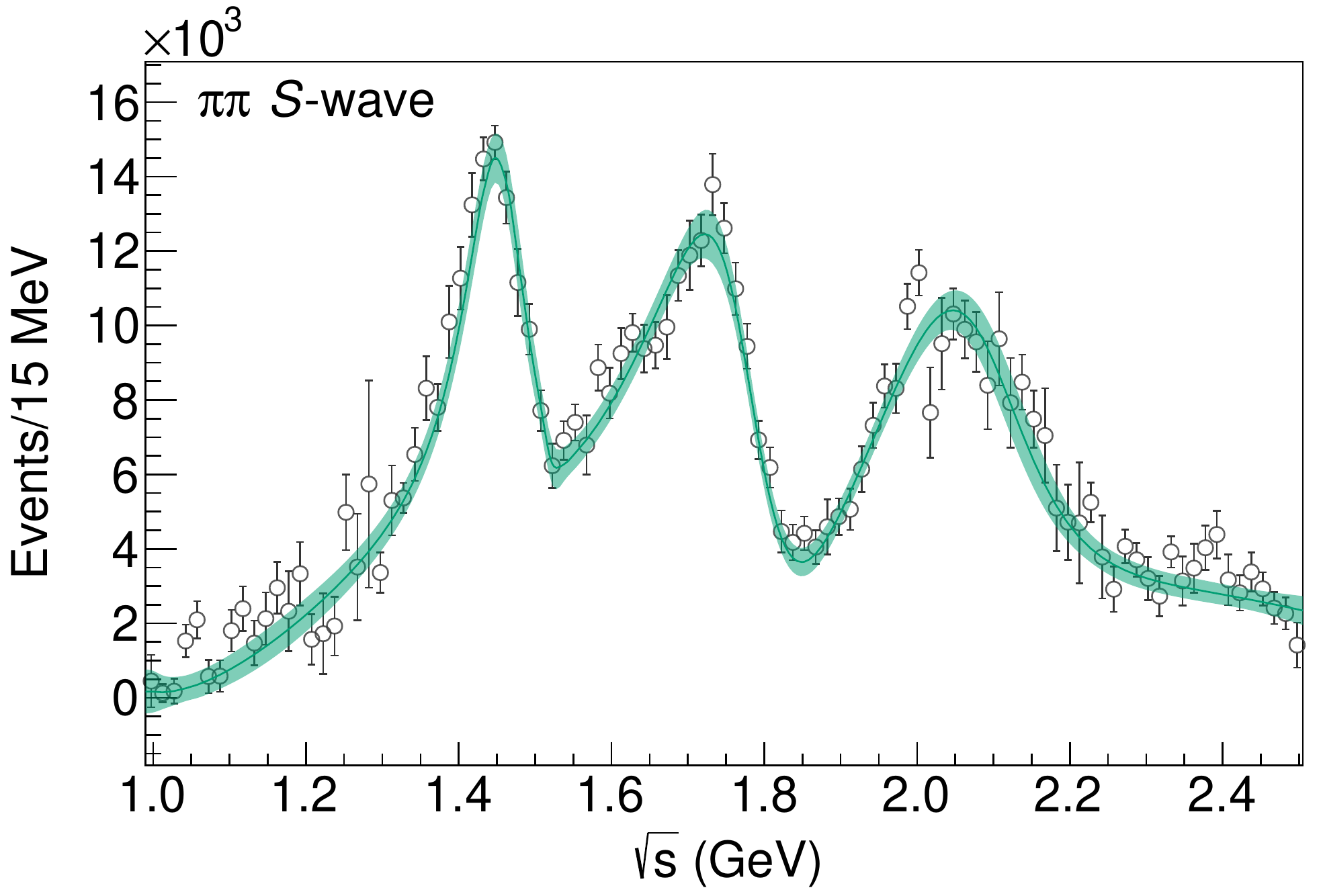} \includegraphics[width=0.32\textwidth]{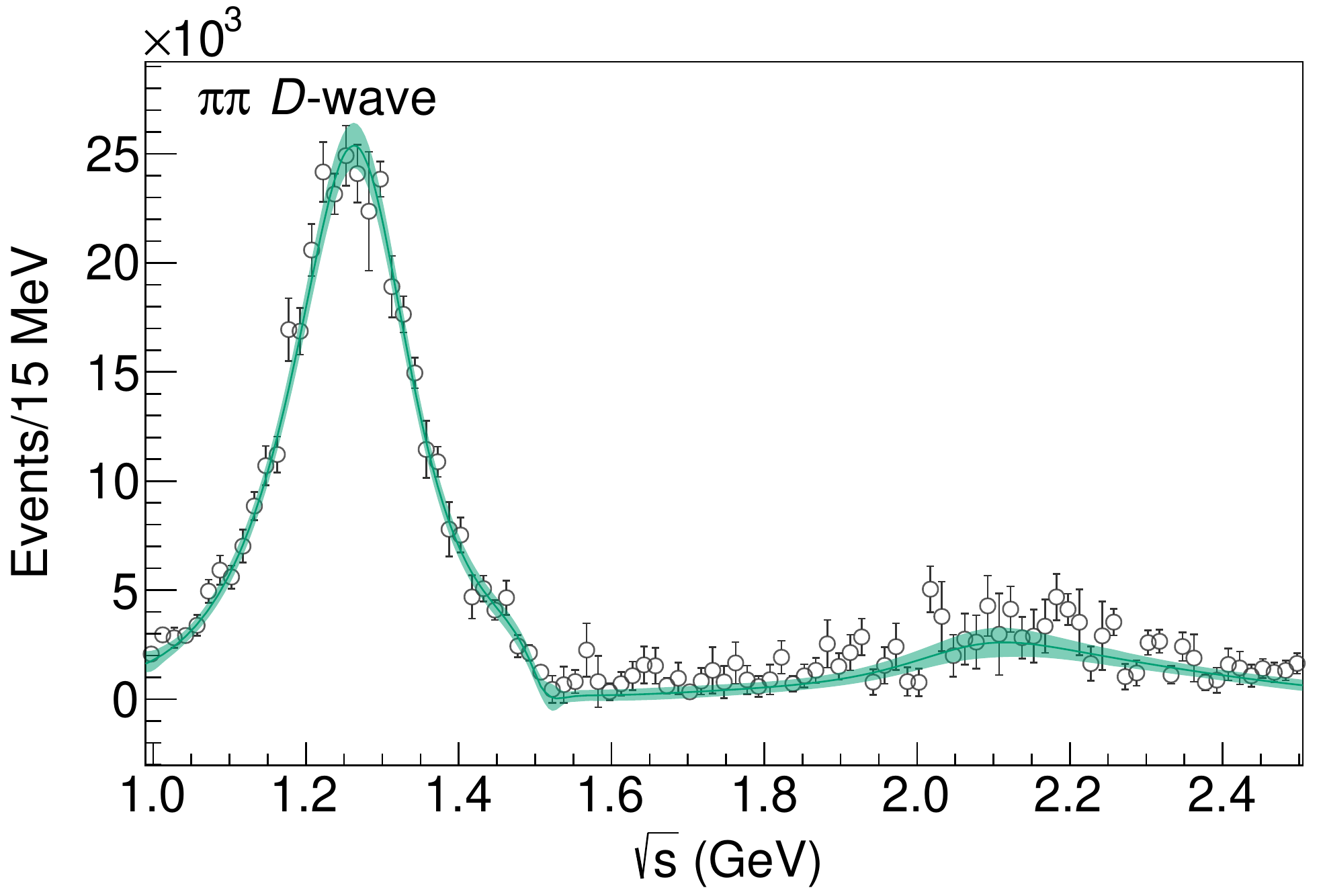} \includegraphics[width=0.32\textwidth]{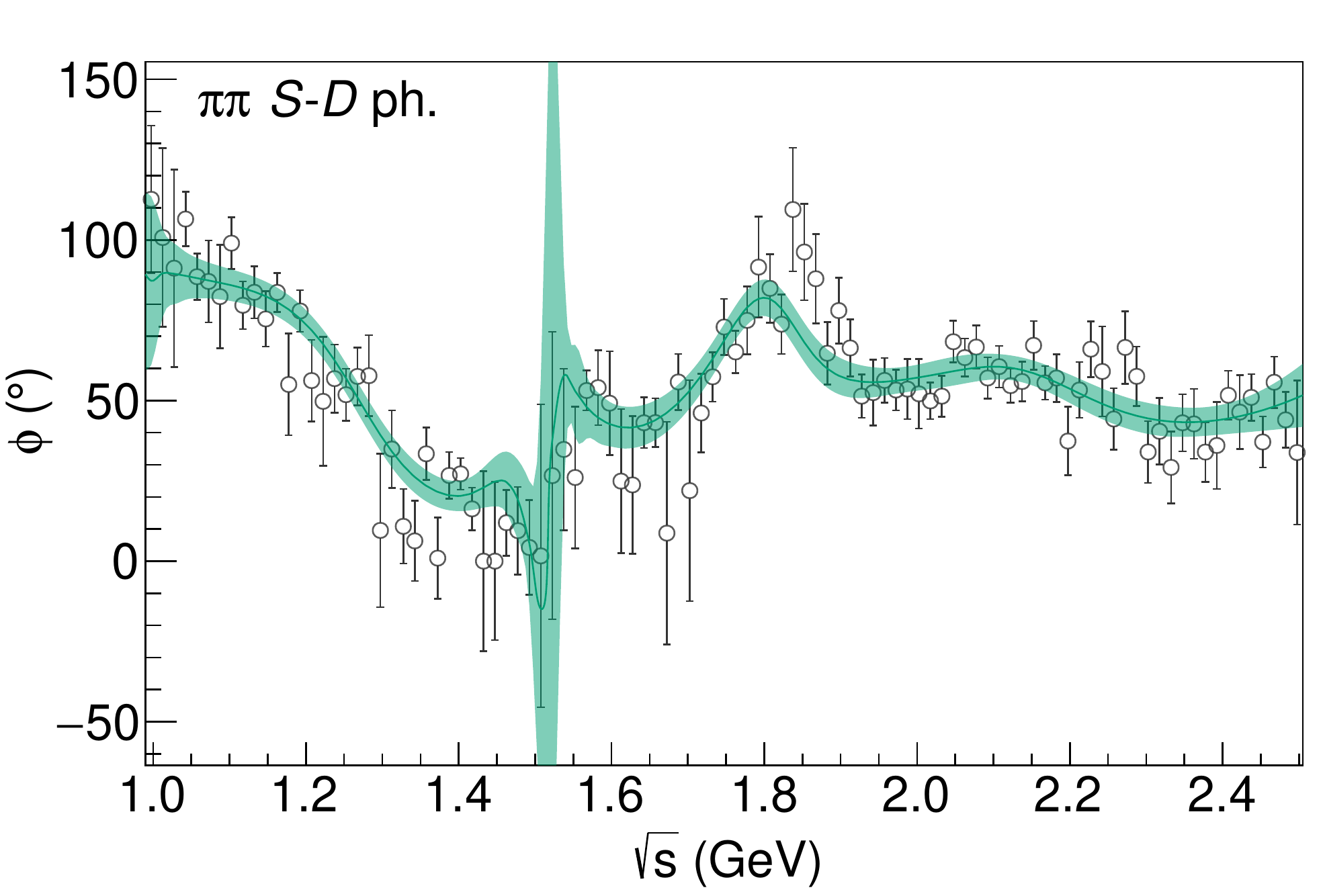}
\includegraphics[width=0.32\textwidth]{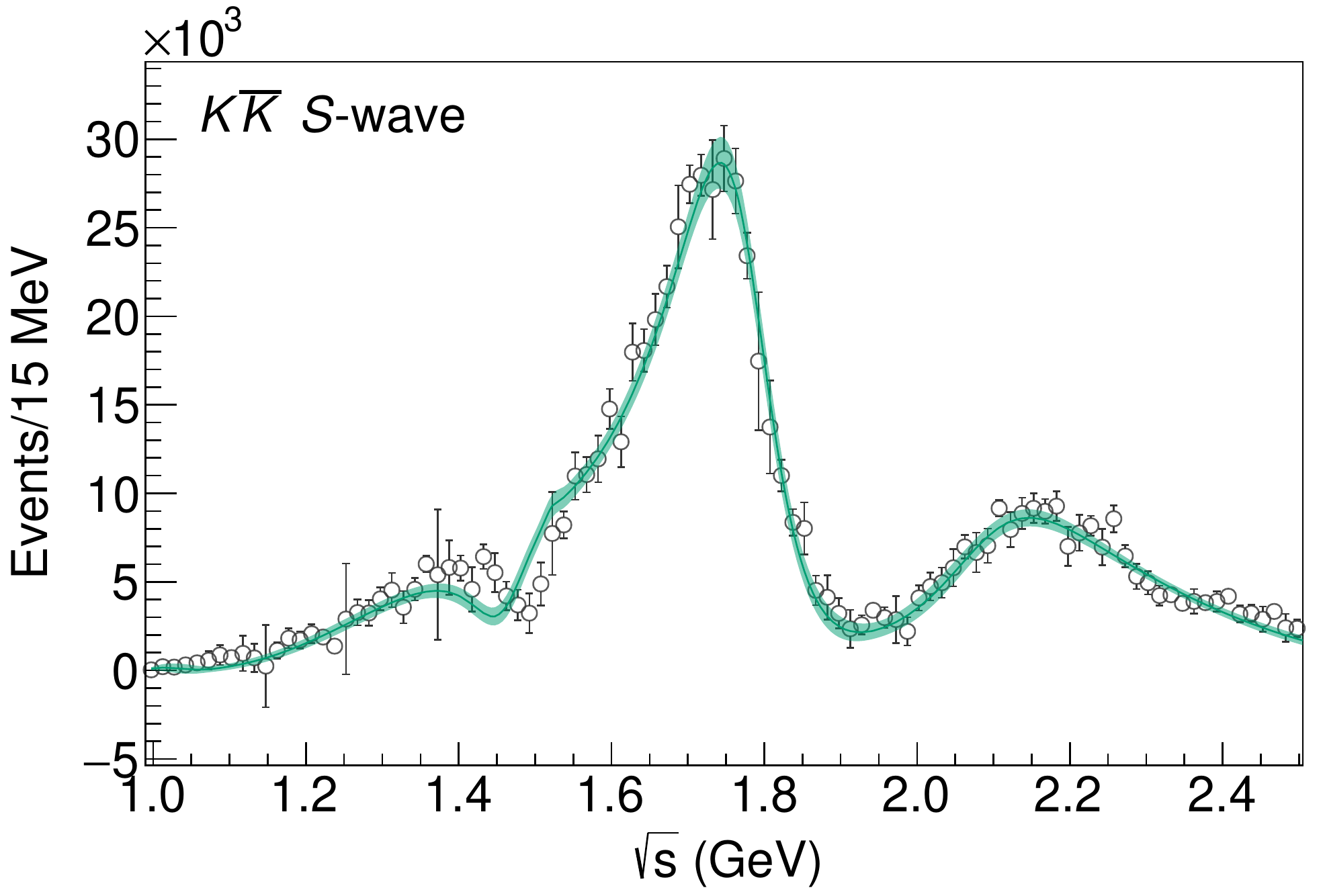} \includegraphics[width=0.32\textwidth]{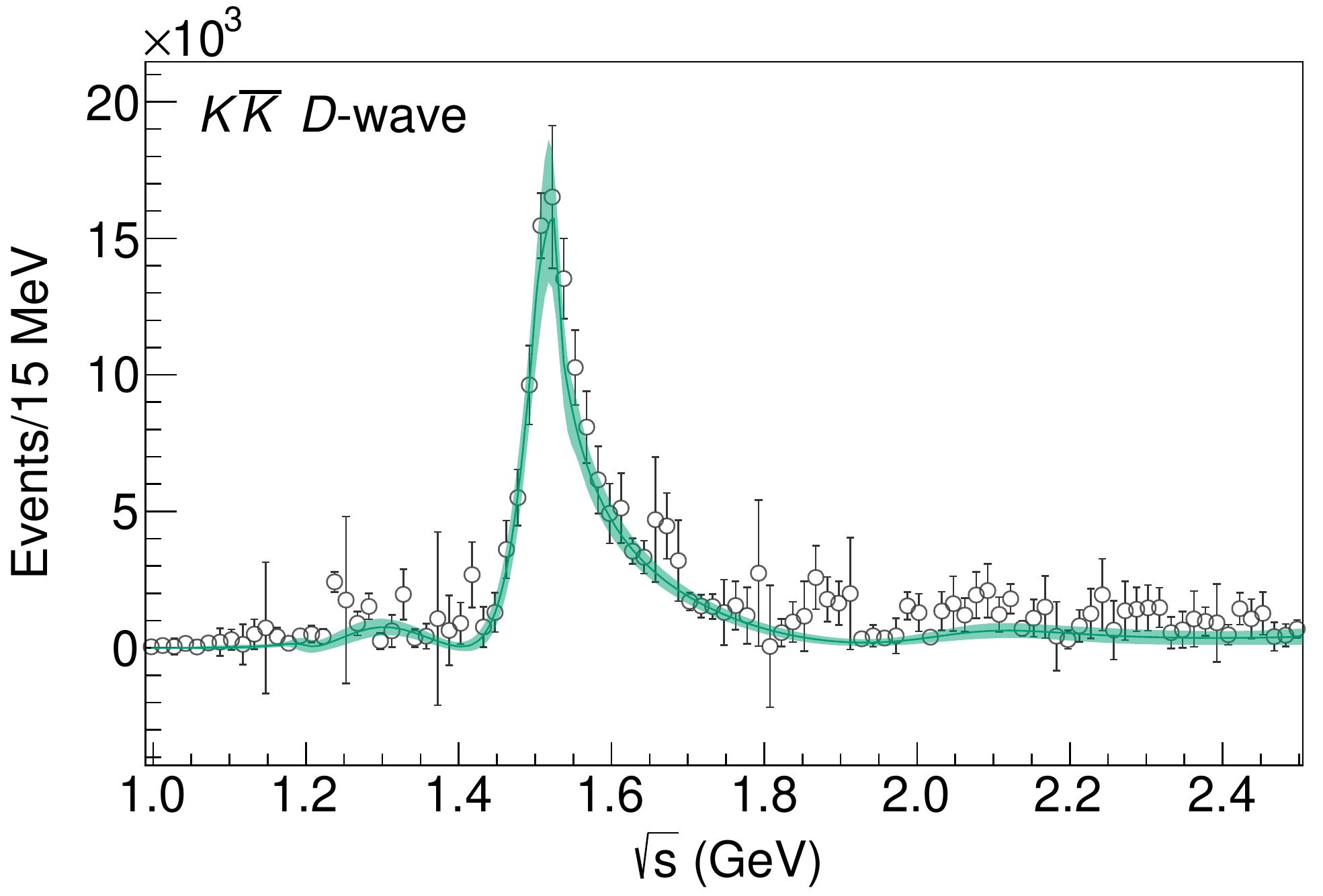} \includegraphics[width=0.32\textwidth]{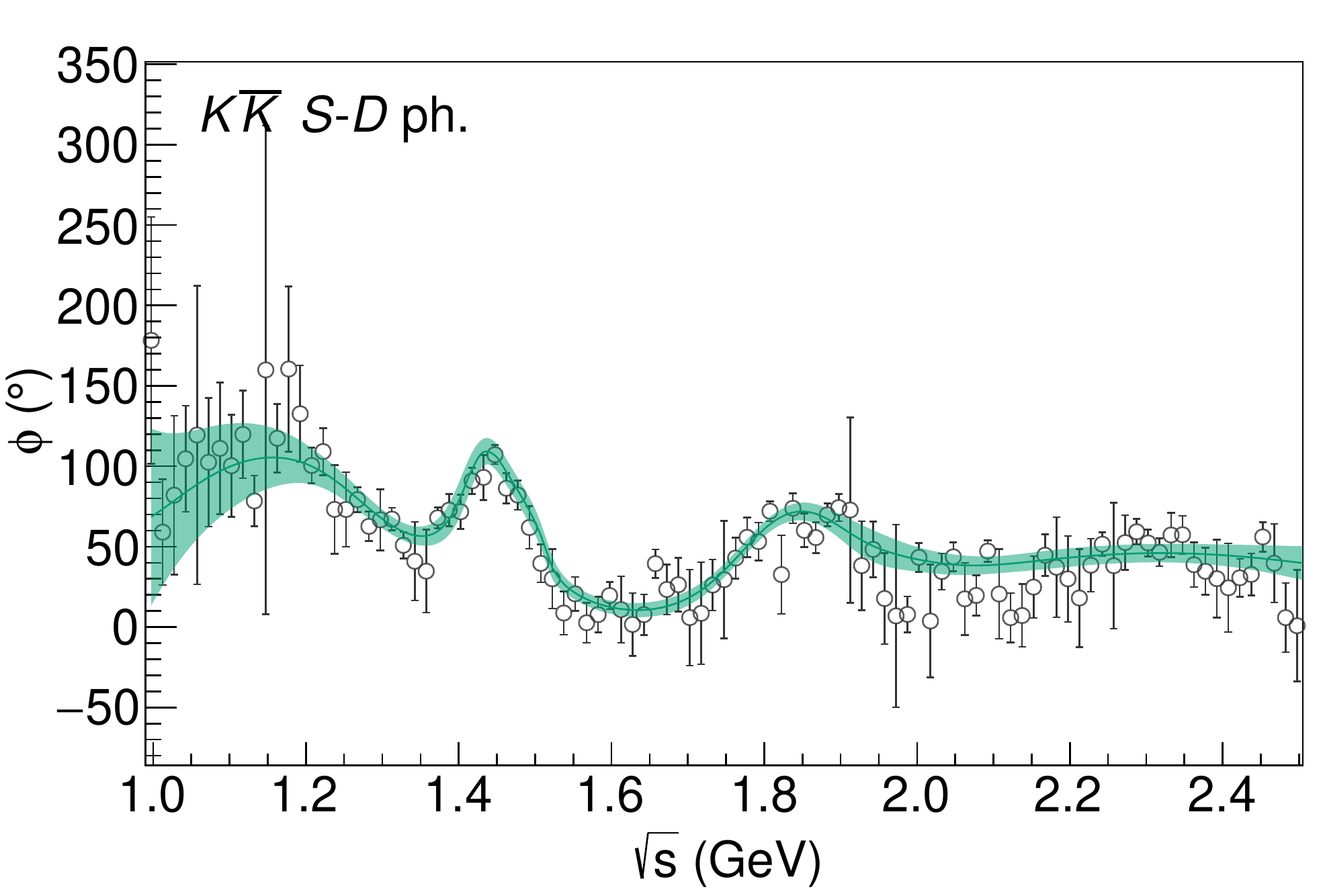}
\end{figure}

\begin{figure}[h]
\centering\includegraphics[width=0.45\textwidth]{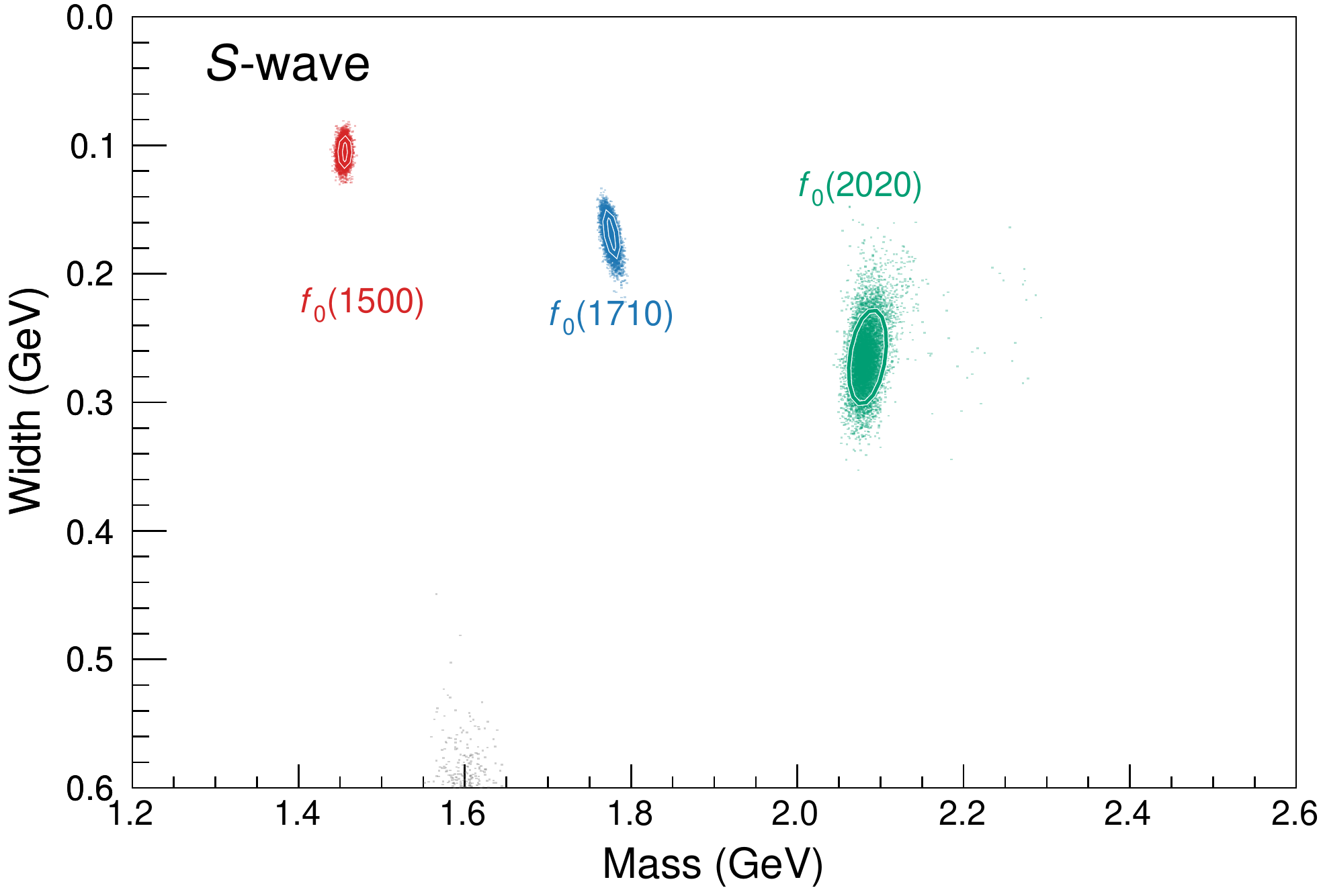} \includegraphics[width=0.45\textwidth]{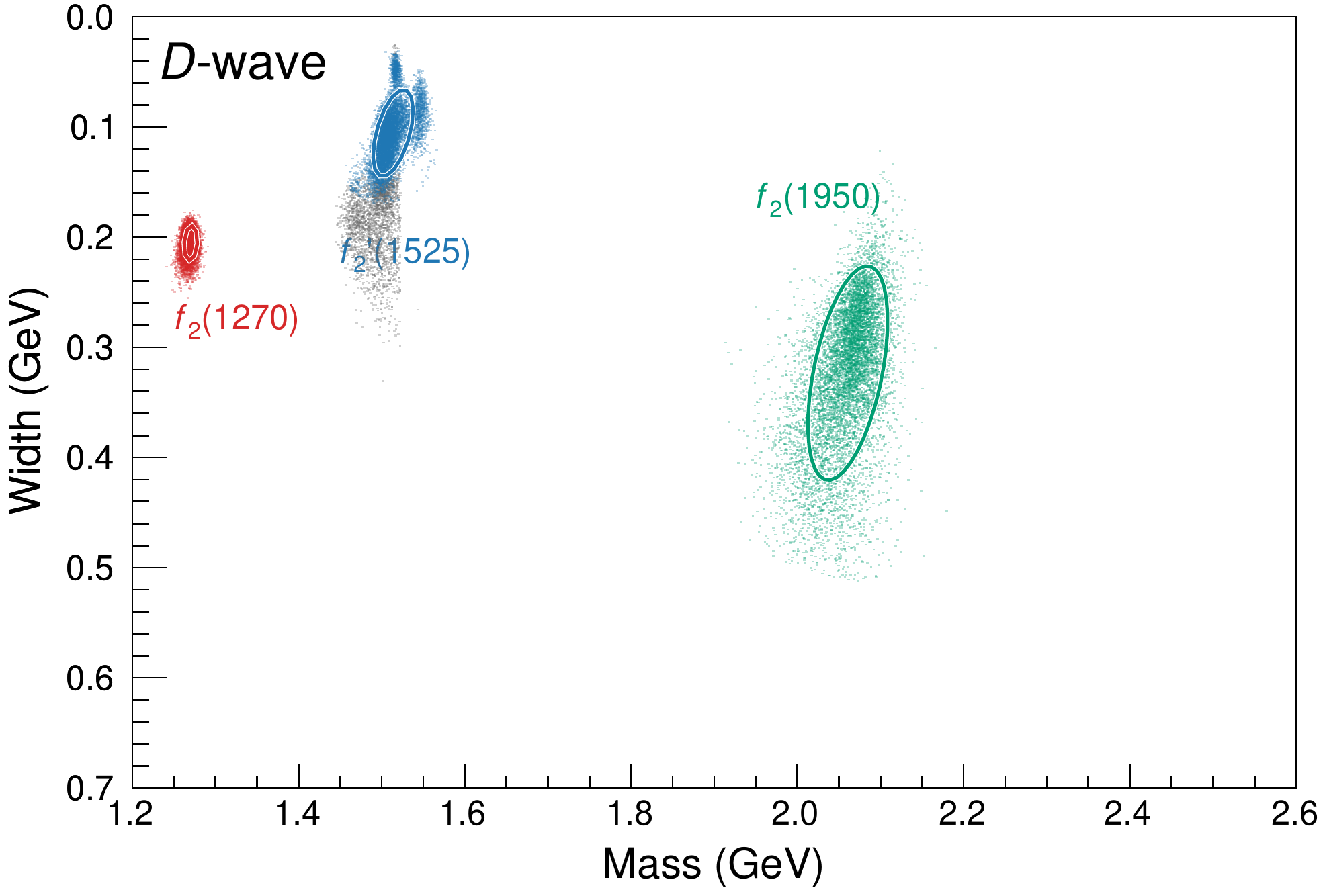}
\end{figure}

\begin{table}[h]
\begin{ruledtabular}
\begin{tabular}{c c c c}
$S$-wave  & $\sqrt{s_p}$ \mevp & $D$-wave & $\sqrt{s_p}$ \mevp \\ \hline
\rule[-0.2cm]{-0.1cm}{.55cm} $f_0(1500)$ &  $(1456 \pm 4) - i (105 \pm 7)/2$  &  $f_2(1270)$ &  $(1270 \pm 5) - i (205 \pm 10)/2$ \\
\rule[-0.2cm]{-0.1cm}{.55cm} $f_0(1710)$ &  $(1776 \pm 5) - i (169 \pm 11)/2$  &  $f_2'(1525)$ &  $(1514 \pm 16) - i (106 \pm 26)/2$ \\
\rule[-0.2cm]{-0.1cm}{.55cm} $f_0(2020)$ &  $(2084 \pm 15) - i (265 \pm 24)/2$  &  $f_2(1950)$ &  $(2061 \pm 32) - i (323 \pm 64)/2$ \\
\end{tabular}
\end{ruledtabular}
\end{table}

\clearpage

\subsection{$\left[K^J(s)^{-1}\right]^\text{CDD} \quad\Big/\quad \omega(s)_\text{scaled} \quad\Big/\quad \rho N^J_{ki}(s')_\text{Q-model} \quad\Big/\quad s_L = 0.6\gevsq$}
\label{subsec:inputcddcm7newc3_bootstrap-out}

\begin{table}[h]
\begin{ruledtabular}
\begin{tabular}{c c c c c c}
\multicolumn{3}{c}{$\pi \pi$ channel}  & \multicolumn{3}{c}{$ K  \bar K$ channel} \\ \hline
\rule[-0.2cm]{-0.1cm}{.55cm} $a^{S,\pi\pi}_0$ &$2011.105$ & $2041.445 \pm 199.672$ & $a^{S,K\bar K}_0$ & $-2104.490$ & $-2188.425 \pm 301.709$ \\
\rule[-0.2cm]{-0.1cm}{.55cm} $a^{S,\pi\pi}_1$ &$-1401.874$ & $-1434.049 \pm 199.323$ & $a^{S,K\bar K}_1$ & $6943.877$ & $7027.954 \pm 638.998$ \\
\rule[-0.2cm]{-0.1cm}{.55cm} $a^{S,\pi\pi}_2$ &$-1722.938$ & $-1711.462 \pm 336.030$ & $a^{S,K\bar K}_2$ & $-7896.936$ & $-7906.077 \pm 1008.859$ \\
\rule[-0.2cm]{-0.1cm}{.55cm} $a^{S,\pi\pi}_3$ &$1991.570$ & $2040.567 \pm 304.295$ & $a^{S,K\bar K}_3$ & $5000.622$ & $5037.468 \pm 799.560$ \\
\hline
\rule[-0.2cm]{-0.1cm}{.55cm} $a^{D,\pi\pi}_0$ &$210.226$ & $180.235 \pm 49.535$ & $a^{D,K\bar K}_0$ & $202.815$ & $214.049 \pm 33.157$ \\
\rule[-0.2cm]{-0.1cm}{.55cm} $a^{D,\pi\pi}_1$ &$110.799$ & $103.064 \pm 54.253$ & $a^{D,K\bar K}_1$ & $-289.486$ & $-288.357 \pm 36.789$ \\
\rule[-0.2cm]{-0.1cm}{.55cm} $a^{D,\pi\pi}_2$ &$-505.600$ & $-446.319 \pm 127.424$ & $a^{D,K\bar K}_2$ & $301.675$ & $285.861 \pm 57.802$ \\
\rule[-0.2cm]{-0.1cm}{.55cm} $a^{D,\pi\pi}_3$ &$377.720$ & $324.931 \pm 104.589$ & $a^{D,K\bar K}_3$ & $-38.537$ & $-22.177 \pm 53.894$ \\
\end{tabular}
\end{ruledtabular}
\end{table}

\begin{table}[h]
\begin{ruledtabular}
\begin{tabular}{c c c c c c c c c}
\multicolumn{3}{c}{$S$-wave couplings $\pi \pi$ channel}  & \multicolumn{3}{c}{$S$-wave couplings $K  \bar K$ channel} & \multicolumn{3}{c}{$S$-wave couplings $\rho\rho$ channel} \\ \hline
\rule[-0.2cm]{-0.1cm}{.55cm} $g^{S,1}_{\pi\pi}$ & $-0.027$ & $-0.037 \pm 0.033$ & $g^{S,1}_{K\bar K}$ & $0.426$ & $0.411 \pm 0.036$ & $g^{S,1}_{\rho\rho}$ & $-0.381$ & $-0.323 \pm 0.090$ \\
\rule[-0.2cm]{-0.1cm}{.55cm} $g^{S,2}_{\pi\pi}$ & $0.660$ & $0.733 \pm 0.104$ & $g^{S,2}_{K\bar K}$ & $0.725$ & $0.571 \pm 0.260$ & $g^{S,2}_{\rho\rho}$ & $0.733$ & $0.691 \pm 0.087$ \\
\rule[-0.2cm]{-0.1cm}{.55cm} $g^{S,3}_{\pi\pi}$ & $0.452$ & $0.092 \pm 0.361$ & $g^{S,3}_{K\bar K}$ & $-0.147$ & $-0.467 \pm 0.343$ & $g^{S,3}_{\rho\rho}$ & $0.166$ & $-0.084 \pm 0.238$ \\
\hline 
\multicolumn{3}{c}{$D$-wave couplings $\pi \pi$ channel}  & \multicolumn{3}{c}{$D$-wave couplings $K  \bar K$ channel} & \multicolumn{3}{c}{$D$-wave couplings $\rho\rho$ channel} \\ \hline
\rule[-0.2cm]{-0.1cm}{.55cm} $g^{D,1}_{\pi\pi}$ & $0.815$ & $0.829 \pm 0.023$ & $g^{D,1}_{K\bar K}$ & $0.909$ & $0.880 \pm 0.059$ & $g^{D,1}_{\rho\rho}$ & $1.256$ & $1.264 \pm 0.035$ \\
\rule[-0.2cm]{-0.1cm}{.55cm} $g^{D,2}_{\pi\pi}$ & $-1.370$ & $-1.347 \pm 0.101$ & $g^{D,2}_{K\bar K}$ & $0.831$ & $0.889 \pm 0.136$ & $g^{D,2}_{\rho\rho}$ & $-1.185$ & $-1.323 \pm 0.241$ \\
\rule[-0.2cm]{-0.1cm}{.55cm} $g^{D,3}_{\pi\pi}$ & $0.672$ & $0.480 \pm 0.379$ & $g^{D,3}_{K\bar K}$ & $-0.943$ & $-0.820 \pm 0.229$ & $g^{D,3}_{\rho\rho}$ & $2.061$ & $1.869 \pm 0.340$ \\
\end{tabular}
\vspace{2em}
\begin{tabular}{c c c c c c}
\multicolumn{3}{c}{$S$-wave background}  & \multicolumn{3}{c}{$D$-wave background} \\ \hline
\rule[-0.2cm]{-0.1cm}{.55cm} $c^{S,0}_{\pi\pi,\pi\pi}$ &$-2.331$ & $-2.474 \pm 0.140$ & $c^{D,0}_{\pi\pi,\pi\pi}$ &$-1.302$ & $-1.274 \pm 0.143$ \\
\rule[-0.2cm]{-0.1cm}{.55cm} $c^{S,0}_{\pi\pi,K\bar K}$ &$3.129$ & $3.301 \pm 0.186$ & $c^{D,0}_{\pi\pi,K\bar K}$ &$0.404$ & $0.472 \pm 0.160$ \\
\rule[-0.2cm]{-0.1cm}{.55cm} $c^{S,0}_{\pi\pi,\rho\rho}$ &$0$ & $0$ & $c^{D,0}_{\pi\pi,\rho\rho}$ &$0$ & $0$ \\
\rule[-0.2cm]{-0.1cm}{.55cm} $c^{S,0}_{K\bar K,K\bar K}$ &$4.240$ & $4.434 \pm 0.301$ & $c^{D,0}_{K\bar K,K\bar K}$ &$1.596$ & $1.491 \pm 0.270$ \\
\rule[-0.2cm]{-0.1cm}{.55cm} $c^{S,0}_{K\bar K,\rho\rho}$ &$0$ & $0$ & $c^{D,0}_{K\bar K,\rho\rho}$ &$0$ & $0$ \\
\rule[-0.2cm]{-0.1cm}{.55cm} $c^{S,0}_{\rho\rho,\rho\rho}$ &$1.586$ & $1.556 \pm 0.098$ & $c^{D,0}_{\rho\rho,\rho\rho}$ &$3.639$ & $3.406 \pm 0.518$ \\
\rule[-0.2cm]{-0.1cm}{.55cm} $c^{S,1}_{\pi\pi,\pi\pi}$ &$1.065$ & $1.142 \pm 0.066$ & $c^{D,1}_{\pi\pi,\pi\pi}$ &$-0.038$ & $0.022 \pm 0.102$ \\
\rule[-0.2cm]{-0.1cm}{.55cm} $c^{S,1}_{\pi\pi,K\bar K}$ &$1.228$ & $1.322 \pm 0.089$ & $c^{D,1}_{\pi\pi,K\bar K}$ &$0.114$ & $0.050 \pm 0.110$ \\
\rule[-0.2cm]{-0.1cm}{.55cm} $c^{S,1}_{\pi\pi,\rho\rho}$ &$0$ & $0$ & $c^{D,1}_{\pi\pi,\rho\rho}$ &$0$ & $0$ \\
\rule[-0.2cm]{-0.1cm}{.55cm} $c^{S,1}_{K\bar K,K\bar K}$ &$1.646$ & $1.730 \pm 0.151$ & $c^{D,1}_{K\bar K,K\bar K}$ &$-0.320$ & $-0.271 \pm 0.110$ \\
\rule[-0.2cm]{-0.1cm}{.55cm} $c^{S,1}_{K\bar K,\rho\rho}$ &$0$ & $0$ & $c^{D,1}_{K\bar K,\rho\rho}$ &$0$ & $0$ \\
\rule[-0.2cm]{-0.1cm}{.55cm} $c^{S,1}_{\rho\rho,\rho\rho}$ &$-0.747$ & $-0.731 \pm 0.047$ & $c^{D,1}_{\rho\rho,\rho\rho}$ &$0$ & $0$ \\
\end{tabular}
\end{ruledtabular}
\end{table}

\begin{figure}[h]
\centering\includegraphics[width=0.32\textwidth]{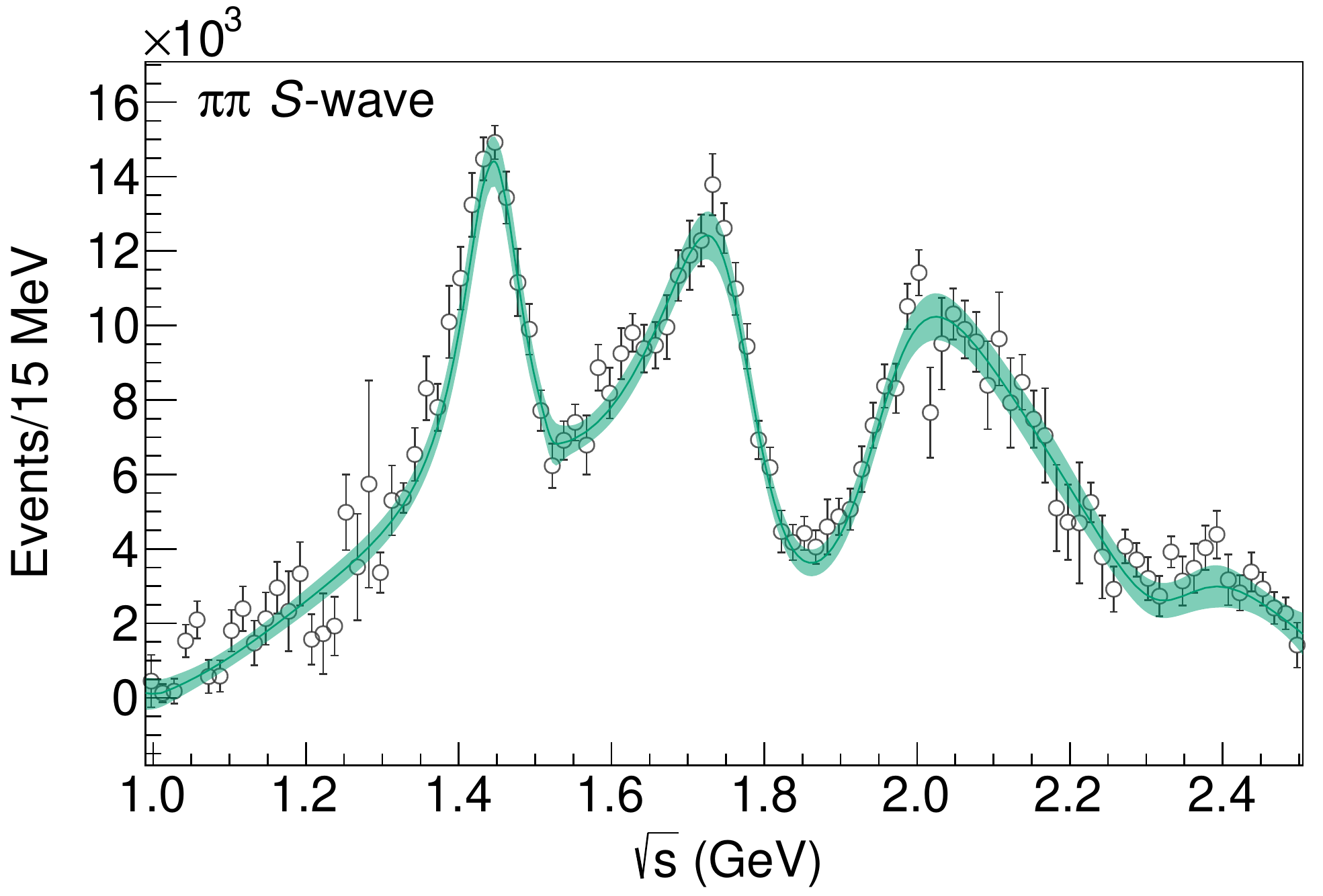} \includegraphics[width=0.32\textwidth]{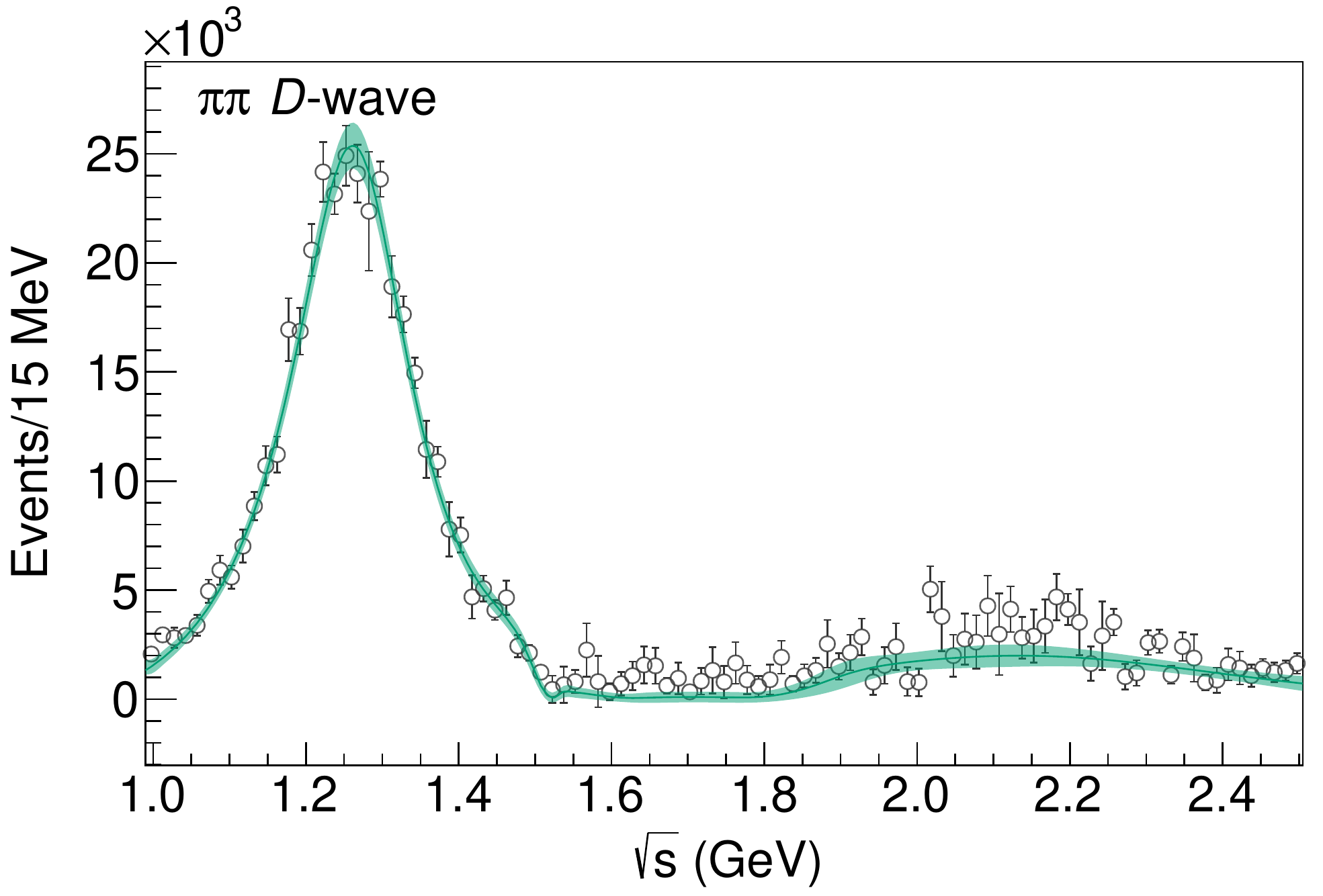} \includegraphics[width=0.32\textwidth]{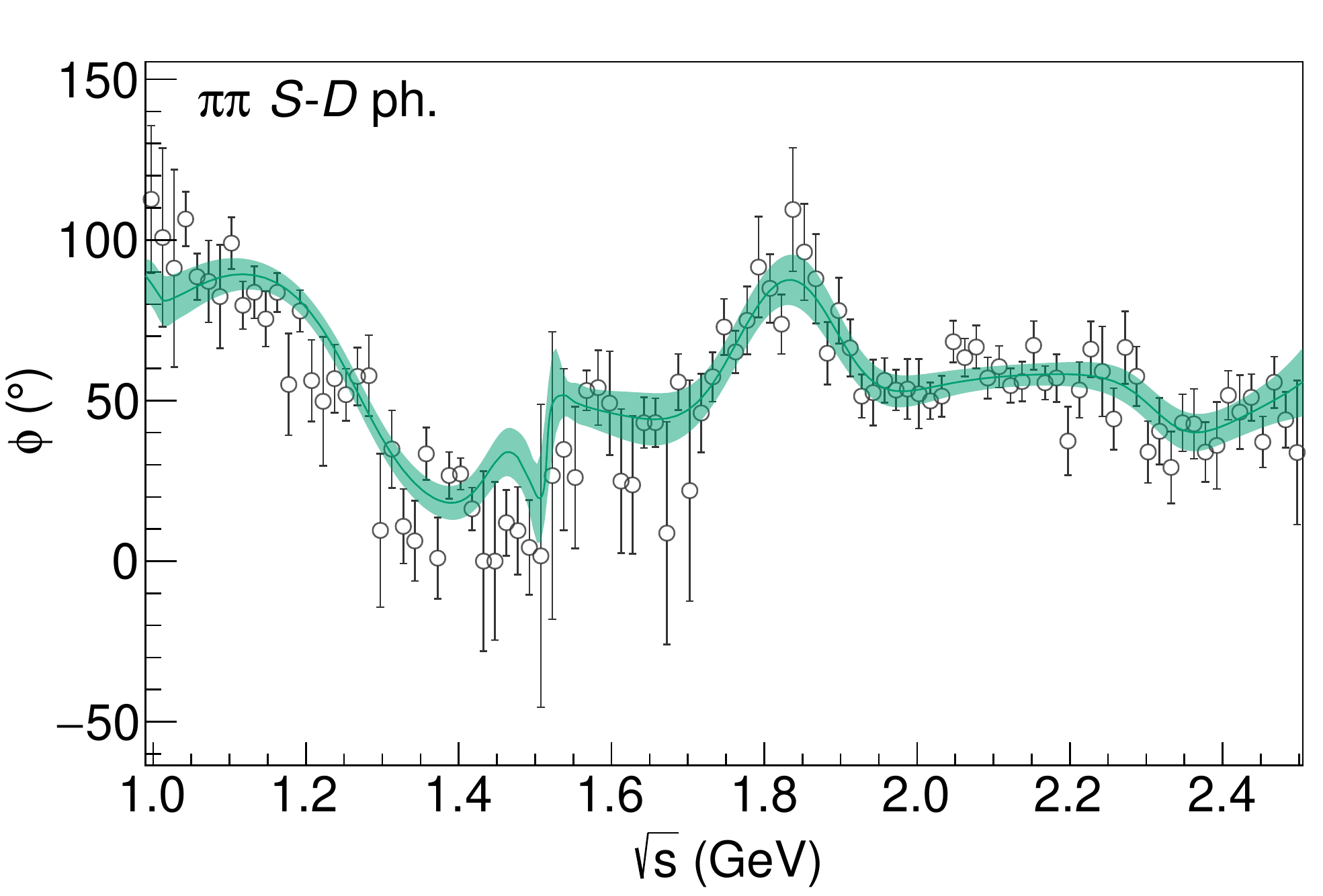}
\includegraphics[width=0.32\textwidth]{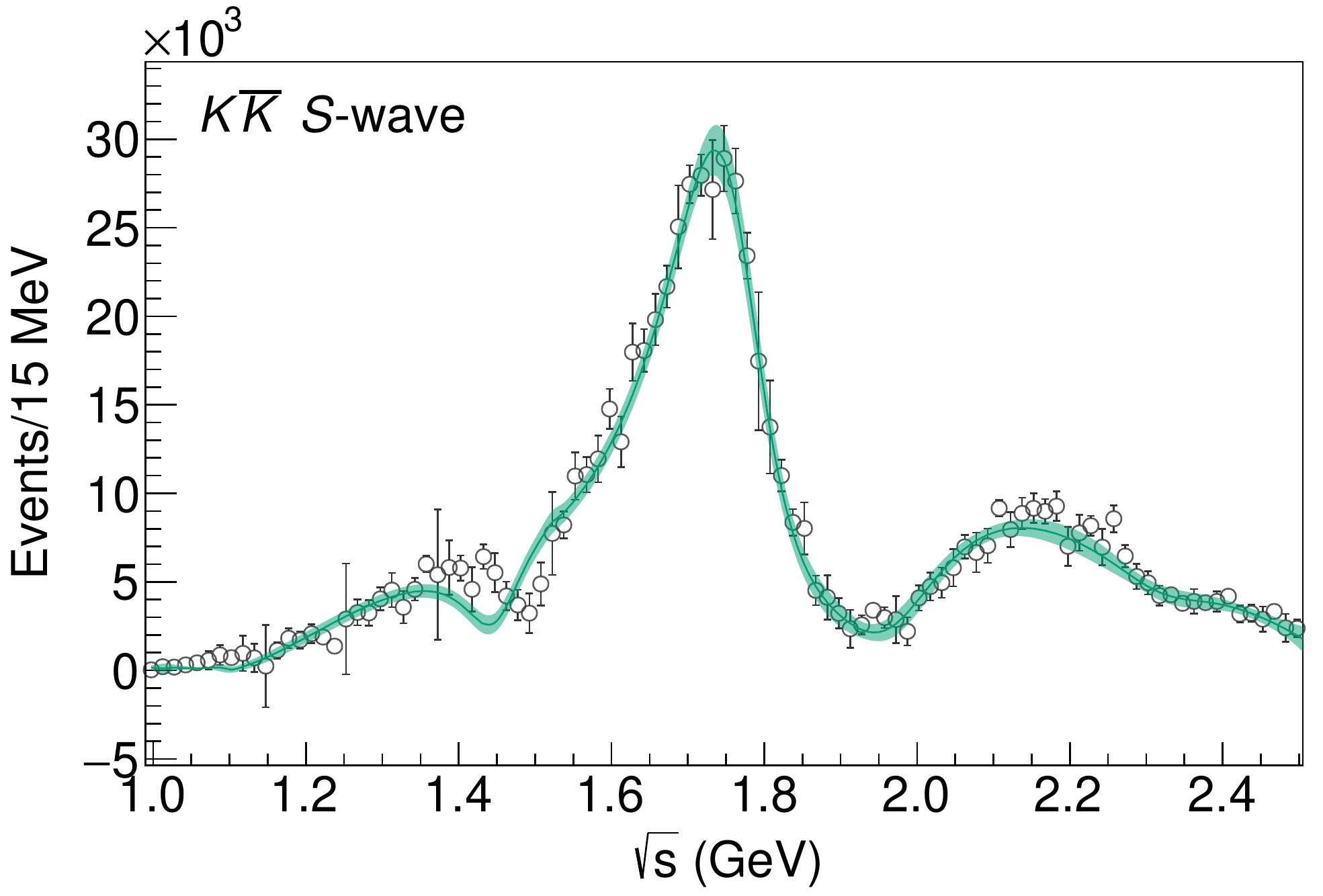} \includegraphics[width=0.32\textwidth]{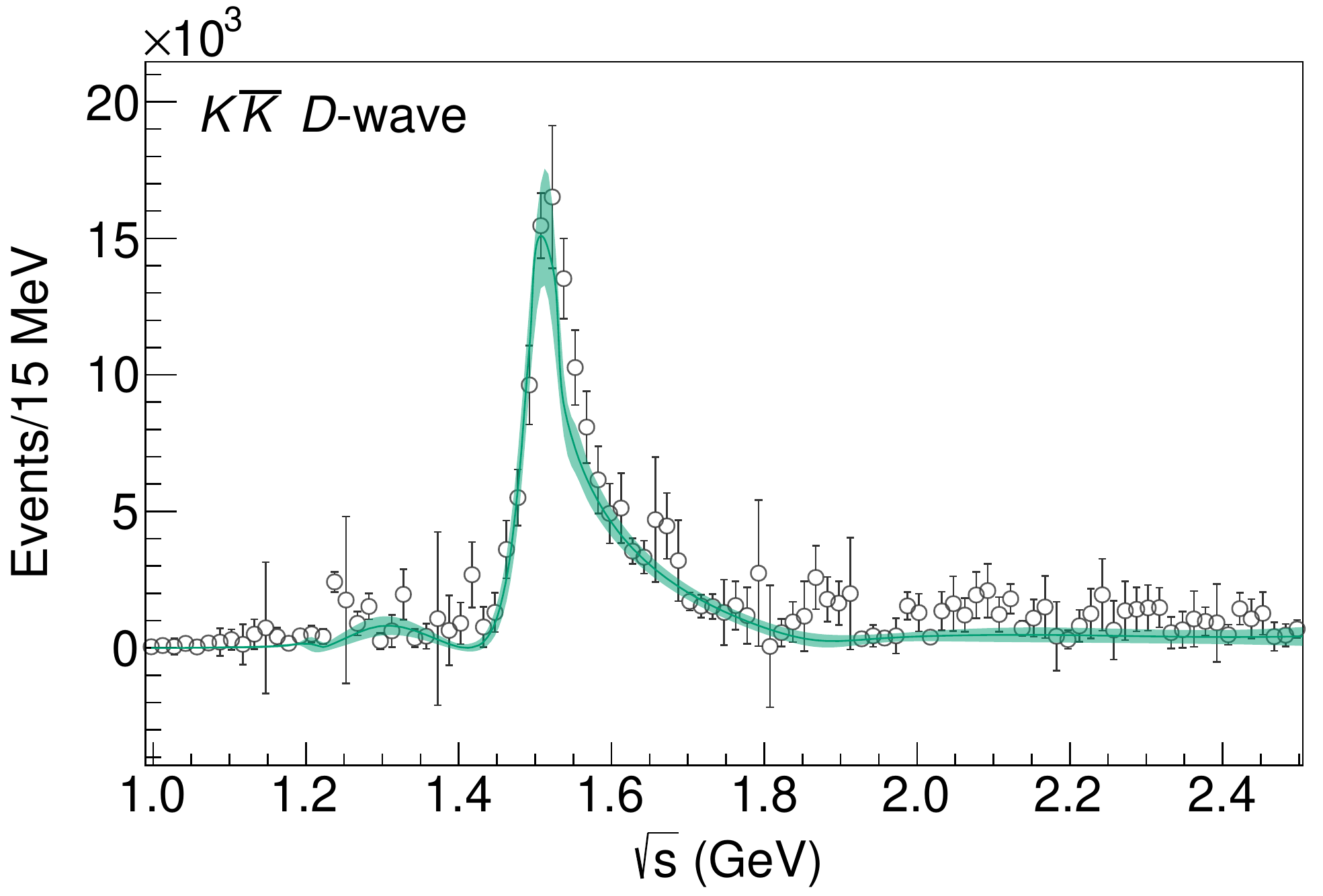} \includegraphics[width=0.32\textwidth]{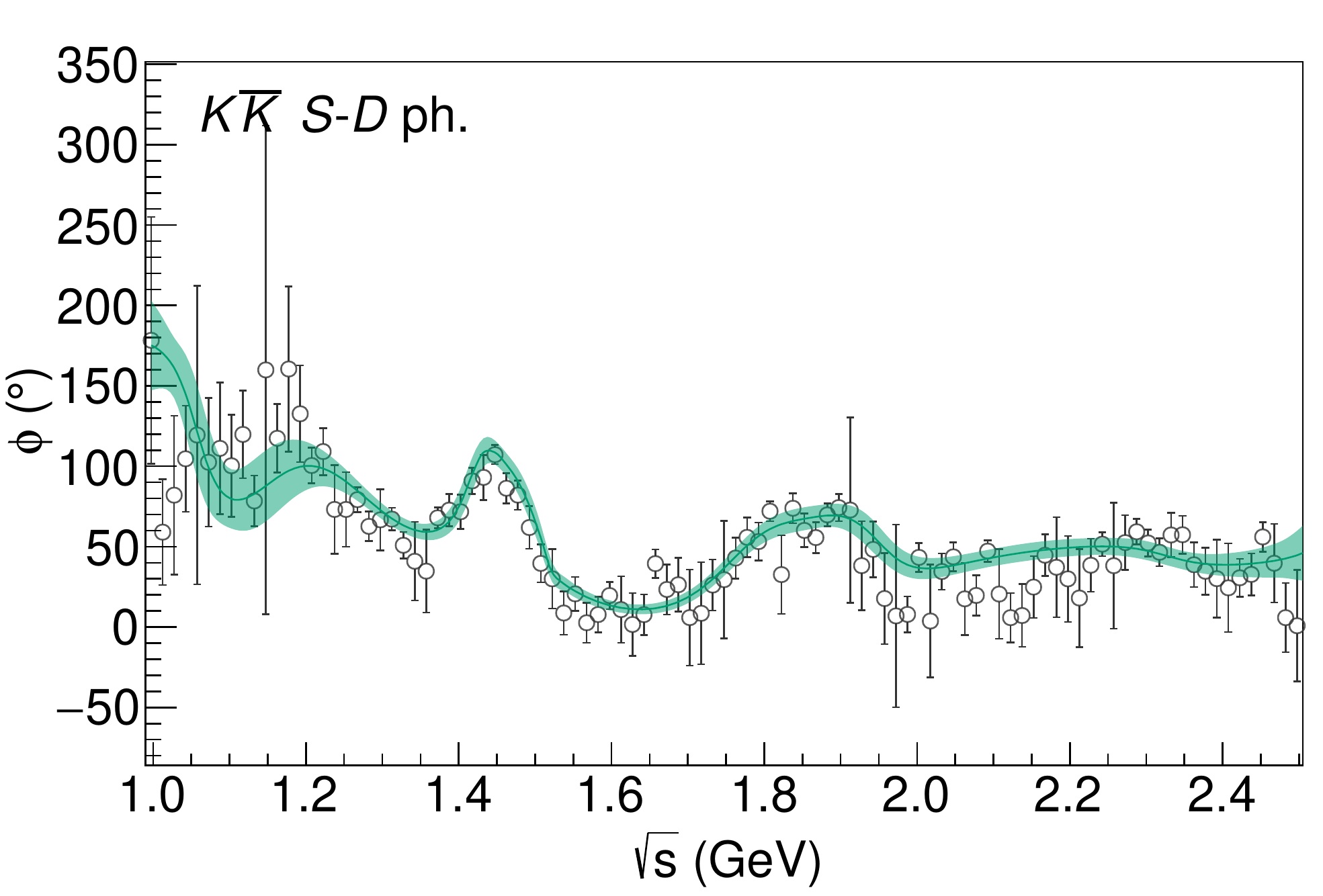}
\end{figure}

\begin{figure}[h]
\centering\includegraphics[width=0.45\textwidth]{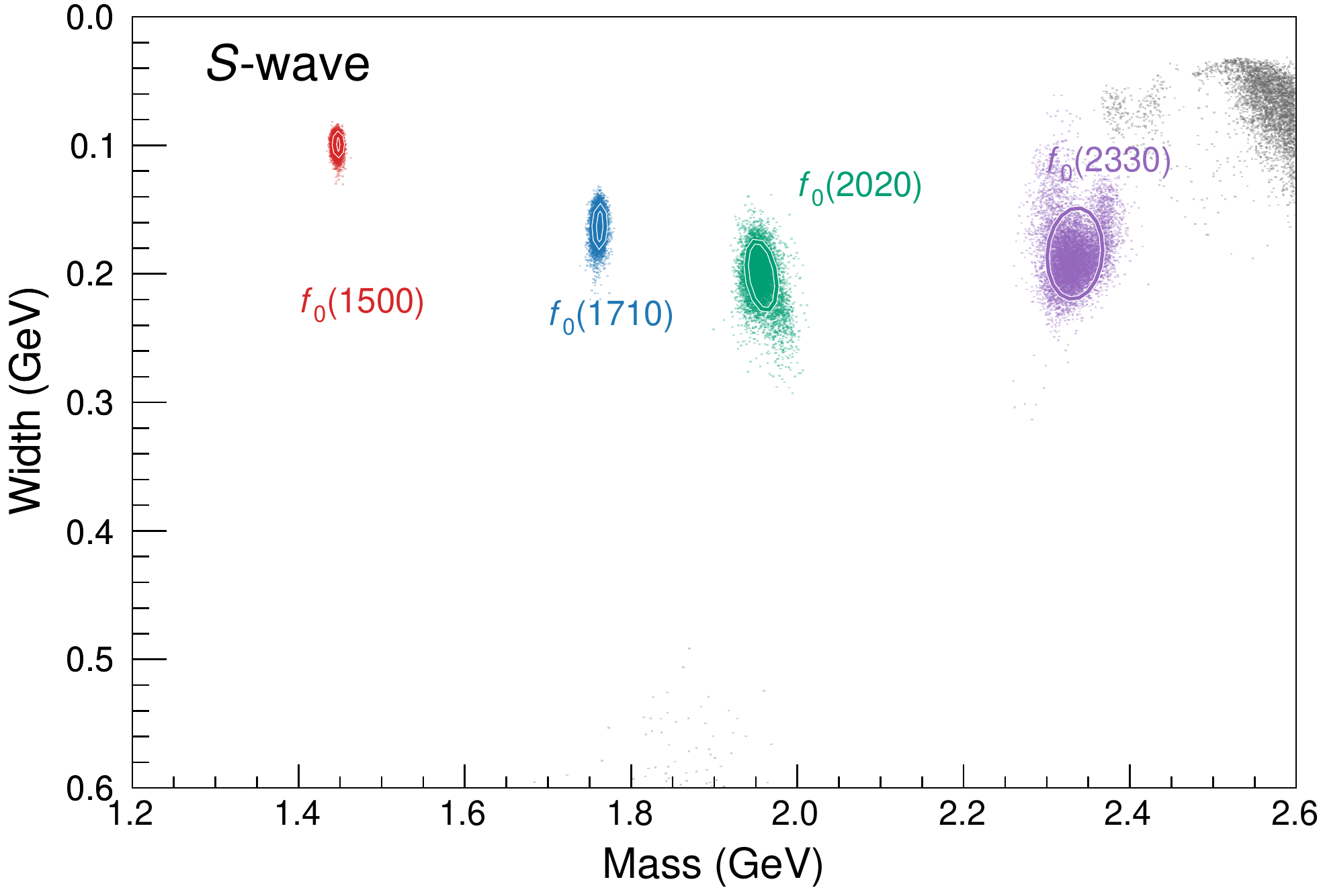} \includegraphics[width=0.45\textwidth]{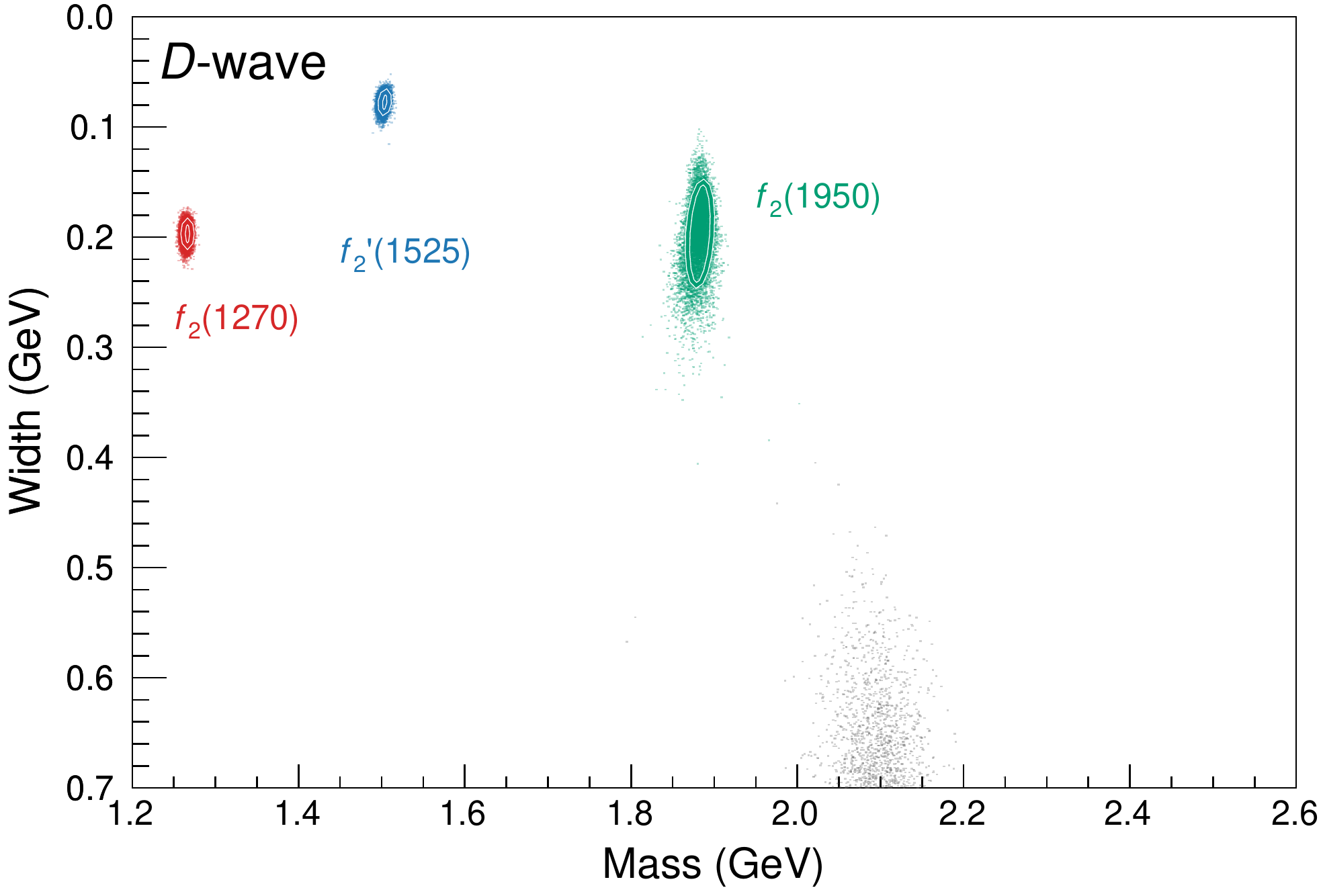}
\end{figure}

\begin{table}[h]
\begin{ruledtabular}
\begin{tabular}{c c c c}
$S$-wave  & $\sqrt{s_p}$ \mevp & $D$-wave & $\sqrt{s_p}$ \mevp \\ \hline
\rule[-0.2cm]{-0.1cm}{.55cm} $f_0(1500)$ &  $(1448 \pm 3) - i (99 \pm 5)/2$  &  $f_2(1270)$ &  $(1266 \pm 4) - i (197 \pm 7)/2$ \\
\rule[-0.2cm]{-0.1cm}{.55cm} $f_0(1710)$ &  $(1762 \pm 4) - i (163 \pm 10)/2$  &  $f_2'(1525)$ &  $(1504 \pm 4) - i (77 \pm 6)/2$ \\
\rule[-0.2cm]{-0.1cm}{.55cm} $f_0(2020)$ &  $(1957 \pm 12) - i (201 \pm 18)/2$  &  $f_2(1950)$ &  $(1883 \pm 10) - i (197 \pm 31)/2$ \\
\rule[-0.2cm]{-0.1cm}{.55cm} $f_0(2330)$ &  $(2334 \pm 22) - i (184 \pm 23)/2$  &   &   \\
\end{tabular}
\end{ruledtabular}
\end{table}

\clearpage

\subsection{$\left[K^J(s)^{-1}\right]^\text{CDD} \quad\Big/\quad \omega(s)_\text{pole} \quad\Big/\quad \rho N^J_{ki}(s')_\text{nominal}\,,\,\alpha = 0  \quad\Big/\quad s_L = 0$}
\label{subsec:inputcddcm9newc3_bootstrap-out}

\begin{table}[h]
\begin{ruledtabular}
\begin{tabular}{c c c c c c}
\multicolumn{3}{c}{$\pi \pi$ channel}  & \multicolumn{3}{c}{$ K  \bar K$ channel} \\ \hline
\rule[-0.2cm]{-0.1cm}{.55cm} $a^{S,\pi\pi}_0$ &$510.938$ & $543.985 \pm 99.170$ & $a^{S,K\bar K}_0$ & $941.510$ & $1005.002 \pm 187.049$ \\
\rule[-0.2cm]{-0.1cm}{.55cm} $a^{S,\pi\pi}_1$ &$929.856$ & $995.931 \pm 189.167$ & $a^{S,K\bar K}_1$ & $1796.670$ & $1929.720 \pm 360.442$ \\
\rule[-0.2cm]{-0.1cm}{.55cm} $a^{S,\pi\pi}_2$ &$292.581$ & $303.695 \pm 61.558$ & $a^{S,K\bar K}_2$ & $698.726$ & $724.303 \pm 130.454$ \\
\rule[-0.2cm]{-0.1cm}{.55cm} $a^{S,\pi\pi}_3$ &$-143.562$ & $-161.427 \pm 52.995$ & $a^{S,K\bar K}_3$ & $-182.237$ & $-218.616 \pm 103.124$ \\
\hline
\rule[-0.2cm]{-0.1cm}{.55cm} $a^{D,\pi\pi}_0$ &$-0.170$ & $-0.082 \pm 0.853$ & $a^{D,K\bar K}_0$ & $4.705$ & $5.130 \pm 1.336$ \\
\rule[-0.2cm]{-0.1cm}{.55cm} $a^{D,\pi\pi}_1$ &$-3.471$ & $-3.428 \pm 0.915$ & $a^{D,K\bar K}_1$ & $6.541$ & $7.045 \pm 1.590$ \\
\rule[-0.2cm]{-0.1cm}{.55cm} $a^{D,\pi\pi}_2$ &$1.087$ & $0.799 \pm 0.737$ & $a^{D,K\bar K}_2$ & $-2.844$ & $-2.722 \pm 0.868$ \\
\rule[-0.2cm]{-0.1cm}{.55cm} $a^{D,\pi\pi}_3$ &$-0.198$ & $-0.283 \pm 0.487$ & $a^{D,K\bar K}_3$ & $-0.893$ & $-0.995 \pm 0.602$ \\
\end{tabular}
\end{ruledtabular}
\end{table}

\begin{table}[h]
\begin{ruledtabular}
\begin{tabular}{c c c c c c c c c}
\multicolumn{3}{c}{$S$-wave couplings $\pi \pi$ channel}  & \multicolumn{3}{c}{$S$-wave couplings $K  \bar K$ channel} & \multicolumn{3}{c}{$S$-wave couplings $\rho\rho$ channel} \\ \hline
\rule[-0.2cm]{-0.1cm}{.55cm} $g^{S,1}_{\pi\pi}$ & $0.209$ & $0.261 \pm 0.140$ & $g^{S,1}_{K\bar K}$ & $1.218$ & $1.200 \pm 0.234$ & $g^{S,1}_{\rho\rho}$ & $0.713$ & $0.689 \pm 0.228$ \\
\rule[-0.2cm]{-0.1cm}{.55cm} $g^{S,2}_{\pi\pi}$ & $2.125$ & $2.423 \pm 0.533$ & $g^{S,2}_{K\bar K}$ & $5.977$ & $6.721 \pm 0.997$ & $g^{S,2}_{\rho\rho}$ & $-2.331$ & $-1.728 \pm 0.771$ \\
\rule[-0.2cm]{-0.1cm}{.55cm} $g^{S,3}_{\pi\pi}$ & $2.496$ & $2.105 \pm 0.538$ & $g^{S,3}_{K\bar K}$ & $4.818$ & $3.747 \pm 1.476$ & $g^{S,3}_{\rho\rho}$ & $1.173$ & $1.036 \pm 0.295$ \\
\hline 
\multicolumn{3}{c}{$D$-wave couplings $\pi \pi$ channel}  & \multicolumn{3}{c}{$D$-wave couplings $K  \bar K$ channel} & \multicolumn{3}{c}{$D$-wave couplings $\rho\rho$ channel} \\ \hline
\rule[-0.2cm]{-0.1cm}{.55cm} $g^{D,1}_{\pi\pi}$ & $-2.127$ & $-2.633 \pm 1.013$ & $g^{D,1}_{K\bar K}$ & $-1.110$ & $-1.197 \pm 0.241$ & $g^{D,1}_{\rho\rho}$ & $0.371$ & $0.290 \pm 0.204$ \\
\rule[-0.2cm]{-0.1cm}{.55cm} $g^{D,2}_{\pi\pi}$ & $0.779$ & $0.776 \pm 0.474$ & $g^{D,2}_{K\bar K}$ & $3.257$ & $3.212 \pm 0.181$ & $g^{D,2}_{\rho\rho}$ & $-0.236$ & $-0.231 \pm 0.090$ \\
\rule[-0.2cm]{-0.1cm}{.55cm} $g^{D,3}_{\pi\pi}$ & $-11.473$ & $-11.247 \pm 8.349$ & $g^{D,3}_{K\bar K}$ & $-6.087$ & $-5.168 \pm 3.454$ & $g^{D,3}_{\rho\rho}$ & $-3.125$ & $-2.581 \pm 1.328$ \\
\end{tabular}
\vspace{2em}
\begin{tabular}{c c c c c c}
\multicolumn{3}{c}{$S$-wave background}  & \multicolumn{3}{c}{$D$-wave background} \\ \hline
\rule[-0.2cm]{-0.1cm}{.55cm} $c^{S,0}_{\pi\pi,\pi\pi}$ &$3.046$ & $3.158 \pm 0.317$ & $c^{D,0}_{\pi\pi,\pi\pi}$ &$-8.984$ & $-20.512 \pm 45.501$ \\
\rule[-0.2cm]{-0.1cm}{.55cm} $c^{S,0}_{\pi\pi,K\bar K}$ &$3.989$ & $4.166 \pm 0.447$ & $c^{D,0}_{\pi\pi,K\bar K}$ &$4.732$ & $0.904 \pm 16.047$ \\
\rule[-0.2cm]{-0.1cm}{.55cm} $c^{S,0}_{\pi\pi,\rho\rho}$ &$0$ & $0$ & $c^{D,0}_{\pi\pi,\rho\rho}$ &$-25.612$ & $-22.526 \pm 11.294$ \\
\rule[-0.2cm]{-0.1cm}{.55cm} $c^{S,0}_{K\bar K,K\bar K}$ &$5.335$ & $5.621 \pm 0.645$ & $c^{D,0}_{K\bar K,K\bar K}$ &$21.301$ & $19.342 \pm 7.393$ \\
\rule[-0.2cm]{-0.1cm}{.55cm} $c^{S,0}_{K\bar K,\rho\rho}$ &$0$ & $0$ & $c^{D,0}_{K\bar K,\rho\rho}$ &$-6.836$ & $-4.858 \pm 4.765$ \\
\rule[-0.2cm]{-0.1cm}{.55cm} $c^{S,0}_{\rho\rho,\rho\rho}$ &$1.259$ & $1.107 \pm 0.272$ & $c^{D,0}_{\rho\rho,\rho\rho}$ &$0.489$ & $0.939 \pm 2.101$ \\
\rule[-0.2cm]{-0.1cm}{.55cm} $c^{S,1}_{\pi\pi,\pi\pi}$ &$-1.529$ & $-1.604 \pm 0.196$ & $c^{D,1}_{\pi\pi,\pi\pi}$ &$-25.360$ & $-20.502 \pm 17.680$ \\
\rule[-0.2cm]{-0.1cm}{.55cm} $c^{S,1}_{\pi\pi,K\bar K}$ &$-1.765$ & $-1.880 \pm 0.305$ & $c^{D,1}_{\pi\pi,K\bar K}$ &$-14.444$ & $-11.906 \pm 7.583$ \\
\rule[-0.2cm]{-0.1cm}{.55cm} $c^{S,1}_{\pi\pi,\rho\rho}$ &$0$ & $0$ & $c^{D,1}_{\pi\pi,\rho\rho}$ &$0$ & $0$ \\
\rule[-0.2cm]{-0.1cm}{.55cm} $c^{S,1}_{K\bar K,K\bar K}$ &$1.956$ & $2.120 \pm 0.525$ & $c^{D,1}_{K\bar K,K\bar K}$ &$-8.864$ & $-7.494 \pm 4.503$ \\
\rule[-0.2cm]{-0.1cm}{.55cm} $c^{S,1}_{K\bar K,\rho\rho}$ &$0$ & $0$ & $c^{D,1}_{K\bar K,\rho\rho}$ &$0$ & $0$ \\
\rule[-0.2cm]{-0.1cm}{.55cm} $c^{S,1}_{\rho\rho,\rho\rho}$ &$0$ & $0$ & $c^{D,1}_{\rho\rho,\rho\rho}$ &$0$ & $0$ \\
\end{tabular}
\end{ruledtabular}
\end{table}

\begin{figure}[h]
\centering\includegraphics[width=0.32\textwidth]{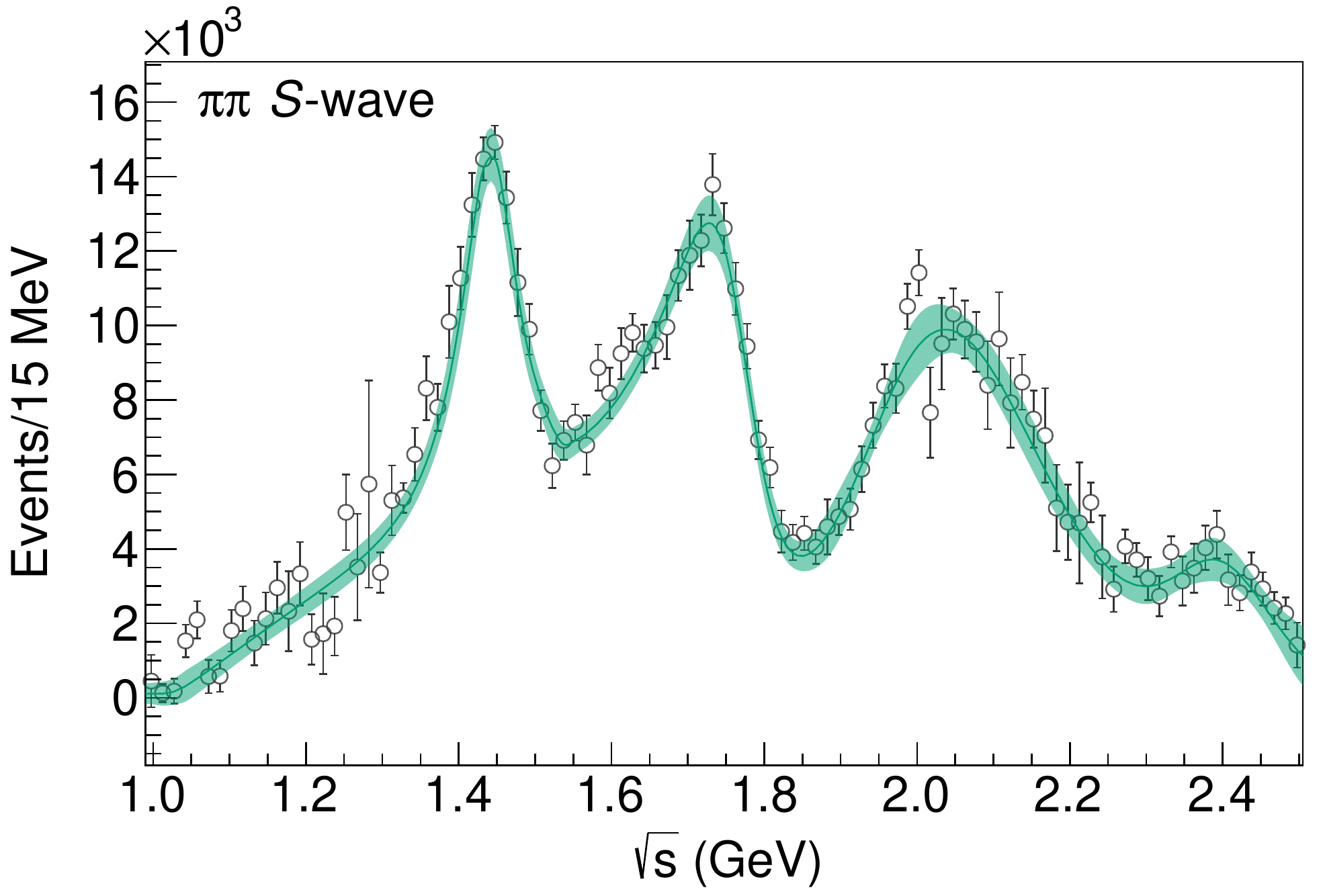} \includegraphics[width=0.32\textwidth]{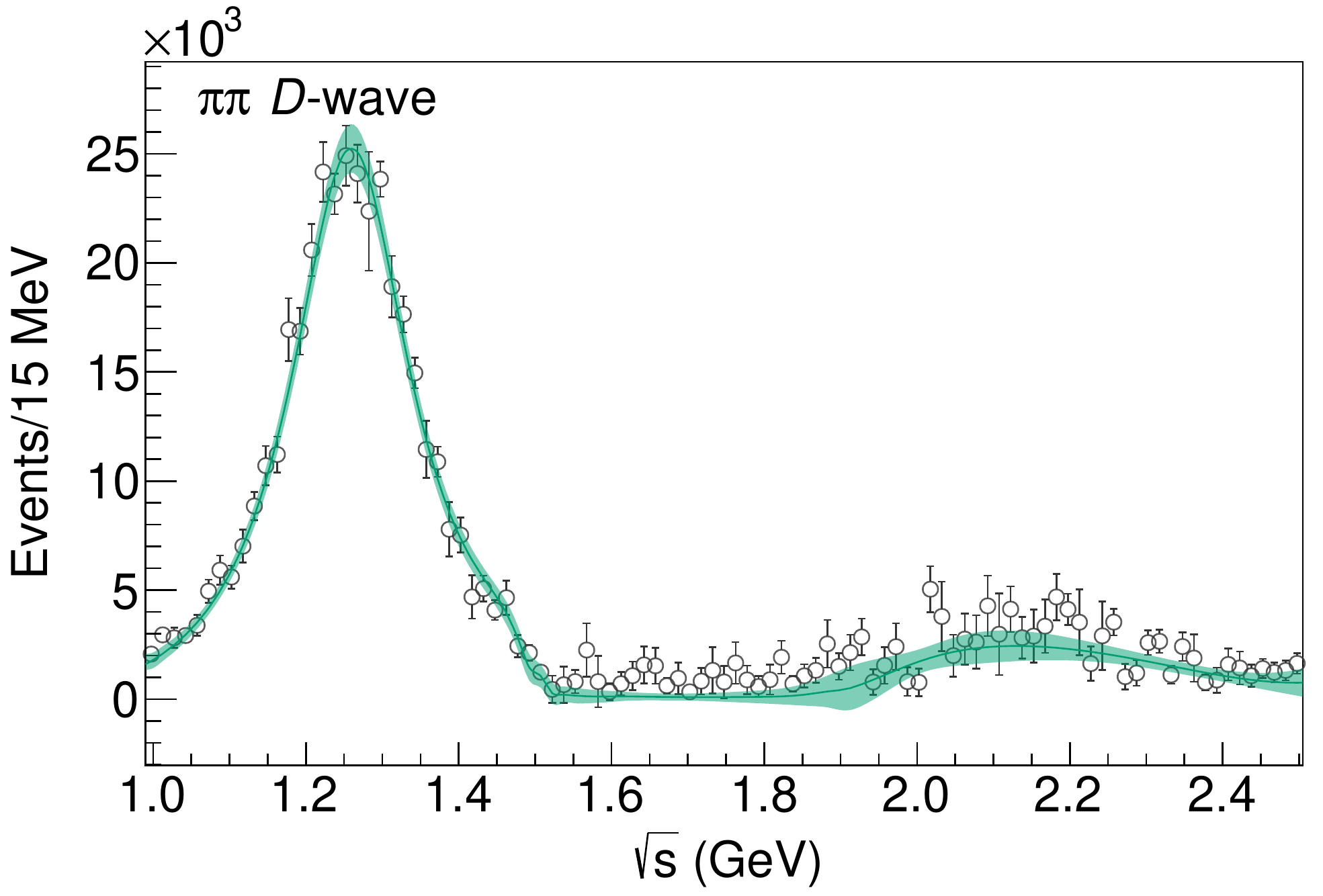} \includegraphics[width=0.32\textwidth]{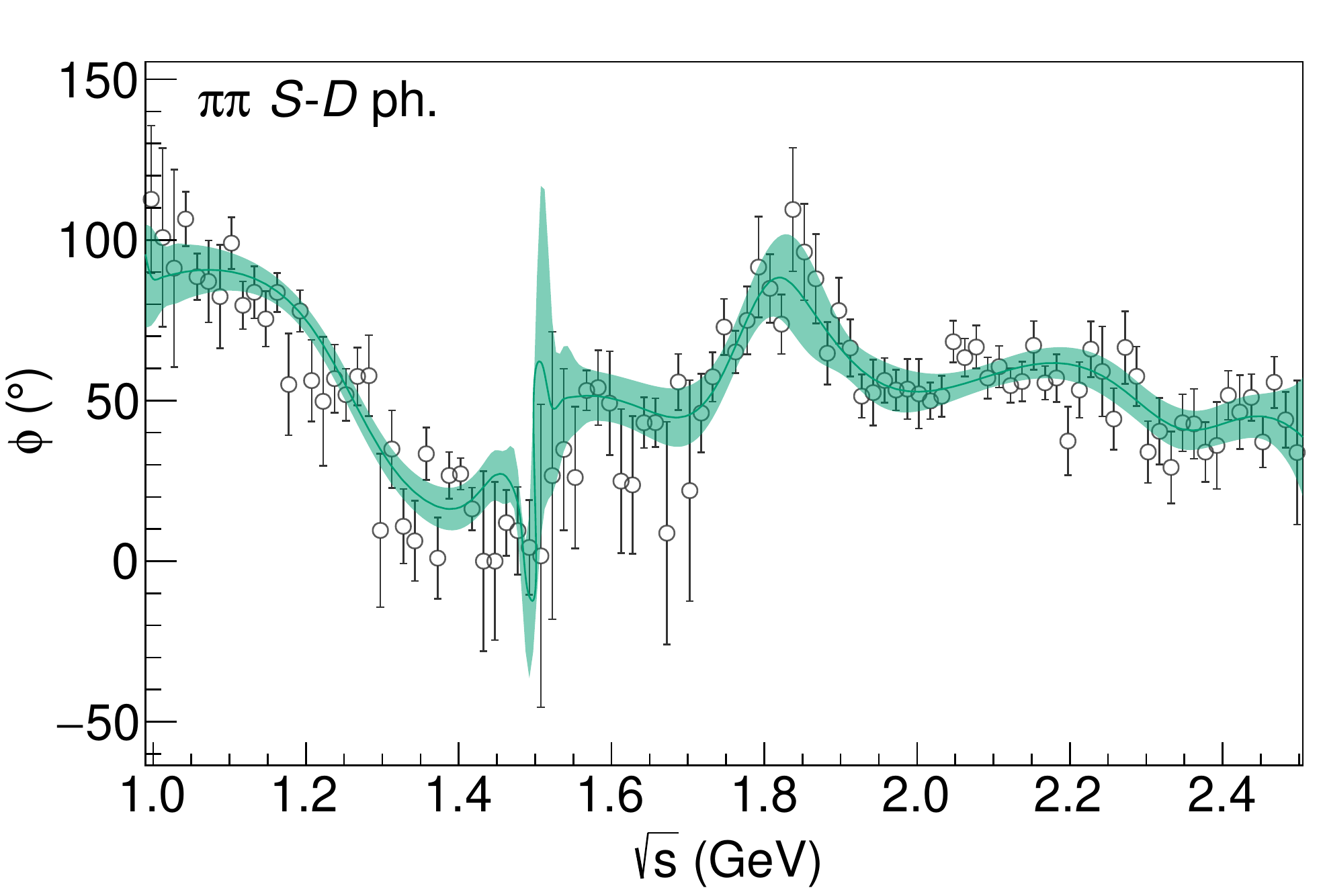}
\includegraphics[width=0.32\textwidth]{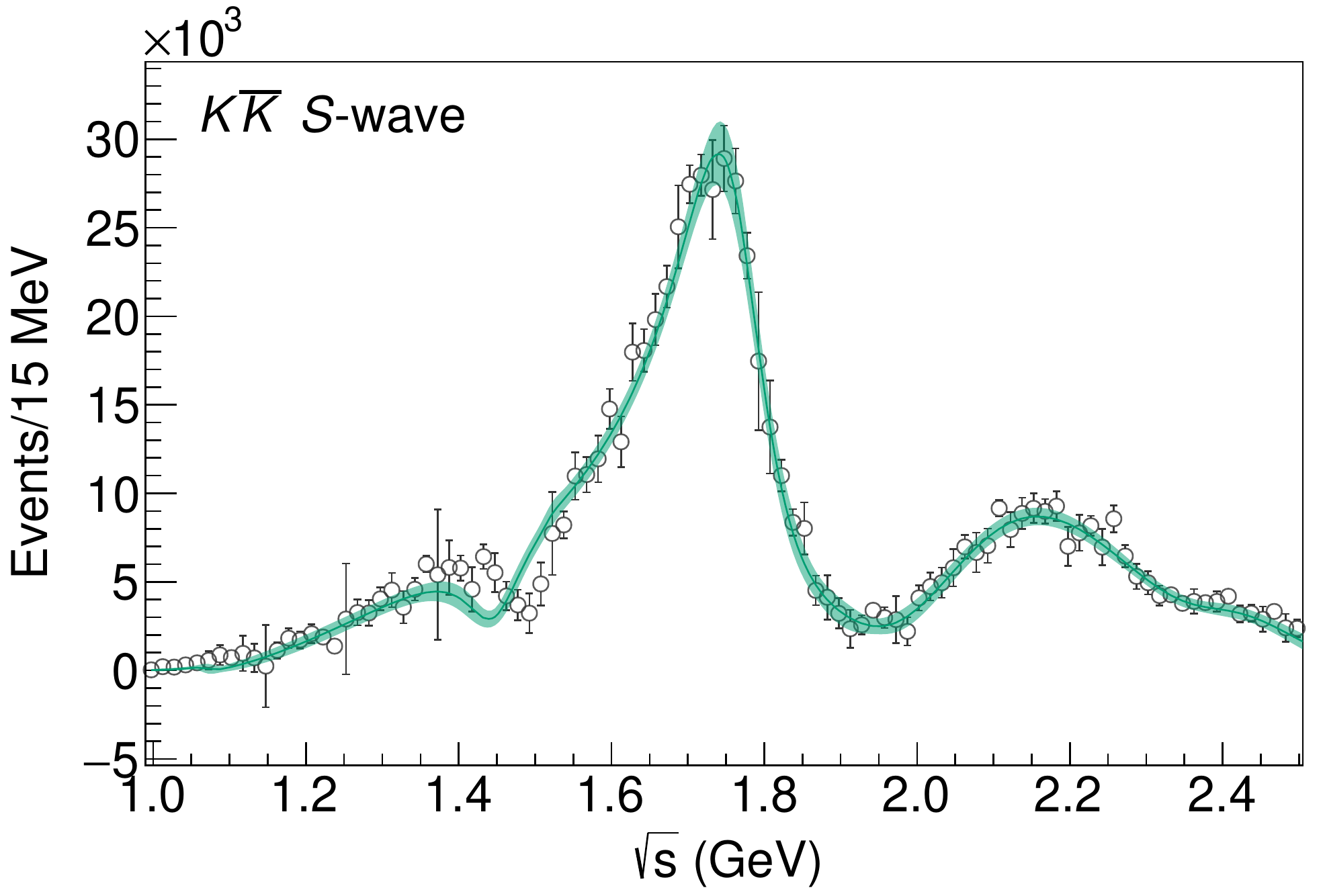} \includegraphics[width=0.32\textwidth]{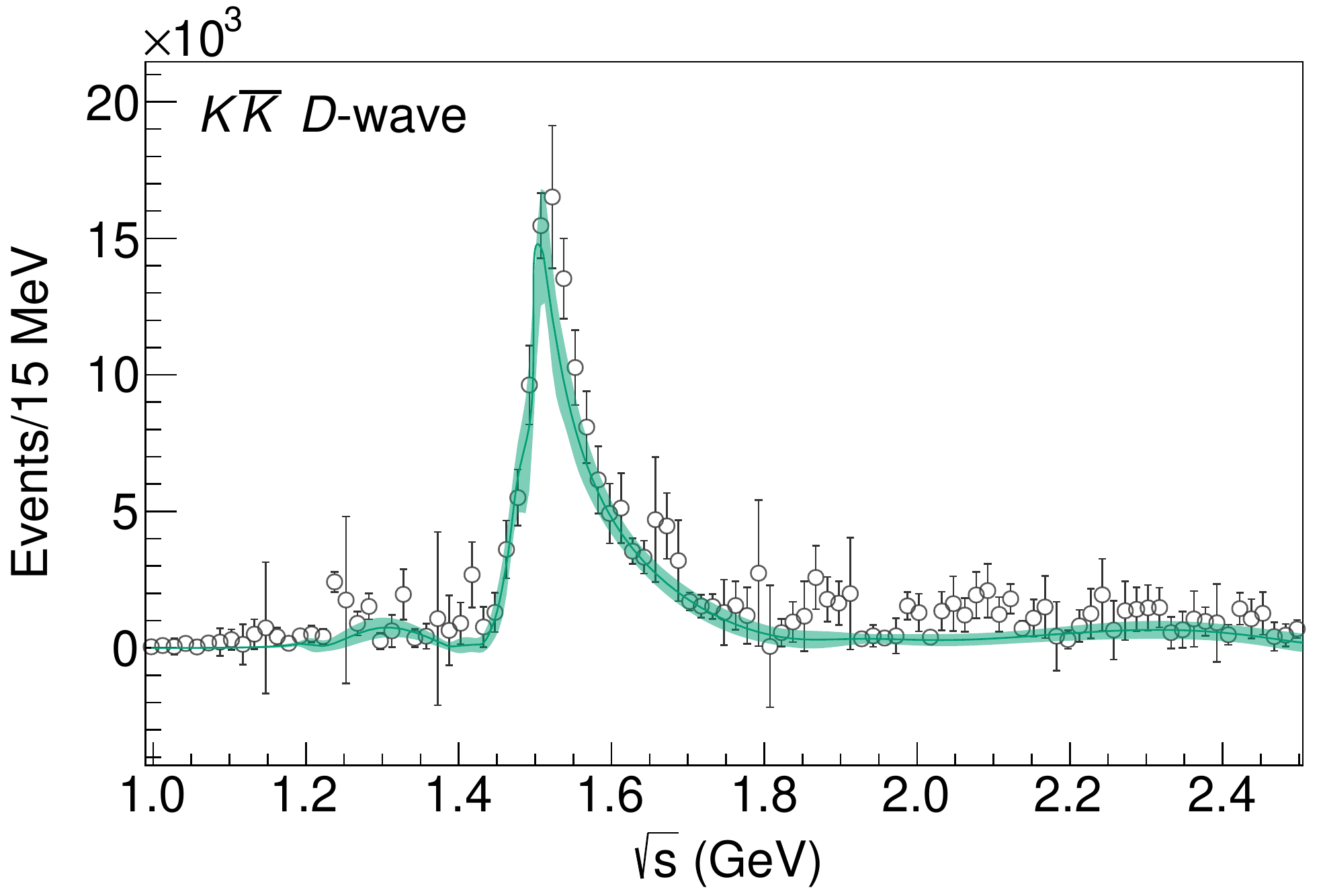} \includegraphics[width=0.32\textwidth]{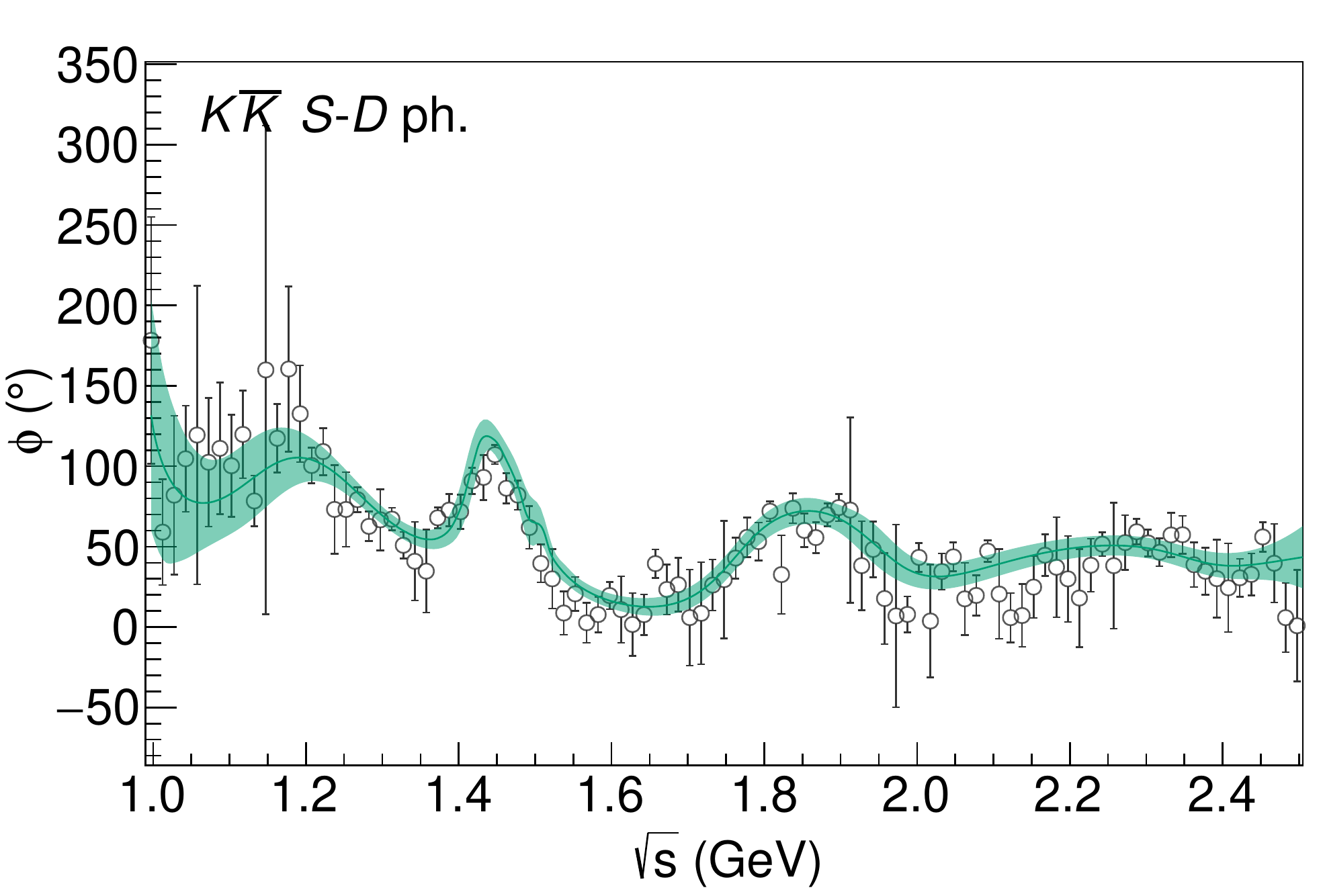}
\end{figure}

\begin{figure}[h]
\centering\includegraphics[width=0.45\textwidth]{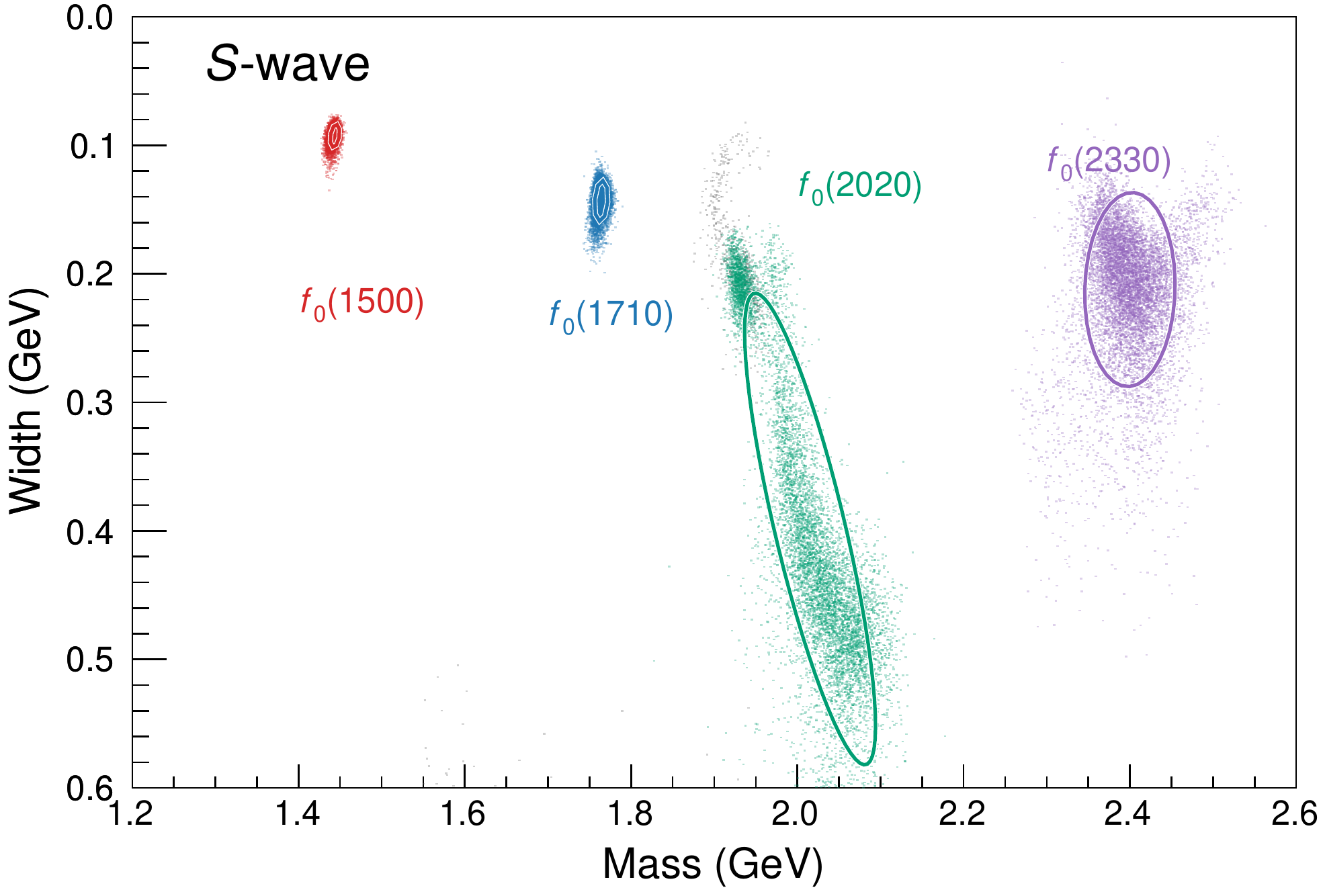} \includegraphics[width=0.45\textwidth]{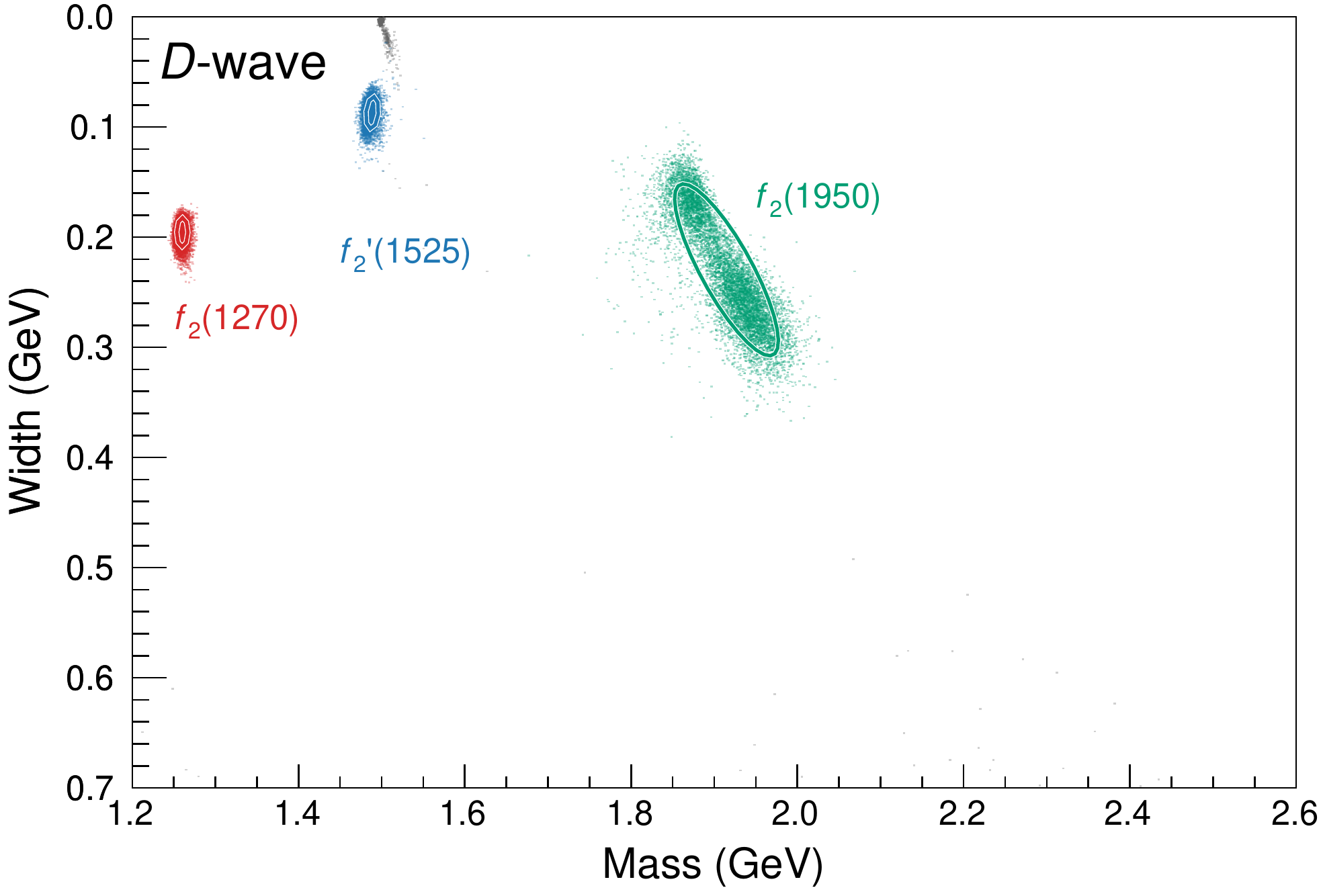}
\end{figure}

\begin{table}[h]
\begin{ruledtabular}
\begin{tabular}{c c c c}
$S$-wave  & $\sqrt{s_p}$ \mevp & $D$-wave & $\sqrt{s_p}$ \mevp \\ \hline
\rule[-0.2cm]{-0.1cm}{.55cm} $f_0(1500)$ &  $(1444 \pm 4) - i (93 \pm 6)/2$  &  $f_2(1270)$ &  $(1260 \pm 4) - i (196 \pm 9)/2$ \\
\rule[-0.2cm]{-0.1cm}{.55cm} $f_0(1710)$ &  $(1764 \pm 5) - i (143 \pm 10)/2$  &  $f_2'(1525)$ &  $(1489 \pm 5) - i (87 \pm 10)/2$ \\
\rule[-0.2cm]{-0.1cm}{.55cm} $f_0(2020)$ &  $(2015 \pm 52) - i (399 \pm 121)/2$  &  $f_2(1950)$ &  $(1915 \pm 41) - i (230 \pm 51)/2$ \\
\rule[-0.2cm]{-0.1cm}{.55cm} $f_0(2330)$ &  $(2400 \pm 36) - i (212 \pm 50)/2$  &   &   \\
\end{tabular}
\end{ruledtabular}
\end{table}

\clearpage

\subsection{$\left[K^J(s)^{-1}\right]^\text{CDD} \quad\Big/\quad \omega(s)_\text{pole+scaled} \quad\Big/\quad \rho N^J_{ki}(s')_\text{nominal}\,,\,\alpha = 0  \quad\Big/\quad s_L = 0$}
\label{subsec:inputcddcmnewc3_bootstrap-out}

\begin{table}[h]
\begin{ruledtabular}
\begin{tabular}{c c c c c c}
\multicolumn{3}{c}{$\pi \pi$ channel}  & \multicolumn{3}{c}{$ K  \bar K$ channel} \\ \hline
\rule[-0.2cm]{-0.1cm}{.55cm} $a^{S,\pi\pi}_0$ &$1754.855$ & $1813.268 \pm 136.054$ & $a^{S,K\bar K}_0$ & $2713.399$ & $2837.206 \pm 278.564$ \\
\rule[-0.2cm]{-0.1cm}{.55cm} $a^{S,\pi\pi}_1$ &$938.765$ & $966.293 \pm 106.127$ & $a^{S,K\bar K}_1$ & $2073.418$ & $2282.138 \pm 271.188$ \\
\rule[-0.2cm]{-0.1cm}{.55cm} $a^{S,\pi\pi}_2$ &$-4384.837$ & $-4430.155 \pm 348.498$ & $a^{S,K\bar K}_2$ & $-7233.067$ & $-7589.865 \pm 726.720$ \\
\rule[-0.2cm]{-0.1cm}{.55cm} $a^{S,\pi\pi}_3$ &$4169.372$ & $4336.171 \pm 329.659$ & $a^{S,K\bar K}_3$ & $7001.418$ & $7509.033 \pm 719.424$ \\
\hline
\rule[-0.2cm]{-0.1cm}{.55cm} $a^{D,\pi\pi}_0$ &$330.656$ & $-196.738 \pm 573.178$ & $a^{D,K\bar K}_0$ & $331.662$ & $26.652 \pm 447.626$ \\
\rule[-0.2cm]{-0.1cm}{.55cm} $a^{D,\pi\pi}_1$ &$-287.942$ & $452.262 \pm 816.725$ & $a^{D,K\bar K}_1$ & $-404.531$ & $25.615 \pm 631.880$ \\
\rule[-0.2cm]{-0.1cm}{.55cm} $a^{D,\pi\pi}_2$ &$113.164$ & $-181.131 \pm 292.034$ & $a^{D,K\bar K}_2$ & $344.428$ & $169.413 \pm 243.348$ \\
\rule[-0.2cm]{-0.1cm}{.55cm} $a^{D,\pi\pi}_3$ &$0$ & $0$ & $a^{D,K\bar K}_3$ & $0$ & $0$ \\
\end{tabular}
\end{ruledtabular}
\end{table}

\begin{table}[h]
\begin{ruledtabular}
\begin{tabular}{c c c c c c c c c}
\multicolumn{3}{c}{$S$-wave couplings $\pi \pi$ channel}  & \multicolumn{3}{c}{$S$-wave couplings $K  \bar K$ channel} & \multicolumn{3}{c}{$S$-wave couplings $\rho\rho$ channel} \\ \hline
\rule[-0.2cm]{-0.1cm}{.55cm} $g^{S,1}_{\pi\pi}$ & $2.434$ & $1.527 \pm 0.996$ & $g^{S,1}_{K\bar K}$ & $1.861$ & $0.830 \pm 0.809$ & $g^{S,1}_{\rho\rho}$ & $1.157$ & $0.840 \pm 0.236$ \\
\rule[-0.2cm]{-0.1cm}{.55cm} $g^{S,2}_{\pi\pi}$ & $0.193$ & $0.210 \pm 0.160$ & $g^{S,2}_{K\bar K}$ & $-0.332$ & $-0.282 \pm 0.290$ & $g^{S,2}_{\rho\rho}$ & $-0.052$ & $-0.060 \pm 0.238$ \\
\rule[-0.2cm]{-0.1cm}{.55cm} $g^{S,3}_{\pi\pi}$ & $4.591$ & $4.698 \pm 1.172$ & $g^{S,3}_{K\bar K}$ & $4.786$ & $4.843 \pm 0.670$ & $g^{S,3}_{\rho\rho}$ & $1.005$ & $0.858 \pm 0.106$ \\
\hline 
\multicolumn{3}{c}{$D$-wave couplings $\pi \pi$ channel}  & \multicolumn{3}{c}{$D$-wave couplings $K  \bar K$ channel} & \multicolumn{3}{c}{$D$-wave couplings $\rho\rho$ channel} \\ \hline
\rule[-0.2cm]{-0.1cm}{.55cm} $g^{D,1}_{\pi\pi}$ & $-0.588$ & $-0.654 \pm 0.128$ & $g^{D,1}_{K\bar K}$ & $-0.179$ & $-0.190 \pm 0.071$ & $g^{D,1}_{\rho\rho}$ & $0$ & $0$ \\
\rule[-0.2cm]{-0.1cm}{.55cm} $g^{D,2}_{\pi\pi}$ & $0.015$ & $-0.000 \pm 0.037$ & $g^{D,2}_{K\bar K}$ & $0.714$ & $0.748 \pm 0.076$ & $g^{D,2}_{\rho\rho}$ & $-0.176$ & $0.209 \pm 0.864$ \\
\rule[-0.2cm]{-0.1cm}{.55cm} $g^{D,3}_{\pi\pi}$ & $-0.851$ & $-0.791 \pm 0.151$ & $g^{D,3}_{K\bar K}$ & $1.479$ & $1.387 \pm 0.394$ & $g^{D,3}_{\rho\rho}$ & $-1.951$ & $-3.624 \pm 3.068$ \\
\hline 
\multicolumn{9}{c}{$S$-wave bare masses}   \\ \hline
\rule[-0.2cm]{-0.1cm}{.55cm} $m^2_{S,1}$ & $0.002$ & $0.873 \pm 0.819$ & $m^2_{S,2}$ & $2.098$ & $2.168 \pm 0.201$ & $m^2_{S,3}$ & $5.924$ & $6.108 \pm 0.729$ \\
\hline 
\multicolumn{9}{c}{$D$-wave bare masses}   \\ \hline
\rule[-0.2cm]{-0.1cm}{.55cm} $m^2_{D,1}$ & $1.261$ & $1.216 \pm 0.084$ & $m^2_{D,2}$ & $2.295$ & $2.312 \pm 0.022$ & $m^2_{D,3}$ & $3.760$ & $3.793 \pm 0.335$ \\
\end{tabular}
\vspace{2em}
\begin{tabular}{c c c c c c}
\multicolumn{3}{c}{$S$-wave background}  & \multicolumn{3}{c}{$D$-wave background} \\ \hline
\rule[-0.2cm]{-0.1cm}{.55cm} $c^{S,0}_{\pi\pi,\pi\pi}$ &$-2.518$ & $-2.759 \pm 0.421$ & $c^{D,0}_{\pi\pi,\pi\pi}$ &$-0.159$ & $0.037 \pm 0.339$ \\
\rule[-0.2cm]{-0.1cm}{.55cm} $c^{S,0}_{\pi\pi,K\bar K}$ &$4.076$ & $4.439 \pm 0.240$ & $c^{D,0}_{\pi\pi,K\bar K}$ &$0.439$ & $0.361 \pm 0.199$ \\
\rule[-0.2cm]{-0.1cm}{.55cm} $c^{S,0}_{\pi\pi,\rho\rho}$ &$0$ & $0$ & $c^{D,0}_{\pi\pi,\rho\rho}$ &$0$ & $0$ \\
\rule[-0.2cm]{-0.1cm}{.55cm} $c^{S,0}_{K\bar K,K\bar K}$ &$-5.323$ & $-5.832 \pm 0.339$ & $c^{D,0}_{K\bar K,K\bar K}$ &$0.566$ & $0.538 \pm 0.375$ \\
\rule[-0.2cm]{-0.1cm}{.55cm} $c^{S,0}_{K\bar K,\rho\rho}$ &$0$ & $0$ & $c^{D,0}_{K\bar K,\rho\rho}$ &$0$ & $0$ \\
\rule[-0.2cm]{-0.1cm}{.55cm} $c^{S,0}_{\rho\rho,\rho\rho}$ &$1.039$ & $0.965 \pm 0.184$ & $c^{D,0}_{\rho\rho,\rho\rho}$ &$-27.282$ & $-154.044 \pm 336.754$ \\
\rule[-0.2cm]{-0.1cm}{.55cm} $c^{S,1}_{\pi\pi,\pi\pi}$ &$0.085$ & $0.007 \pm 0.292$ & $c^{D,1}_{\pi\pi,\pi\pi}$ &$-0.107$ & $-0.137 \pm 0.087$ \\
\rule[-0.2cm]{-0.1cm}{.55cm} $c^{S,1}_{\pi\pi,K\bar K}$ &$-1.585$ & $-1.750 \pm 0.105$ & $c^{D,1}_{\pi\pi,K\bar K}$ &$0.048$ & $0.055 \pm 0.096$ \\
\rule[-0.2cm]{-0.1cm}{.55cm} $c^{S,1}_{\pi\pi,\rho\rho}$ &$0$ & $0$ & $c^{D,1}_{\pi\pi,\rho\rho}$ &$0$ & $0$ \\
\rule[-0.2cm]{-0.1cm}{.55cm} $c^{S,1}_{K\bar K,K\bar K}$ &$-2.420$ & $-2.675 \pm 0.203$ & $c^{D,1}_{K\bar K,K\bar K}$ &$-0.159$ & $-0.149 \pm 0.133$ \\
\rule[-0.2cm]{-0.1cm}{.55cm} $c^{S,1}_{K\bar K,\rho\rho}$ &$0$ & $0$ & $c^{D,1}_{K\bar K,\rho\rho}$ &$0$ & $0$ \\
\rule[-0.2cm]{-0.1cm}{.55cm} $c^{S,1}_{\rho\rho,\rho\rho}$ &$0.522$ & $0.488 \pm 0.052$ & $c^{D,1}_{\rho\rho,\rho\rho}$ &$7.036$ & $42.133 \pm 93.118$ \\
\end{tabular}
\end{ruledtabular}
\end{table}

\begin{figure}[h]
\centering\includegraphics[width=0.32\textwidth]{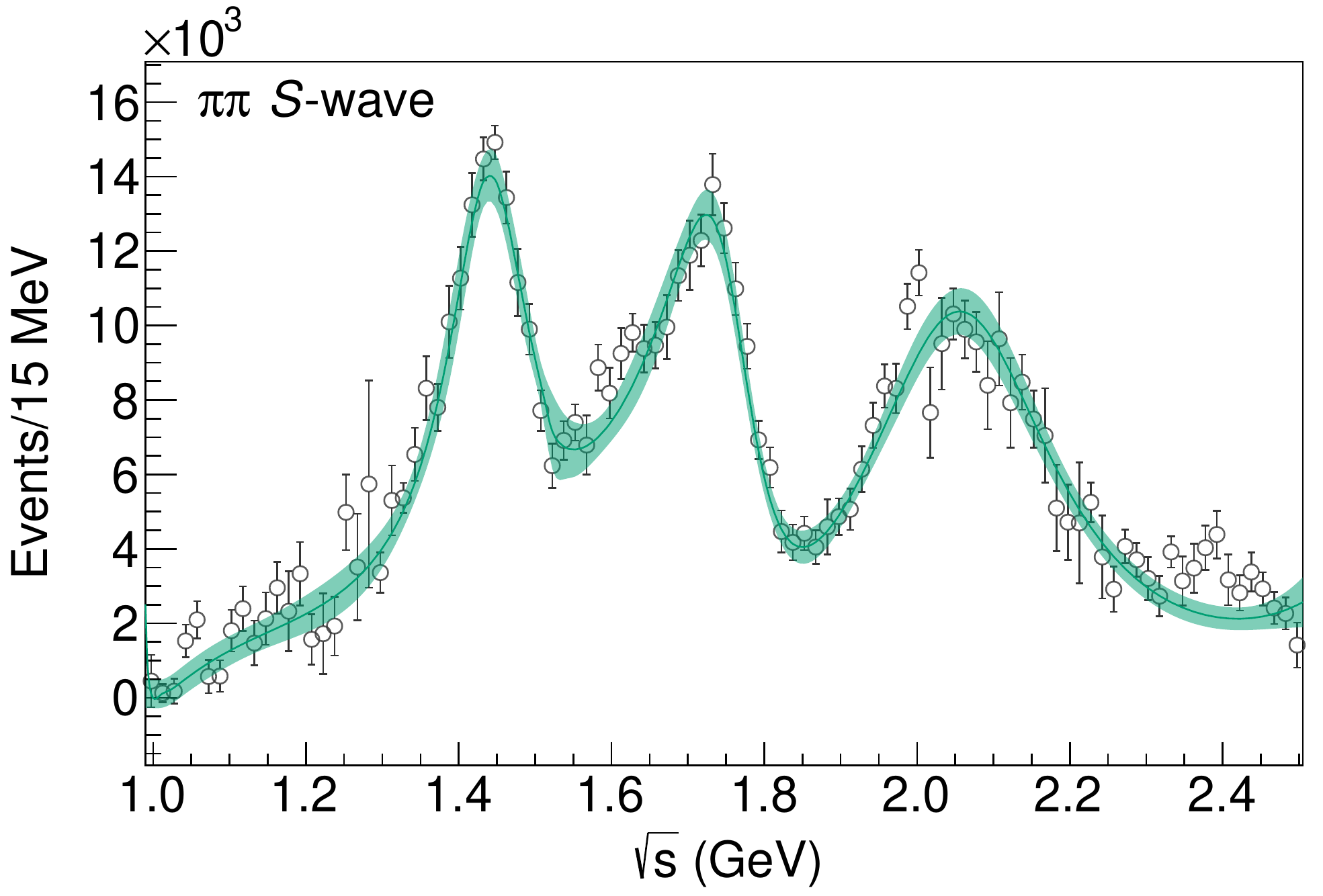} \includegraphics[width=0.32\textwidth]{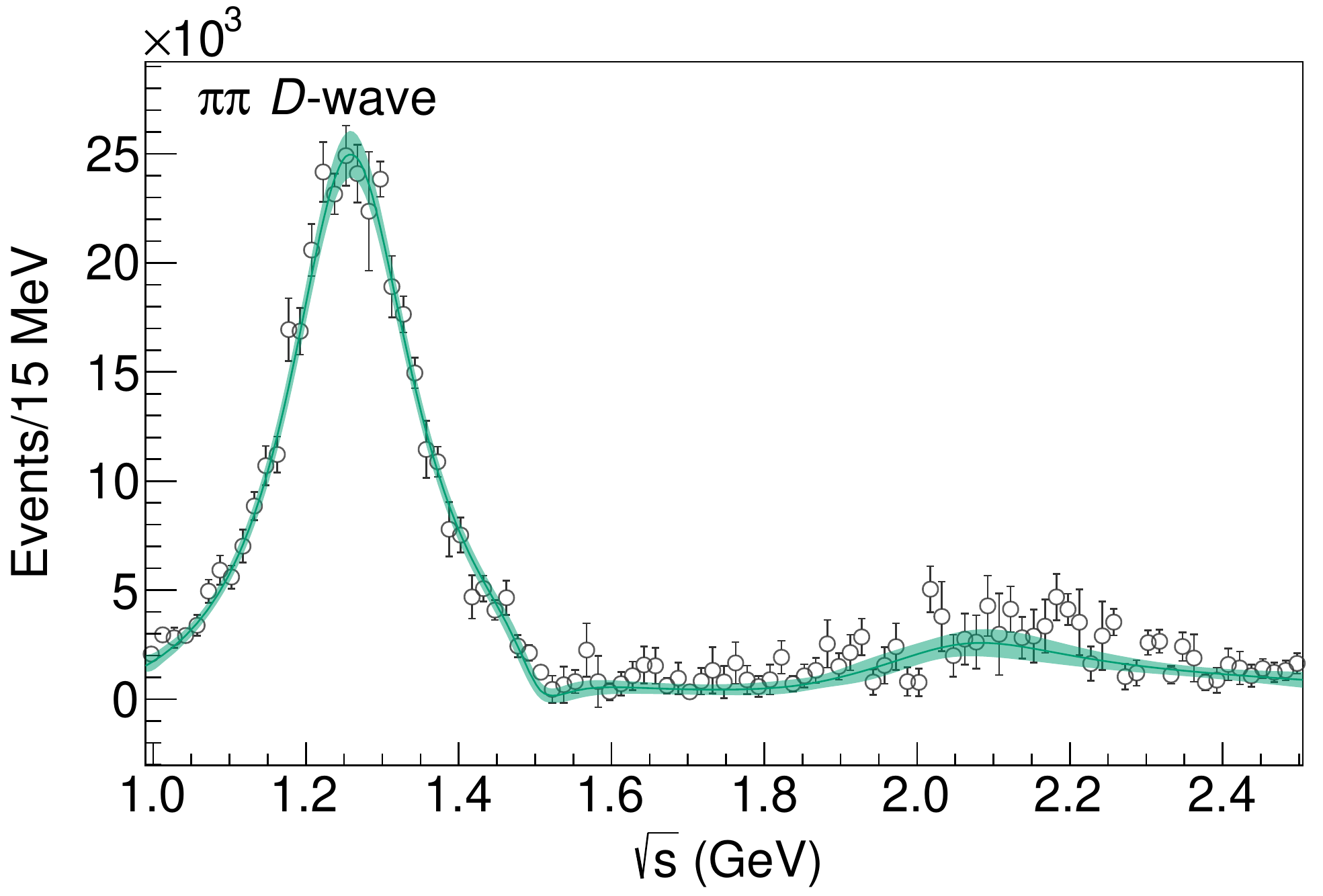} \includegraphics[width=0.32\textwidth]{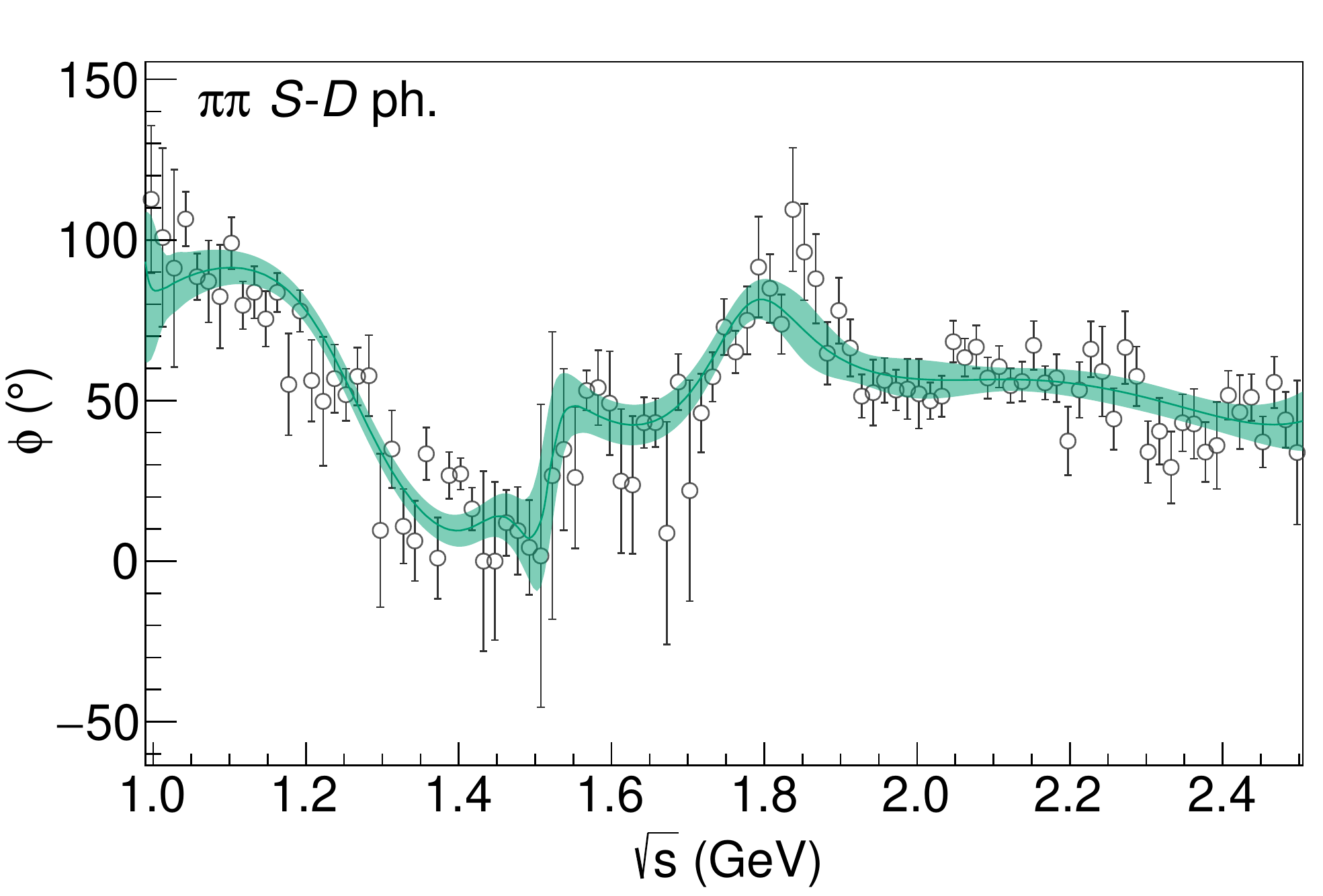}
\includegraphics[width=0.32\textwidth]{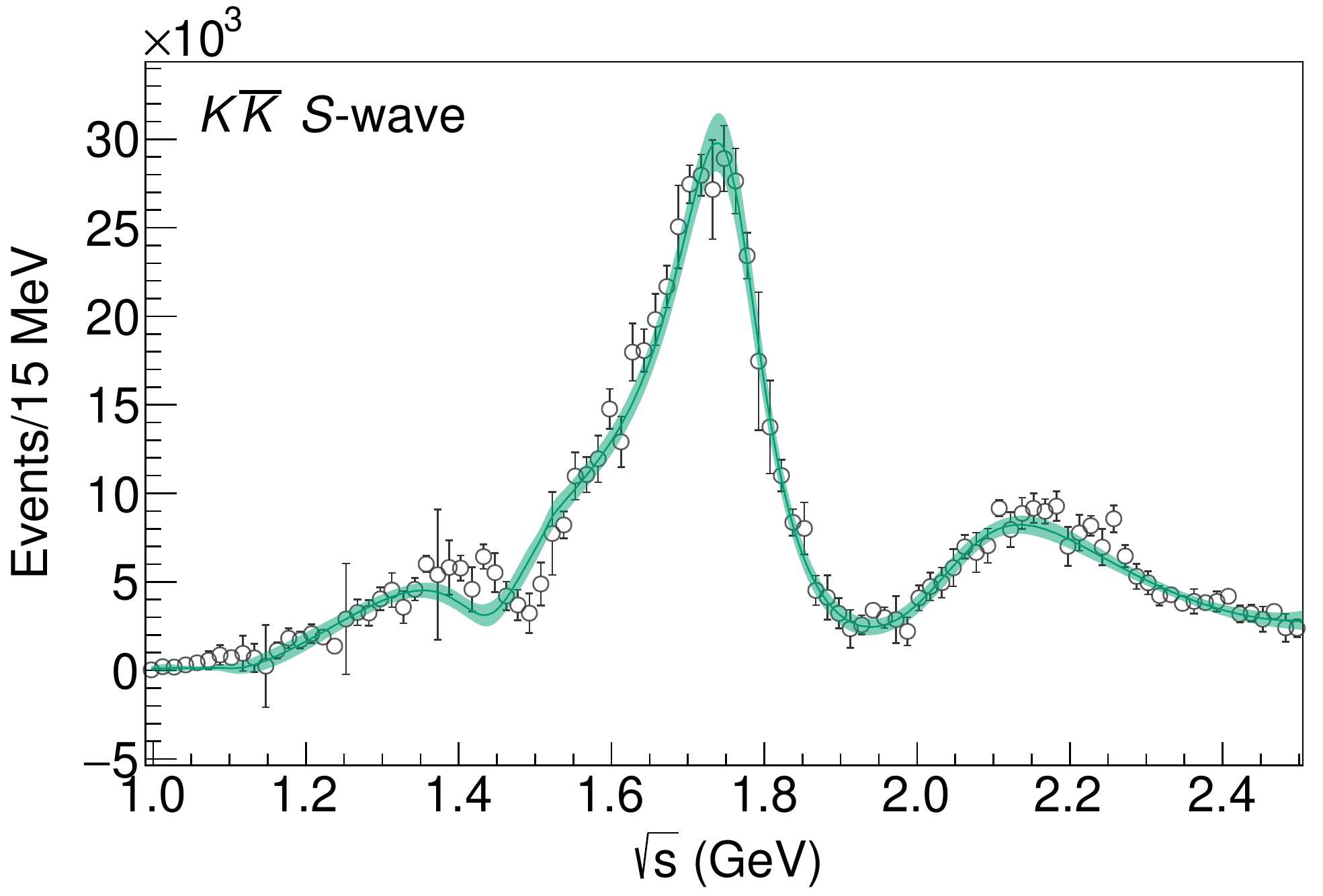} \includegraphics[width=0.32\textwidth]{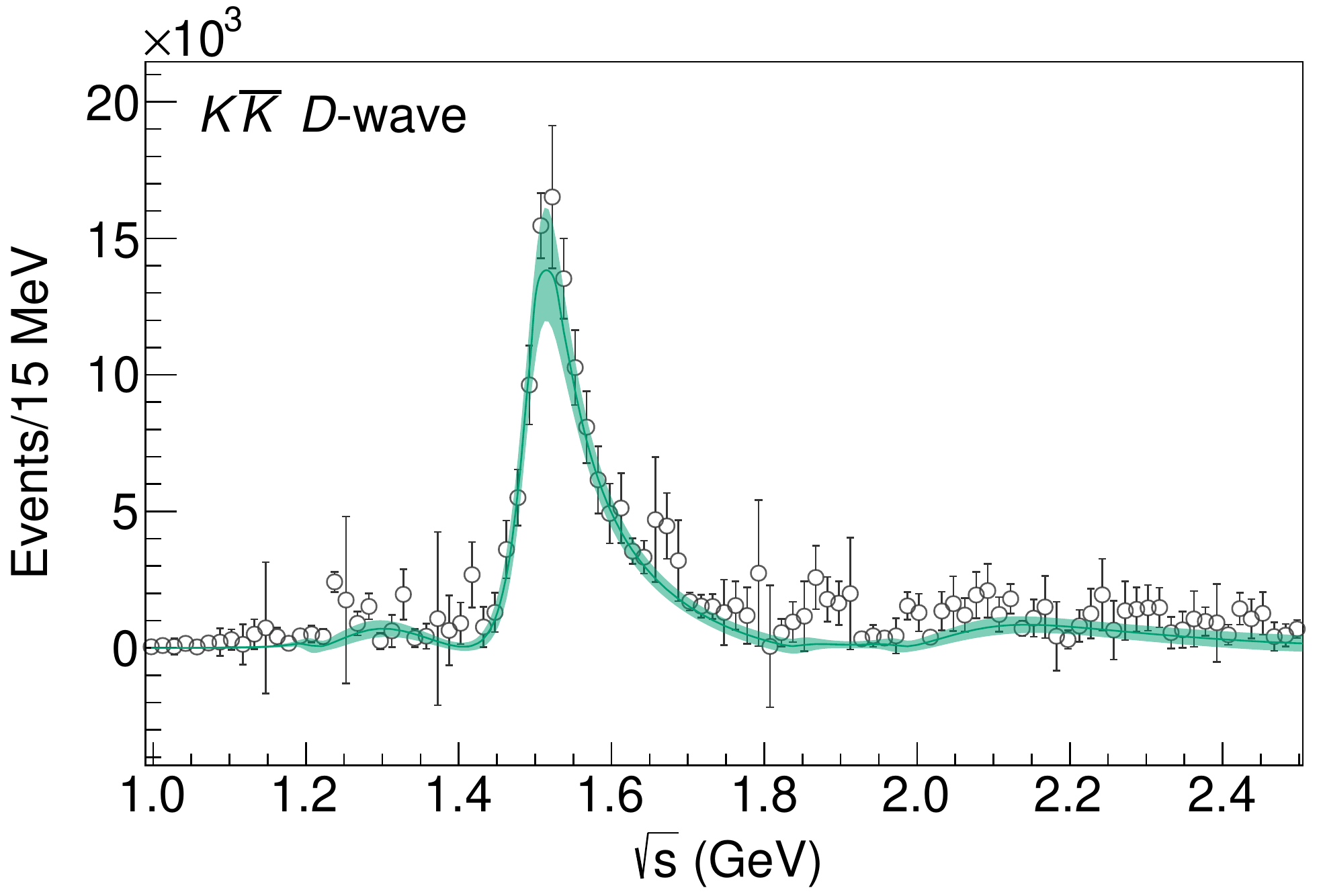} \includegraphics[width=0.32\textwidth]{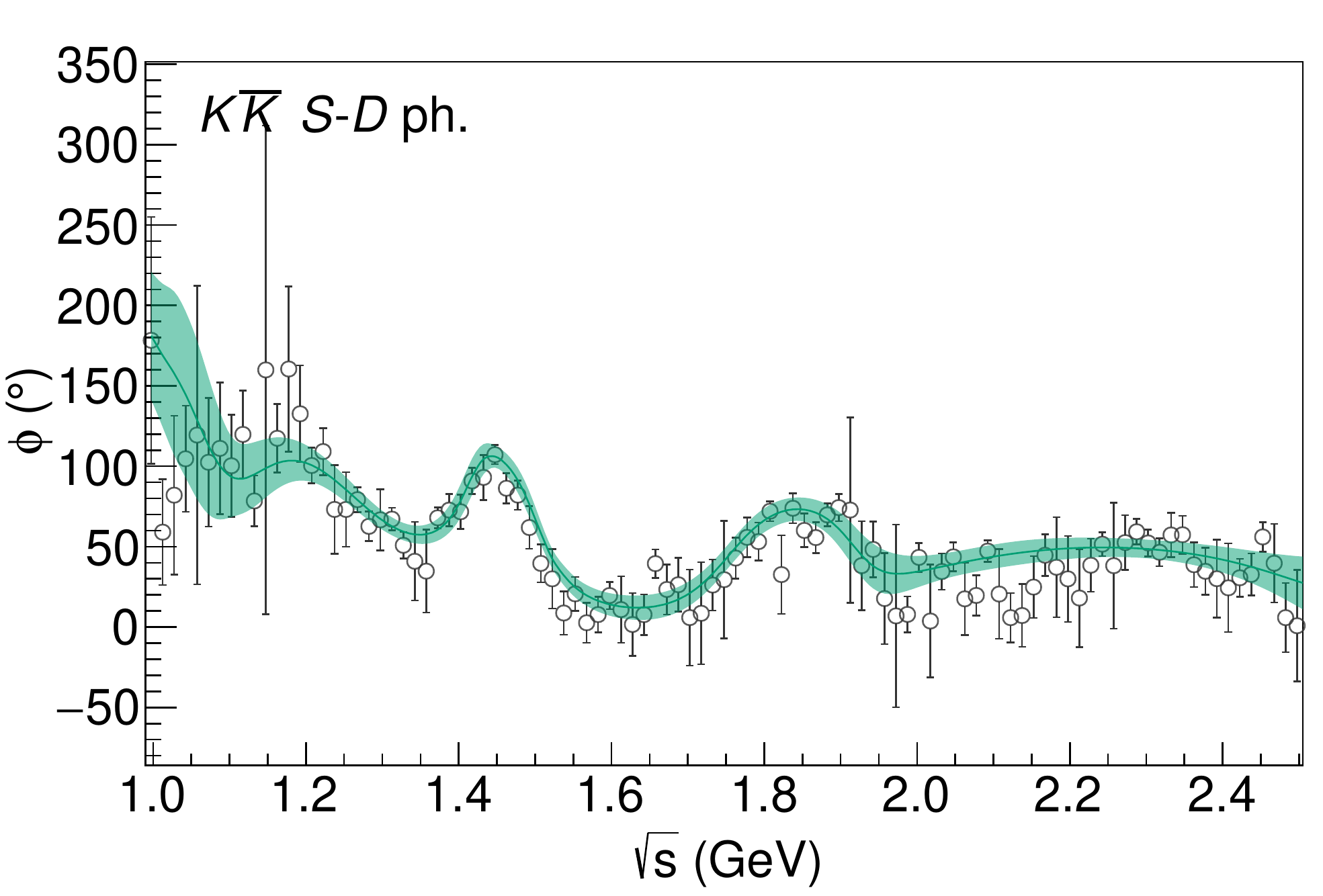}
\end{figure}

\begin{figure}[h]
\centering\includegraphics[width=0.45\textwidth]{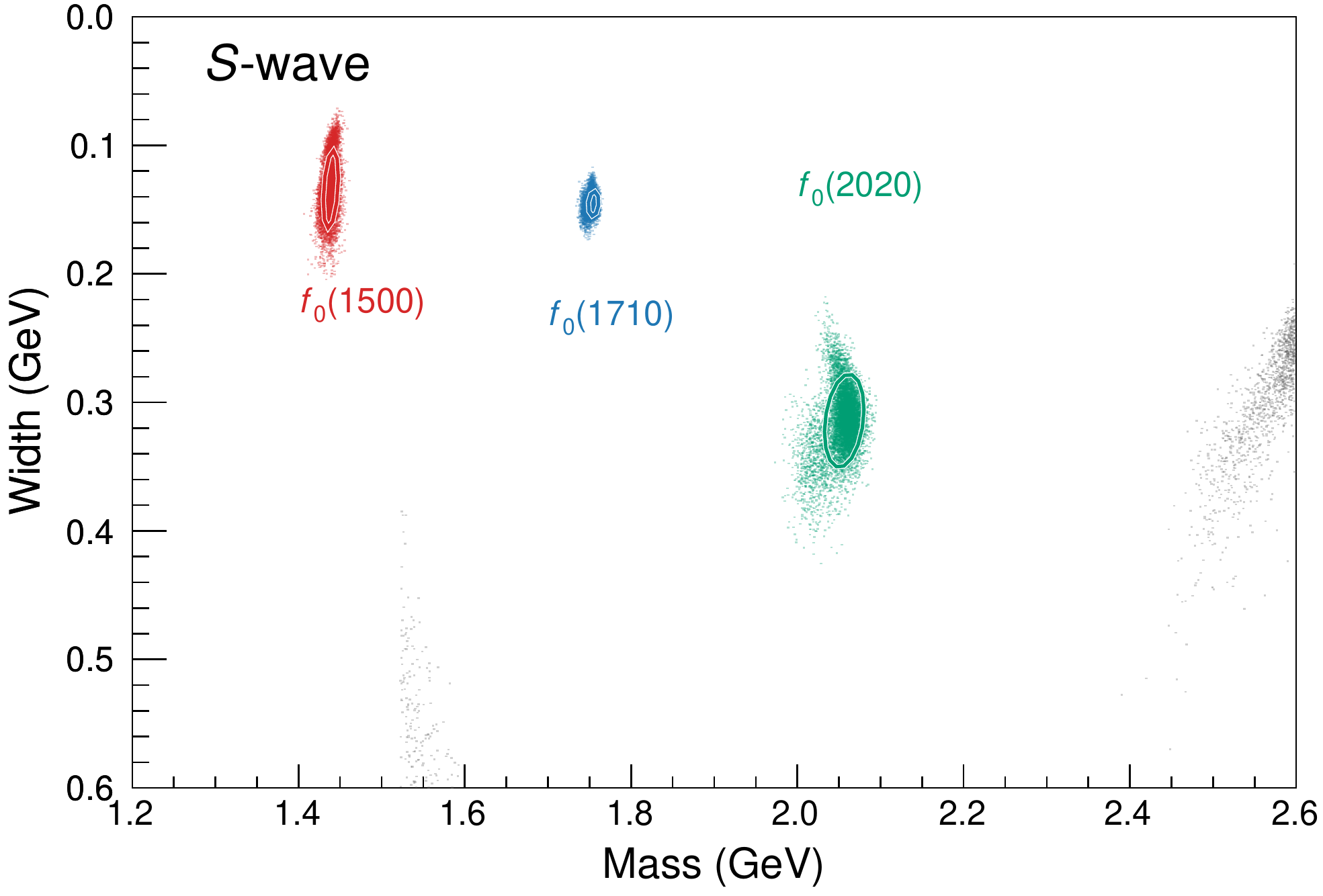} \includegraphics[width=0.45\textwidth]{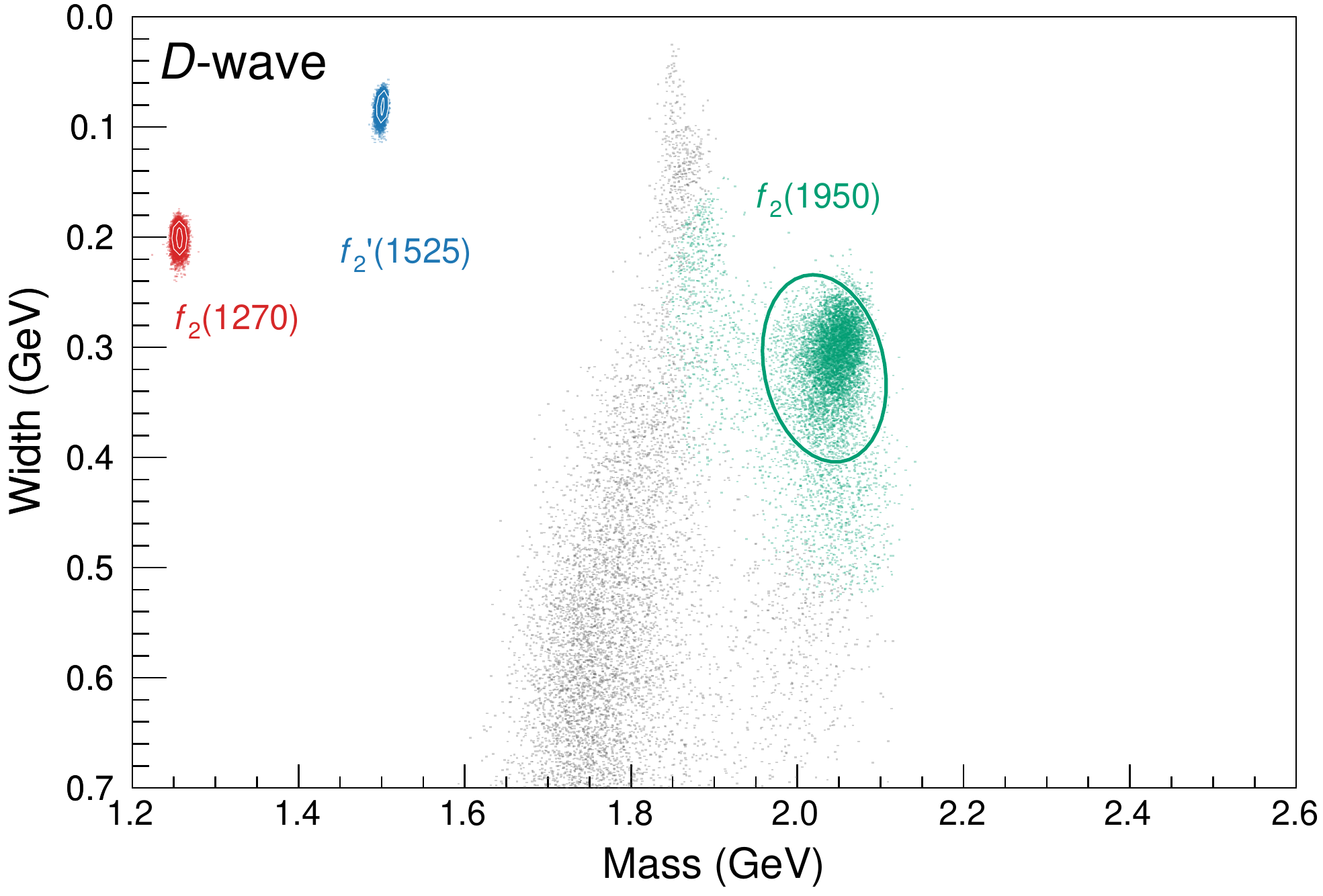}
\end{figure}

\begin{table}[h]
\begin{ruledtabular}
\begin{tabular}{c c c c}
$S$-wave  & $\sqrt{s_p}$ \mevp & $D$-wave & $\sqrt{s_p}$ \mevp \\ \hline
\rule[-0.2cm]{-0.1cm}{.55cm} $f_0(1500)$ &  $(1439 \pm 6) - i (134 \pm 20)/2$  &  $f_2(1270)$ &  $(1257 \pm 4) - i (201 \pm 8)/2$ \\
\rule[-0.2cm]{-0.1cm}{.55cm} $f_0(1710)$ &  $(1755 \pm 4) - i (145 \pm 6)/2$  &  $f_2'(1525)$ &  $(1501 \pm 3) - i (81 \pm 7)/2$ \\
\rule[-0.2cm]{-0.1cm}{.55cm} $f_0(2020)$ &  $(2056 \pm 16) - i (314 \pm 24)/2$  &  $f_2(1950)$ &  $(2032 \pm 49) - i (319 \pm 56)/2$ \\
\end{tabular}
\end{ruledtabular}
\end{table}

\clearpage

\subsection{$K^J(s) \quad\Big/\quad \omega(s)_\text{pole+scaled} \quad\Big/\quad \rho N^J_{ki}(s')_\text{nominal}\,,\,\alpha = 0  \quad\Big/\quad s_L = 0$}
\label{subsec:inputstartcm11new2c3_bootstrap-out}

\begin{table}[h]
\begin{ruledtabular}
\begin{tabular}{c c c c c c}
\multicolumn{3}{c}{$\pi \pi$ channel}  & \multicolumn{3}{c}{$ K  \bar K$ channel} \\ \hline
\rule[-0.2cm]{-0.1cm}{.55cm} $a^{S,\pi\pi}_0$ &$-2173.729$ & $-2196.868 \pm 78.733$ & $a^{S,K\bar K}_0$ & $1927.495$ & $1938.219 \pm 63.447$ \\
\rule[-0.2cm]{-0.1cm}{.55cm} $a^{S,\pi\pi}_1$ &$3272.052$ & $3301.354 \pm 122.306$ & $a^{S,K\bar K}_1$ & $-2996.838$ & $-3009.397 \pm 100.868$ \\
\rule[-0.2cm]{-0.1cm}{.55cm} $a^{S,\pi\pi}_2$ &$-1553.726$ & $-1561.688 \pm 58.187$ & $a^{S,K\bar K}_2$ & $1354.327$ & $1356.176 \pm 49.282$ \\
\rule[-0.2cm]{-0.1cm}{.55cm} $a^{S,\pi\pi}_3$ &$361.790$ & $360.535 \pm 17.280$ & $a^{S,K\bar K}_3$ & $-287.158$ & $-285.375 \pm 14.004$ \\
\hline
\rule[-0.2cm]{-0.1cm}{.55cm} $a^{D,\pi\pi}_0$ &$109.759$ & $111.857 \pm 6.750$ & $a^{D,K\bar K}_0$ & $-26.964$ & $-30.346 \pm 8.700$ \\
\rule[-0.2cm]{-0.1cm}{.55cm} $a^{D,\pi\pi}_1$ &$-2.844$ & $-1.076 \pm 8.895$ & $a^{D,K\bar K}_1$ & $-135.324$ & $-138.644 \pm 17.232$ \\
\rule[-0.2cm]{-0.1cm}{.55cm} $a^{D,\pi\pi}_2$ &$71.095$ & $72.804 \pm 7.401$ & $a^{D,K\bar K}_2$ & $-14.887$ & $-18.525 \pm 8.882$ \\
\rule[-0.2cm]{-0.1cm}{.55cm} $a^{D,\pi\pi}_3$ &$0$ & $0$ & $a^{D,K\bar K}_3$ & $0$ & $0$ \\
\end{tabular}
\end{ruledtabular}
\end{table}

\begin{table}[h]
\begin{ruledtabular}
\begin{tabular}{c c c c c c c c c}
\multicolumn{3}{c}{$S$-wave couplings $\pi \pi$ channel}  & \multicolumn{3}{c}{$S$-wave couplings $K  \bar K$ channel} & \multicolumn{3}{c}{$S$-wave couplings $\rho\rho$ channel} \\ \hline
\rule[-0.2cm]{-0.1cm}{.55cm} $g^{S,1}_{\pi\pi}$ & $-6.058$ & $-6.051 \pm 0.030$ & $g^{S,1}_{K\bar K}$ & $-7.247$ & $-7.251 \pm 0.030$ & $g^{S,1}_{\rho\rho}$ & $1.956$ & $1.928 \pm 0.090$ \\
\rule[-0.2cm]{-0.1cm}{.55cm} $g^{S,2}_{\pi\pi}$ & $-0.236$ & $-0.242 \pm 0.030$ & $g^{S,2}_{K\bar K}$ & $-1.092$ & $-1.087 \pm 0.050$ & $g^{S,2}_{\rho\rho}$ & $-0.233$ & $-0.252 \pm 0.075$ \\
\rule[-0.2cm]{-0.1cm}{.55cm} $g^{S,3}_{\pi\pi}$ & $8.295$ & $8.288 \pm 0.046$ & $g^{S,3}_{K\bar K}$ & $12.908$ & $12.908 \pm 0.058$ & $g^{S,3}_{\rho\rho}$ & $-1.049$ & $-1.028 \pm 0.094$ \\
\hline 
\multicolumn{3}{c}{$D$-wave couplings $\pi \pi$ channel}  & \multicolumn{3}{c}{$D$-wave couplings $K  \bar K$ channel} & \multicolumn{3}{c}{$D$-wave couplings $\rho\rho$ channel} \\ \hline
\rule[-0.2cm]{-0.1cm}{.55cm} $g^{D,1}_{\pi\pi}$ & $0.155$ & $0.160 \pm 0.046$ & $g^{D,1}_{K\bar K}$ & $1.027$ & $1.026 \pm 0.062$ & $g^{D,1}_{\rho\rho}$ & $-0.344$ & $-0.415 \pm 0.173$ \\
\rule[-0.2cm]{-0.1cm}{.55cm} $g^{D,2}_{\pi\pi}$ & $0.908$ & $0.904 \pm 0.014$ & $g^{D,2}_{K\bar K}$ & $-0.250$ & $-0.239 \pm 0.035$ & $g^{D,2}_{\rho\rho}$ & $0$ & $0$ \\
\rule[-0.2cm]{-0.1cm}{.55cm} $g^{D,3}_{\pi\pi}$ & $0.852$ & $0.851 \pm 0.069$ & $g^{D,3}_{K\bar K}$ & $0.698$ & $0.705 \pm 0.149$ & $g^{D,3}_{\rho\rho}$ & $1.134$ & $1.120 \pm 0.223$ \\
\hline 
\multicolumn{9}{c}{$S$-wave bare masses}   \\ \hline
\rule[-0.2cm]{-0.1cm}{.55cm} $m^2_{S,1}$ & $0.006$ & $0.007 \pm 0.011$ & $m^2_{S,2}$ & $3.184$ & $3.177 \pm 0.028$ & $m^2_{S,3}$ & $7.548$ & $7.546 \pm 0.038$ \\
\hline 
\multicolumn{9}{c}{$D$-wave bare masses}   \\ \hline
\rule[-0.2cm]{-0.1cm}{.55cm} $m^2_{D,1}$ & $2.330$ & $2.332 \pm 0.014$ & $m^2_{D,2}$ & $1.667$ & $1.669 \pm 0.013$ & $m^2_{D,3}$ & $2.894$ & $2.878 \pm 0.068$ \\
\end{tabular}
\vspace{2em}
\begin{tabular}{c c c c c c}
\multicolumn{3}{c}{$S$-wave background}  & \multicolumn{3}{c}{$D$-wave background} \\ \hline
\rule[-0.2cm]{-0.1cm}{.55cm} $c^{S,0}_{\pi\pi,\pi\pi}$ &$16.190$ & $16.190 \pm 0.062$ & $c^{D,0}_{\pi\pi,\pi\pi}$ &$-1.170$ & $-1.185 \pm 0.126$ \\
\rule[-0.2cm]{-0.1cm}{.55cm} $c^{S,0}_{\pi\pi,K\bar K}$ &$18.308$ & $18.308 \pm 0.062$ & $c^{D,0}_{\pi\pi,K\bar K}$ &$-1.012$ & $-1.006 \pm 0.139$ \\
\rule[-0.2cm]{-0.1cm}{.55cm} $c^{S,0}_{\pi\pi,\rho\rho}$ &$0$ & $0$ & $c^{D,0}_{\pi\pi,\rho\rho}$ &$0$ & $0$ \\
\rule[-0.2cm]{-0.1cm}{.55cm} $c^{S,0}_{K\bar K,K\bar K}$ &$19.239$ & $19.250 \pm 0.091$ & $c^{D,0}_{K\bar K,K\bar K}$ &$0.395$ & $0.417 \pm 0.174$ \\
\rule[-0.2cm]{-0.1cm}{.55cm} $c^{S,0}_{K\bar K,\rho\rho}$ &$0$ & $0$ & $c^{D,0}_{K\bar K,\rho\rho}$ &$0$ & $0$ \\
\rule[-0.2cm]{-0.1cm}{.55cm} $c^{S,0}_{\rho\rho,\rho\rho}$ &$-15.811$ & $-15.867 \pm 2.310$ & $c^{D,0}_{\rho\rho,\rho\rho}$ &$-11.365$ & $-17.953 \pm 15.273$ \\
\rule[-0.2cm]{-0.1cm}{.55cm} $c^{S,1}_{\pi\pi,\pi\pi}$ &$-6.274$ & $-6.278 \pm 0.032$ & $c^{D,1}_{\pi\pi,\pi\pi}$ &$0.211$ & $0.213 \pm 0.028$ \\
\rule[-0.2cm]{-0.1cm}{.55cm} $c^{S,1}_{\pi\pi,K\bar K}$ &$-9.197$ & $-9.197 \pm 0.037$ & $c^{D,1}_{\pi\pi,K\bar K}$ &$0.168$ & $0.168 \pm 0.036$ \\
\rule[-0.2cm]{-0.1cm}{.55cm} $c^{S,1}_{\pi\pi,\rho\rho}$ &$0$ & $0$ & $c^{D,1}_{\pi\pi,\rho\rho}$ &$0$ & $0$ \\
\rule[-0.2cm]{-0.1cm}{.55cm} $c^{S,1}_{K\bar K,K\bar K}$ &$-13.371$ & $-13.367 \pm 0.056$ & $c^{D,1}_{K\bar K,K\bar K}$ &$0.005$ & $0.002 \pm 0.040$ \\
\rule[-0.2cm]{-0.1cm}{.55cm} $c^{S,1}_{K\bar K,\rho\rho}$ &$0$ & $0$ & $c^{D,1}_{K\bar K,\rho\rho}$ &$0$ & $0$ \\
\rule[-0.2cm]{-0.1cm}{.55cm} $c^{S,1}_{\rho\rho,\rho\rho}$ &$6.031$ & $6.059 \pm 0.730$ & $c^{D,1}_{\rho\rho,\rho\rho}$ &$3.923$ & $5.832 \pm 4.521$ \\
\end{tabular}
\end{ruledtabular}
\end{table}

\begin{figure}[h]
\centering\includegraphics[width=0.32\textwidth]{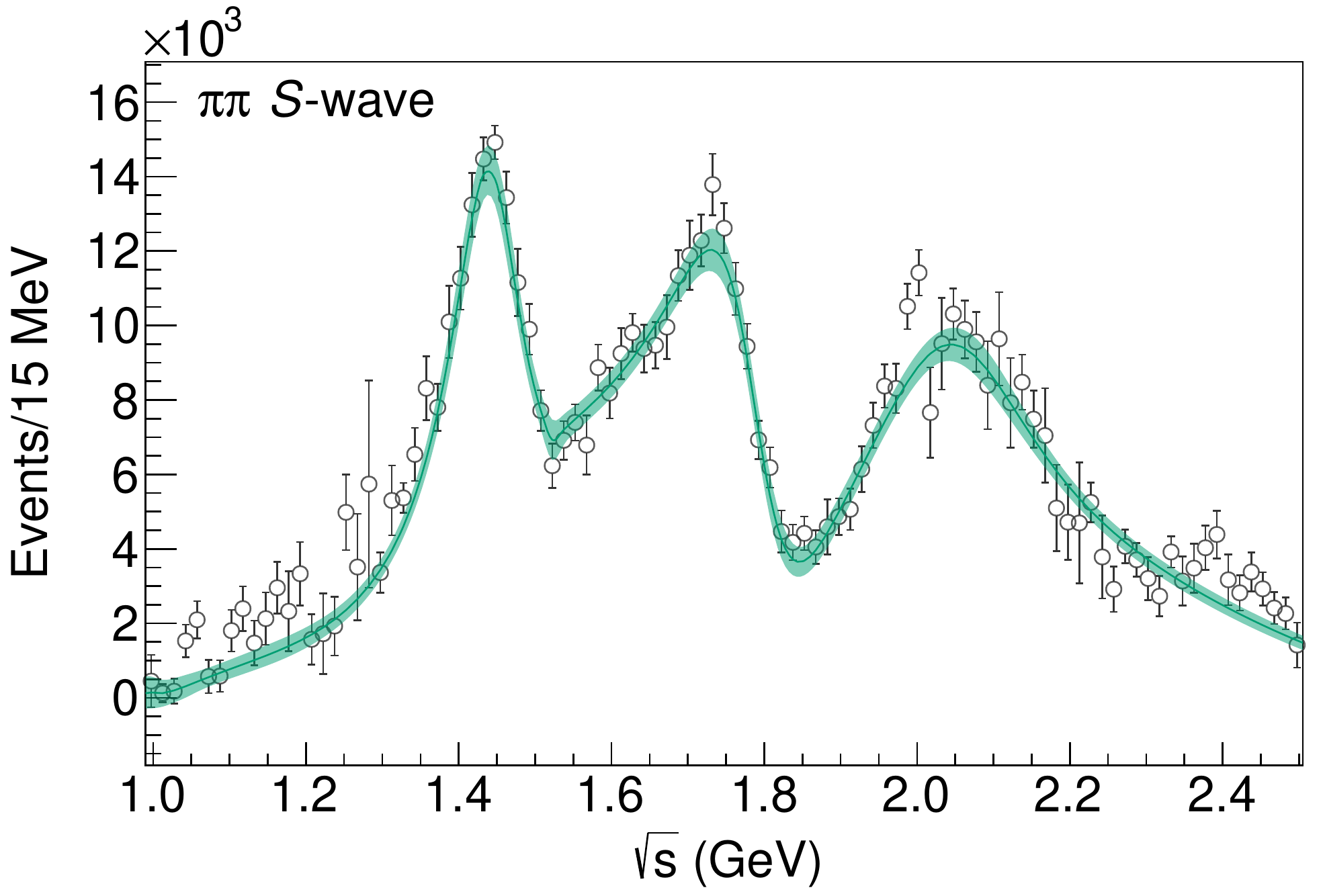} \includegraphics[width=0.32\textwidth]{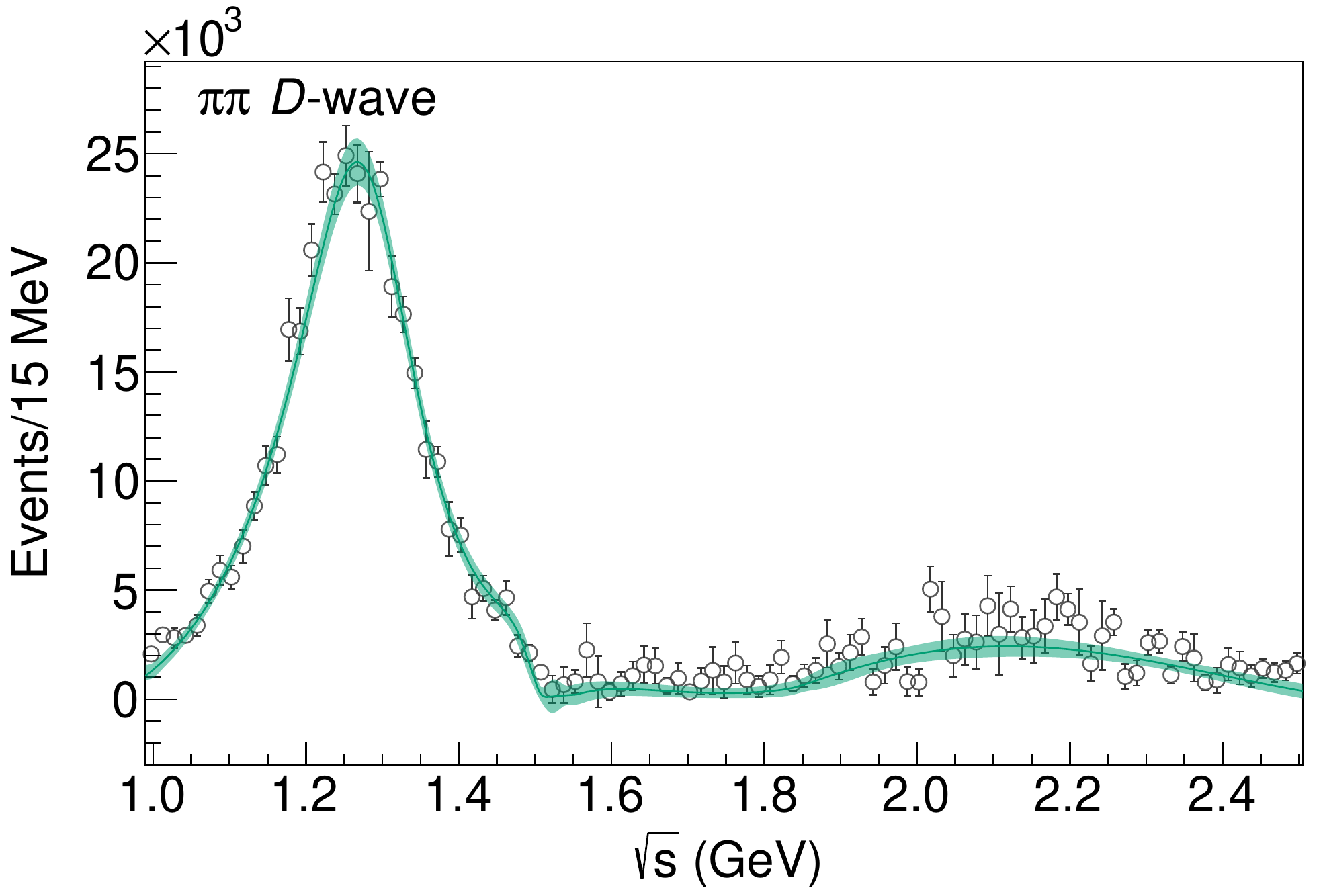} \includegraphics[width=0.32\textwidth]{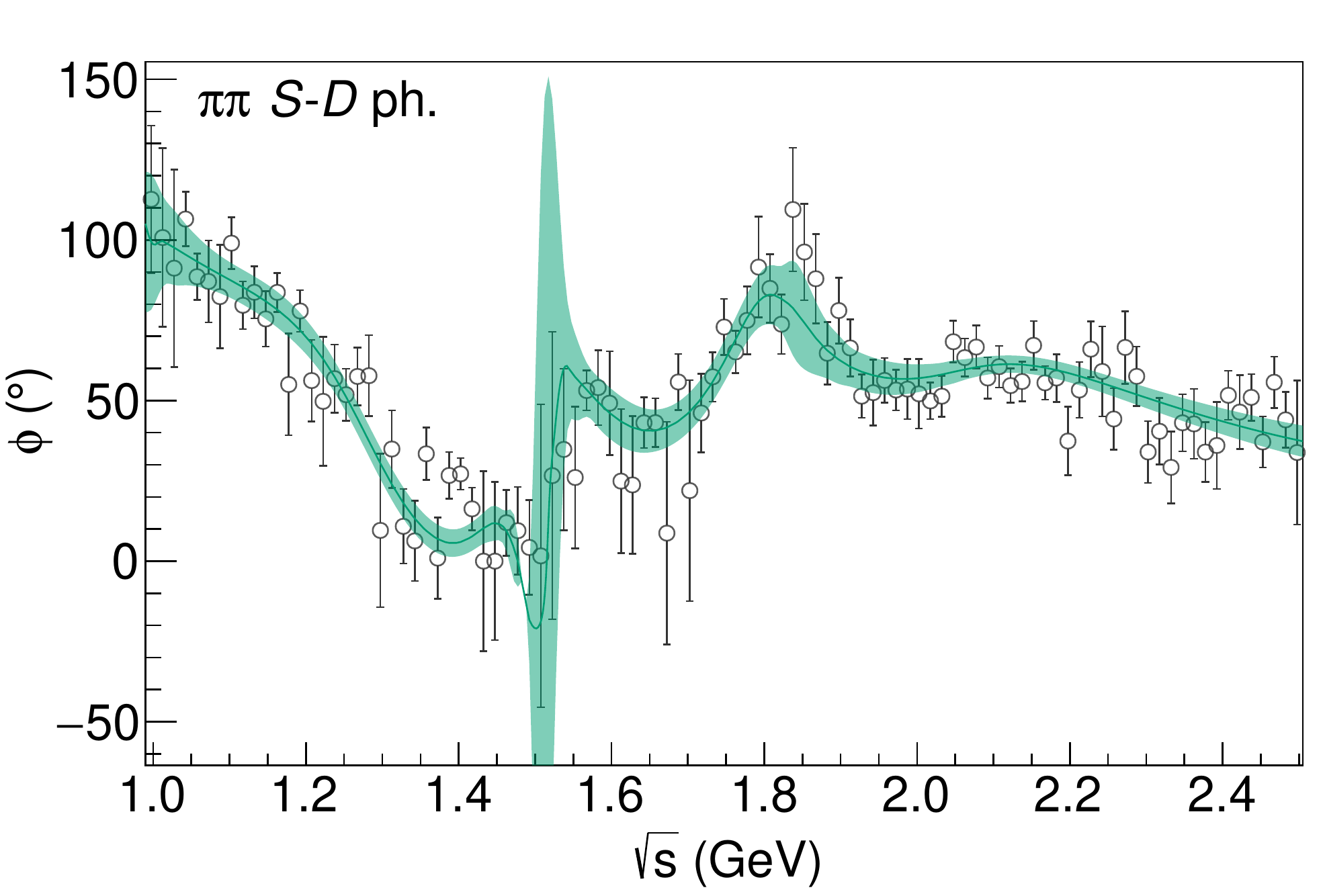}
\includegraphics[width=0.32\textwidth]{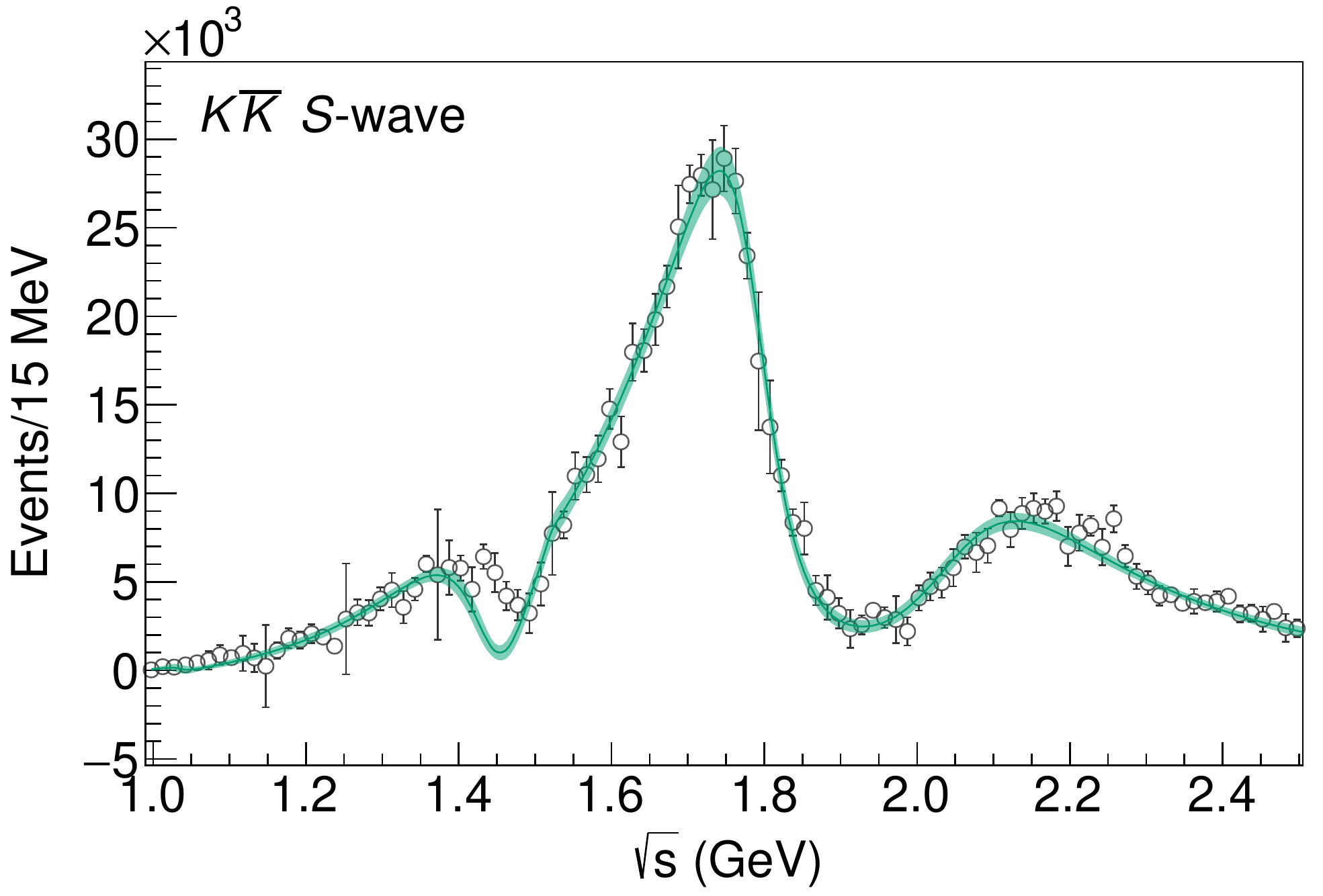} \includegraphics[width=0.32\textwidth]{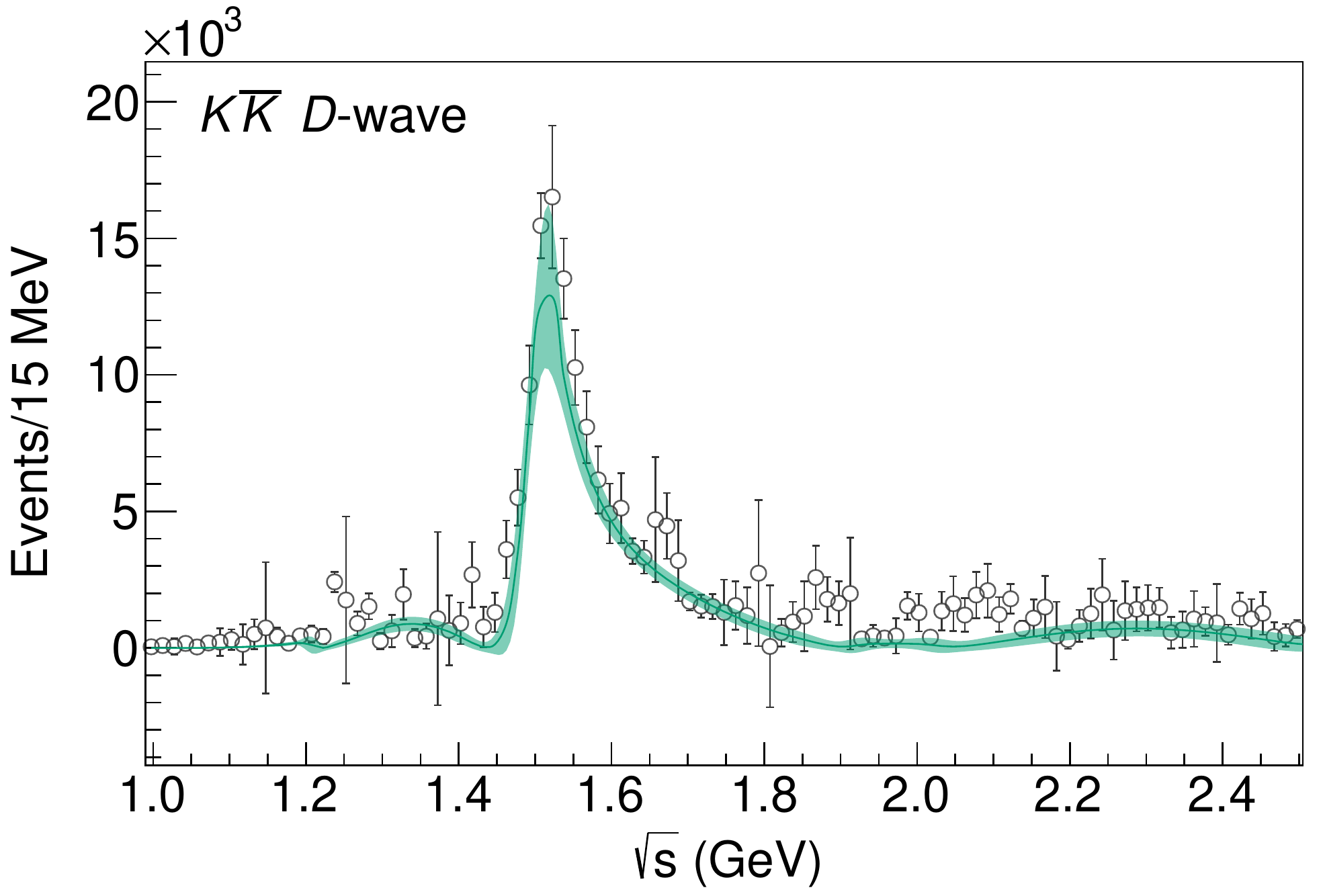} \includegraphics[width=0.32\textwidth]{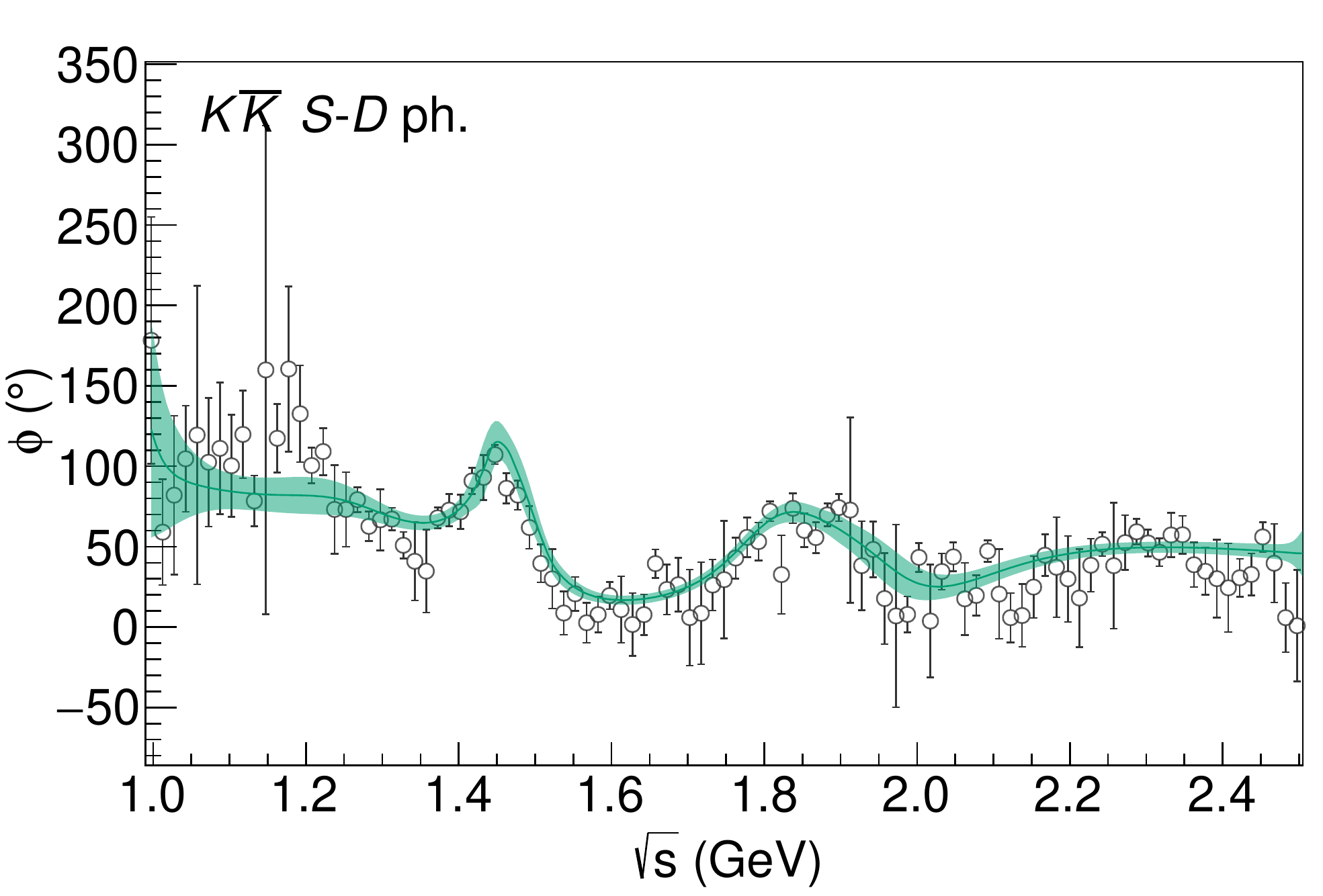}
\end{figure}

\begin{figure}[h]
\centering\includegraphics[width=0.45\textwidth]{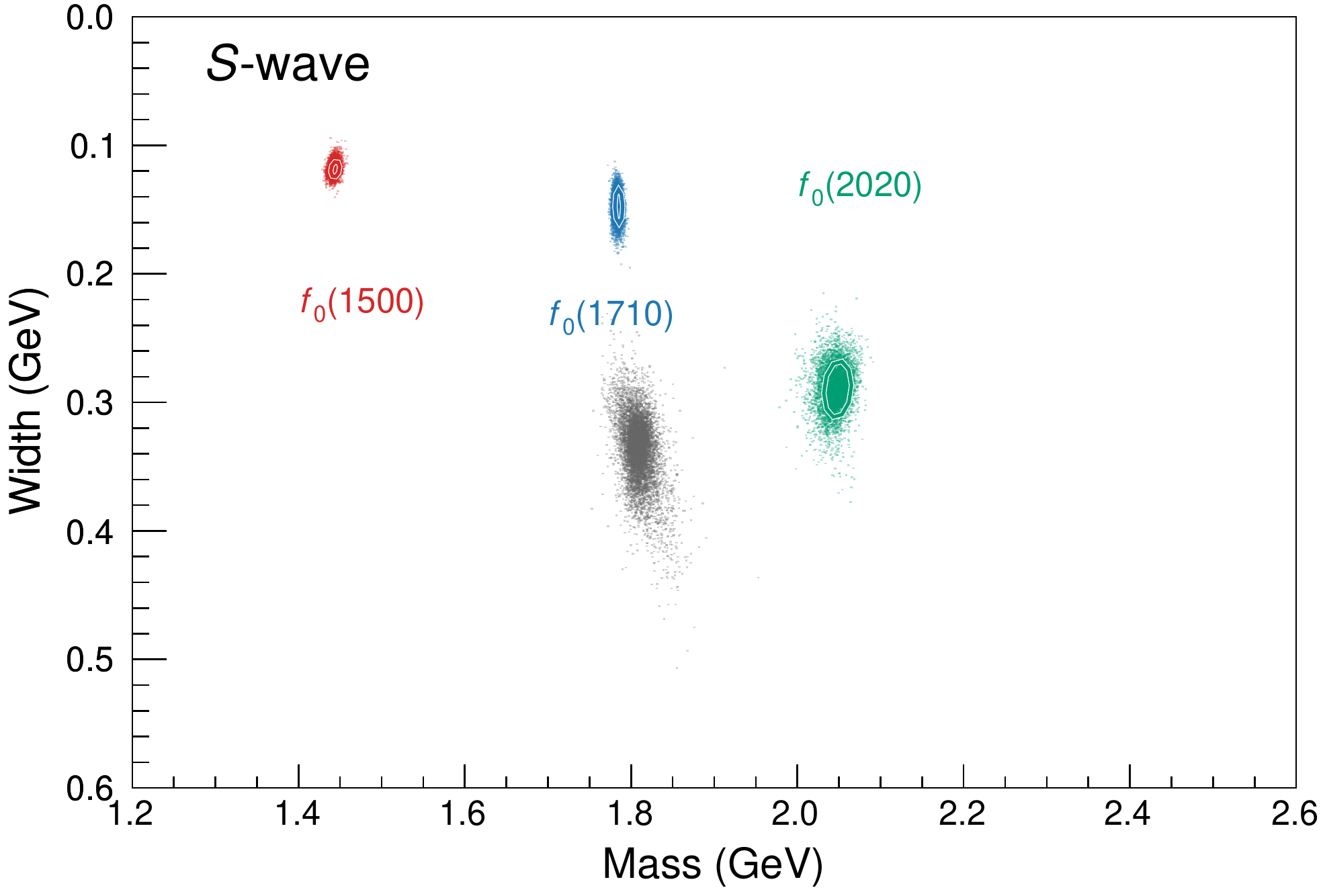} \includegraphics[width=0.45\textwidth]{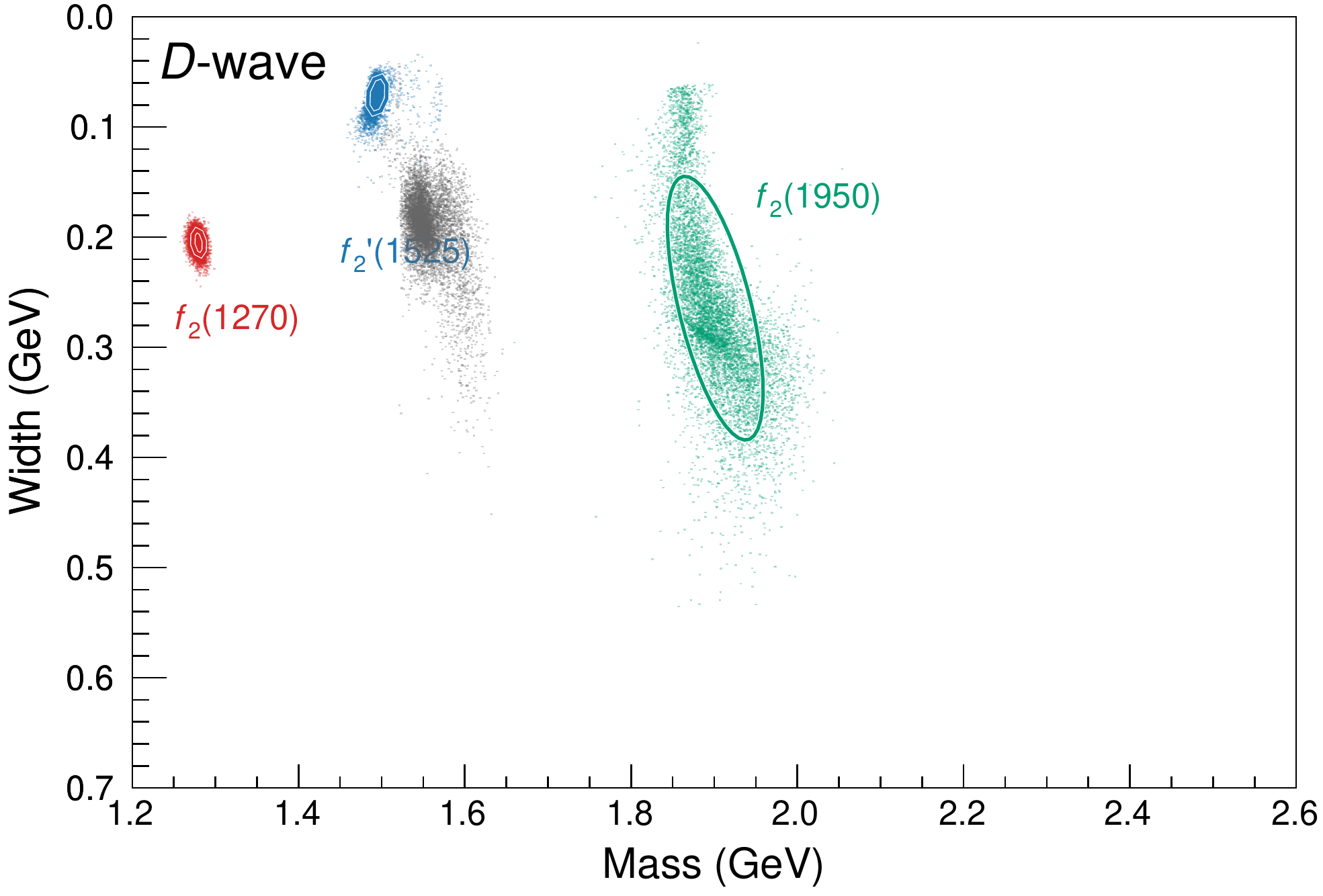}
\end{figure}

\begin{table}[h]
\begin{ruledtabular}
\begin{tabular}{c c c c}
$S$-wave  & $\sqrt{s_p}$ \mevp & $D$-wave & $\sqrt{s_p}$ \mevp \\ \hline
\rule[-0.2cm]{-0.1cm}{.55cm} $f_0(1500)$ &  $(1444 \pm 4) - i (119 \pm 4)/2$  &  $f_2(1270)$ &  $(1280 \pm 5) - i (205 \pm 8)/2$ \\
\rule[-0.2cm]{-0.1cm}{.55cm} $f_0(1710)$ &  $(1785 \pm 3) - i (148 \pm 9)/2$  &  $f_2'(1525)$ &  $(1495 \pm 8) - i (71 \pm 11)/2$ \\
\rule[-0.2cm]{-0.1cm}{.55cm} $f_0(2020)$ &  $(2048 \pm 11) - i (290 \pm 14)/2$  &  $f_2(1950)$ &  $(1901 \pm 38) - i (264 \pm 79)/2$ \\
\end{tabular}
\end{ruledtabular}
\end{table}

\clearpage

\subsection{$K^J(s) \quad\Big/\quad \omega(s)_\text{pole+scaled} \quad\Big/\quad \rho N^J_{ki}(s')_\text{Q-model} \quad\Big/\quad s_L = 0$}
\label{subsec:inputstartcm133c3_bootstrap-out}

\begin{table}[h]
\begin{ruledtabular}
\begin{tabular}{c c c c c c}
\multicolumn{3}{c}{$\pi \pi$ channel}  & \multicolumn{3}{c}{$ K  \bar K$ channel} \\ \hline
\rule[-0.2cm]{-0.1cm}{.55cm} $a^{S,\pi\pi}_0$ &$-932.715$ & $-987.763 \pm 124.584$ & $a^{S,K\bar K}_0$ & $-1667.292$ & $-1727.113 \pm 152.361$ \\
\rule[-0.2cm]{-0.1cm}{.55cm} $a^{S,\pi\pi}_1$ &$1483.985$ & $1567.741 \pm 195.201$ & $a^{S,K\bar K}_1$ & $2545.060$ & $2635.865 \pm 235.279$ \\
\rule[-0.2cm]{-0.1cm}{.55cm} $a^{S,\pi\pi}_2$ &$-731.004$ & $-767.311 \pm 91.181$ & $a^{S,K\bar K}_2$ & $-1171.917$ & $-1209.463 \pm 103.175$ \\
\rule[-0.2cm]{-0.1cm}{.55cm} $a^{S,\pi\pi}_3$ &$205.755$ & $212.457 \pm 20.734$ & $a^{S,K\bar K}_3$ & $251.251$ & $257.448 \pm 20.547$ \\
\hline
\rule[-0.2cm]{-0.1cm}{.55cm} $a^{D,\pi\pi}_0$ &$-5.172$ & $-5.196 \pm 0.692$ & $a^{D,K\bar K}_0$ & $5.847$ & $5.783 \pm 0.752$ \\
\rule[-0.2cm]{-0.1cm}{.55cm} $a^{D,\pi\pi}_1$ &$-11.454$ & $-11.498 \pm 1.141$ & $a^{D,K\bar K}_1$ & $9.946$ & $9.834 \pm 1.217$ \\
\rule[-0.2cm]{-0.1cm}{.55cm} $a^{D,\pi\pi}_2$ &$-6.142$ & $-6.180 \pm 0.639$ & $a^{D,K\bar K}_2$ & $4.692$ & $4.628 \pm 0.765$ \\
\rule[-0.2cm]{-0.1cm}{.55cm} $a^{D,\pi\pi}_3$ &$-1.515$ & $-1.533 \pm 0.237$ & $a^{D,K\bar K}_3$ & $0.650$ & $0.609 \pm 0.302$ \\
\end{tabular}
\end{ruledtabular}
\end{table}

\begin{table}[h]
\begin{ruledtabular}
\begin{tabular}{c c c c c c c c c}
\multicolumn{3}{c}{$S$-wave couplings $\pi \pi$ channel}  & \multicolumn{3}{c}{$S$-wave couplings $K  \bar K$ channel} & \multicolumn{3}{c}{$S$-wave couplings $\rho\rho$ channel} \\ \hline
\rule[-0.2cm]{-0.1cm}{.55cm} $g^{S,1}_{\pi\pi}$ & $-6.338$ & $-6.367 \pm 0.095$ & $g^{S,1}_{K\bar K}$ & $4.369$ & $4.446 \pm 0.145$ & $g^{S,1}_{\rho\rho}$ & $2.155$ & $2.181 \pm 0.163$ \\
\rule[-0.2cm]{-0.1cm}{.55cm} $g^{S,2}_{\pi\pi}$ & $-0.901$ & $-0.938 \pm 0.086$ & $g^{S,2}_{K\bar K}$ & $0.545$ & $0.562 \pm 0.062$ & $g^{S,2}_{\rho\rho}$ & $-0.646$ & $-0.645 \pm 0.061$ \\
\rule[-0.2cm]{-0.1cm}{.55cm} $g^{S,3}_{\pi\pi}$ & $-8.789$ & $-8.873 \pm 0.355$ & $g^{S,3}_{K\bar K}$ & $10.425$ & $10.433 \pm 0.273$ & $g^{S,3}_{\rho\rho}$ & $1.470$ & $1.469 \pm 0.133$ \\
\hline 
\multicolumn{3}{c}{$D$-wave couplings $\pi \pi$ channel}  & \multicolumn{3}{c}{$D$-wave couplings $K  \bar K$ channel} & \multicolumn{3}{c}{$D$-wave couplings $\rho\rho$ channel} \\ \hline
\rule[-0.2cm]{-0.1cm}{.55cm} $g^{D,1}_{\pi\pi}$ & $0.737$ & $0.769 \pm 0.186$ & $g^{D,1}_{K\bar K}$ & $3.629$ & $3.654 \pm 0.116$ & $g^{D,1}_{\rho\rho}$ & $0$ & $0$ \\
\rule[-0.2cm]{-0.1cm}{.55cm} $g^{D,2}_{\pi\pi}$ & $-3.526$ & $-4.314 \pm 2.168$ & $g^{D,2}_{K\bar K}$ & $-5.282$ & $-6.385 \pm 2.967$ & $g^{D,2}_{\rho\rho}$ & $17.600$ & $16.958 \pm 3.053$ \\
\rule[-0.2cm]{-0.1cm}{.55cm} $g^{D,3}_{\pi\pi}$ & $5.980$ & $9.030 \pm 10.091$ & $g^{D,3}_{K\bar K}$ & $72.146$ & $72.730 \pm 2.555$ & $g^{D,3}_{\rho\rho}$ & $3.232$ & $3.589 \pm 1.337$ \\
\hline 
\multicolumn{9}{c}{$S$-wave bare masses}   \\ \hline
\rule[-0.2cm]{-0.1cm}{.55cm} $m^2_{S,1}$ & $0.163$ & $0.146 \pm 0.052$ & $m^2_{S,2}$ & $3.219$ & $3.219 \pm 0.039$ & $m^2_{S,3}$ & $6.839$ & $6.836 \pm 0.102$ \\
\hline 
\multicolumn{9}{c}{$D$-wave bare masses}   \\ \hline
\rule[-0.2cm]{-0.1cm}{.55cm} $m^2_{D,1}$ & $2.497$ & $2.500 \pm 0.016$ & $m^2_{D,2}$ & $7.033$ & $7.005 \pm 0.872$ & $m^2_{D,3}$ & $18.114$ & $17.687 \pm 1.163$ \\
\end{tabular}
\vspace{2em}
\begin{tabular}{c c c c c c}
\multicolumn{3}{c}{$S$-wave background}  & \multicolumn{3}{c}{$D$-wave background} \\ \hline
\rule[-0.2cm]{-0.1cm}{.55cm} $c^{S,0}_{\pi\pi,\pi\pi}$ &$15.046$ & $14.677 \pm 1.054$ & $c^{D,0}_{\pi\pi,\pi\pi}$ &$14.392$ & $2.850 \pm 35.312$ \\
\rule[-0.2cm]{-0.1cm}{.55cm} $c^{S,0}_{\pi\pi,K\bar K}$ &$-8.388$ & $-8.265 \pm 0.701$ & $c^{D,0}_{\pi\pi,K\bar K}$ &$-36.041$ & $-54.850 \pm 55.743$ \\
\rule[-0.2cm]{-0.1cm}{.55cm} $c^{S,0}_{\pi\pi,\rho\rho}$ &$0$ & $0$ & $c^{D,0}_{\pi\pi,\rho\rho}$ &$0$ & $0$ \\
\rule[-0.2cm]{-0.1cm}{.55cm} $c^{S,0}_{K\bar K,K\bar K}$ &$1.729$ & $1.850 \pm 0.764$ & $c^{D,0}_{K\bar K,K\bar K}$ &$-269.583$ & $-285.371 \pm 38.587$ \\
\rule[-0.2cm]{-0.1cm}{.55cm} $c^{S,0}_{K\bar K,\rho\rho}$ &$0$ & $0$ & $c^{D,0}_{K\bar K,\rho\rho}$ &$0$ & $0$ \\
\rule[-0.2cm]{-0.1cm}{.55cm} $c^{S,0}_{\rho\rho,\rho\rho}$ &$0.147$ & $0.165 \pm 1.293$ & $c^{D,0}_{\rho\rho,\rho\rho}$ &$-31.902$ & $-31.312 \pm 9.572$ \\
\rule[-0.2cm]{-0.1cm}{.55cm} $c^{S,1}_{\pi\pi,\pi\pi}$ &$-8.312$ & $-8.359 \pm 0.217$ & $c^{D,1}_{\pi\pi,\pi\pi}$ &$2.431$ & $1.641 \pm 3.236$ \\
\rule[-0.2cm]{-0.1cm}{.55cm} $c^{S,1}_{\pi\pi,K\bar K}$ &$7.822$ & $7.856 \pm 0.151$ & $c^{D,1}_{\pi\pi,K\bar K}$ &$5.792$ & $3.949 \pm 6.888$ \\
\rule[-0.2cm]{-0.1cm}{.55cm} $c^{S,1}_{\pi\pi,\rho\rho}$ &$0$ & $0$ & $c^{D,1}_{\pi\pi,\rho\rho}$ &$0$ & $0$ \\
\rule[-0.2cm]{-0.1cm}{.55cm} $c^{S,1}_{K\bar K,K\bar K}$ &$-7.635$ & $-7.653 \pm 0.124$ & $c^{D,1}_{K\bar K,K\bar K}$ &$-20.919$ & $-23.637 \pm 8.572$ \\
\rule[-0.2cm]{-0.1cm}{.55cm} $c^{S,1}_{K\bar K,\rho\rho}$ &$0$ & $0$ & $c^{D,1}_{K\bar K,\rho\rho}$ &$0$ & $0$ \\
\rule[-0.2cm]{-0.1cm}{.55cm} $c^{S,1}_{\rho\rho,\rho\rho}$ &$1.133$ & $1.111 \pm 0.470$ & $c^{D,1}_{\rho\rho,\rho\rho}$ &$-12.643$ & $-11.460 \pm 3.568$ \\
\end{tabular}
\end{ruledtabular}
\end{table}

\begin{figure}[h]
\centering\includegraphics[width=0.32\textwidth]{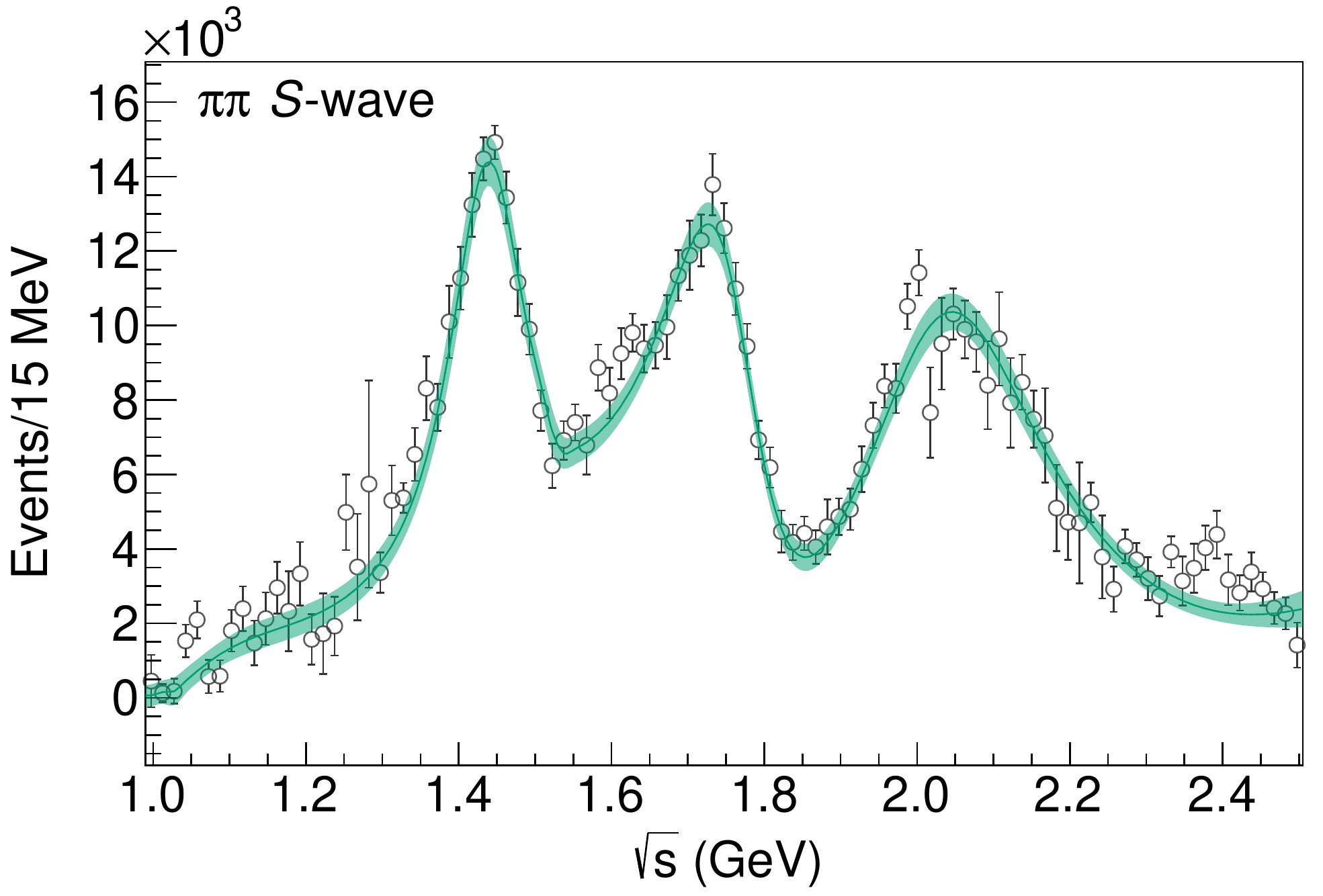} \includegraphics[width=0.32\textwidth]{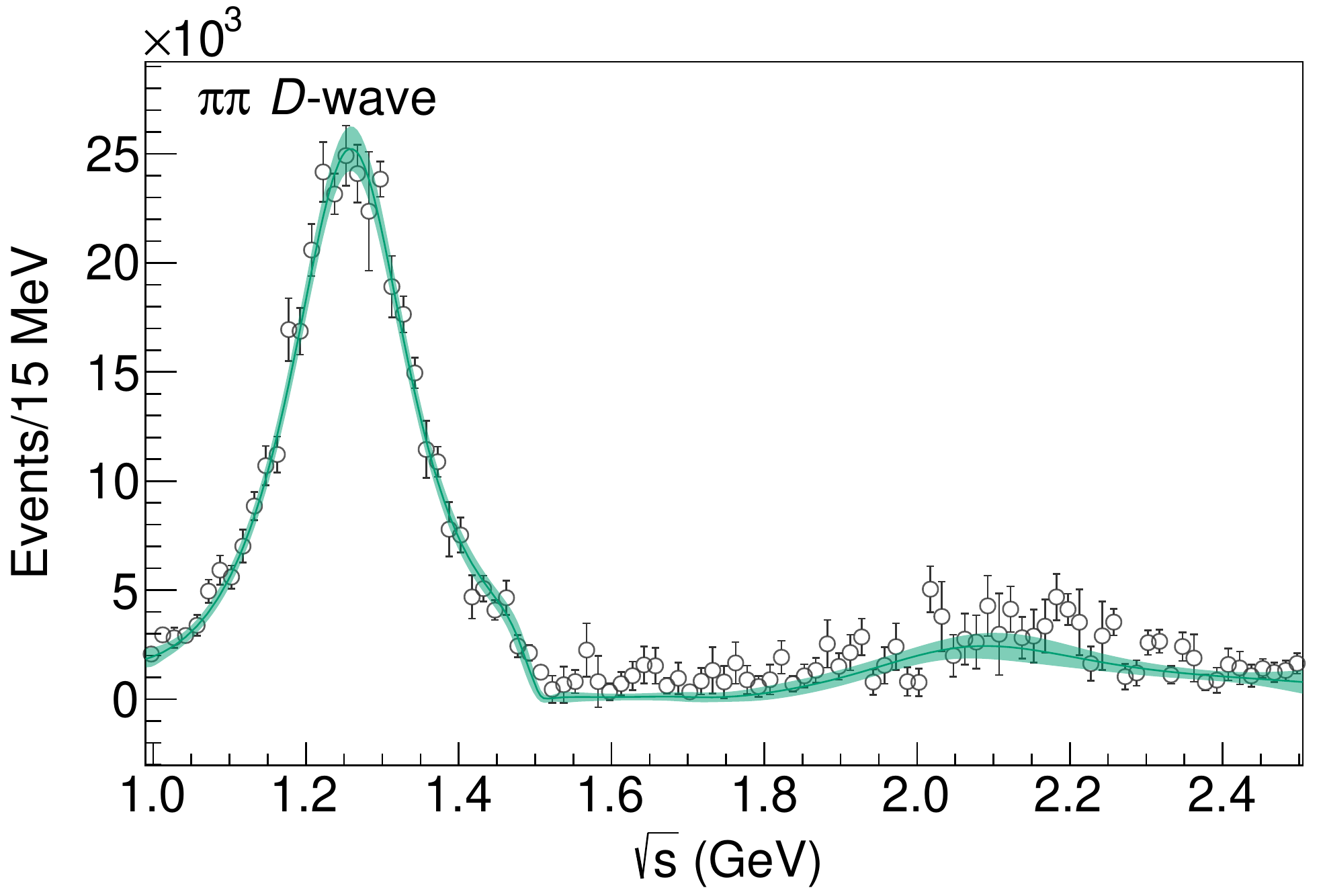} \includegraphics[width=0.32\textwidth]{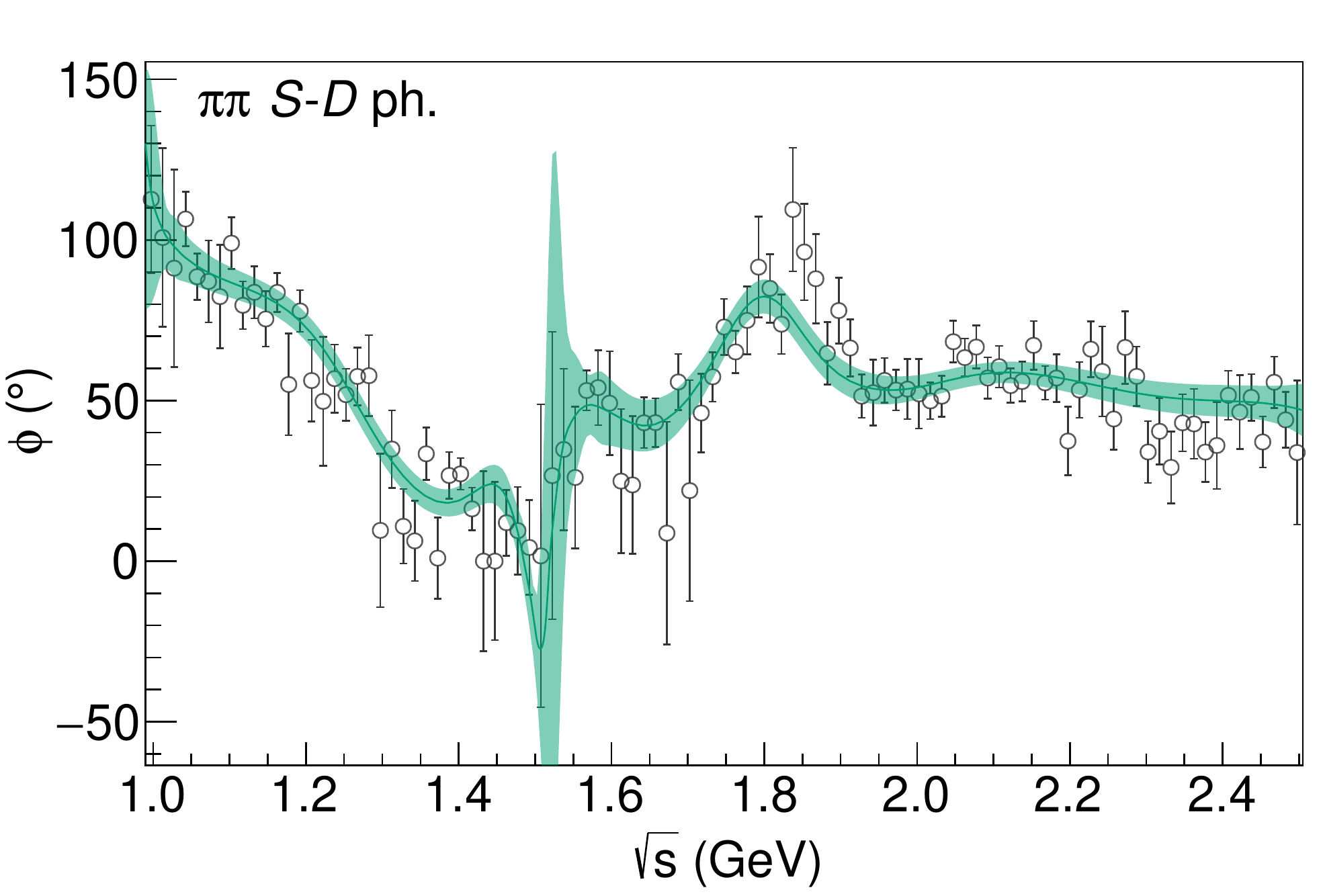}
\includegraphics[width=0.32\textwidth]{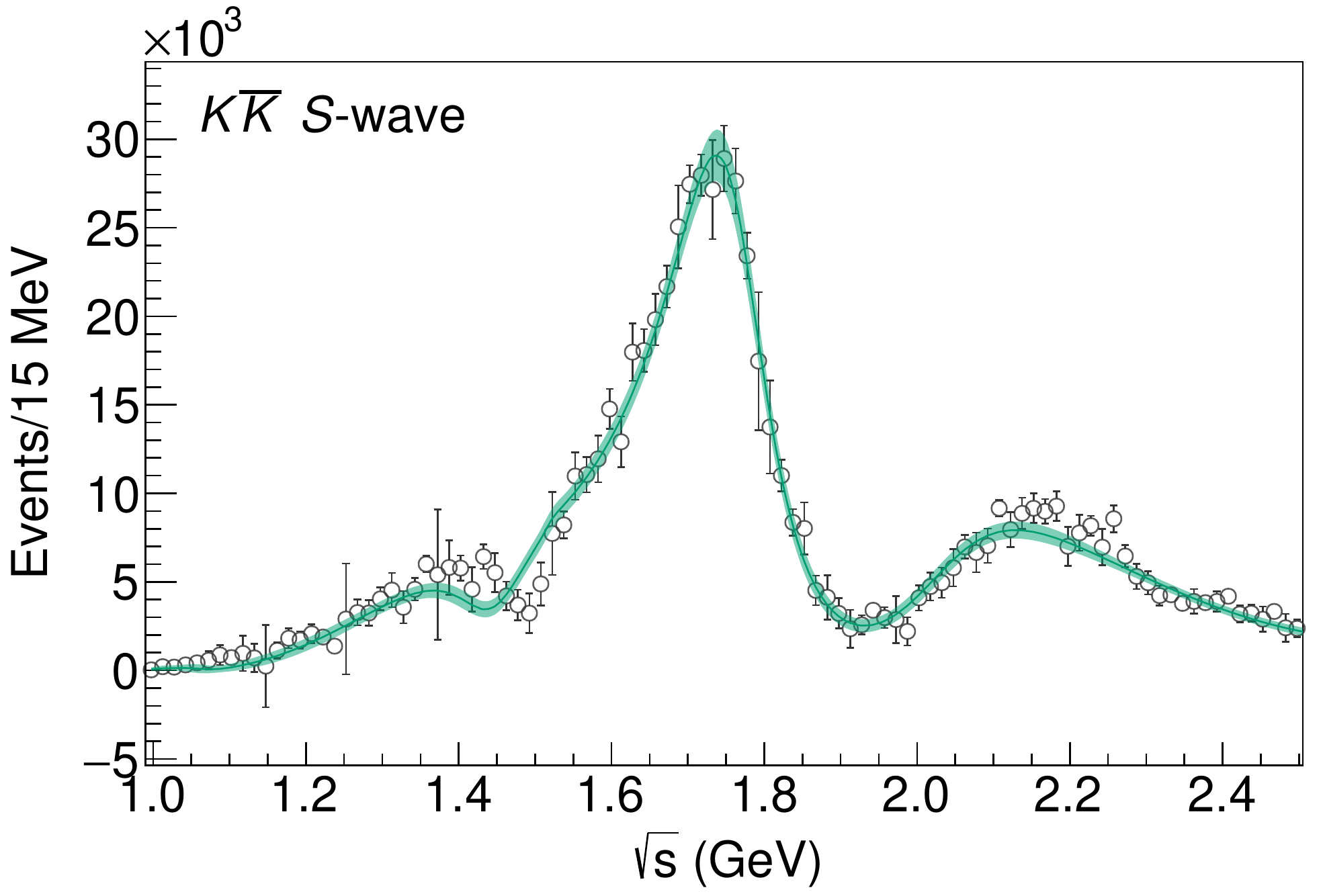} \includegraphics[width=0.32\textwidth]{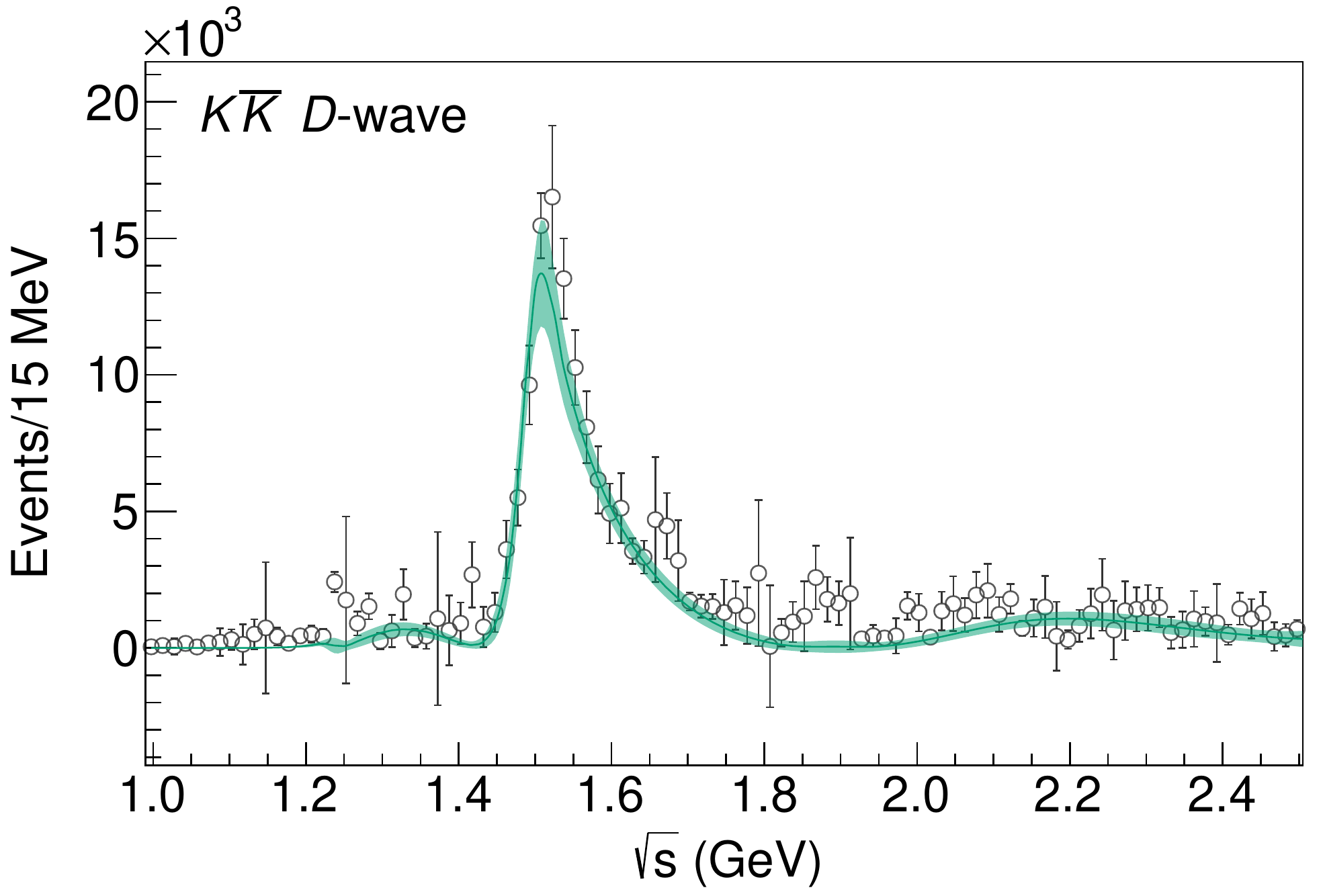} \includegraphics[width=0.32\textwidth]{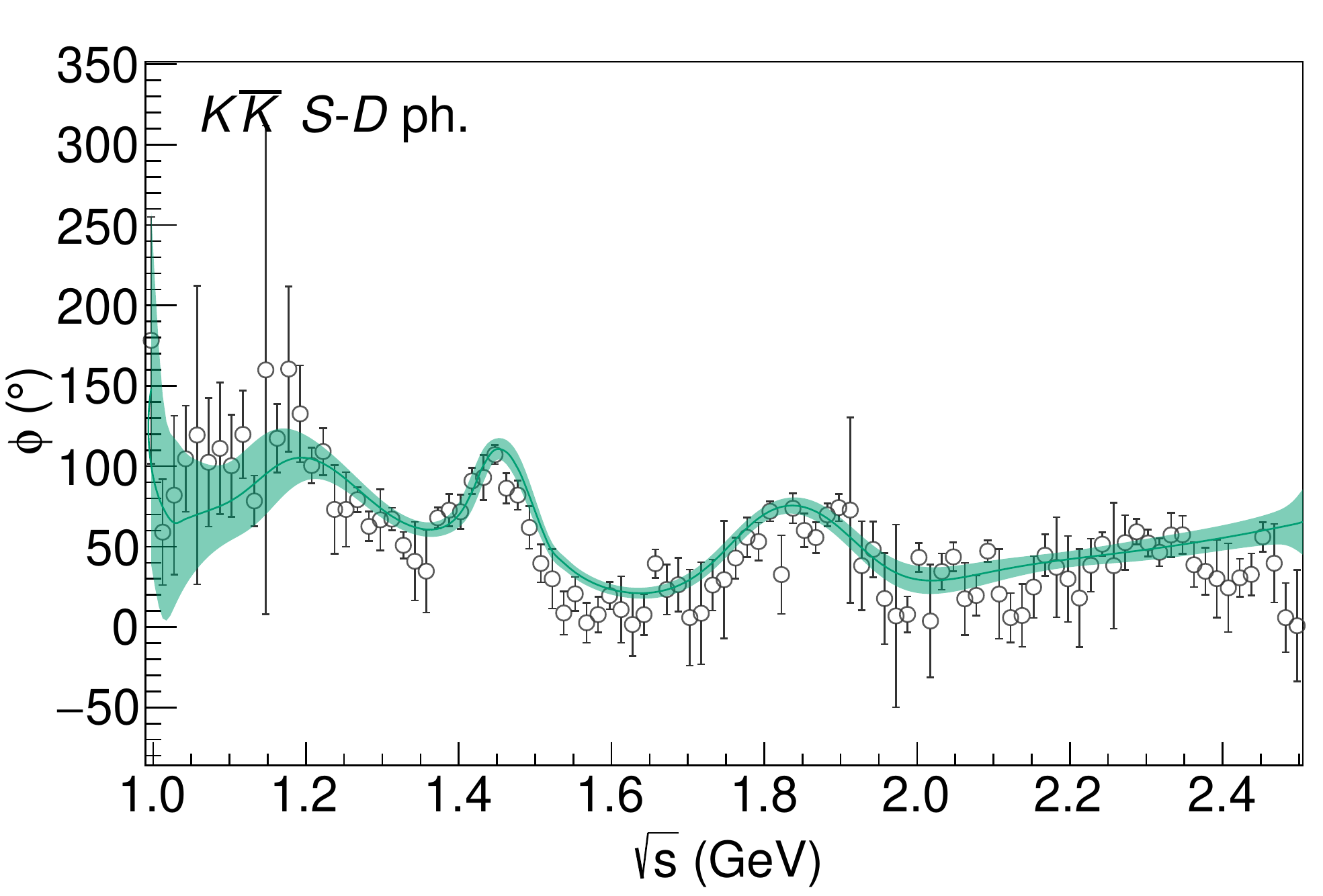}
\end{figure}

\begin{figure}[h]
\centering\includegraphics[width=0.45\textwidth]{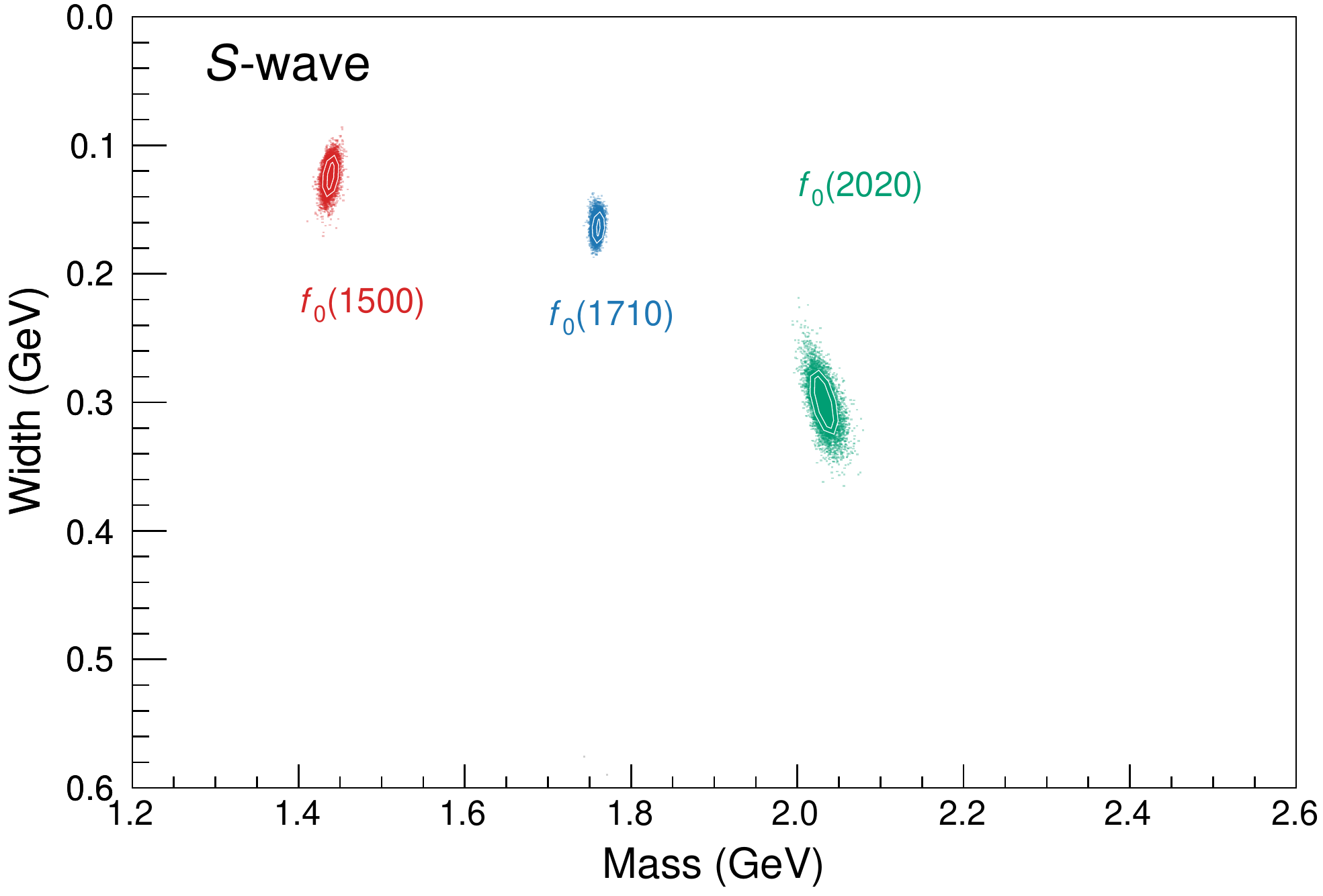} \includegraphics[width=0.45\textwidth]{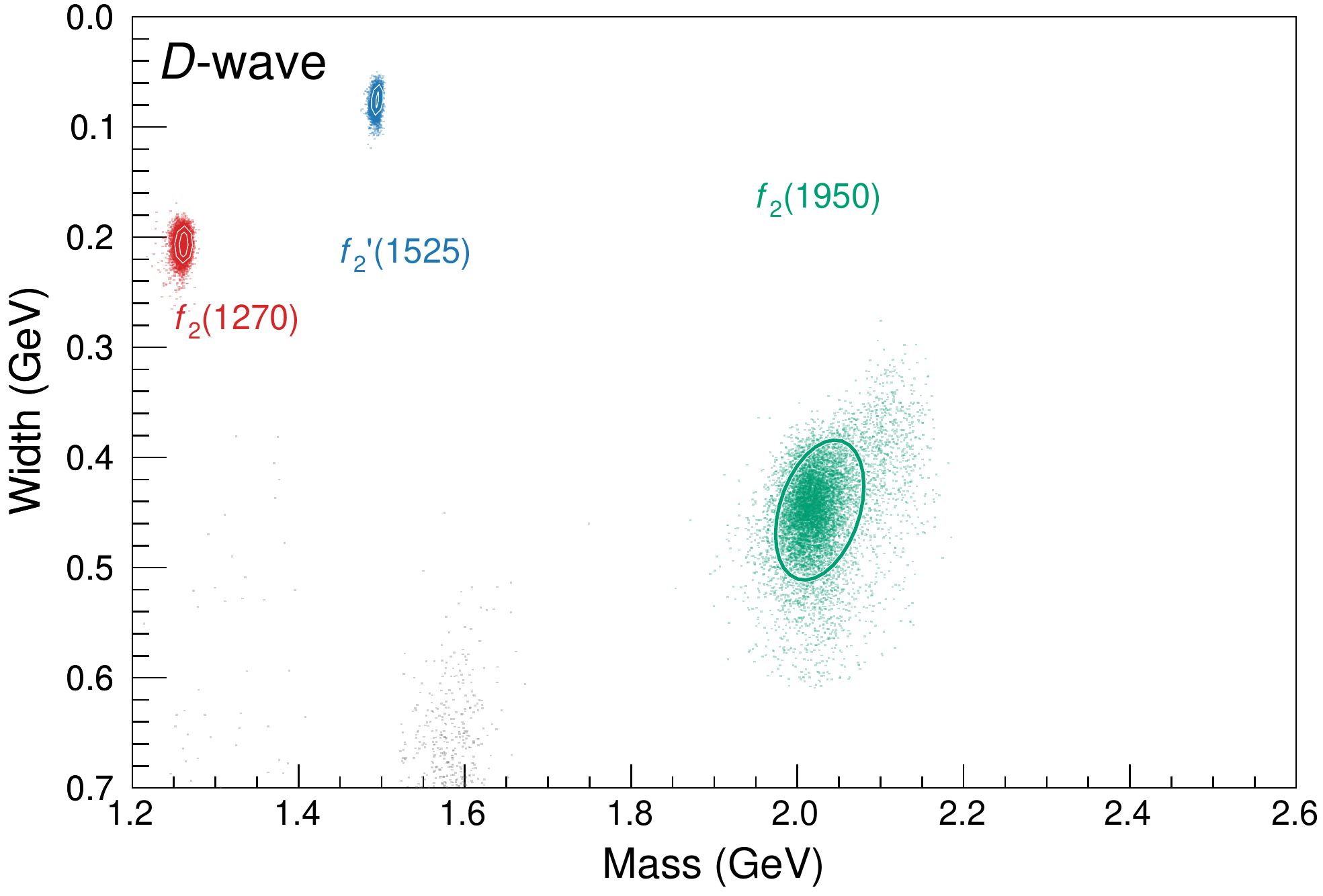}
\end{figure}

\begin{table}[h]
\begin{ruledtabular}
\begin{tabular}{c c c c}
$S$-wave  & $\sqrt{s_p}$ \mevp & $D$-wave & $\sqrt{s_p}$ \mevp \\ \hline
\rule[-0.2cm]{-0.1cm}{.55cm} $f_0(1500)$ &  $(1439 \pm 5) - i (124 \pm 9)/2$  &  $f_2(1270)$ &  $(1262 \pm 5) - i (207 \pm 9)/2$ \\
\rule[-0.2cm]{-0.1cm}{.55cm} $f_0(1710)$ &  $(1761 \pm 3) - i (164 \pm 6)/2$  &  $f_2'(1525)$ &  $(1495 \pm 3) - i (75 \pm 7)/2$ \\
\rule[-0.2cm]{-0.1cm}{.55cm} $f_0(2020)$ &  $(2032 \pm 10) - i (300 \pm 15)/2$  &  $f_2(1950)$ &  $(2027 \pm 35) - i (448 \pm 42)/2$ \\
\end{tabular}
\end{ruledtabular}
\end{table}

\clearpage

\subsection{$K^J(s) \quad\Big/\quad \omega(s)_\text{pole} \quad\Big/\quad \rho N^J_{ki}(s')_\text{Q-model} \quad\Big/\quad s_L = 0$}
\label{subsec:inputstartcm15new2c3_bootstrap-out}

\input{tabs-supp-material/numerator-table-inputstartcm15new2c3_bootstrap-out}

\input{tabs-supp-material/denominator-table-inputstartcm15new2c3_bootstrap-out}

\begin{figure}[h]
\centering\includegraphics[width=0.32\textwidth]{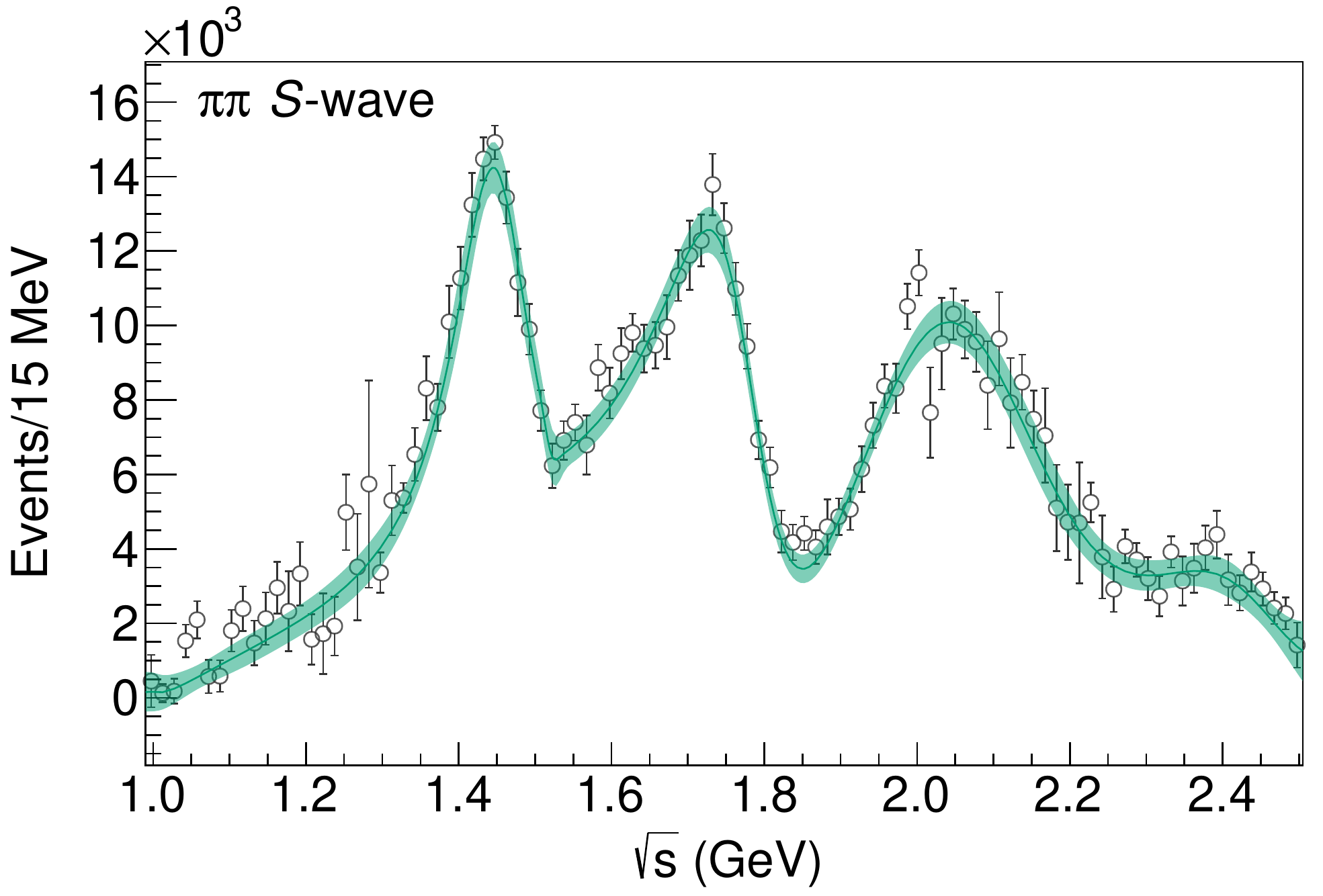} \includegraphics[width=0.32\textwidth]{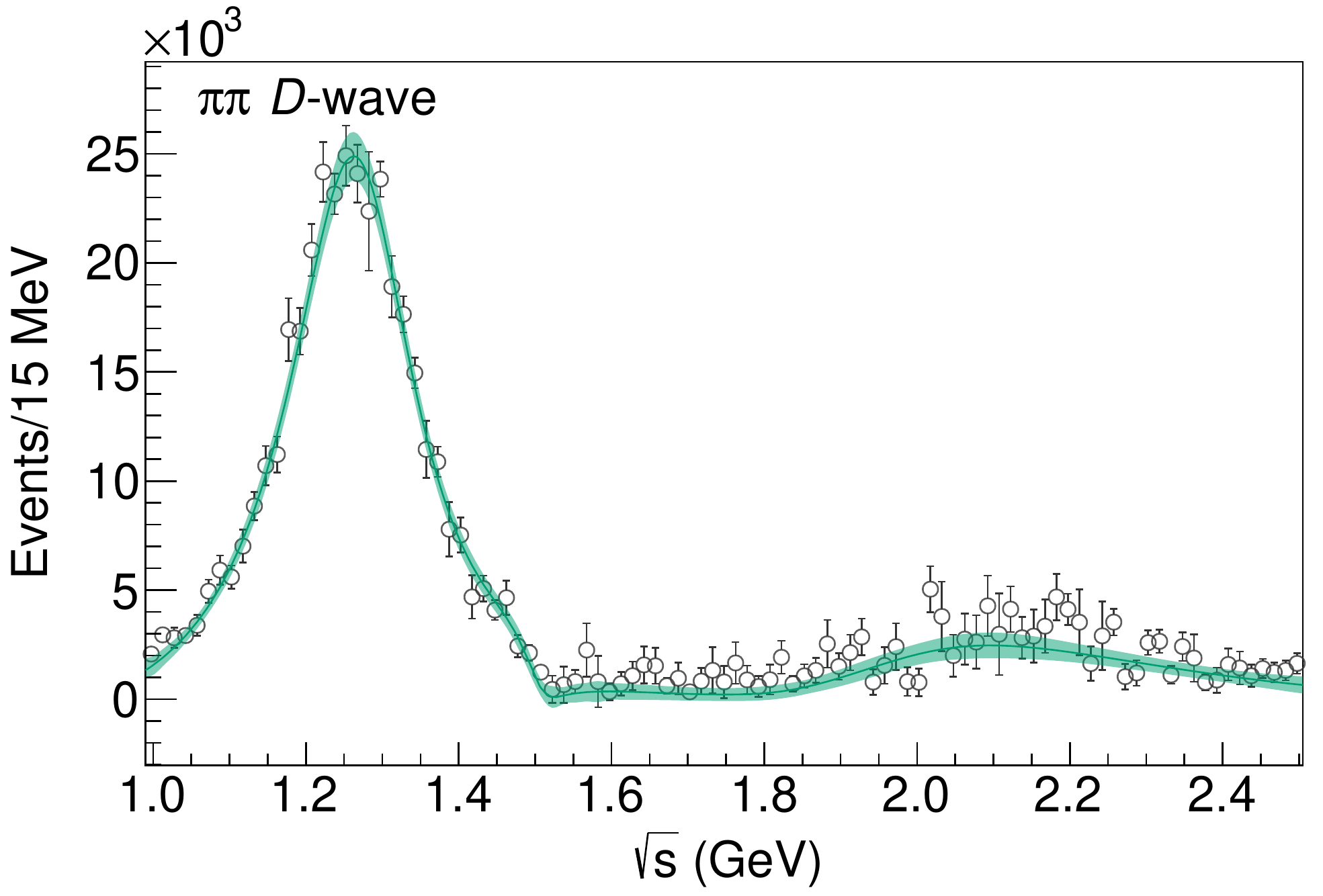} \includegraphics[width=0.32\textwidth]{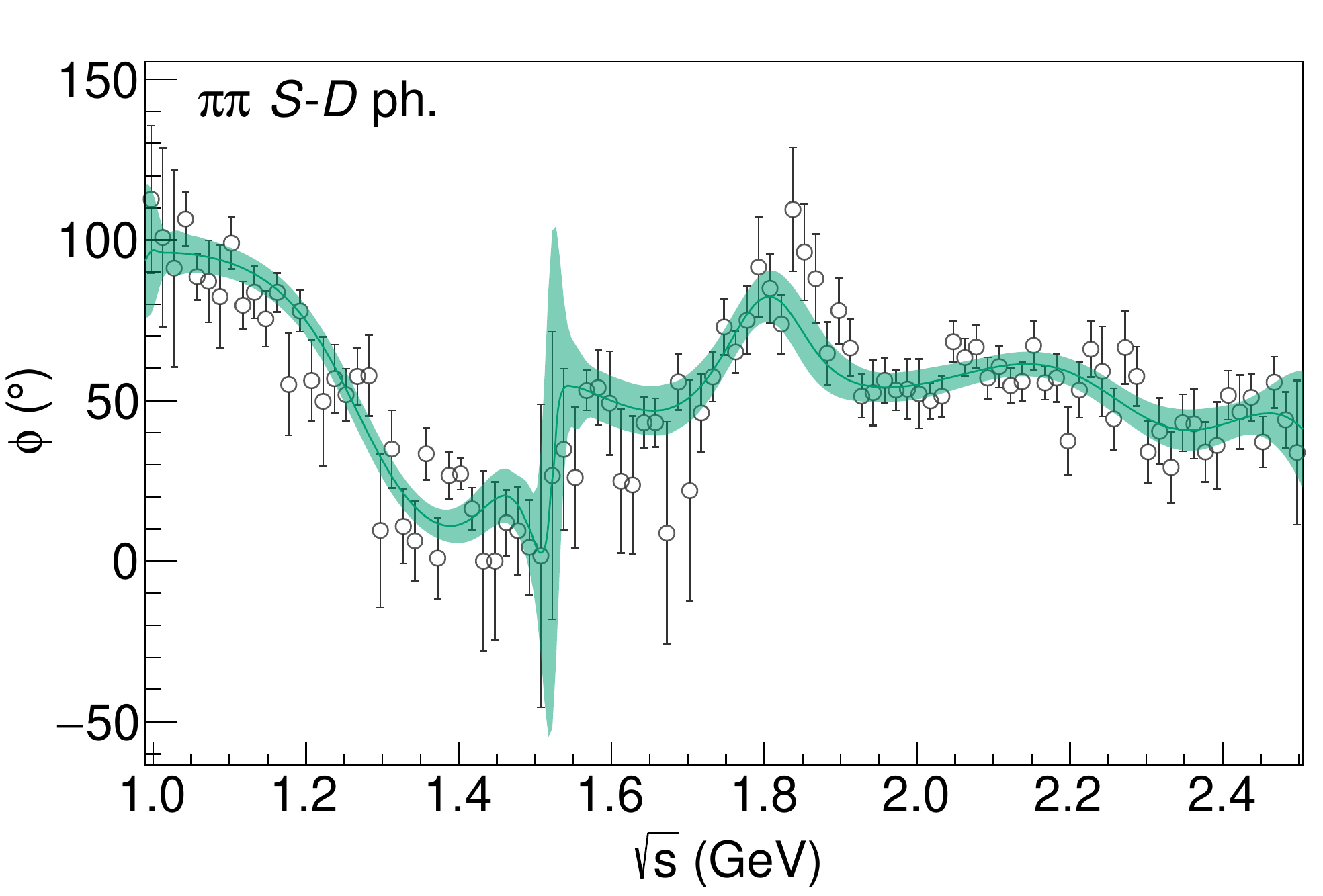}
\includegraphics[width=0.32\textwidth]{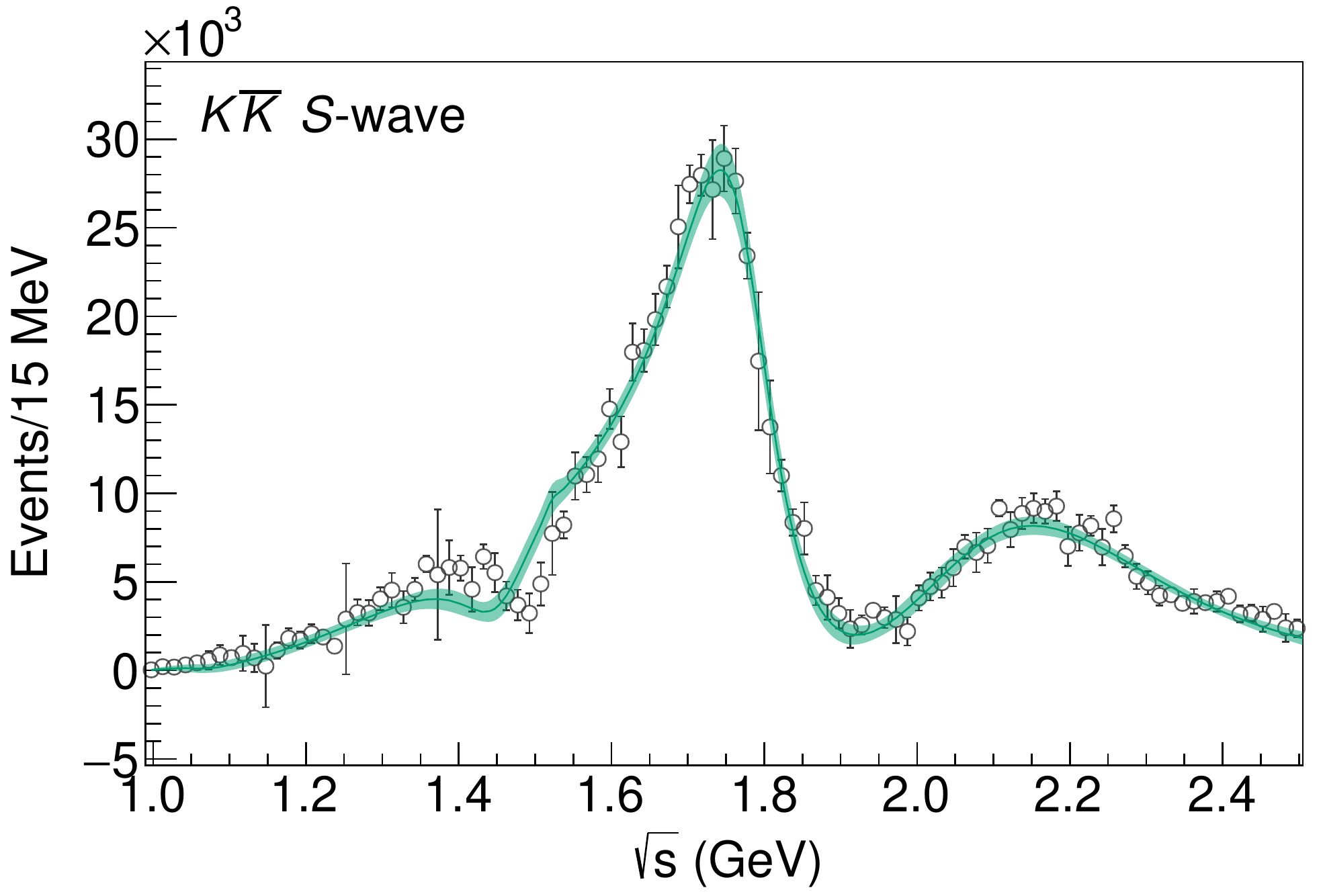} \includegraphics[width=0.32\textwidth]{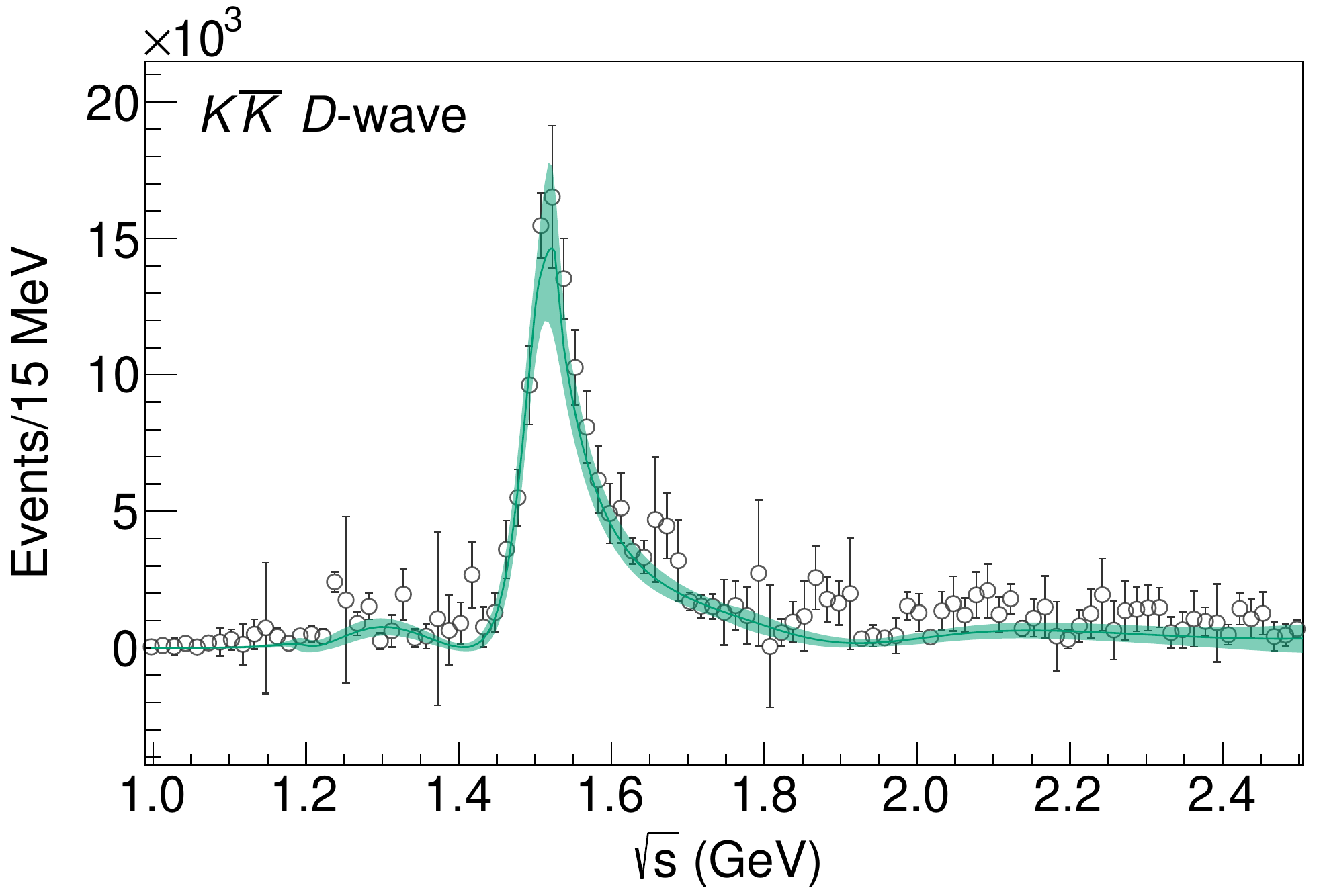} \includegraphics[width=0.32\textwidth]{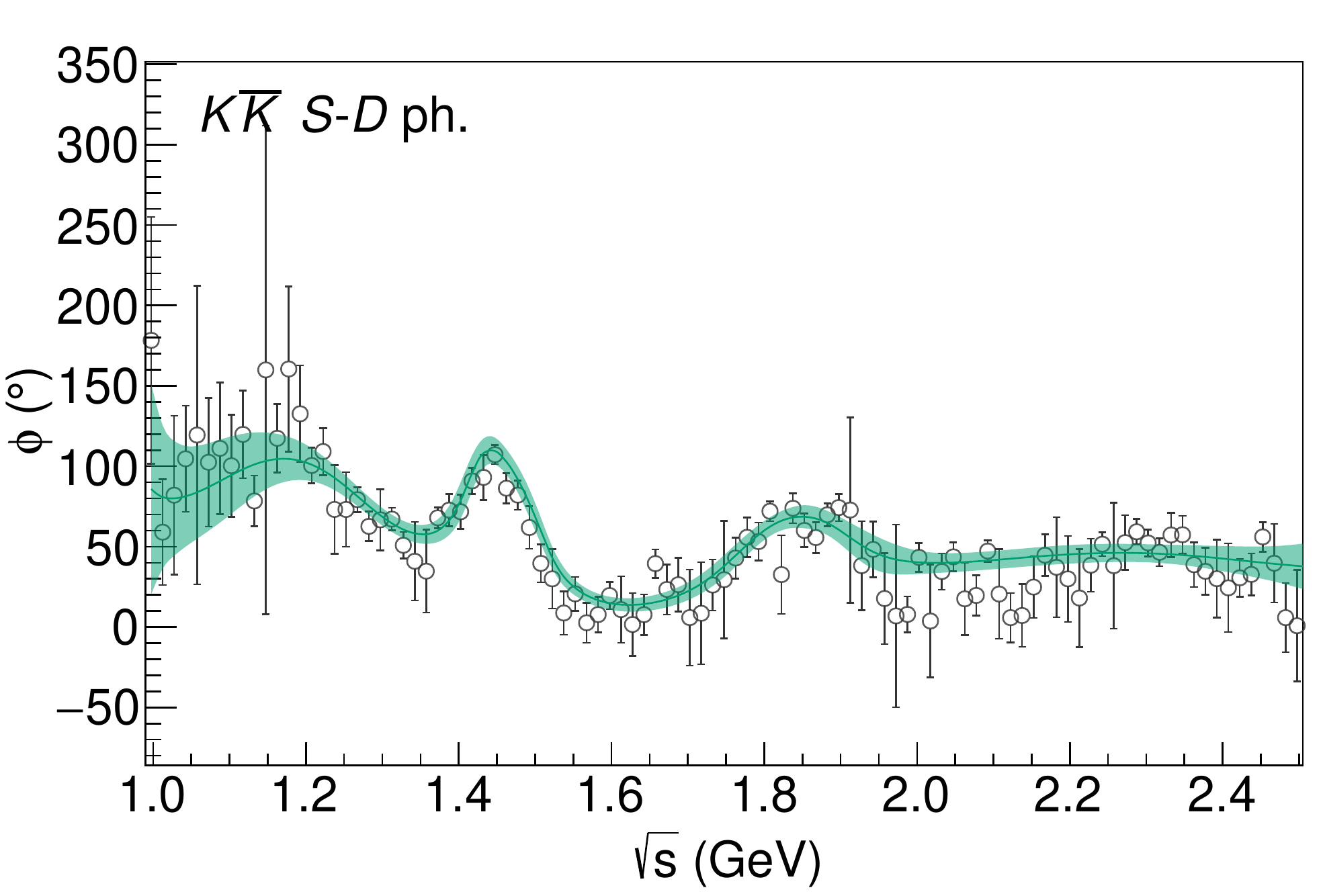}
\end{figure}

\begin{figure}[h]
\centering\includegraphics[width=0.45\textwidth]{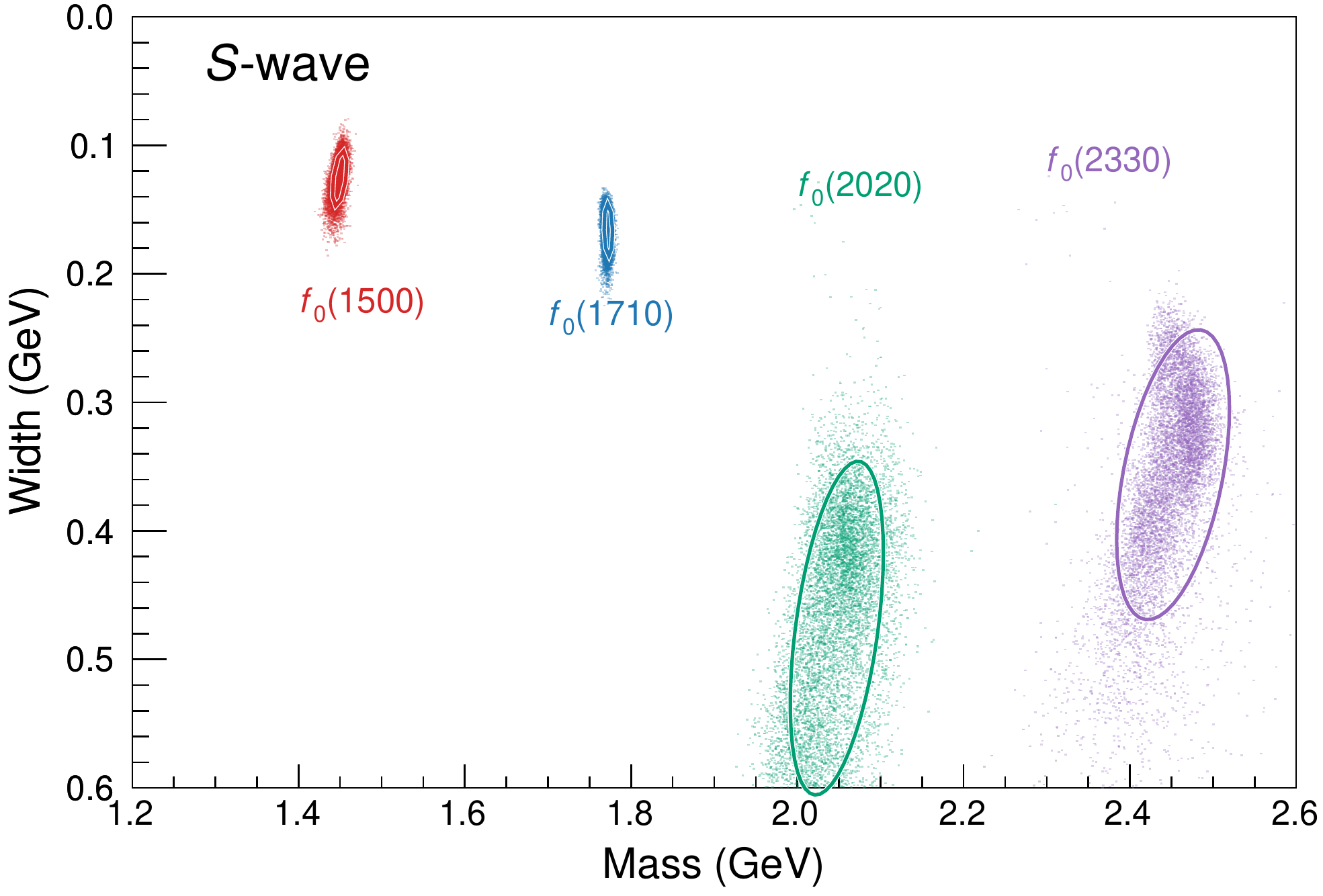} \includegraphics[width=0.45\textwidth]{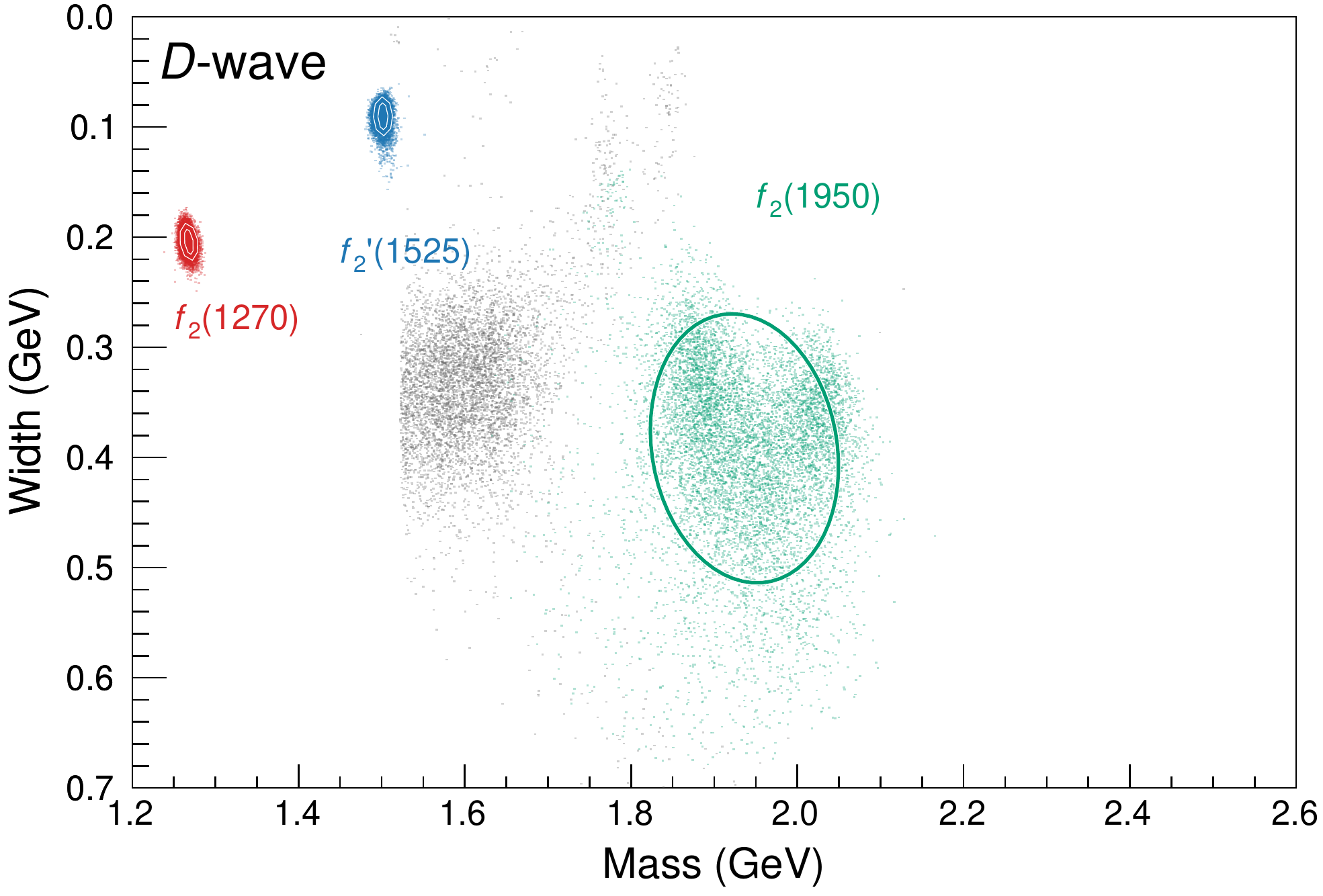}
\end{figure}

\input{tabs-supp-material/poles-inputstartcm15new2c3_bootstrap-out}

\clearpage

\subsection{$K^J(s) \quad\Big/\quad \omega(s)_\text{pole} \quad\Big/\quad \rho N^J_{ki}(s')_\text{Q-model} \quad\Big/\quad s_L = 0.6\gevsq$}
\label{subsec:inputstartcm3c3_bootstrap-out}

\input{tabs-supp-material/numerator-table-inputstartcm3c3_bootstrap-out}

\input{tabs-supp-material/denominator-table-inputstartcm3c3_bootstrap-out}

\begin{figure}[h]
\centering\includegraphics[width=0.32\textwidth]{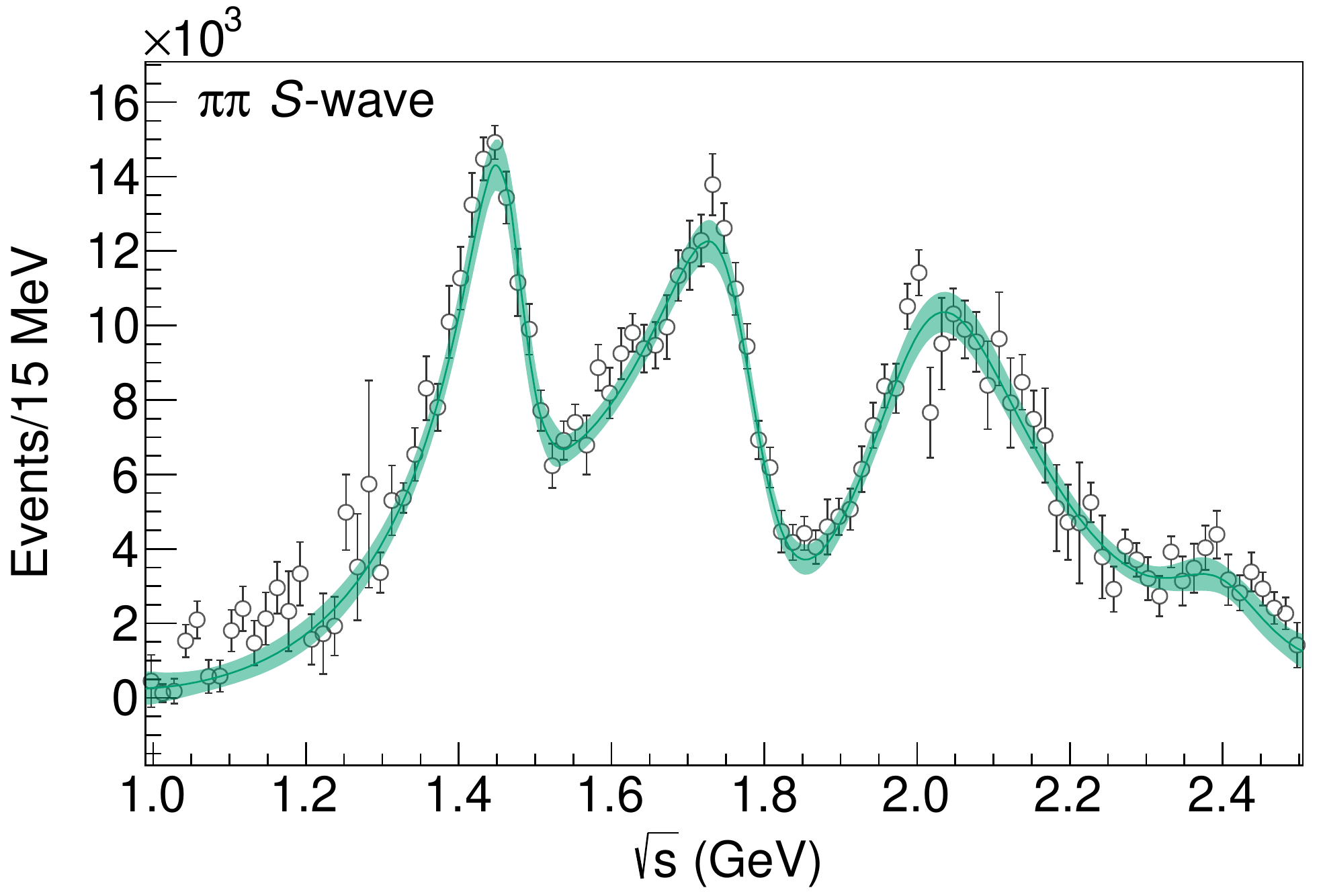} \includegraphics[width=0.32\textwidth]{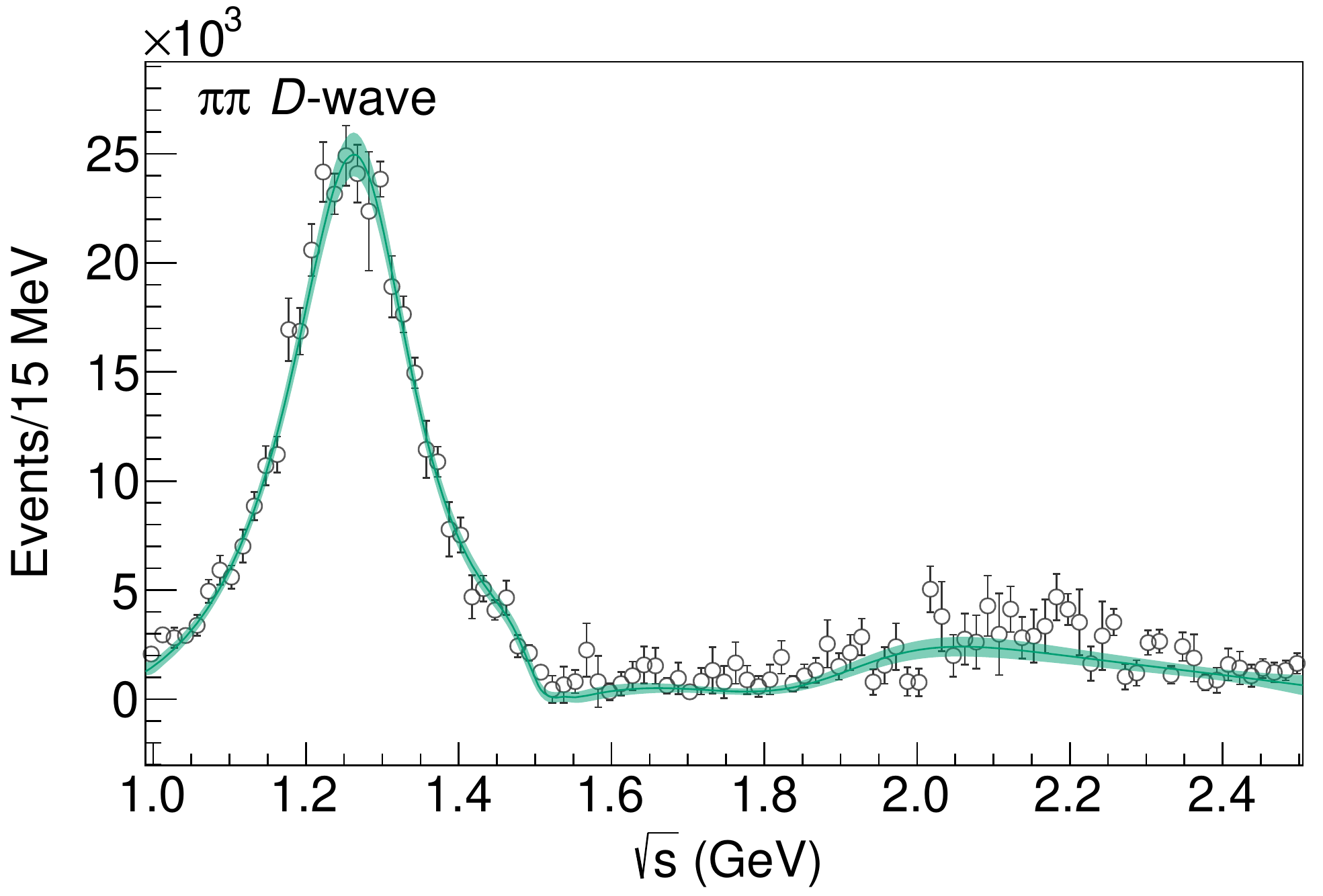} \includegraphics[width=0.32\textwidth]{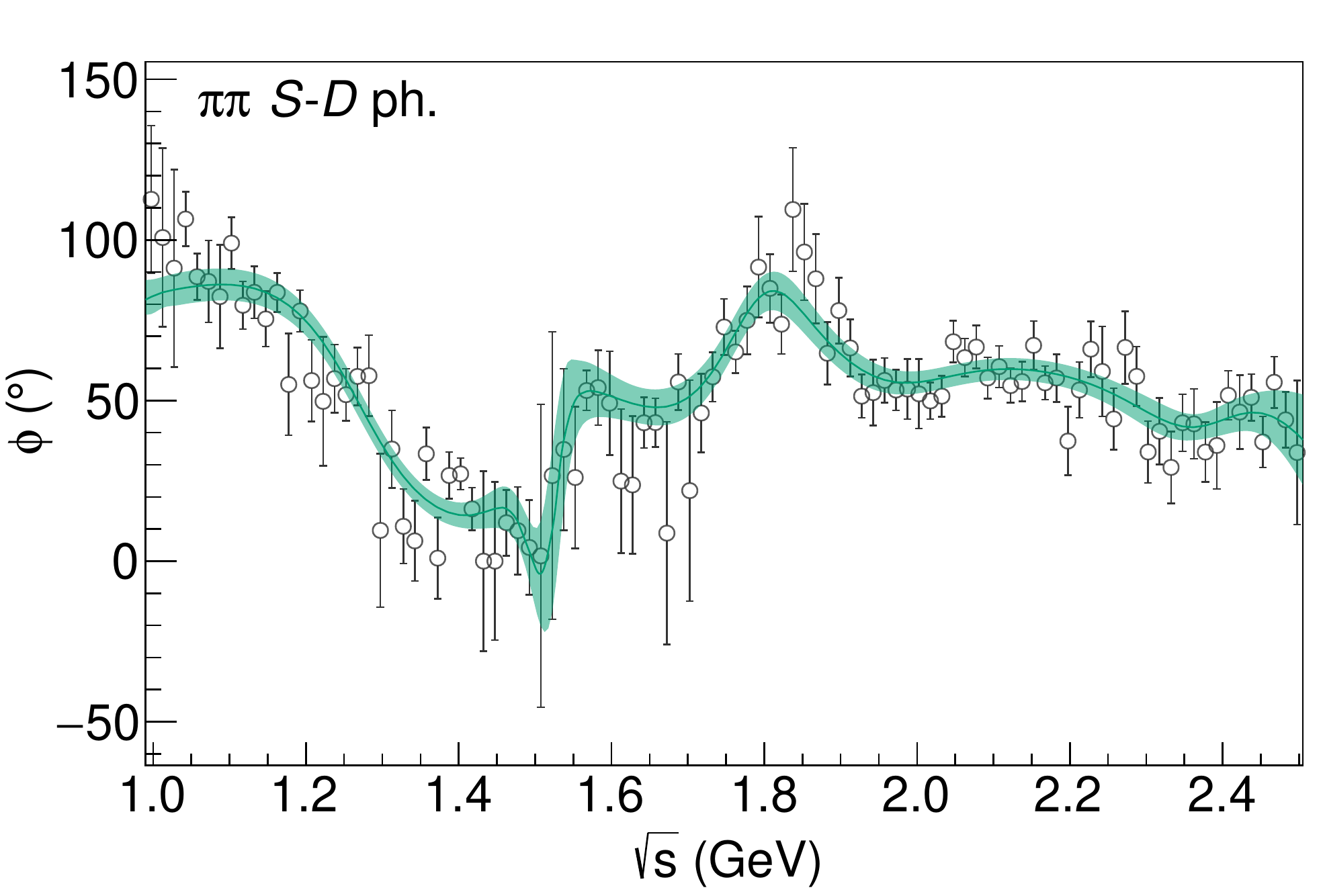}
\includegraphics[width=0.32\textwidth]{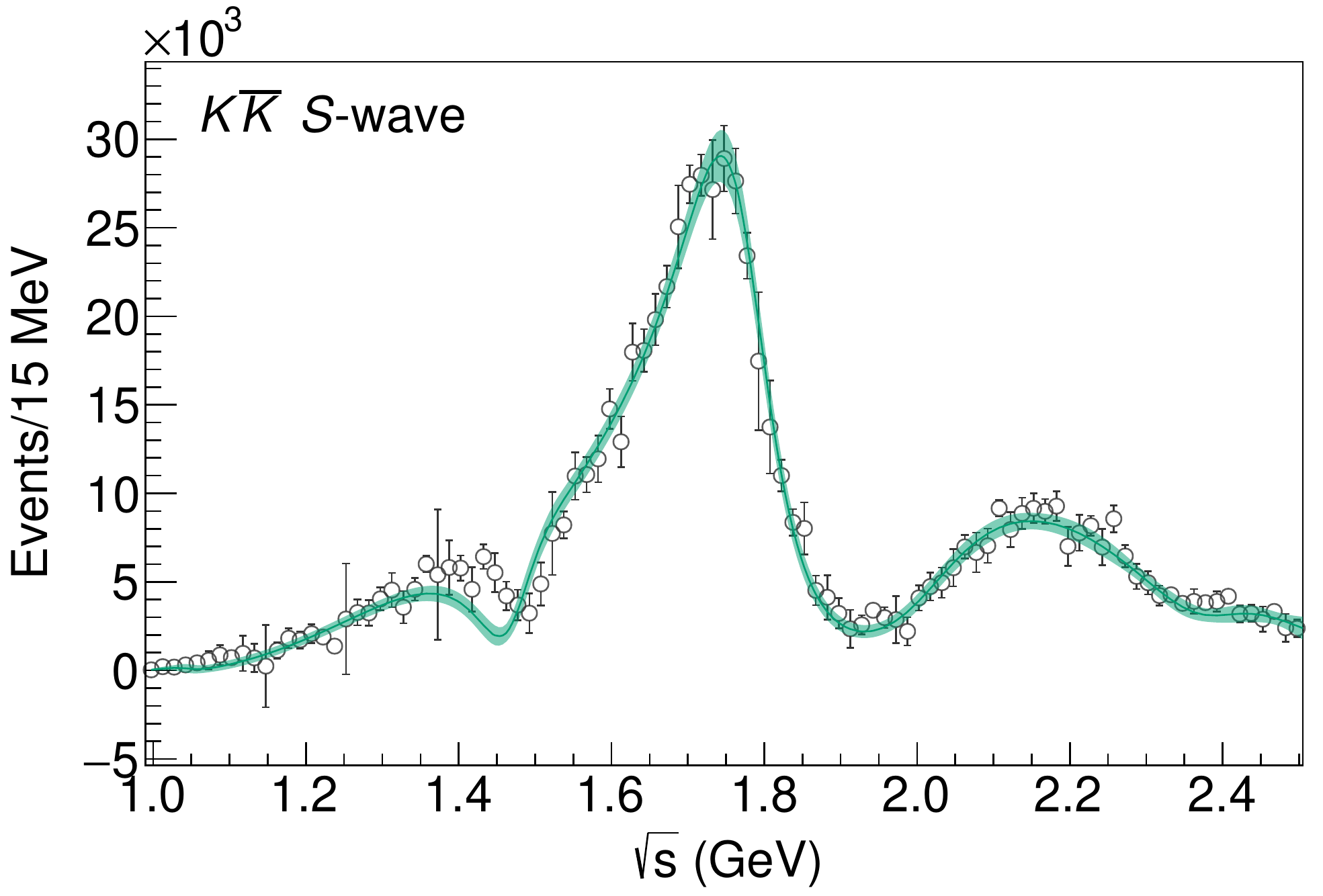} \includegraphics[width=0.32\textwidth]{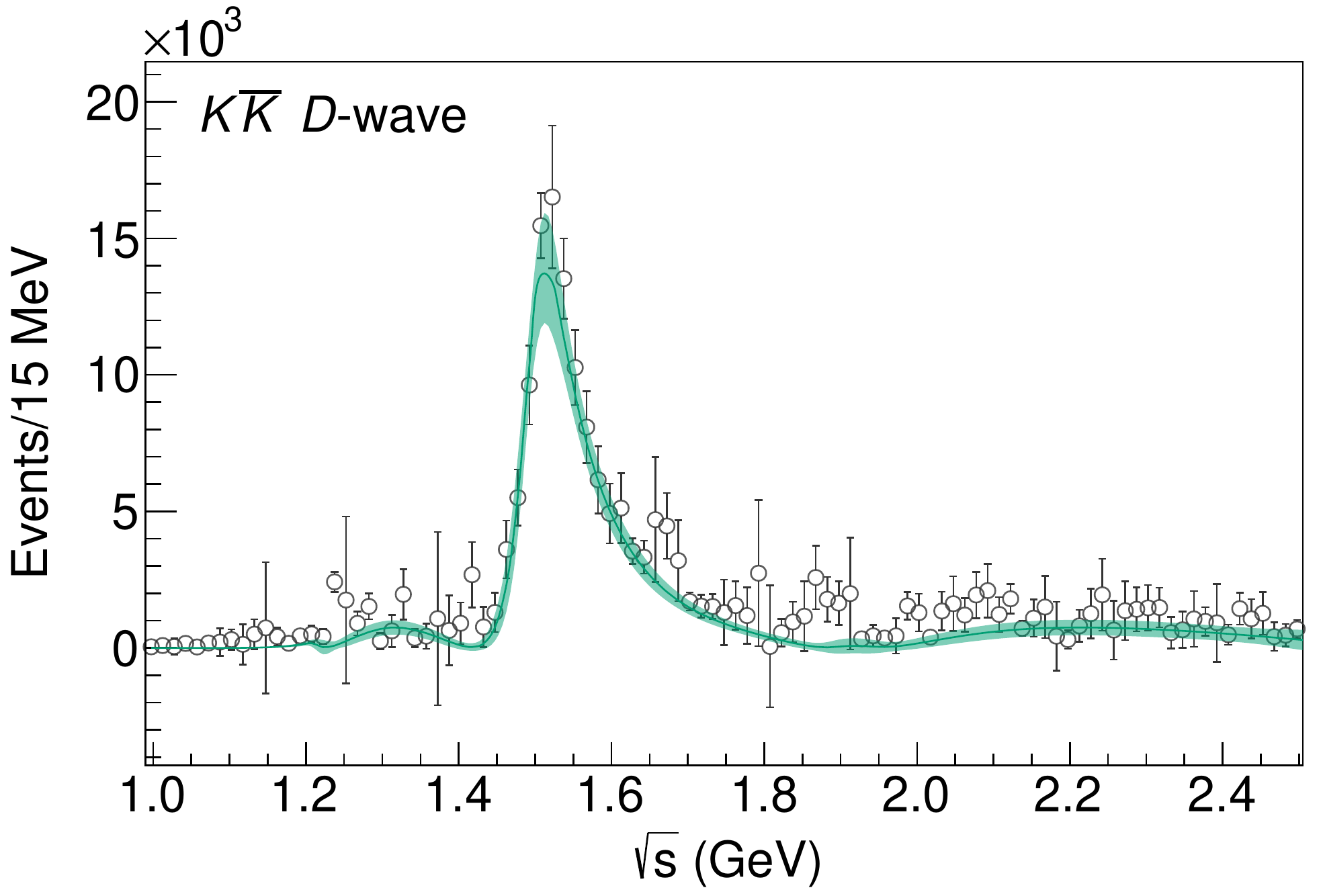} \includegraphics[width=0.32\textwidth]{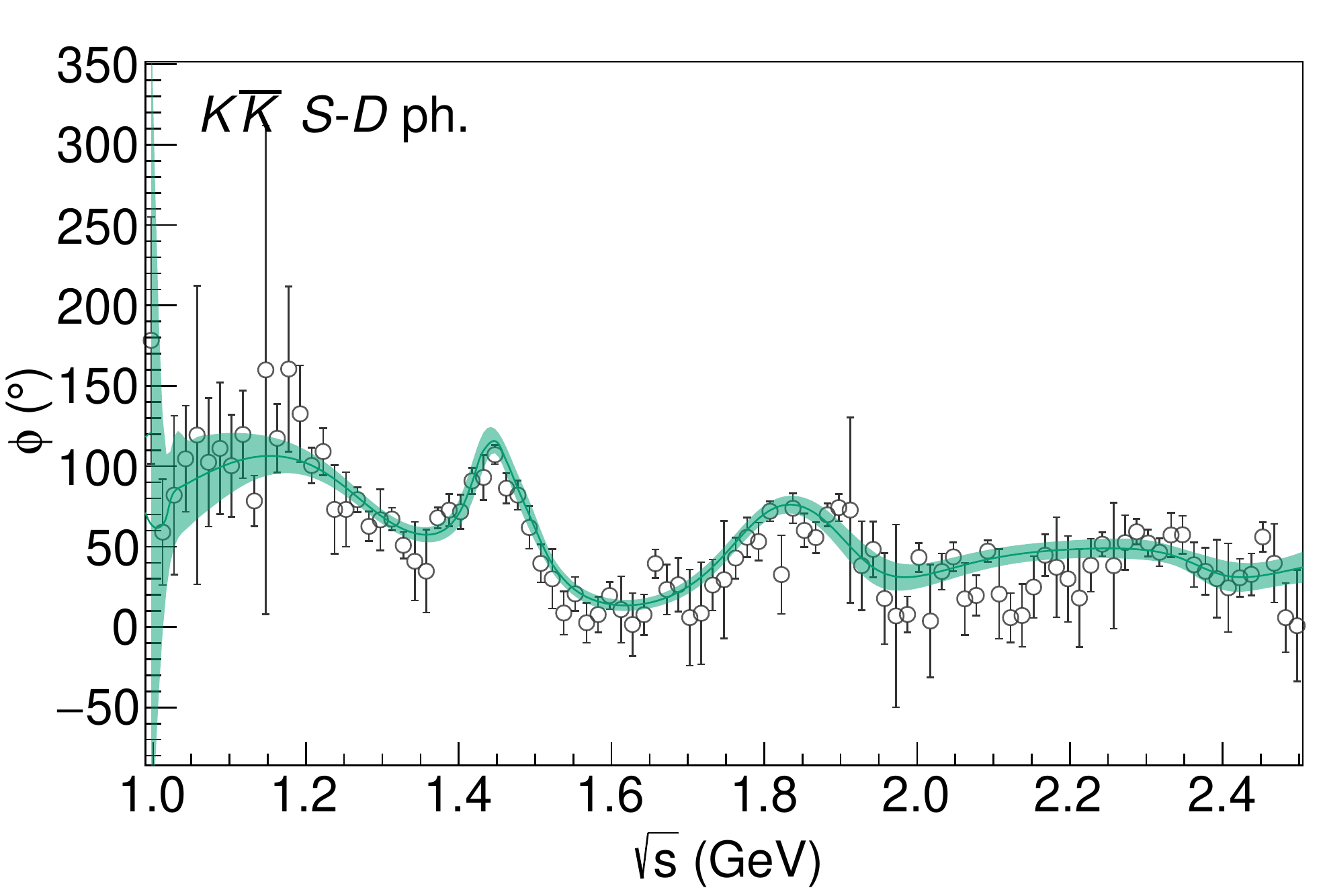}
\end{figure}

\begin{figure}[h]
\centering\includegraphics[width=0.45\textwidth]{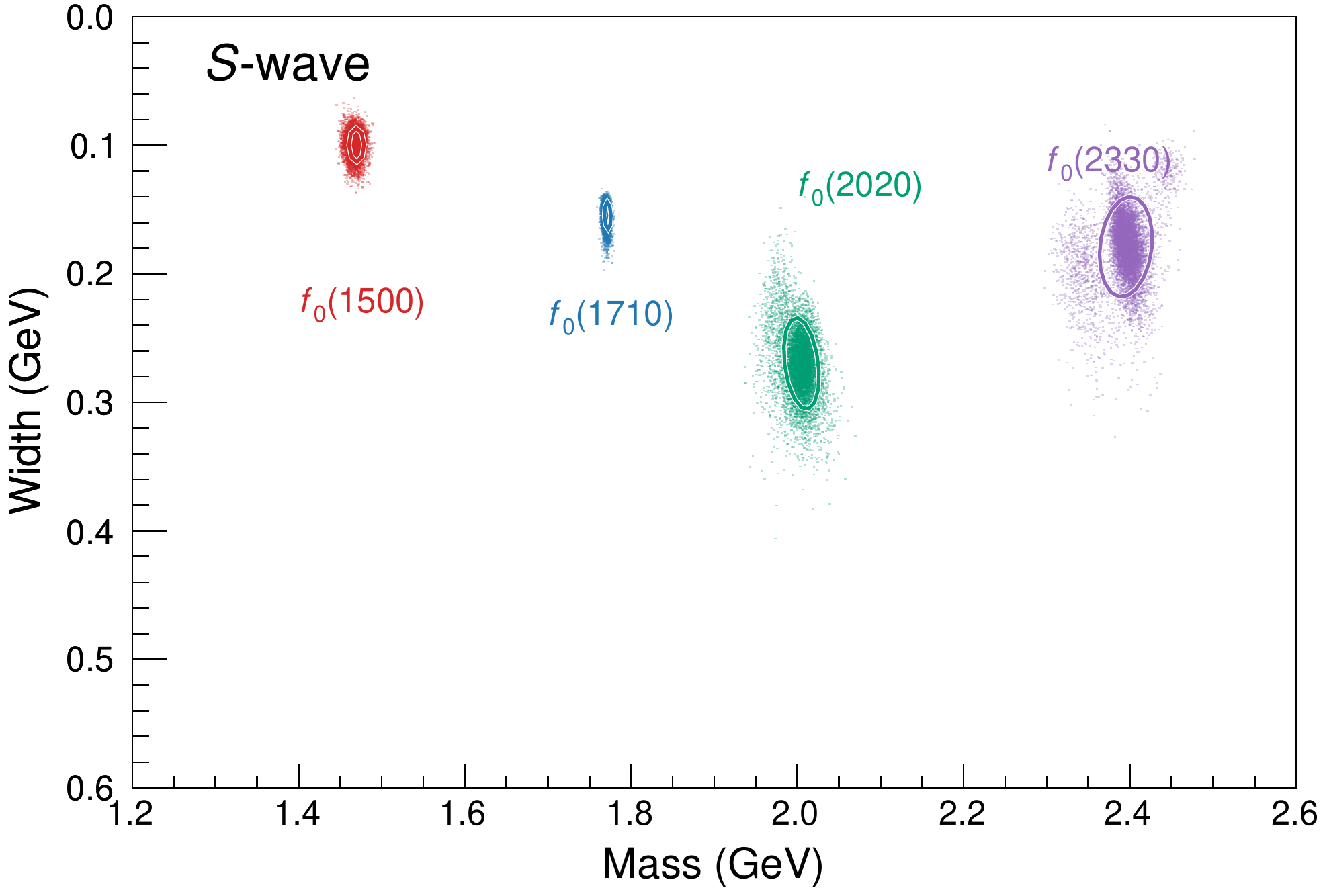} \includegraphics[width=0.45\textwidth]{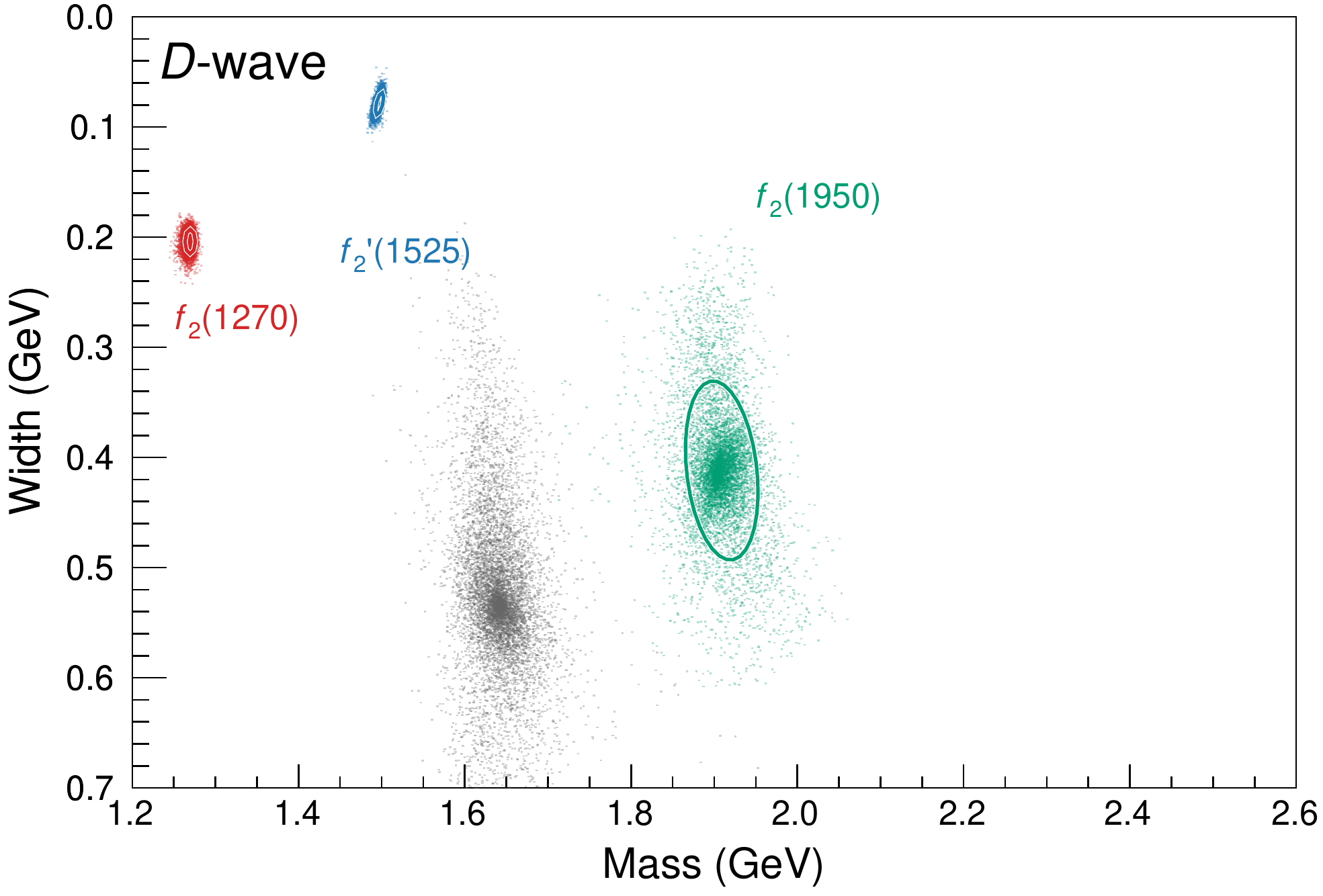}
\end{figure}

\input{tabs-supp-material/poles-inputstartcm3c3_bootstrap-out}

\clearpage

\subsection{$K^J(s) \quad\Big/\quad \omega(s)_\text{scaled} \quad\Big/\quad \rho N^J_{ki}(s')_\text{Q-model} \quad\Big/\quad s_L = 0.6\gevsq$}
\label{subsec:inputstartcm5newc3_bootstrap-out}

\input{tabs-supp-material/numerator-table-inputstartcm5newc3_bootstrap-out}

\input{tabs-supp-material/denominator-table-inputstartcm5newc3_bootstrap-out}

\begin{figure}[h]
\centering\includegraphics[width=0.32\textwidth]{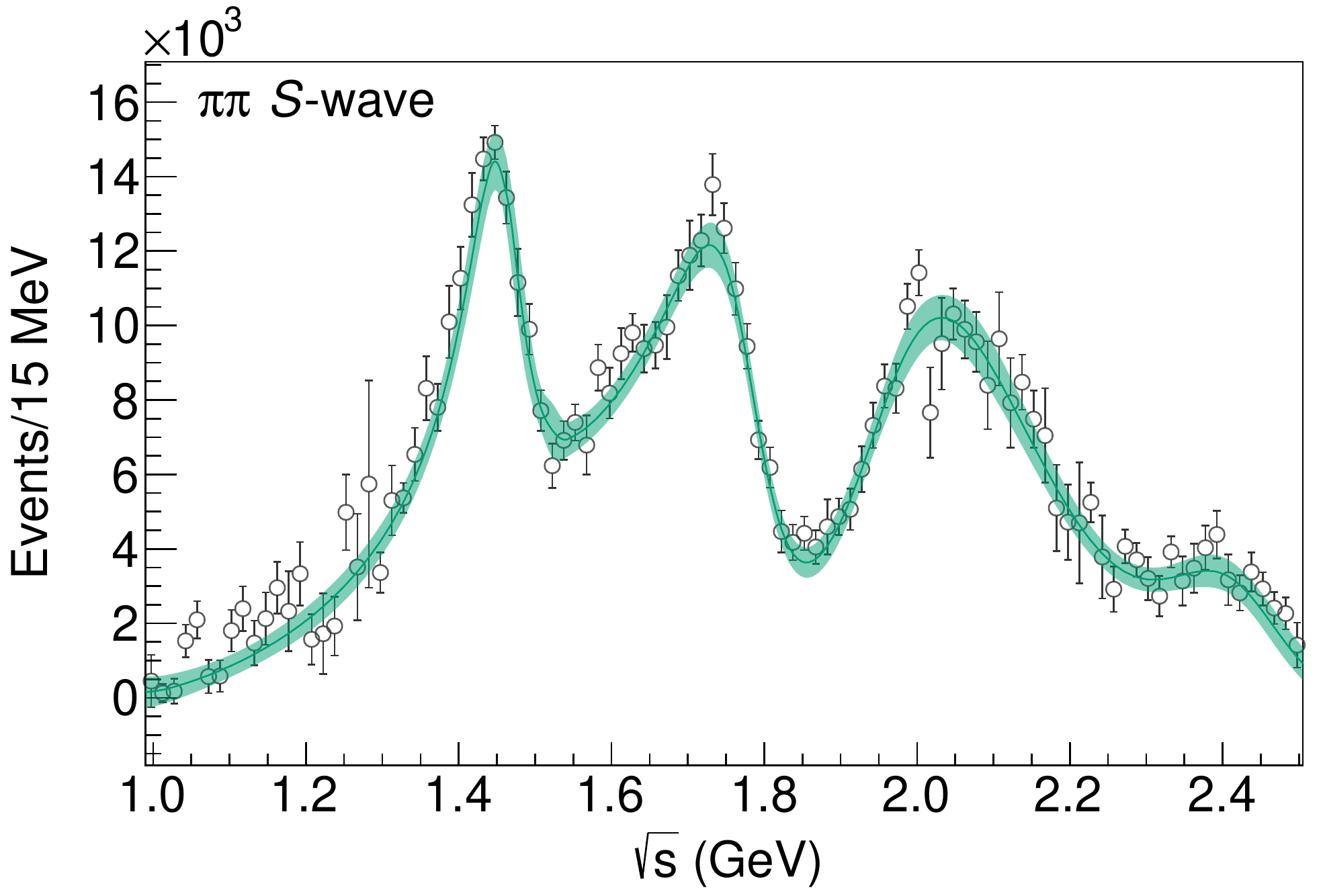} \includegraphics[width=0.32\textwidth]{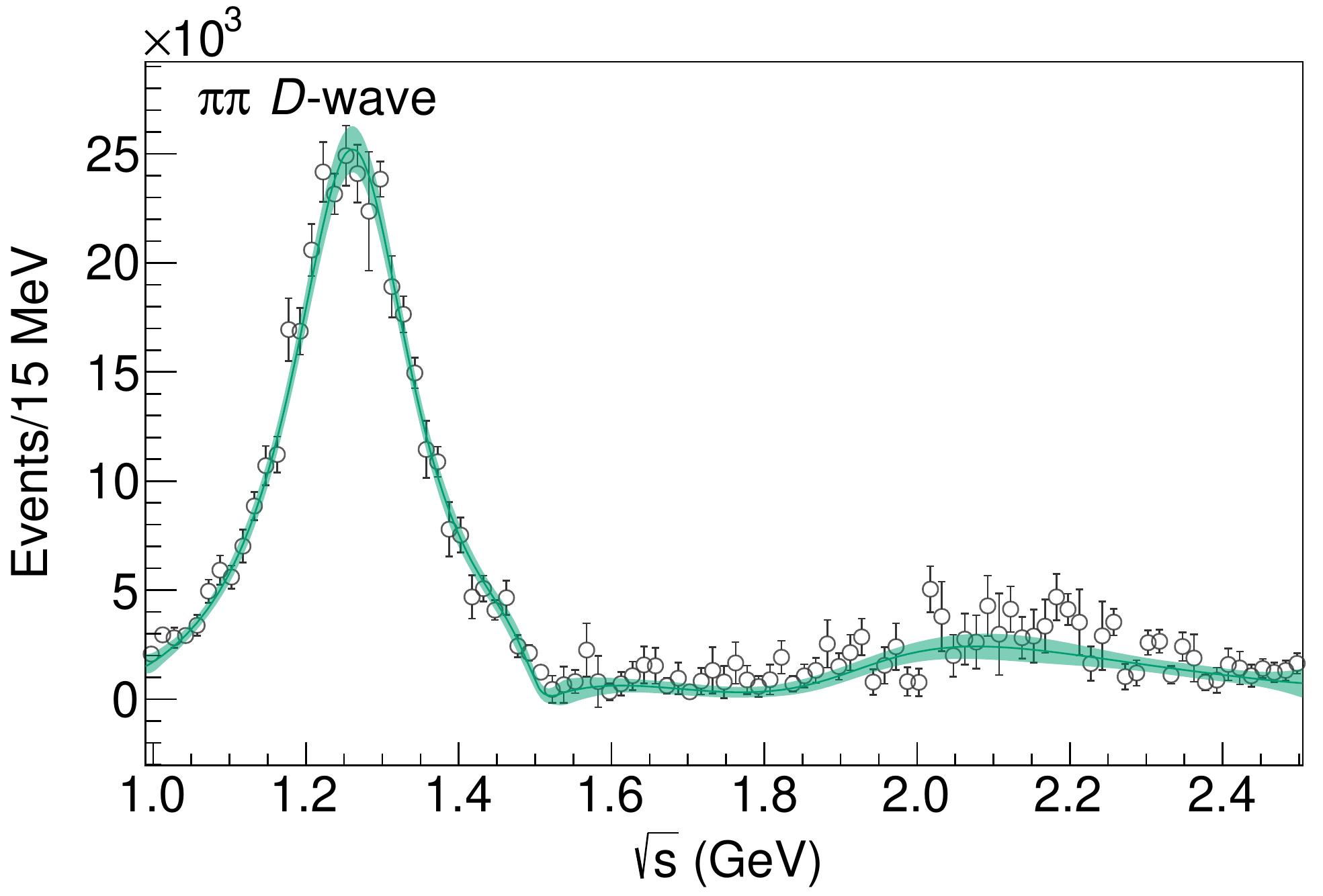} \includegraphics[width=0.32\textwidth]{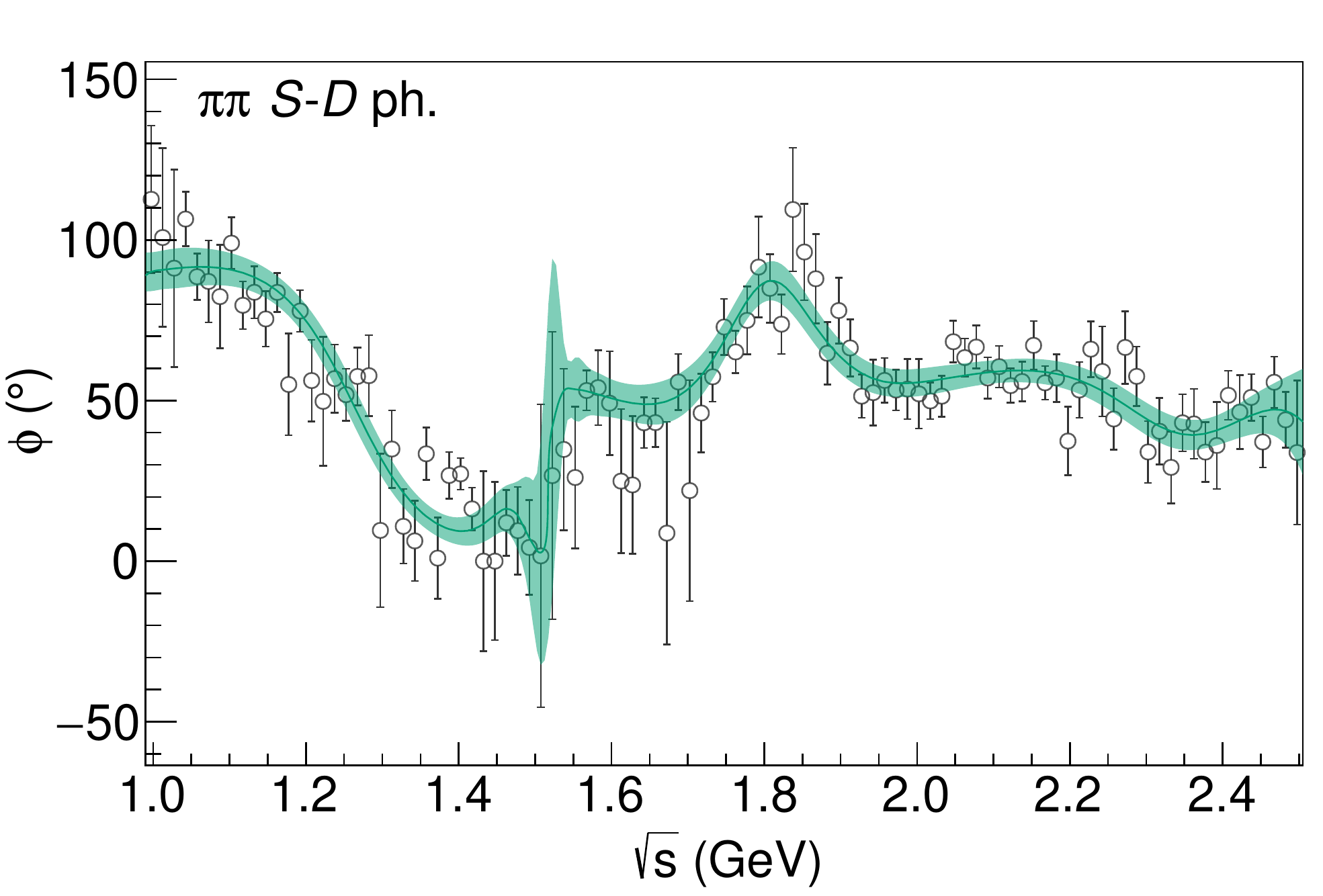}
\includegraphics[width=0.32\textwidth]{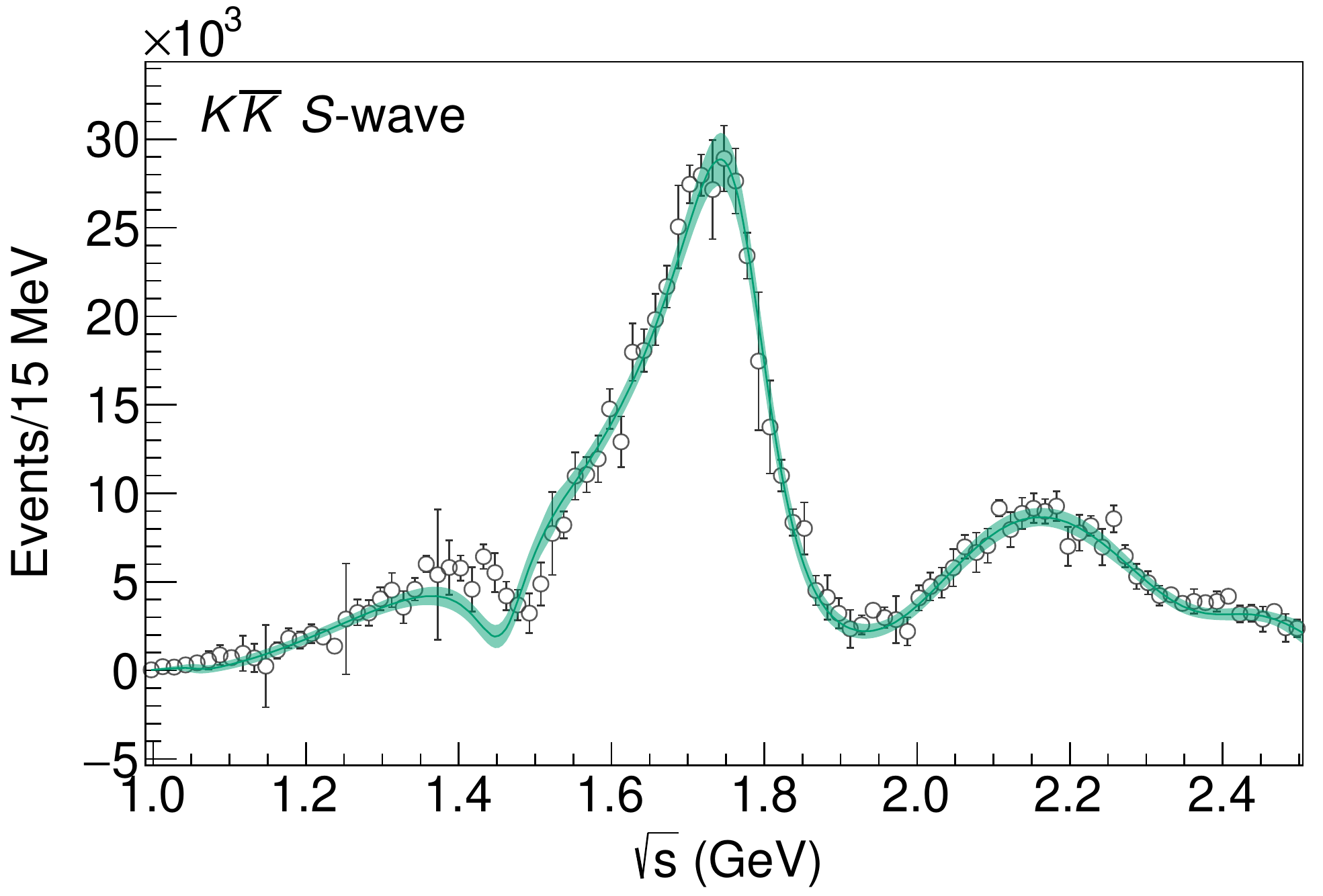} \includegraphics[width=0.32\textwidth]{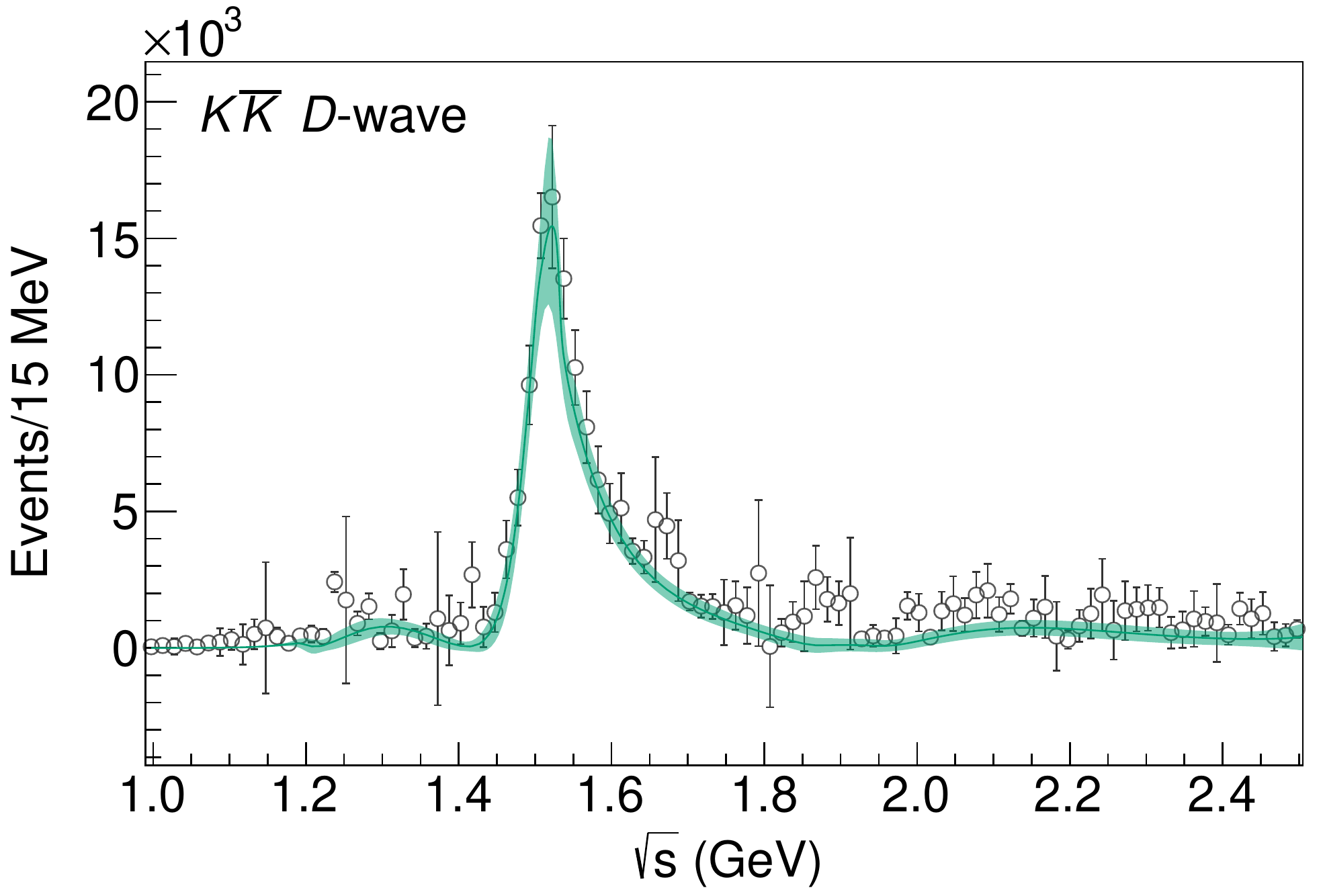} \includegraphics[width=0.32\textwidth]{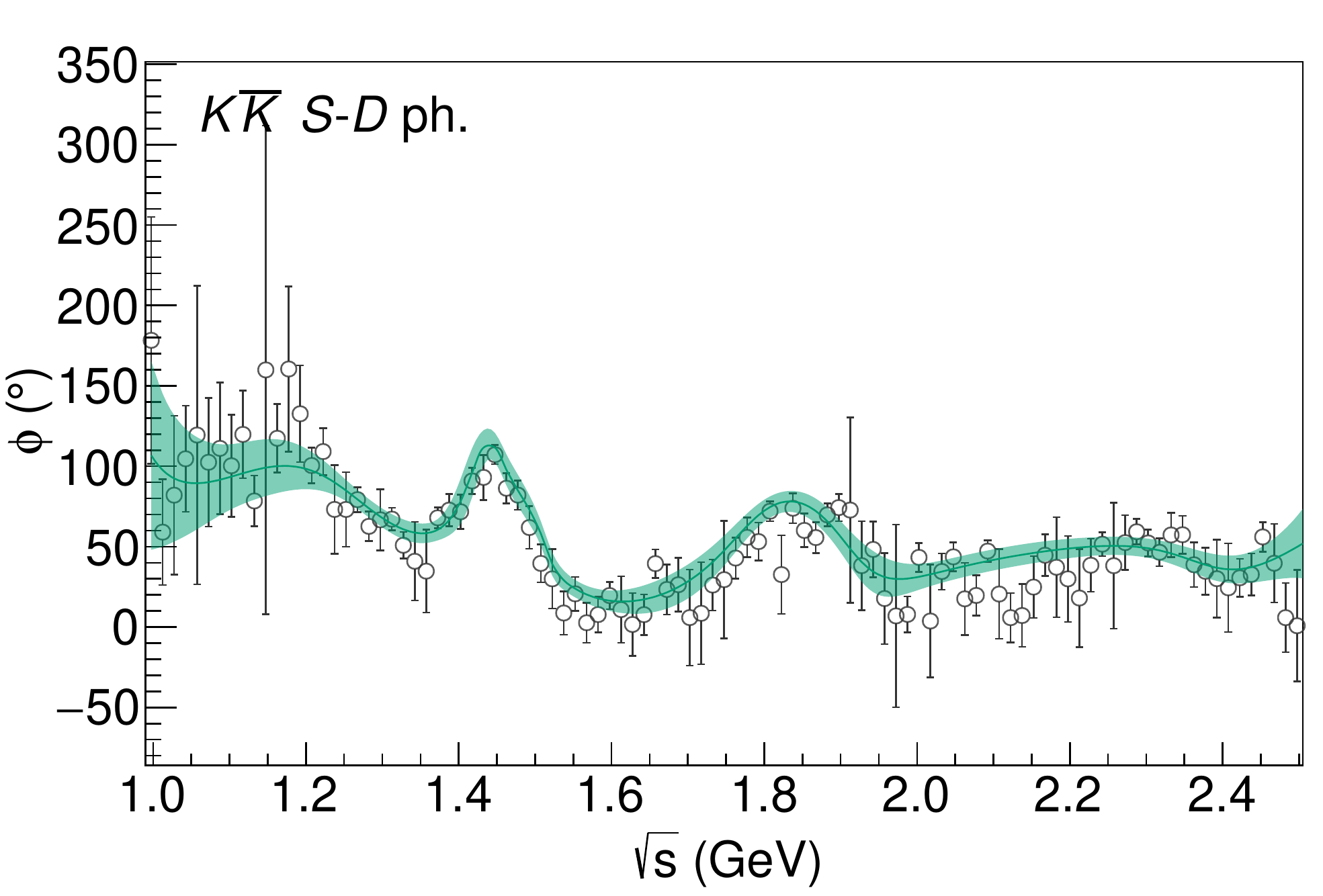}
\end{figure}

\begin{figure}[h]
\centering\includegraphics[width=0.45\textwidth]{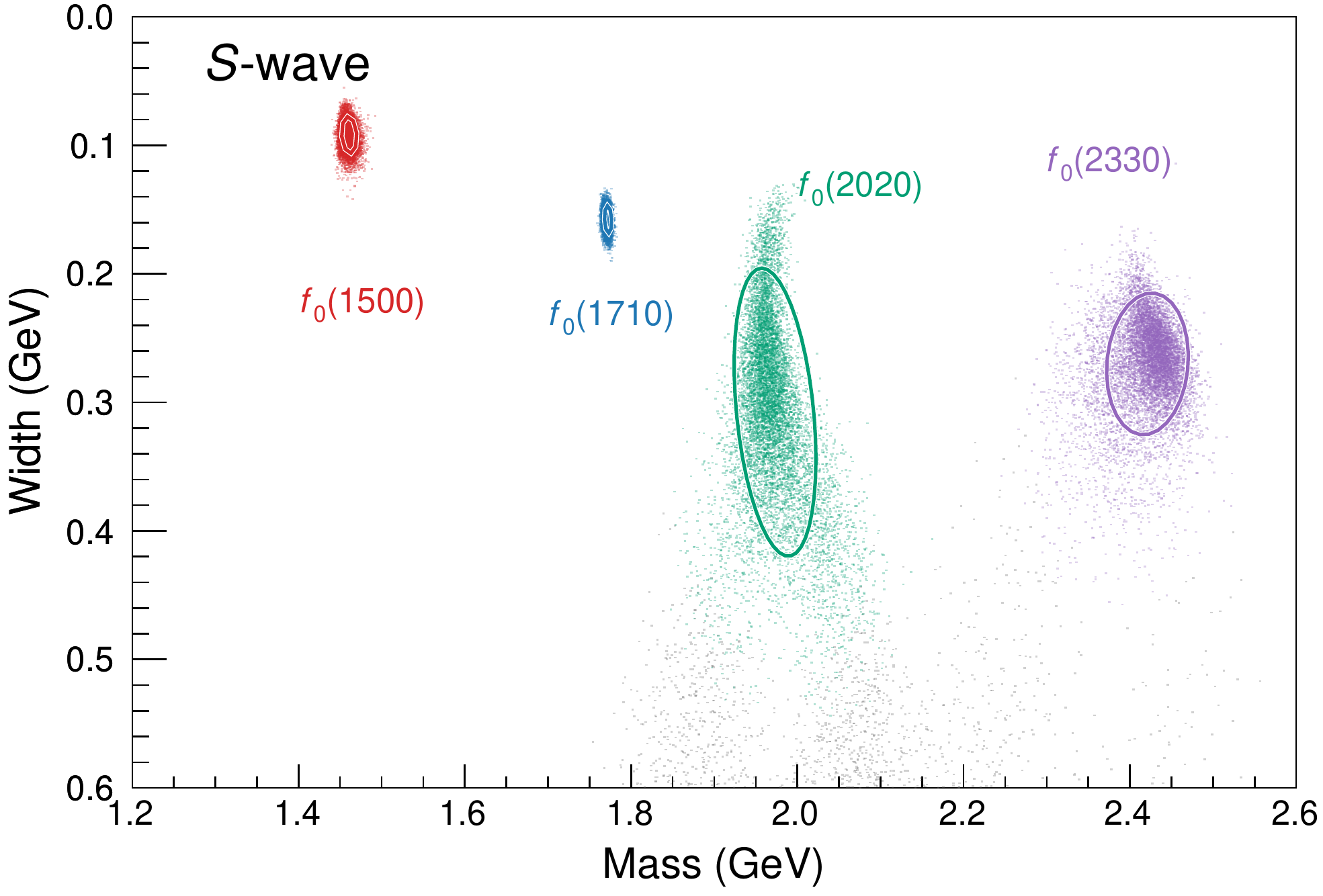} \includegraphics[width=0.45\textwidth]{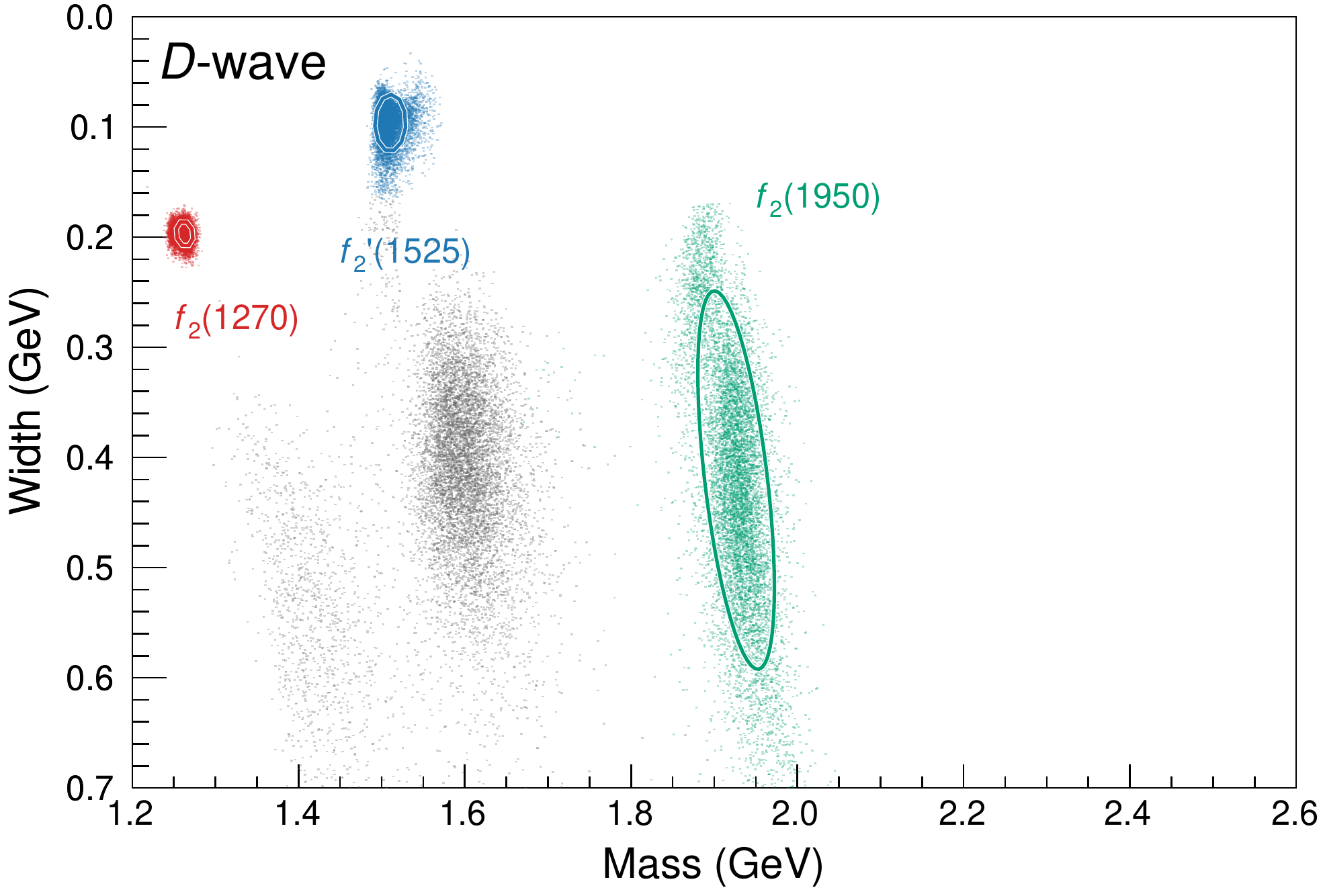}
\end{figure}

\input{tabs-supp-material/poles-inputstartcm5newc3_bootstrap-out}

\clearpage

\subsection{$K^J(s) \quad\Big/\quad \omega(s)_\text{scaled} \quad\Big/\quad \rho N^J_{ki}(s')_\text{Q-model} \quad\Big/\quad s_L = 0$}
\label{subsec:inputstartcm6new2c3_bootstrap-out}

\input{tabs-supp-material/numerator-table-inputstartcm6new2c3_bootstrap-out}

\input{tabs-supp-material/denominator-table-inputstartcm6new2c3_bootstrap-out}

\begin{figure}[h]
\centering\includegraphics[width=0.32\textwidth]{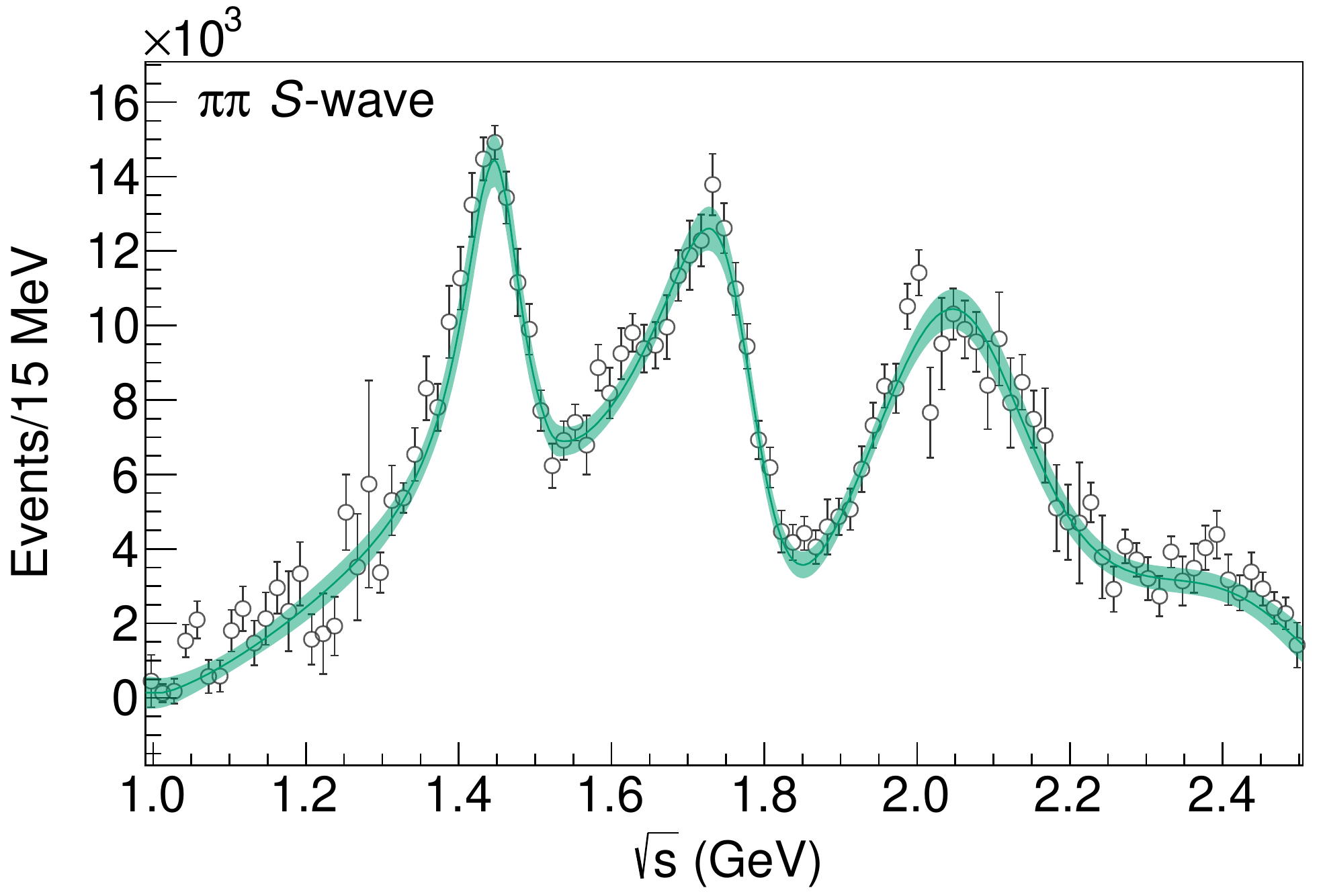} \includegraphics[width=0.32\textwidth]{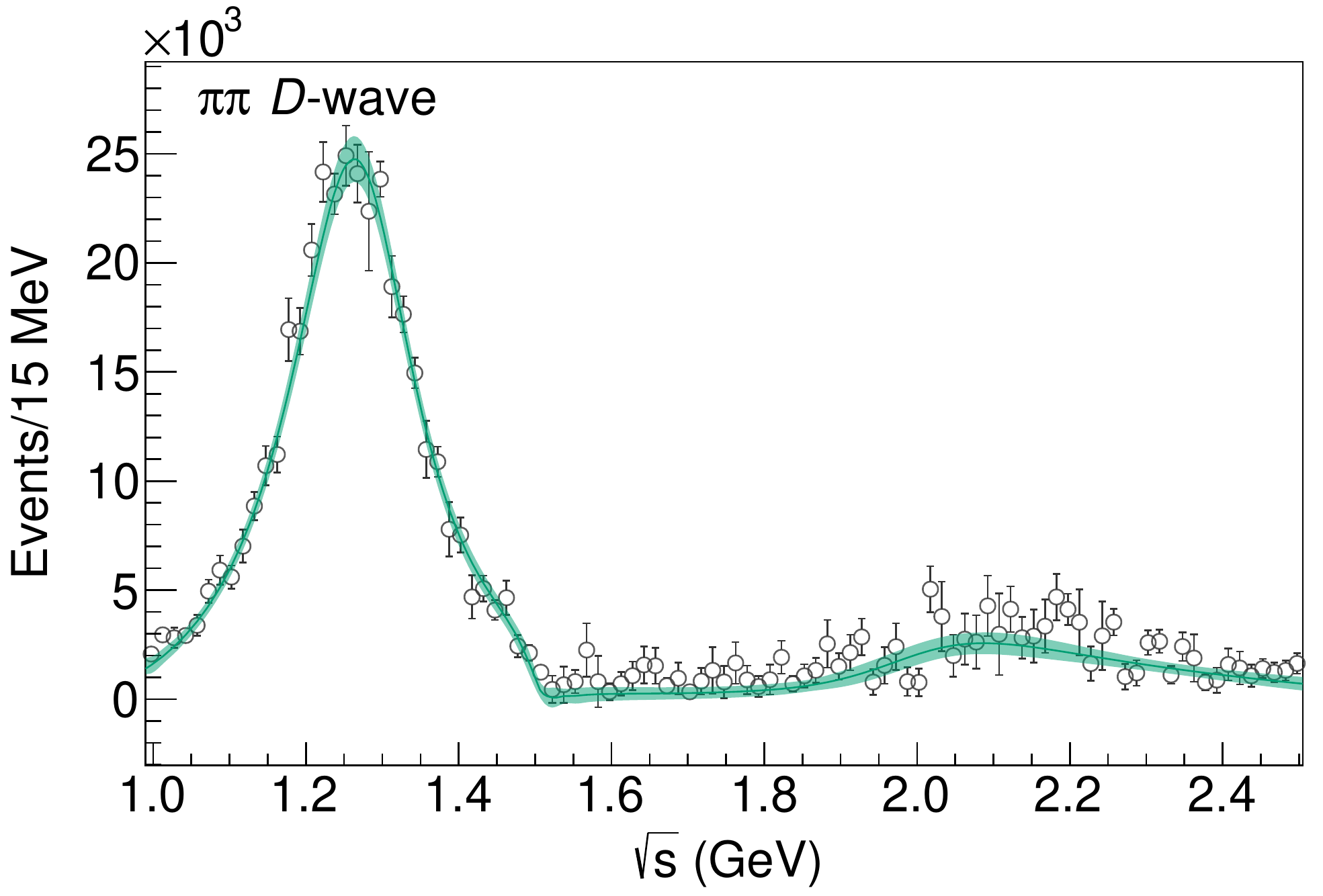} \includegraphics[width=0.32\textwidth]{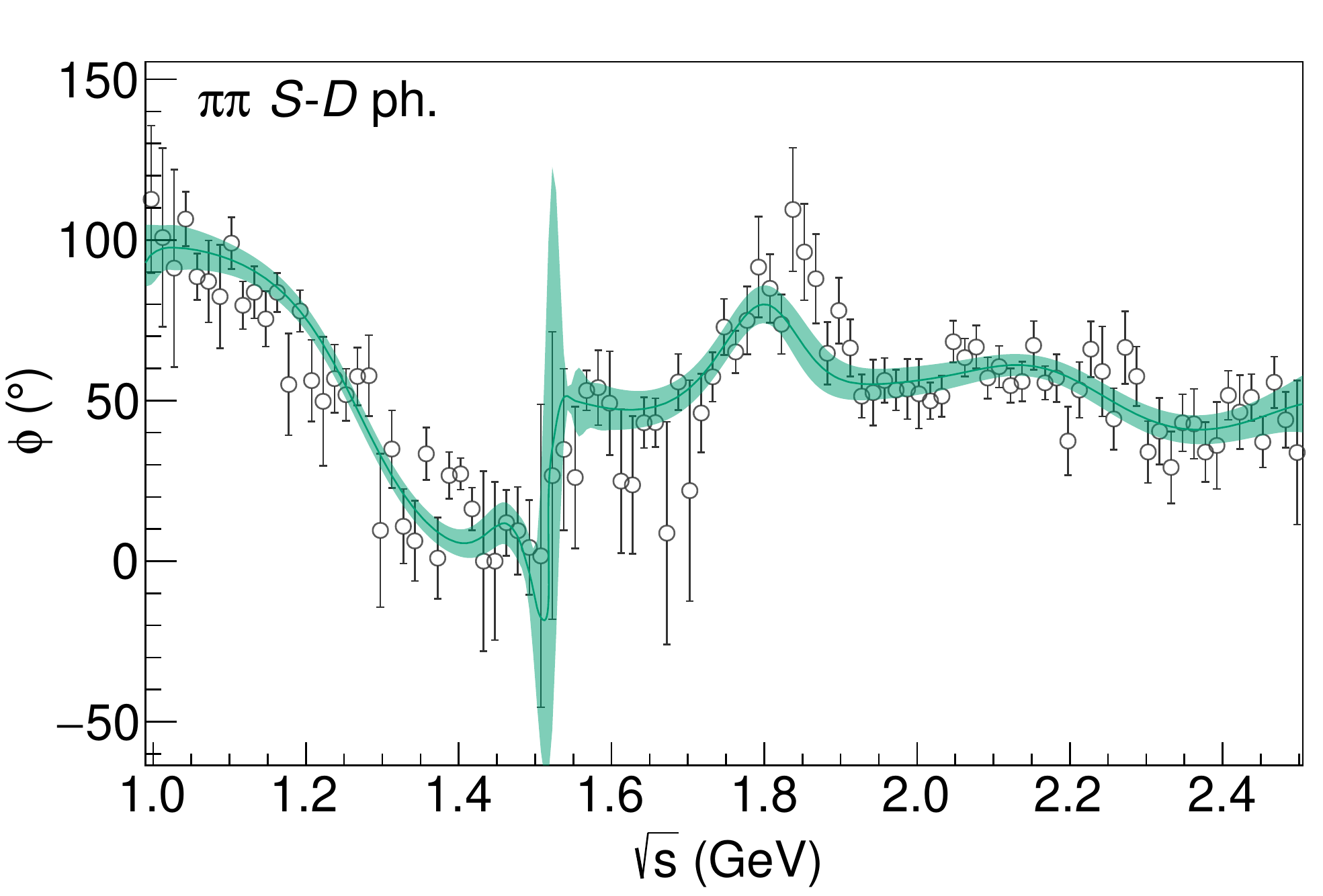}
\includegraphics[width=0.32\textwidth]{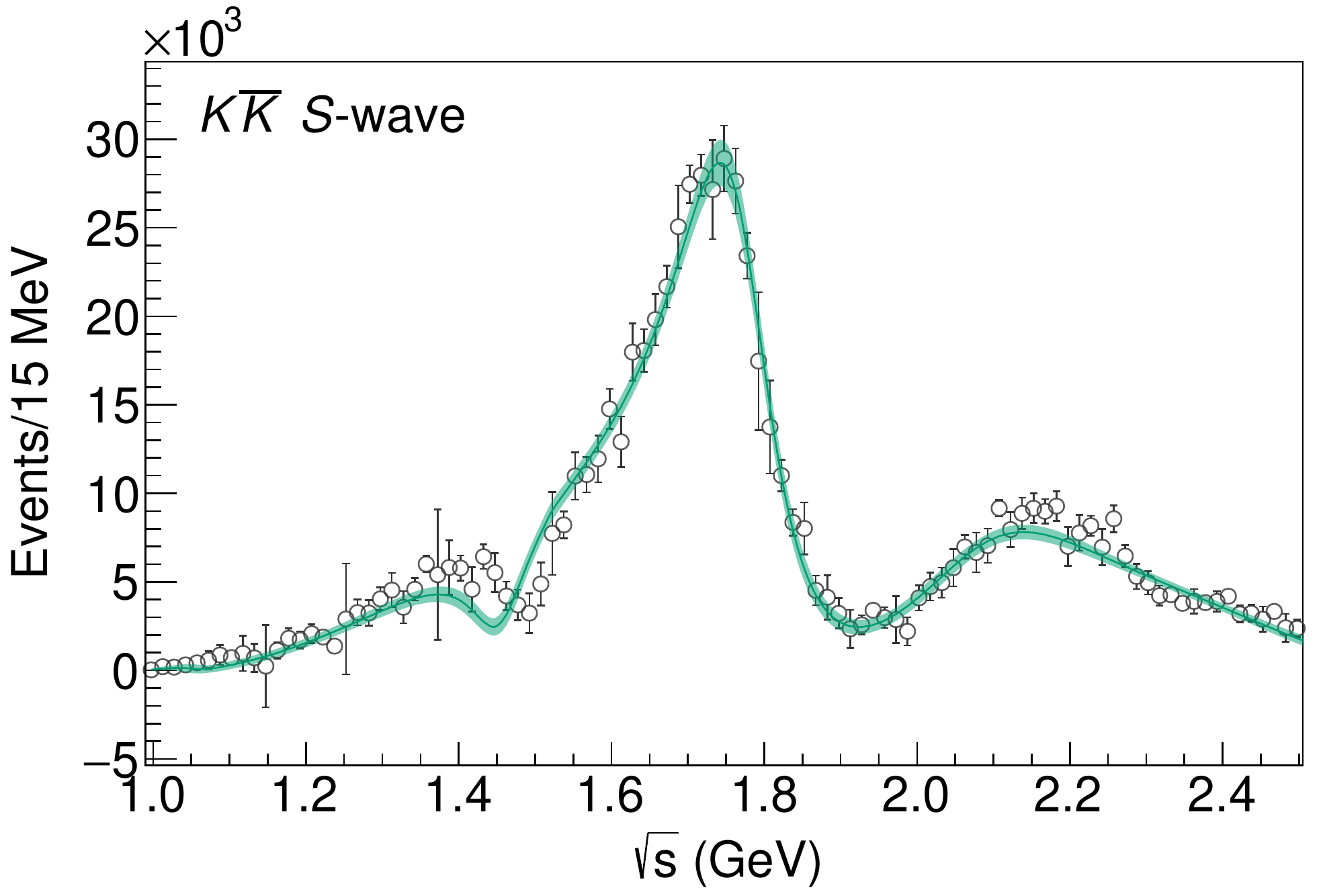} \includegraphics[width=0.32\textwidth]{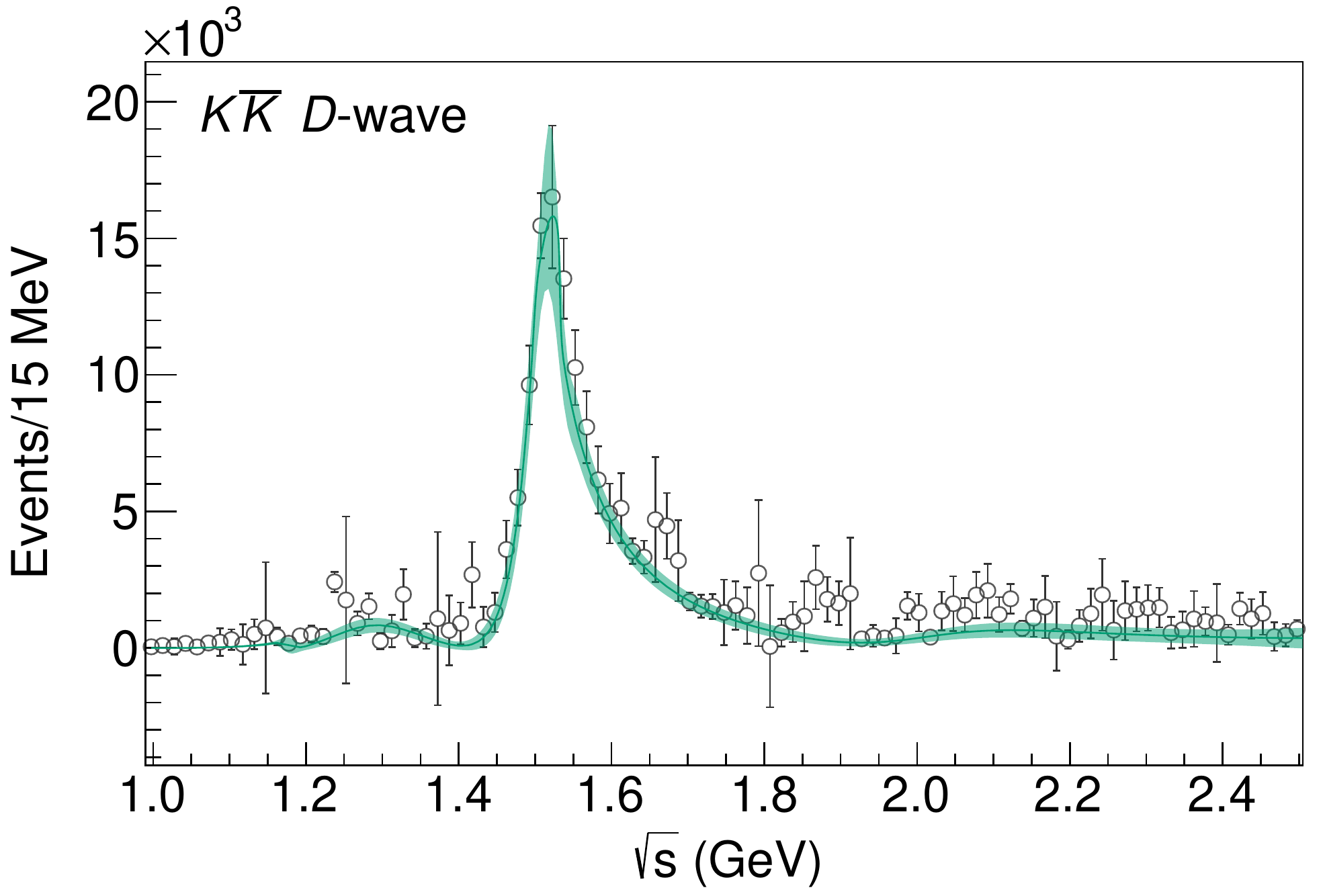} \includegraphics[width=0.32\textwidth]{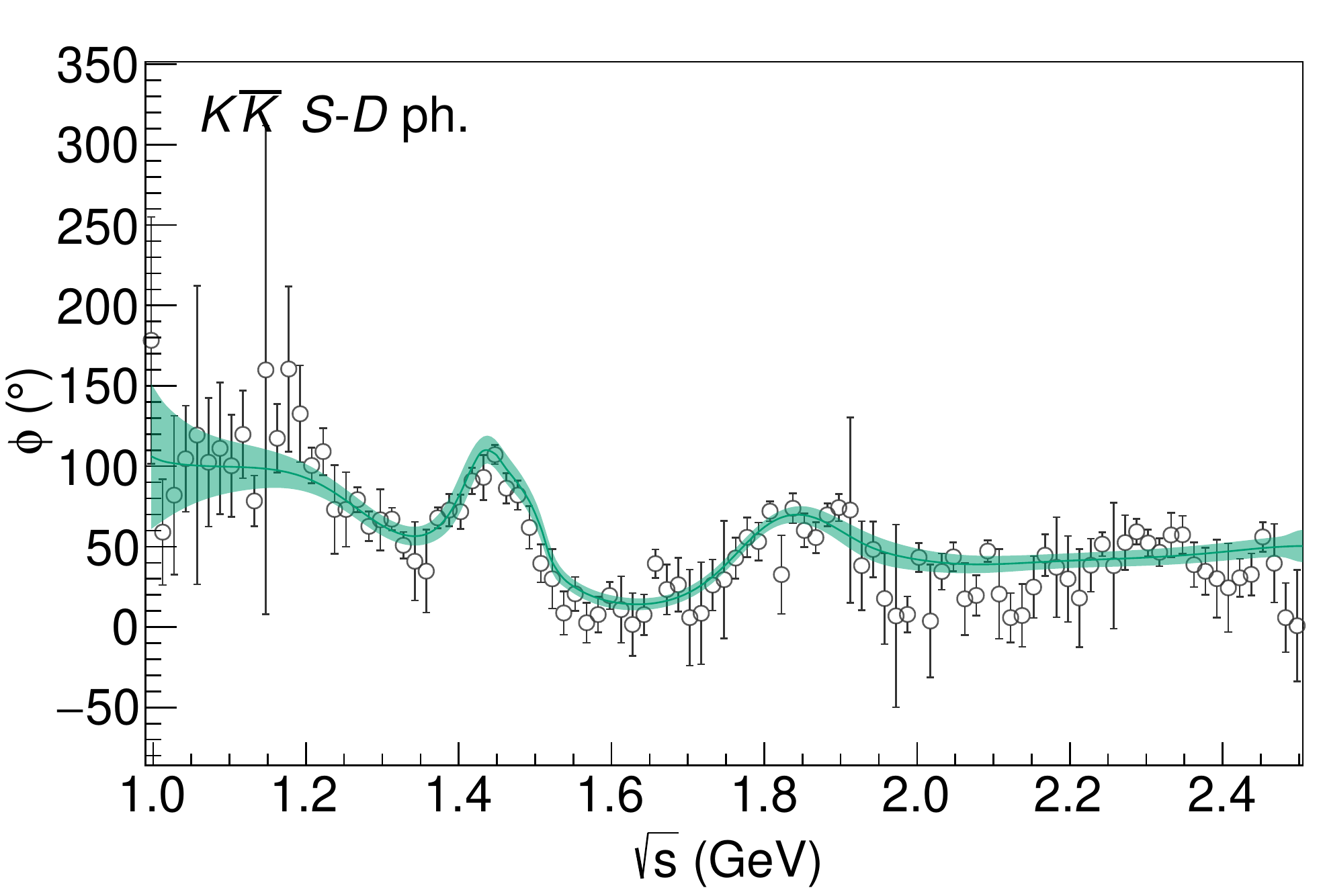}
\end{figure}

\begin{figure}[h]
\centering\includegraphics[width=0.45\textwidth]{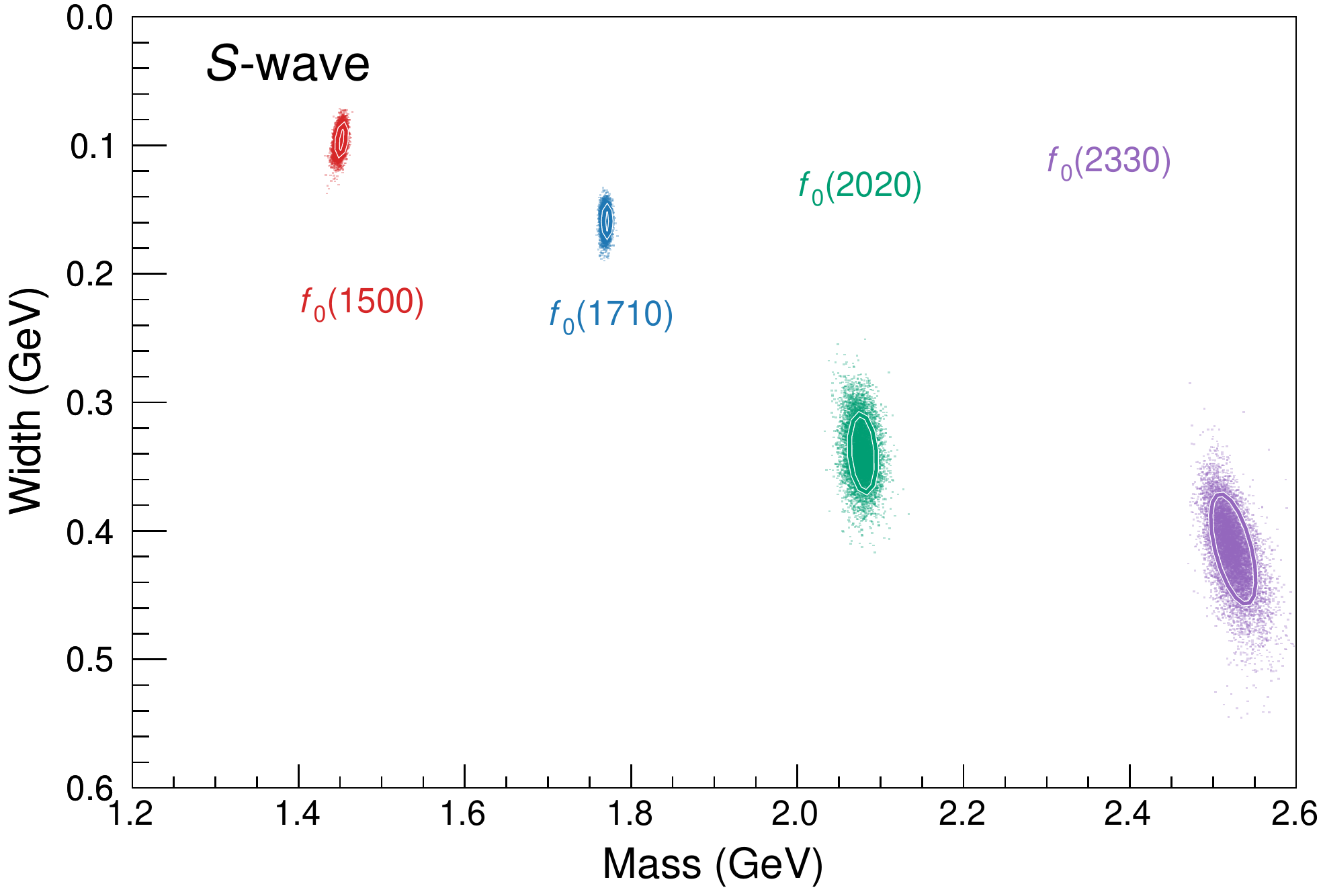} \includegraphics[width=0.45\textwidth]{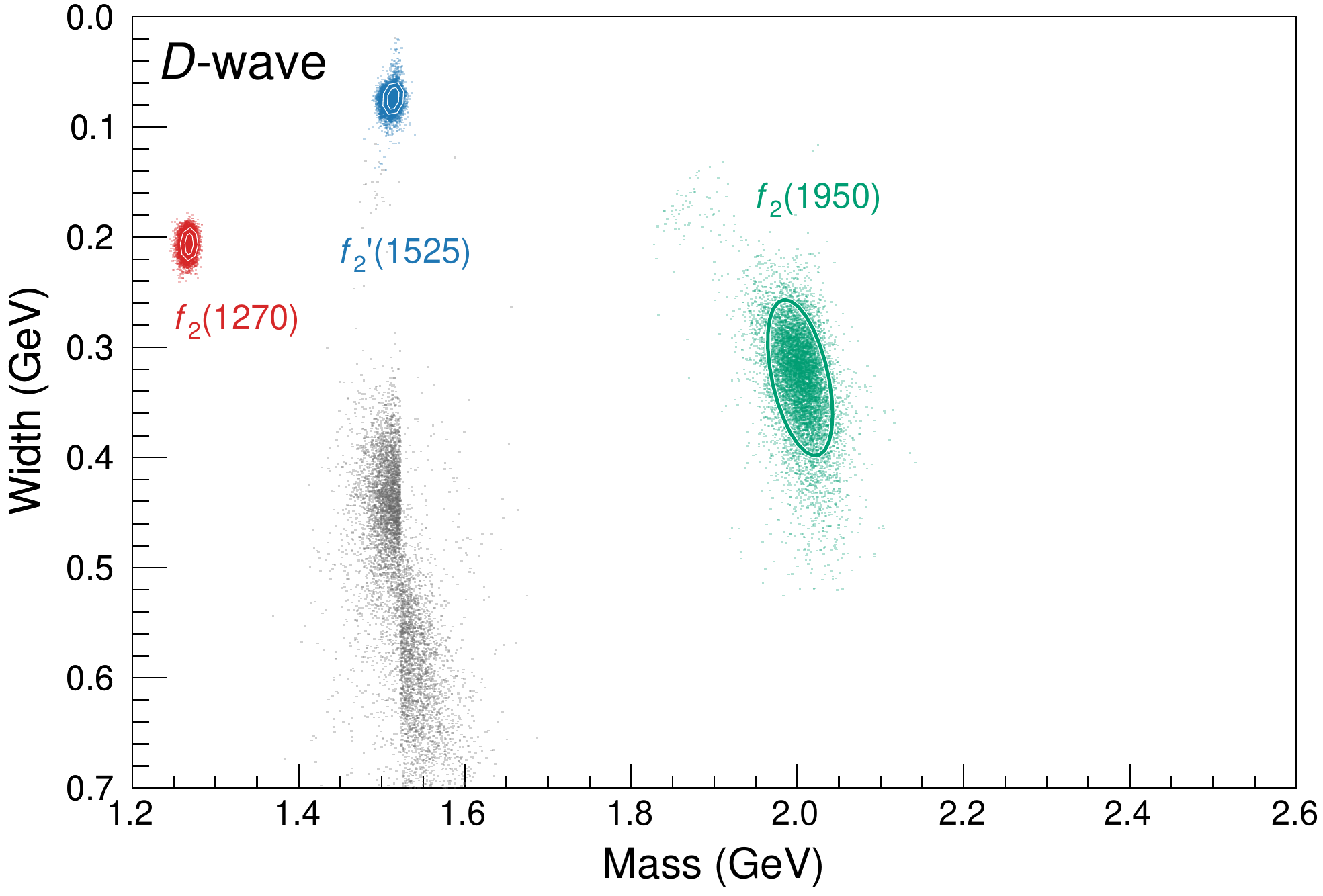}
\end{figure}

\input{tabs-supp-material/poles-inputstartcm6new2c3_bootstrap-out}

\clearpage

\subsection{$K^J(s) \quad\Big/\quad \omega(s)_\text{pole+scaled} \quad\Big/\quad \rho N^J_{ki}(s')_\text{Q-model} \quad\Big/\quad s_L = 0.6\gevsq$}
\label{subsec:inputstartcmnew2c3_bootstrap-out}

\input{tabs-supp-material/numerator-table-inputstartcmnew2c3_bootstrap-out}

\input{tabs-supp-material/denominator-table-inputstartcmnew2c3_bootstrap-out}

\begin{figure}[h]
\centering\includegraphics[width=0.32\textwidth]{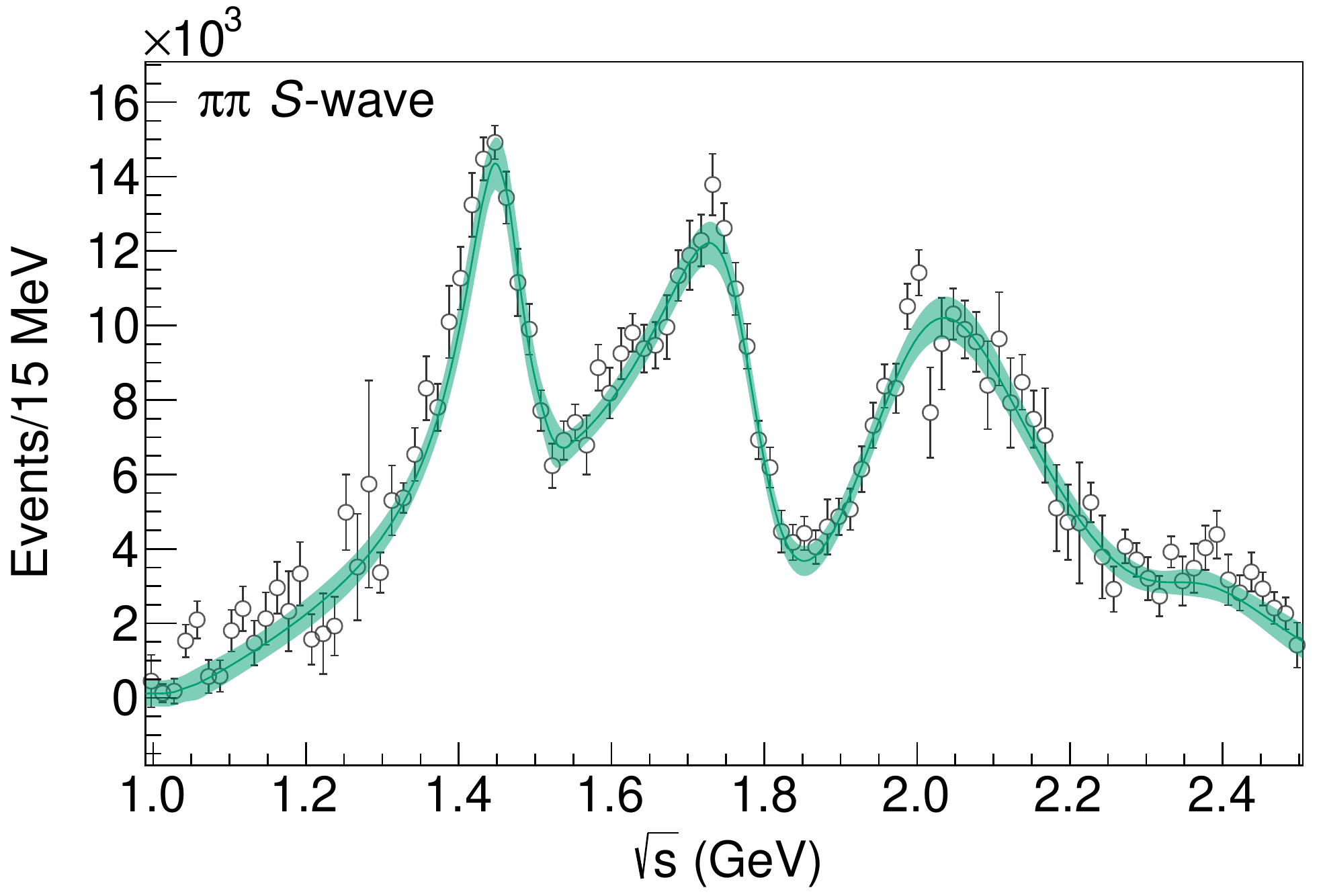} \includegraphics[width=0.32\textwidth]{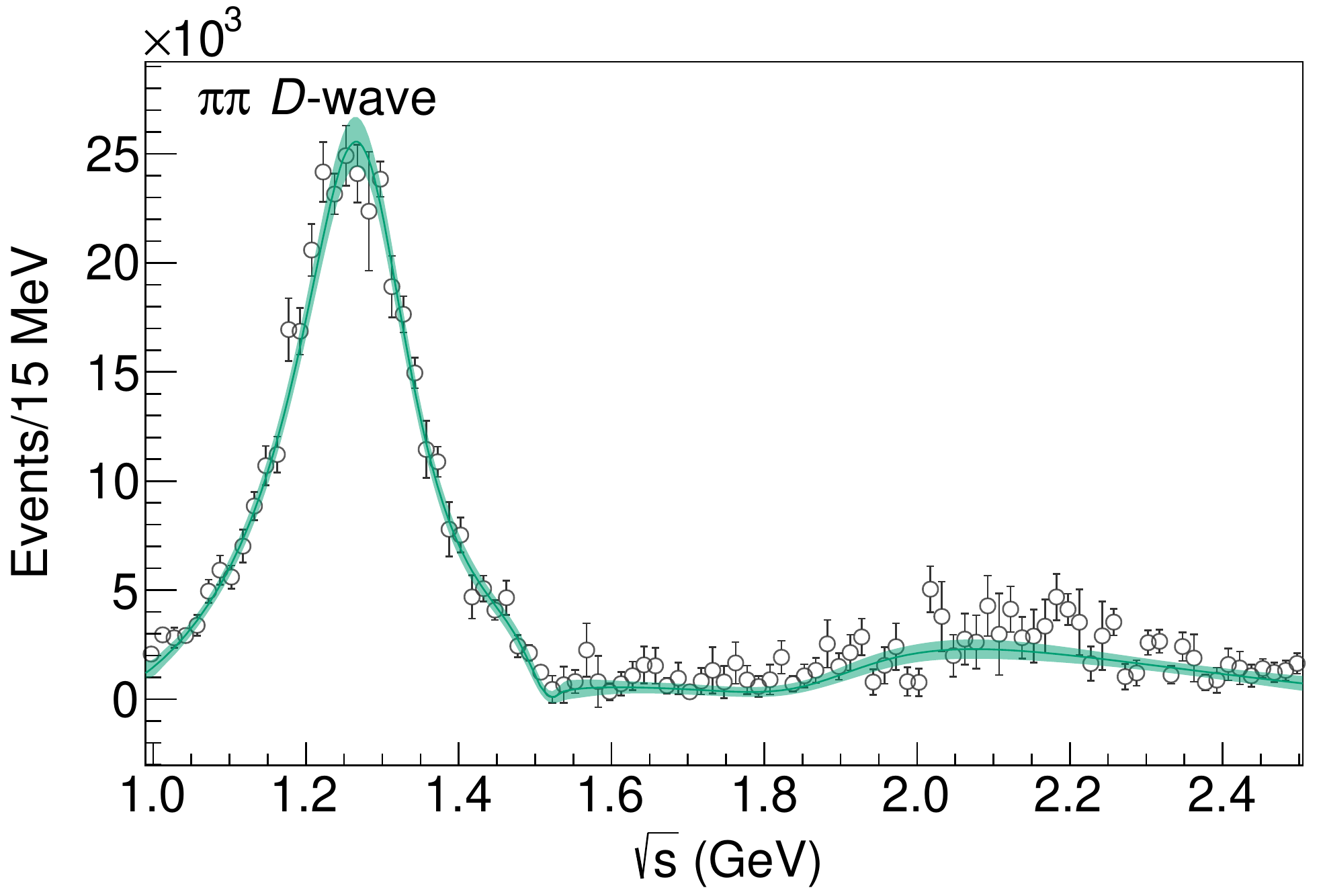} \includegraphics[width=0.32\textwidth]{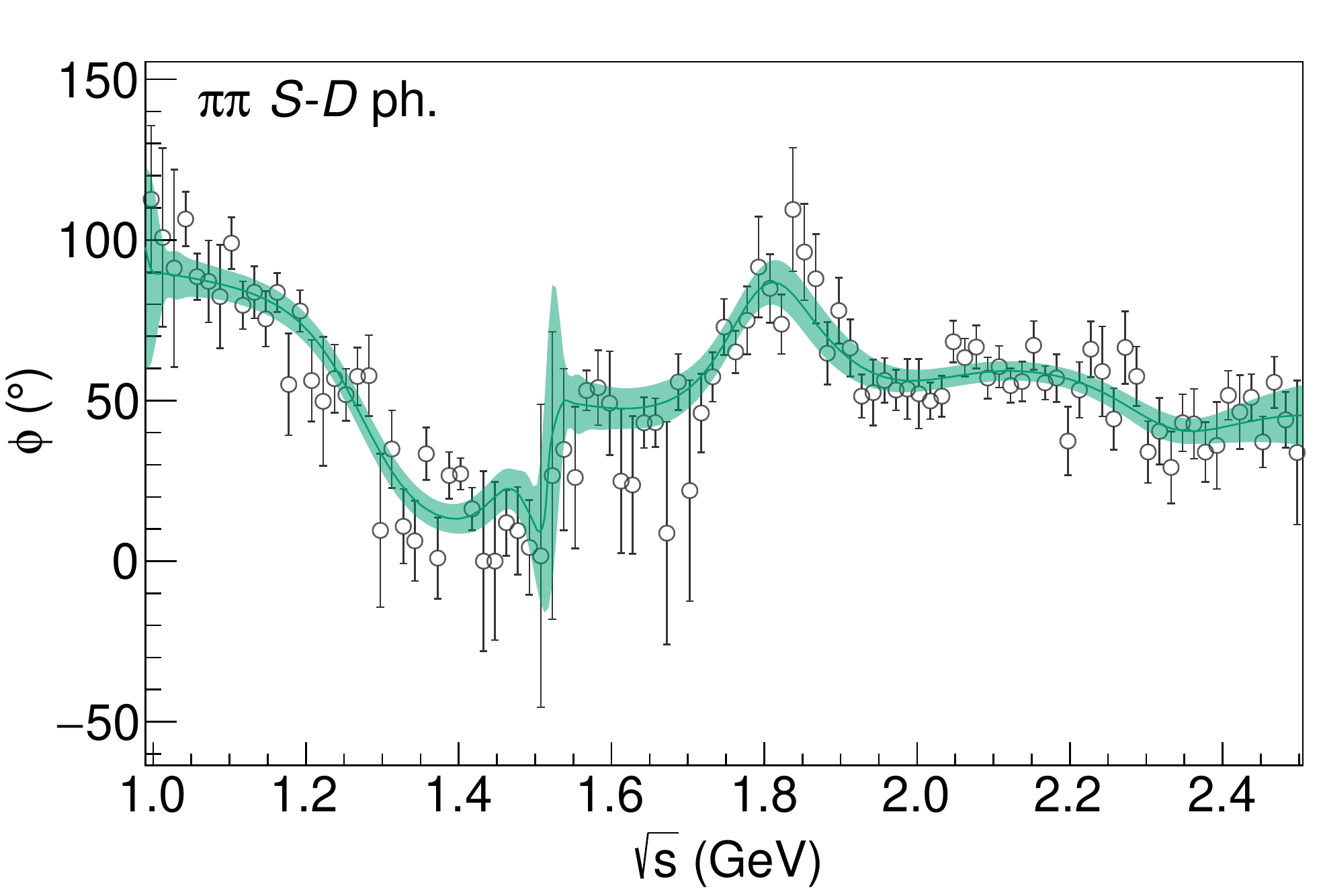}
\includegraphics[width=0.32\textwidth]{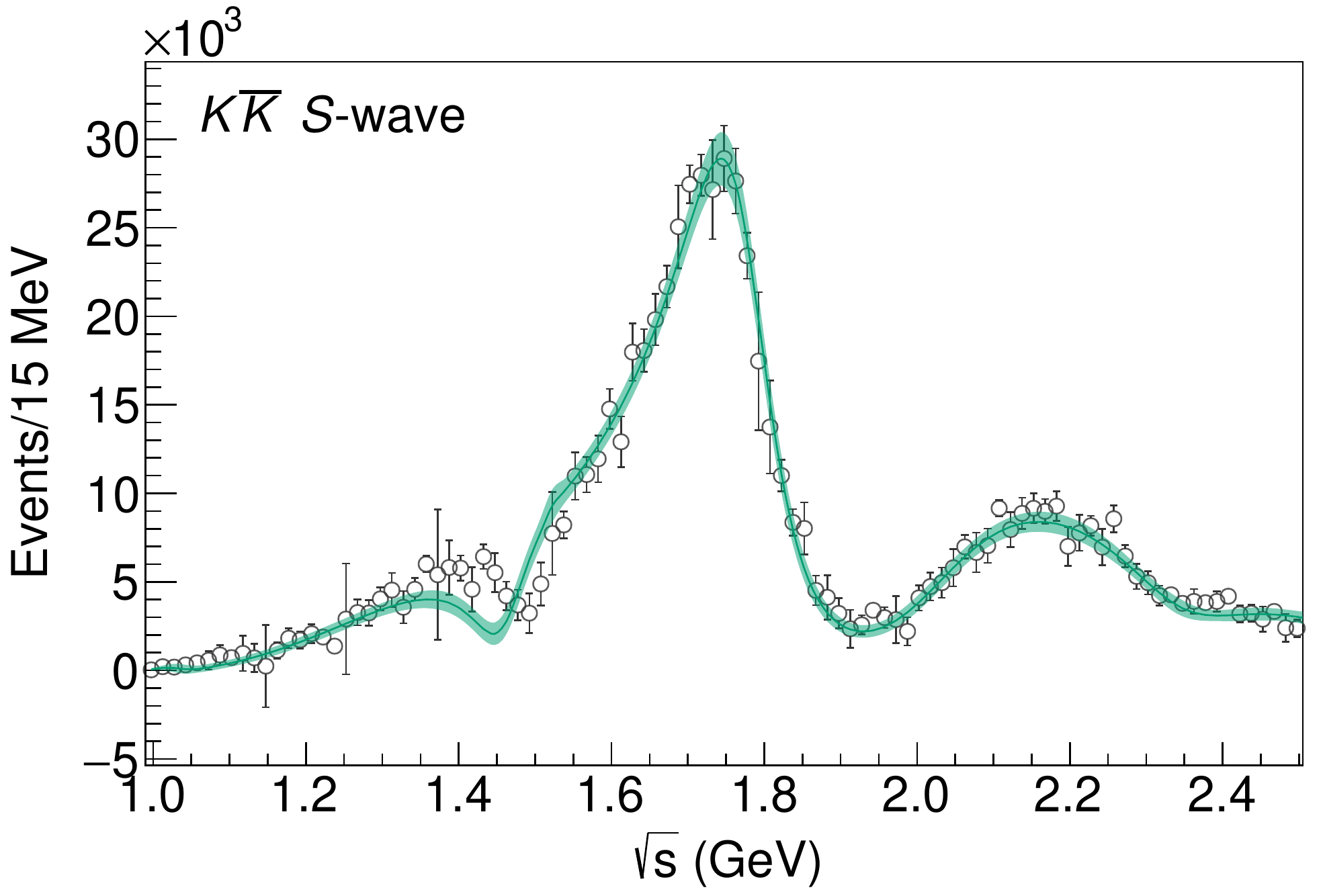} \includegraphics[width=0.32\textwidth]{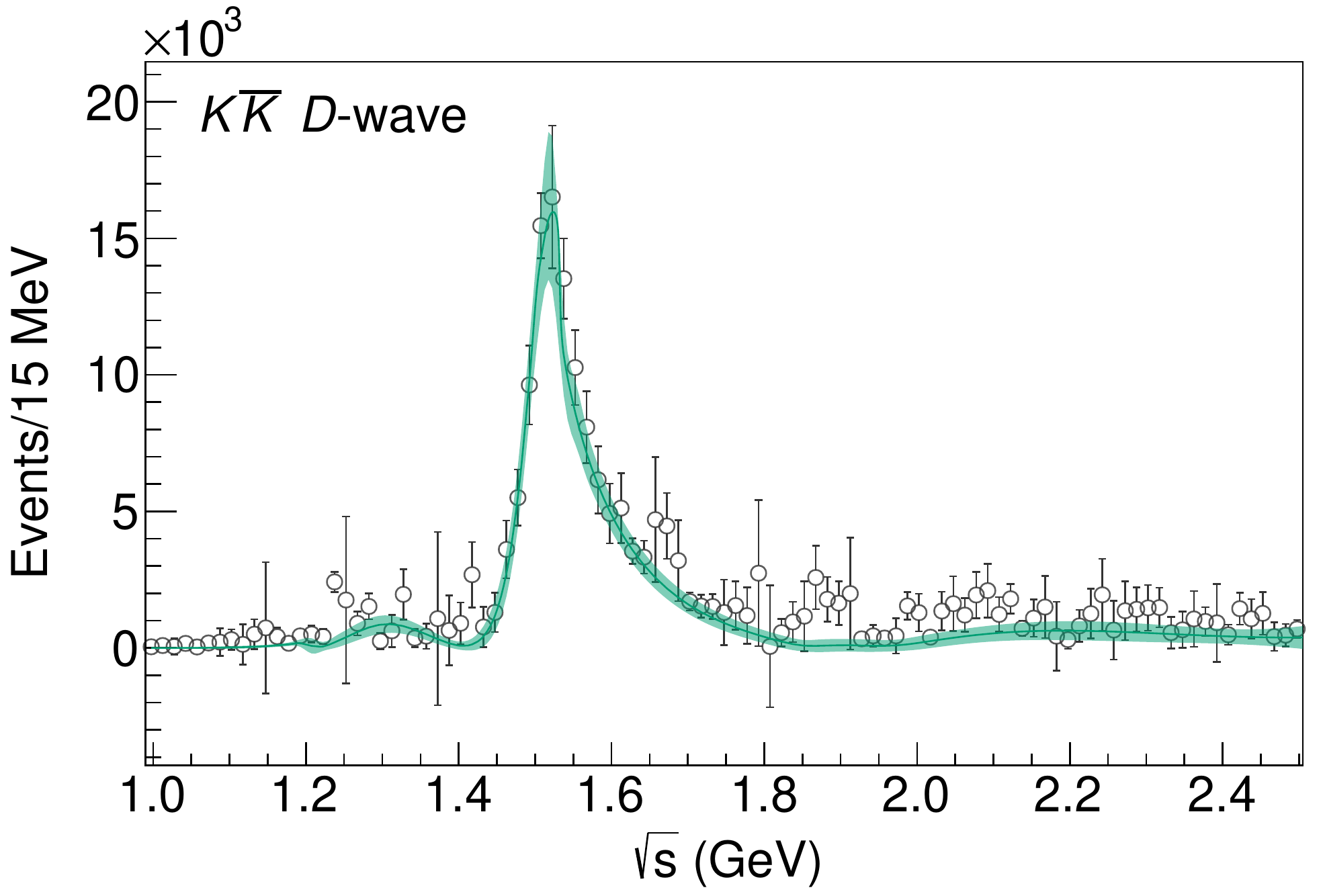} \includegraphics[width=0.32\textwidth]{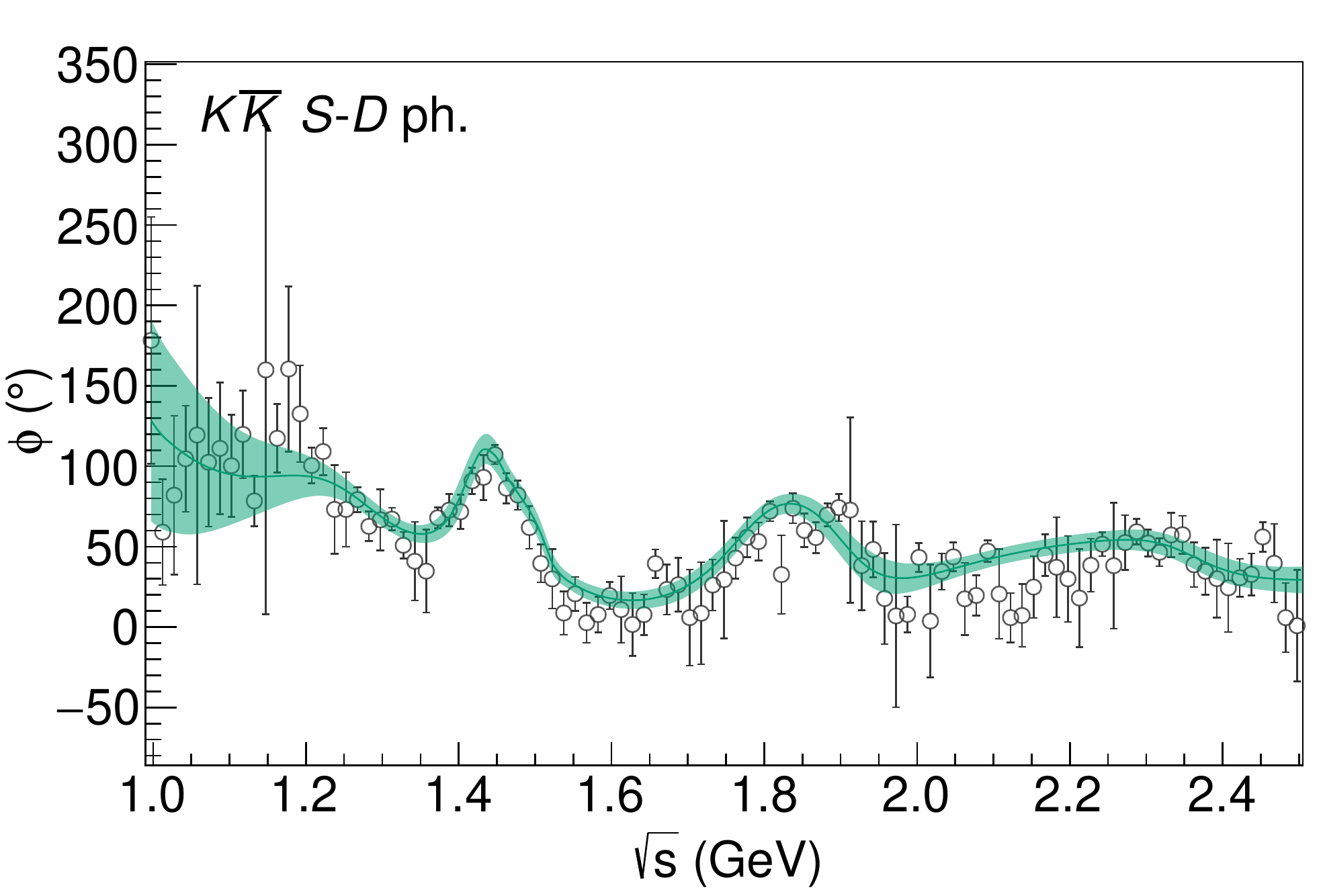}
\end{figure}

\begin{figure}[h]
\centering\includegraphics[width=0.45\textwidth]{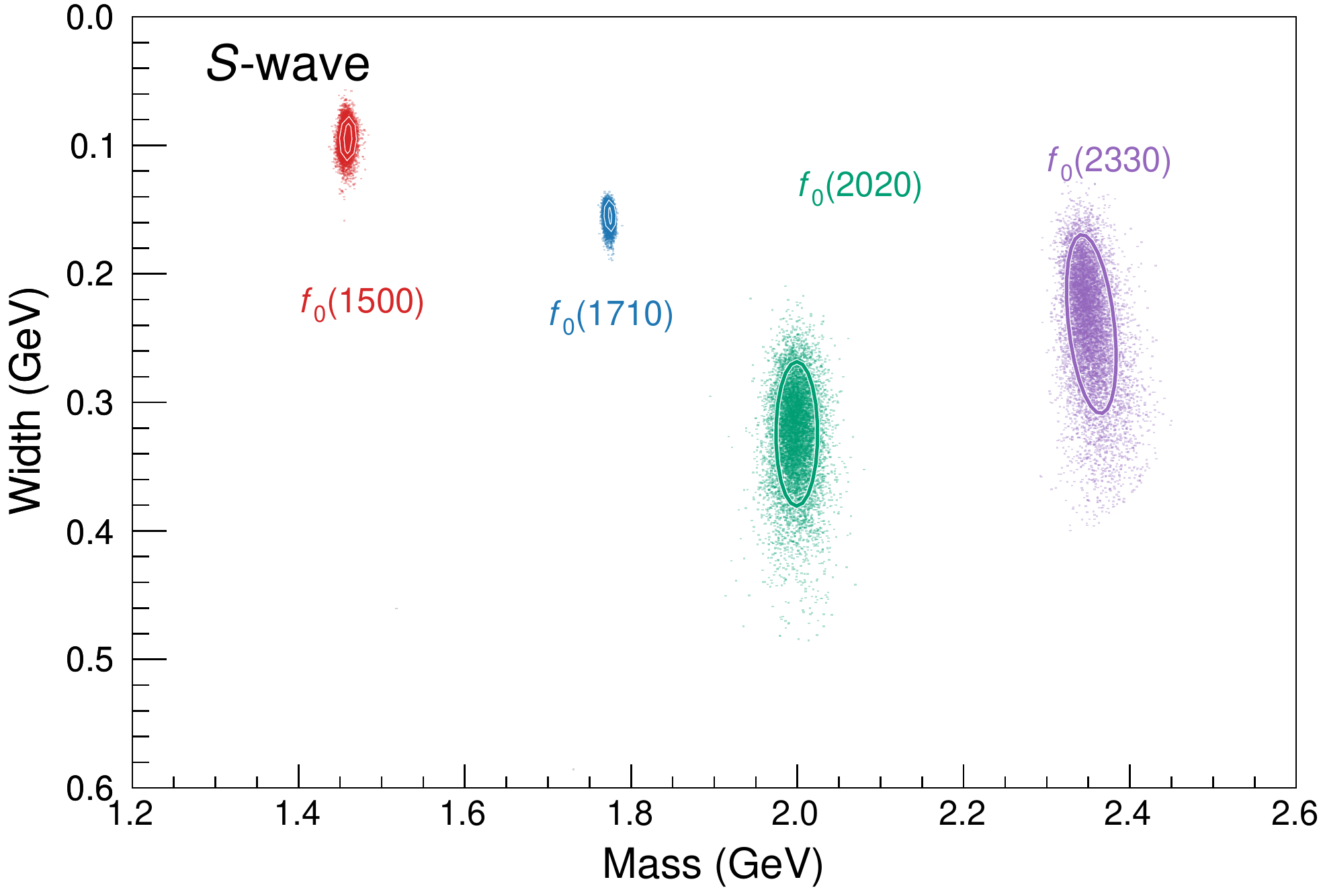} \includegraphics[width=0.45\textwidth]{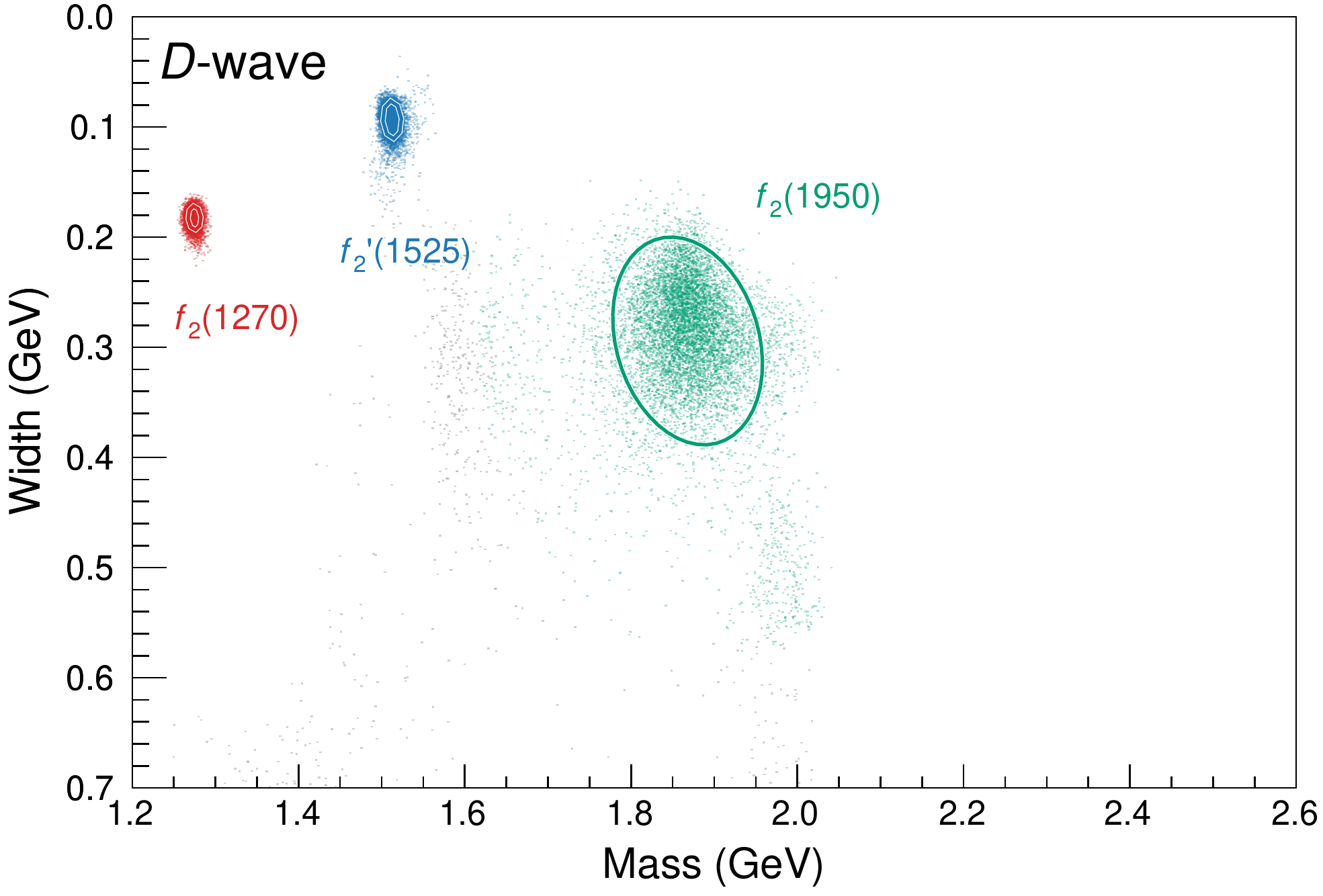}
\end{figure}

\input{tabs-supp-material/poles-inputstartcmnew2c3_bootstrap-out}

\end{document}